\documentclass[12pt,oneside]{book}
\usepackage{latexsym}
\usepackage{multibox}
\usepackage{amsmath,amsfonts,amsbsy}
\usepackage[dvips]{graphicx,epsfig}

\csname @addtoreset\endcsname{equation}{section}

\def\miso{\mathfrak{iso}}
\def\mso{\mathfrak{so}}
\def\msu{\mathfrak{su}}
\def\msp{\mathfrak{sp}}
\def\musp{\mathfrak{usp}}
\def\msl{\mathfrak{sl}}

\def\msp{\mathfrak{sp}}

\def\mho{\mathfrak{ho}}
\def\mg{\mathfrak{g}}
\def\mh{\mathfrak{h}}
\def\mD{\mathfrak{D}}
\def\mV{\mathfrak{V}}

\def\mC{\mathfrak{C}}
\def\mN{\mathfrak{N}}
\def\mI{\mathfrak{I}}

\def\mg{\mathfrak{g}}
\def\mR{\mathfrak{R}}

\def\Real{{\mathbb R}}
\def\Comp{{\mathbb C}}
\def\integ{{\mathbb Z}}
\def\1{1\hspace{-4pt}1}
\def\j1{\widetilde{1\hspace{-4pt}1}}

\def\Tr{{\mathrm{Tr}}}

\def\Str{{\mathrm{Str}}}
\def\Ad{{\mathrm{Ad}}}
\def\Ac{{\mathrm{Ac}}}

\def\bec{\begin{center}}
\def\ec{\end{center}}
\def\a{\alpha} \def\ad{\dot{\a}} \def\hA{{\widehat A}}

\def\b{\beta} \def\bd{\dot{\b}} 
\def\c{\gamma} \def\cd{\dot{\c}}
\def\C{\Gamma}
\def\d{\delta} \def\dd{\dot{\d}}
\def\D{\Delta}
\def\e{\epsilon} \def\he{\hat{\epsilon}}

\def\F{\Phi}

\def\k{\kappa}
\def\l{\lambda}
\def\L{\Lambda}
\def\m{\mu}
\def\n{\nu}
\def\r{\rho}
\def\s{\sigma}
\def\S{\Sigma}
\def\t{\tau}
\def\th{\theta} 

\def\x{\xi}
\def\y{\eta}

\def\O{\Omega}
\def\o{\omega}
\def\pb{{\bar\pi}}

\def\sb{{\bar\s}}

\def\cG{{\cal G}}
\def\span#1{\mathrm{span}_{#1}}
\def\cc{{\cal C}}
\def\cL{{\cal L}}
\def\cD{{\cal D}}
\def\cF{{\cal F}}
\def\cO{{\cal O}}
\def\cA{{\cal A}}
\def\cM{{\cal M}}
\def\cN{{\cal N}}

\def\cS{{\cal S}}

\def\cI{{\cal I}}
\def\cU{{\cal U}}

\def\cA{{\cal A}}
\def\hA{\widehat A}
\def\hD{\widehat D}
\def\hPhi{\widehat \Phi}
\def\hS{\widehat S}
\def\hM{\widehat M}
\def\he{\widehat \e}

\def\hmho{\widehat{\mho}}
\def\hF{\widehat{F}}

\def\hd{\widehat{d}}

\def\yb{{\bar y}}
\def\zb{{\bar z}}

\def\ob{{\bar \o}}

\def\ts{\tilde{\s}}

\def\cO{{\cal O}}

\def\ra{\rightarrow}

\def\del{\partial}

\let\la=\label

\def\nn{\nonumber}
\newcommand{\eq}[1]{(\ref{#1})}

\newcommand{\w}[1]{\\[0.#1cm]}
\def\be{\begin{equation}}
\def\ee{\end{equation}}
\def\bea{\begin{eqnarray}}
\def\eea{\end{eqnarray}}
\def\ba{\begin{array}}
\def\ea{\end{array}}

\def\mx#1#2#3#4{\left#1\begin{array}{#2} #3 \end{array}\right#4}

\def\ft#1#2{{\textstyle{{\scriptstyle #1}
\over {\scriptstyle #2}}}}

\def\ket#1{|#1\rangle}
\def\bra#1{\langle#1|}

\def\scs#1{\section{\sc \large #1}}
\def\scss#1{\subsection{\sc #1}}
\def\scsss#1{\subsubsection{\sc \small #1}}

\def\ad{\dot\alpha}
\def\bd{\dot\beta}
\def\sb{\bar\sigma}

\def\cross{{}_\times\!\!\!\!{}^\times}

\begin{document}

\pagenumbering{roman}

\thispagestyle{empty}

\vspace*{-1.5truecm}

\begin{center}
{\LARGE UNIVERSIT\`A DEGLI STUDI DI ROMA \\
       ``TOR \rule{0pt}{24pt}VERGATA''} \\
\end{center}

\vspace{0.8cm}

\begin{center}
FACOLT\`A DI SCIENZE MATEMATICHE, FISICHE E NATURALI \\
DOTTORATO DI RICERCA IN FISICA \\
\end{center}

\vspace{0.5cm}
\begin{center}
XX CICLO
\end{center}

\vspace{1cm}

\begin{center}
{\LARGE {\bf On the Algebraic Structure of Higher-Spin \\
\vspace{.5cm}Field Equations and New Exact Solutions}}
\end{center}

\vspace{0.5cm}

\begin{center}
{\large Carlo Iazeolla}\footnote{Present address: {\it\small Scuola
Normale Superiore and INFN, Piazza dei Cavalieri 7, 56126 Pisa,
Italy}}
\end{center}

\begin{center}
{\small{\it{Dipartimento di Fisica, Universit\`{a} di Roma ``Tor
Vergata"}}}\\
     {\it{INFN, Sezione di Roma ``Tor Vergata" }}\\
      {\it{Via della Ricerca Scientifica 1, 00133 Roma, Italy }}\\
\end{center}

\vspace{1cm}
\begin{center}
{\small Ph.D. Thesis} \\
Supervisors: \rule{0pt}{20pt}{Dr. {\em Per
Sundell}} \ and \ {Prof.
{\em Augusto Sagnotti}} \\
{\small (Scuola Normale Superiore, Pisa)}\\
Coordinator:
\rule{0pt}{20pt}{Prof. {\em Piergiorgio Picozza}} \\
\end{center}

\newpage
\begin{center}
{\LARGE {\textbf{Abstract}}}
\end{center}
\vspace{1.5cm}

This Thesis reviews Vasiliev's approach to Higher-Spin Gauge Theory
and contains some original results concerning new exact solutions of
the Vasiliev equations and the representation theory of the
higher-spin algebra.

The review part covers the various formulations of the free theory
as well as Vasiliev's full nonlinear equations, in particular
focusing on their algebraic structure and on their properties in
various space-time signatures.

Then, the original results are presented. First, the 4D Vasiliev
equations are formulated in space-times with signatures $(4-p,p)$
and non-vanishing cosmological constant, and some new exact
solutions are found, depending on continuous and discrete
parameters: (a) an $SO(4-p,p)$-invariant family of solutions; (b)
non-maximally symmetric solutions with vanishing Weyl tensors and
higher-spin gauge fields, that differ from the maximally symmetric
background solutions in the auxiliary field sector; and (c)
solutions of the chiral models with an infinite tower of Weyl
tensors proportional to totally symmetric products of two principal
spinors. These are apparently the first exact 4D solutions with
non-vanishing massless higher-spin fields.
\\
Finally, a generalized harmonic expansion of the Vasiliev's master
zero-form is performed as a map from the associative algebra ${\cal
A}$ of operators on the singleton phase space to representations of
the background isometry algebra that include one-particle states
along with linearized runaway solutions. Such Harish-Chandra modules
are unitarizable in a $\Tr_{\cal A}$-norm rather than in the
standard Killing norm. We also take the first steps towards a
regularization scheme for handling strongly coupled
higher-derivative interactions within this operator formalism.

\newpage \vspace*{16cm}
\begin{flushright}

\vspace*{6cm}

{\large {\em Ai miei genitori, e a Chlo\'e}}
\end{flushright}

 \vspace*{16cm}
\newpage
{\LARGE {\textbf{Acknowledgments}}}

\vspace{1.5cm}

This Thesis work includes most of the research I have done during my
PhD course at the University of Rome ``Tor Vergata''. I wish to
thank the Department of Physics and the Coordinator of the PhD
program Prof. Piergiorgio Picozza for supporting my work. I spent a
consistent part of my PhD years on leave of absence from Tor
Vergata, at CERN and at the Scuola Normale Superiore in Pisa, both
of which I would like to thank for the hospitality and the resources
they made available. I am also grateful to INFN for traveling
support. Having spent my PhD years in various different places has
also considerably enlarged the list of persons that I would like to
thank, and, more generally, that have enriched my life.

First of all, my gratitude goes to my advisor Augusto Sagnotti, who
first introduced and then continued to support me in a fascinating
research field, and always encouraged me to think to research
problems in my own way. I also wish to thank him for the
opportunities he gave me to be exposed to life at the forefront of
research in Theoretical Physics, and to grow up as a researcher.
Furthermore, I am grateful to him for having made possible my stay
at CERN and at the Scuola Normale Superiore in Pisa, and for always
having been very friendly and hospitable. I am also grateful to Per
Sundell, for having followed my research closely, for having taught
me most of what I know on Higher-Spin Gauge Theories, for being very
helpful and amenable to discussions at all hours of the day and
night. He has also been a good friend and adventure-mate in Pisa: we
have shared happy and funny moments during the years of my PhD, and
I appreciated a lot the work we did together. I would like to thank
Ergin Sezgin and Mikhail Vasiliev, with whom I had the pleasure of
collaborating, and of discussing some of the main topics of this
research field. In particular, I am grateful to Mikhail Vasiliev for
having carefully refereed this Thesis, giving me many helpful
suggestions that have improved the final form of this work. Many
thanks also to Xavier Bekaert and Sandrine Cnockaert, for being good
colleagues and good friends, whom it has been a pleasure to work
with. It is also a pleasure to acknowledge stimulating discussions
with Damiano Anselmi, Glenn Barnich, Xavier Bekaert, Nicolas
Boulanger, Andrea Campoleoni, Sandrine Cnockaert, Paul Cook, Johan
Engquist, Dario Francia, Maxim Grigoriev, Francesco Guerrieri, Paul
Heslop, Olaf Hohm, Jihad Mourad, George Papadopoulos, Fabio
Riccioni, Slava Rychkov, Philippe Spindel, Jan Troost, Mirian
Tsulaia, Silvia Vaul\`a, Peter West. I also wish to thank Costas
Bachas, Glenn Barnich, Paul Heslop, Jihad Mourad, Fabio Riccioni,
Gabriele Travaglini and Peter West for having kindly invited me to
present my research in their Institutes. Many thanks also to Massimo
Bianchi and Gianfranco Pradisi for their kind assistance and support
in Tor Vergata.

I am very grateful to all the friends that I have been in touch with
during these years: having been travelling around kept me away from
some of them, but also gave me the opportunity to know many others.
In particular, I would like to thank for their friendship and
sustain Alessandra, Anna, Edoardo, Francesco, Nicolas, Paul, Per,
Sandrine, Sara, Valentina, Zofia. And our shiny skittish Camilla.
With all of them I spent happy hours.

Above all, my deepest gratitude and love goes to my parents, for
their encouragement throughout all my studies and for being close to
me in any circumstance with their visits, words, patience, support
and infinite care; and to Chlo\'e, for her love, for the wonderful
time we have together, and for always being next to me.

This work is dedicated to them.

\pagebreak

\tableofcontents

\chapter{Introduction}

\pagenumbering{arabic} \setcounter{page}{1}

Higher-Spin (HS) Fields have attracted the attention of theoretical
physicists since the very early days of Relativistic Field Theory.
As soon as the importance of space-time isometries was exploited,
and free propagating particles were associated to the solutions of
invariant equations under the corresponding symmetry group, it
became natural to investigate the properties of relativistic fields
of more general types. Indeed, the relativity principle reduced the
classification of linear relativistic wave equations to the
classification of the unitary irreducible representations (UIRs) of
the Poincar\'e group, and the latter where found to be fully
specified by the two quantum numbers of mass and spin
\cite{Wigner:1939cj}. Therefore, the very existence of massive and
massless UIRs of the Poincar\'e group with arbitrary spin (or, more
properly, helicity, in the massless case) was, and keeps on being,
the first motivation for a study of the corresponding field
theories. \\
The massless, discrete-helicity case has received the greatest
attention due to the local symmetry principles associated to it. The
gauge invariance that the field equations acquire in the massless
limit is a signal that they actually involve more variables than the
physical degrees of freedom (dof). In group-theoretical terms, this
means that the space of solutions no longer corresponds to an
irreducible representation of the Poincar\'e group, but rather to an
indecomposable one, in which the unphysical polarizations form an
invariant submodule. The role of the gauge symmetry is precisely
that of factoring out such submodule, thereby defining the gauge
field via an equivalence class.\\
The huge interest in gauge theories of course arises from the fact
that massless lower-spin fields (\emph{i.e.}, spin $s\leq 2$) are
known to describe the fundamental interactions, encoded in the
Standard Model and in Einstein's General Relativity (GR):
electroweak and strong interactions are based on abelian and
non-abelian spin-$1$ gauge fields, that implement $SU(3)\times
SU(2)\times U(1)$ local symmetry, and gravity on a spin-$2$ gauge
field, that essentially implements diffeomorphism invariance. In
Classical and Quantum Field Theory (QFT), local symmetries put
strong restrictions on, or completely determine, the dynamics of
fundamental constituents and the possible interactions between
them - and have indeed led theoretical physicists to spectacular
predictions. \\
By now, the classical lower-spin gauge theories are very well known,
and explain an amazing variety of very different phenomena. But the
history of Physics is, to a good extent, a history of unification:
formerly separated areas of investigation were shown to admit a
description within a single conceptual and formal framework, which
proved to be successful in increasing their predictive power. The
last century achieved a partial success in unifying the fundamental
interactions mediated by spin-$1$ massless particles within the
framework of QFT. However, joining gravity in this technical and
conceptual scheme still faces enormous difficulties. \\
Indeed, a number of very important differences arise already at the
classical level, when one compares gauge fields of spin $1$ and $2$:
first, the former is associated to a gauge parameter $\e(x)$ which
is an arbitrary scalar function of space-time coordinates, and the
corresponding local symmetries are therefore \emph{internal}, while
spin-$2$ gauge fields are associated to a vector gauge parameter
$\e^\m(x)$, which can be identified with an infinitesimal change of
coordinates, \emph{i.e.}, with \emph{space-time} symmetries. This
fact is at the root of many differences - that, for example, account
for the known subtleties in the \emph{tout-court} interpretation of
gravity as a gauge theory in the Yang-Mills sense - that we shall
examine in detail in a broader and more general context. Let us only
stress here the well-known fact that, while consistent
self-interactions of non Abelian spin-$1$ gauge fields demand
finitely many nonlinear terms (up to the quartic order in the
Lagrangian), the same request for a spin-$2$ field actually implies
infinitely many nonlinear corrections of higher and higher order!
Furthermore, the propagators of massless fields in QFT show that the
exchange of \emph{even}-spin massless boson gives rise to
universally attractive static potentials, while \emph{odd}-spin
gauge fields mediate attractive and repulsive static interactions
among particles of unlike and like charges, respectively. Finally, a
QFT of spin-$1$ gauge fields, both abelian and non-abelian, has been
constructed and has been shown to be renormalizable, while gravity
is plagued by non-renormalizable infinities. \\
The fundamental difficulty of the many attempts at combining
internal and space-time symmetries into a bigger, more fundamental
symmetry group, within a purely field-theoretical scheme, were
encoded into the famous Coleman-Mandula theorem \cite{Coleman:ad} in
1967: roughly speaking, it states that, under certain assumptions
which at that time seemed reasonable for any physically relevant
field theory, any S-matrix - governing interactions among quantum
fields - symmetric under an algebra $\mg$ which is bigger than the
direct sum of the Poincar\'e (or at most the conformal) algebra
${\cal P}$ and some internal symmetry algebra (spanned by elements
that commute with the generators of ${\cal P}$) has to be trivial.
However, one very important exception was indeed found a few years
later: supersymmetry was the first instance of a global symmetry
that is not forbidden by the Coleman-Mandula theorem and that
transforms fields of different spin into one another. Such a way out
was found by explicitly evading one of the hypothesis of the
theorem: in particular, by introducing Grassmann-odd generators,
\emph{i.e.}, spin-1/2 fermionic parameters, and constructing a
\emph{graded} symmetry algebra, that acts via commutators and
anti-commutators. The realm of the physically interesting theories
was therefore expanded, and the new landmarks were put down with the
formulation of the Haag-Lopuszanski-Sohnius theorem
\cite{Haag:1974qh}, stating that, under similar assumptions, the
maximal symmetry of a nontrivial S-matrix is the direct sum of the
superconformal algebra and some internal symmetry algebra.\\
In 1976, the discovery of supergravity \cite{Freedman:1976xh}, a
theory with \emph{local} supersymmetry, renewed the interest for
Higher-Spin Gauge Theories (HSGT). Indeed, simple supergravity can
be obtained from the requirement of a consistent gravitational
coupling (a crucial requirement for any matter or gauge field, due
to the universality of the gravitational interaction) for a massless
spin-3/2 field, the gravitino, whose free equation had been written
down already in 1941 \cite{Rarita:mf}. In some sense this was,
historically, the first example of a HSGT, of a theory which is
invariant under a local transformations that rotates fields of
different spins among themselves, albeit a particularly simple one,
for a reason that we shall explain later. Since then, higher spins
are fields with spin $s\geq 5/2$. The contributions of the fermionic
superpartners were actually improving the quantum behavior of the
theory, compared to pure gravity, but the theory still seemed to be
divergent\footnote{More recent investigations \cite{Bern:2006kd},
however, suggest that the divergences of supergravity are not as
severe as were originally thought, and that $\cN=8$ supergravity
might even be finite!}.\\
Therefore, Higher-Spin Gauge Theories correspond to field theories
that include massless fields of various spins $s$ and that are
invariant under more general local symmetry principles associated
with parameters that are spin-$(s-1)$ fields. Although there is, at
present, no experimental evidence for fundamental particles of
higher spin, from a purely field theoretical point of view the study
of a more general case may teach some important lesson on the known
gauge theories, and can lead to a better understanding of the gauge
principle. Moreover, our experience with supergravity suggests that
the quantum theory of a bigger symmetry multiplet, that contains the
spin-2 gauge field together with other fields of different spins,
might behave better than the known gravity theories, and hopefully
be free of infinities.

Nowadays, the dreams of unifications of the fundamental interactions
and of a finite quantum theory of gravity both seem within reach in
String Theory - for the very reason that the fundamental constituent
is not a point particle but a one-dimensional extended object of
length $ l_{s}\sim10^{-33} cm $. The string length, indeed, not only
acts as a natural UV cutoff, but also provides some necessary
conditions for unification in a very natural way, compared to the
case of point-like particles. Indeed, in the stringy framework,
fields of different spins arise as vibration modes of the string,
with the spin naturally carried by Fourier coefficients that carry
indices related to target space-time coordinates. However, such a
framework leads to a much richer set of fields than the familiar
low-spin ones. As a generic vibration of a classical string, with
certain boundary condition, admits an expansion over an infinite set
of harmonics with higher and higher frequency, in the same way the
spectrum of a quantum mechanical relativistic string naturally
involves \emph{infinitely many} excitations that can be identified
with different particles of higher and higher mass and spin. The
length and tension $T$ of the string are related as (in units
$\hbar=c=1$)
\bea l_{s}=\sqrt{2\alpha'}=\frac{1}{\sqrt{\pi T}}\ ,\eea
where $\a'$ is a constant, the so-called \emph{Regge slope}, while
the mass and spin of string excitations are related among themselves
and to the tension as
\bea m^{2}\sim \frac{1}{\alpha'}(s-a)\ , \qquad a \ = \
\left\{\ba{c} 1 \ , \ \textrm{open string}\\ 2 \ , \
\textrm{closed string}\ea \right.\ , \eea
where $m^2$ is the squared mass of string states, $s$ is their spin,
and $a$ a constant. The string tension is typically assumed to be of
the order of the Planck mass,  $ M_{Pl}\sim10^{19} GeV $, and is
therefore so high that, at low energies, only the modes with low
frequency can be excited, that exactly correspond to the massless
fields - spin-1 and spin-2 gauge fields, in the open and closed
string sector, respectively - that mediate the known long-range
interactions\footnote{In the bosonic string spectrum there is also a
tachyon, which is a signal of the instability of the vacuum around
which the theory is expanded. Here we do not comment further on this
subtlety, but just remind the reader that the tachyon is anyway not
present in the spectrum of superstrings. Moreover, let us also
mention here that theories of open strings only are inconsistent,
and necessarily a closed string sector needs to be introduced. In
other words, in some sense String Theory predicts the existence of
gravity.}. Moreover, in the superstring spectrum, whose critical
dimension is $D=10$, a massless spinor with $s=3/2$ is also present.
The low-energy superstring dynamics is therefore well-approximated
by ten-dimensional supergravity, while all higher-spin excitations
have very high masses and essentially decouple. However, an infinite
tower of higher-spin fields is necessarily present in String Theory,
and is crucial for its consistency. Indeed, from a field theoretical
point of view, the finiteness of the Theory, due to the natural
cutoff given by the finite string length, translates into the
inevitable appearance of an infinite set of higher-spin fields, with
certain fine-tuned masses, whose contribution to the quantum
perturbative expansion cancels exactly the divergencies of the
lower-spin sector. A study of the properties of HS fields may
therefore lead to a better understanding of String Theory. This
becomes particularly true when referred to its high-energy regime,
where HS fields can indeed contribute significantly to the dynamics.
Despite many efforts, however, much less is known about strings in
this limit than in the supergravity approximation, but nevertheless
interesting observations have been made about the appearance of some
intriguing symmetry enhancements. Indeed, hints of a higher-spin
symmetry have been found in the study of string scattering
amplitudes in the high-energy limit. This fact, together with our
knowledge of spin-1 gauge theories - according to which the quantum
theory of a massive boson is renormalizable only if the mass is
generated via spontaneous breaking of the gauge symmetries -, led
many to think that String Theory might in fact correspond to a
Higher-Spin Gauge Theory in some spontaneously broken phase. All the
extra symmetries should therefore be recovered in the tensionless
limit, in which the whole tower of string excitations become
massless and each corresponding field mediates new long-range
interactions. If this is the case, the study of HSGT in relation
with the tensionless limit of strings might unravel the true
symmetries on which String Theory is based, and give some guiding
principle for exploring the physics of the so-far elusive
eleven-dimensional M-theory that has been recognized to underlie the
different ten-dimensional Superstring Theories. Furthermore, the
latter are still lacking a background independent formulation, where
space-time is a concept that only emerges from the dynamics of the
various string excitations and is not fixed \emph{a priori}. There
are reasons to believe that this is essentially a technical problem,
rather than an intrinsic weakness of String Theory: in particular,
it might depend on the fact that, as mentioned before, it is only
the low-energy dynamics of strings - that we are accustomed to
describe with tools and concepts derived from the known low-spin
gauge theories - that has been widely explored, while a more
``fundamental'' description may need to take into account the
contribution of HS excitations to the space-time texture. The search
for such a description may also therefore benefit from the
developments of HSGT, since, as we shall see, classical interacting
HS field equations exist that are fully background independent, in a
way that is at the same time compatible with HS gauge symmetries.

As mentioned above, the idea that String Theory is a broken phase of
a more symmetric HSGT has been somehow in the back of many
theoretical physicists' mind, but has never found a truly
quantitative formulation. Nowadays, however, the level reached by
our knowledge of HSGT and of the so-called AdS/CFT correspondence
has enabled to recognize, at least at a kinematical level,
signatures of the huge Higgs mechanism (that some authors called
\emph{La Grande Bouffe}) expected to take place in switching on the
tension, giving mass to all higher-spin states and leaving massless
only the low-spin excitations that mediate the known long-range
interactions. A very important window for the study of both String
Theory and quantum gauge field theories, the holographic AdS/CFT
dualities relate, in their most general version, type IIB
Superstring Theory in space-time geometries that are asymptotically
Anti-de Sitter (AdS) times a compact space to conformal field
theories. More precisely, the AdS/CFT correspondence
\cite{Maldacena:1997re} is a conjectured equivalence, at the level
of partition functions, between the type IIB Superstring Theory on
an $AdS_5\times S^5$ background and the $\cN=4$ Supersymmetric
Yang-Mills Theory (SYM), with $SU(N)$ gauge group, living on the
four-dimensional boundary of $AdS_5$. The equivalence is, in fact, a
\emph{duality}: that is, the string and field theoretic pictures are
two descriptions of the same physics, such that when one is weakly
coupled, the dual one is strongly coupled. Although the4
correspondence is supposed to hold for every regime of both
descriptions, \emph{i.e.}, for every value of the relevant
parameters, it has been mainly tested in the limit of large $N$,
large 't Hooft coupling $\l=g^2_{YM}N$ and high tension $T\sim
\sqrt{\l}$, in which the bulk (string) side is well approximated by
classical IIB supergravity and the boundary theory is a strongly
coupled $\cN=4$ SYM in the planar limit (in which there is now some
evidence that it is an integrable theory). \\
However, the correspondence also offers a powerful framework to
explore the physics of strings in the tensionless limit, that is
dual to a weakly coupled or free boundary theory. In such a limit,
the bulk theory cannot be approximated by supergravity anymore,
since the mass of HS excitations becomes small, and, on the boundary
side, the free SYM admits an infinite number of conserved currents
of all spins. In the spirit of the correspondence, the dual bulk
theory should therefore be a theory of interacting gauge fields of
higher spin, based on some non-abelian infinite-dimensional
extension of the $AdS$ superalgebra. Recent investigations have
indeed found tracks of a HS Higgs mechanism in the fact that,
switching on a small $\l$ for the SYM, all the currents that
correspond to bulk HS fields acquire anomalous dimensions, which is
equivalent to a spontaneous breaking of all local HS bulk symmetries
to the finite-dimensional ones of the supergravity theory. There are
also very interesting issues of integrability, on the two sides of
the correspondence, that show remarkable similarities: particularly
intriguing are, for example, those between the construction of
Yangian algebras \cite{MacKay:2004tc} and HS algebras.

The study of String Theory in the tensionless limit and of the HS
dynamics has also been carried on, to some extent, in the framework
of String Field Theory (SFT), a second quantized approach to
strings. Interacting HSGT and SFT show very interesting
similarities, since they both deal with similar variables (the
\emph{string field} and the \emph{master fields}, that we will
present later on, that both include space-time fields of all spins
in an expansion over higher and higher powers of oscillators), that
have similar associative but noncommutative composition laws (a
$\star$-product that essentially implements an (infinite) matrix
multiplication), and since their field equations share the simple
and elegant form of zero-curvature conditions. Therefore, also
certain methods for finding exact solutions are very common in
spirit in the two settings (and are also common to noncommutative
field theories, in general), as well as the language and main tools. \\
In fact, one expects that SFT in the tensionless limit reduces
somehow to a HSGT. However, there are a number of difficulties in
relating explicitly the two theories. To begin with, important
corners are still missing, on both sides: Closed SFT (CSFT) is,
under many respects, much more complicated than Open SFT (OSFT), and
its present formulation is still incomplete; on the other hand, the
full classical theory of HS fields at all orders in interactions has
only been worked out for a particular class of fields, those
represented by totally symmetric tensors, that is exhaustive, up to
dualities, only in four dimensions. For this reason, interacting
HSGT deals today essentially with the fields that belong to a single
Regge trajectory of the open string spectrum, \emph{i.e.}, obtained
acting on the vacuum with powers of a single oscillator, like
\bea \varphi^{\m_1\ldots \m_s} \ = \ \a_{-1}^{\m_1}
\ldots\a_{-1}^{\m_s}\,\ket{0} \ ,\eea
while ten-dimensional Superstring theories also contain mixed
symmetry tensor excitations, obtained acting on the vacuum with
several oscillators. In $D>4$, such vibrating modes describe degrees
of freedom that are independent from those encoded in totally
symmetric fields, and have therefore necessarily to be taken into
account in the dynamics. It is possible that the tensionless limit
of SFT will be clearer once a more complete formulation of CSFT and
HSGT will be available. \\
However, the study of the tensionless limit of the \emph{free}
string field equations, that can be consistently truncated to
totally symmetric fields, has already given some positive results:
in particular, it has been shown that in such a limit they reduce to
equations for triplets of fields that describe the propagation of
massless fields of all spins and that exactly reduce to the
\emph{unconstrained} free HS field equations of Francia and Sagnotti
that we shall describe in Chapter \ref{free} (see
\cite{Sagnotti:2003qa}, also for generalizations to mixed symmetry
free equations and (A)dS background, and for further references).

\vspace{0.5cm}

\begin{flushleft}
\textit{History and properties of Higher-spin Gauge Theories}
\end{flushleft}

\vspace{0.2cm}

\noindent From what has been said so far, one should appreciate the
importance of a better understanding of HS dynamics, and how HSGT
are connected to many important aspects and open questions of the
current research in String Theory. In particular, interacting HSGT
restricted to the totally symmetric sector, offers a somewhat
simplified framework for the study of certain string excitations
when collapsed to zero mass. Indeed, the research in HS fields has
been carried out since many years independently of String Theory,
and produced results that are very interesting in their own right,
for the reasons mentioned in the beginning of this Introduction.\\
Already in 1939, Fierz and Pauli wrote free equations for massive HS
fields in flat space-times. For bosons, the correct number of
polarizations was found to be propagated by a real, totally
symmetric traceless and divergenceless field $\Phi_{\m_1\ldots\m_s}$
satisfying a massive Klein-Gordon equation. Analogously, massive
spin-(s+1/2) fields were described by a totally symmetric,
$\c$-traceless and divergenceless spinor-tensor
$\Psi_{\m_1\ldots\m_s}$ satisfying a massive Dirac equation. Already
at this level, some difficulty related to the peculiarities of HS
fields was recognized: indeed, limiting here the discussion to
bosonic fields for brevity, while the equations of motion were
evidently a generalization of the Proca construction for the massive
photon, a corresponding variational principle seemed not easy to
find. Only in 1974 Singh and Hagen \cite{Singh:qz, Singh:rc} were
able to cast the free theory of a massive spin-$s$ field in
Lagrangian form, by making use of a set of
auxiliary fields of all spins $s-2, s-1,..., 0$. \\
The massless limit of this formulation was studied for the first
time in 1978 by Fronsdal, for bosons, and by Fang and Fronsdal for
fermions \cite{Fronsdal:1978rb, Fang:1978wz, Fronsdal:1978vb,
Fang:1979hq}. It was found that, as the mass tends to zero, all
auxiliary fields of the massive bosonic Singh-Hagen lagrangian
decouple from the physical field $\Phi_{\m_1\ldots\m_s}$, except
for the one with spin $s-2$. The latter can then be combined with
the spin-$s$ field in a totally symmetric \emph{doubly traceless}
field $\varphi_{\m_1\ldots\m_s}$ satisfying an equation that is a
straightforward generalization of Maxwell's and linearized
Einstein's equations: in particular, it possesses an abelian gauge
symmetry under
\bea \d\varphi_{\m_1\ldots\m_s}(x) \ = \
\del_{\m_1}\e_{\m_2\ldots\m_s}(x)+\textrm{symmetrizations} \ ,
\eea
where the gauge parameter is a totally symmetric \emph{traceless}
tensor field of rank $s-1$. It was also found that such construction
admitted an extension to (A)dS background, that essentially amounted
to replace standard derivatives with (A)dS-covariant ones and to add
a mass-like term that encoded the coupling of the spin-$s$ field
with the constant curvature scalar $R$. \\
The Fronsdal equations marked the beginning of the studies on HSGT.
However, the constraints on the gauge parameter and on the field,
that were crucial for the gauge invariance of the field equations
and the lagrangian, were somehow unsatisfactory signs of an
incomplete formulation; indeed, there are no such algebraic
constraints in the free string field equations. These were the main
motivations for a deeper investigations of the free HS dynamics,
that was carried on mainly by Francia and Sagnotti
\cite{Francia:2002aa, Francia:2002pt}, and Bekaert and Boulanger
\cite{Bekaert:2002dt}: the result was a more general formulation,
that achieves \emph{unconstrained} gauge invariance and, in doing
so, elucidates the geometry of HS gauge fields. For instance, the
resulting Francia-Sagnotti unconstrained equations for free spin-$s$
massless fields are \emph{non-local} generalizations of the Maxwell
and linearized Einstein ones, but this time written in terms of
proper HS curvatures, that were first constructed by de Wit and
Freedman \cite{deWit:pe}. The conventional Fronsdal formulation can
be recovered via a gauge fixing. Moreover, the non-locality is pure
gauge, and can therefore be removed by introducing, for every
spin-$s$ field, a compensator field of rank $s-3$. An important
feature of such equations, among others, is that they allowed a
direct, successful comparison with those of the triplets coming from
free SFT \cite{Sagnotti:2003qa}. More recently, the Francia-Sagnotti
equations, in both their local and non-local forms, were also cast
in a lagrangian form, by making use of the compensator and of an
additional auxiliary field, a spin-$(s-4)$ Lagrange
multiplier\footnote{It should be also mentioned that the free HS
off-shell formulation for symmetric tensors had also been worked out
previously by Pashnev and Tsulaia using BRST techniques: the
spin-$s$ lagrangian, however, involved a number of auxiliary fields
that grew proportionally to $s$.}.

Notwithstanding the many motivations for a careful study of a full
HSGT, for decades there has not been much progress in the
construction of interactions of massless HS fields, among themselves
and with low-spin gauge fields. The ``HS interaction problem'' was
already recognized from the early analysis of Fierz and Pauli, that
studied the coupling of HS fields to the electromagnetic field, and
was later formalized in general no-go theorems - such as those due
to Coleman-Mandula and Haag-Lopuszanski-Sohnius already cited and
another due to Weinberg and Witten \cite{Weinberg:1980kq} - that
seemed to rule out a consistent nontrivial embedding of the
lower-spin symmetries into some bigger symmetry algebra mixing
fields of different spins, therefore forbidding the possibility of
couplings with HS fields that would not break the gauge symmetries.
A special attention was devoted to the coupling to gravity, due to
its universality: the analysis of Aragone-Deser and Aragone-Laroche
\cite{Aragone:1979hx} showed that indeed the coupling of the spin-2
field with a gauge field with spin $s\geq 5/2$ is inconsistent with
the gauge symmetries, and this essentially because the presence of
``too many'' indices attached to a massless field inevitably leads
to the appearance of the Weyl tensor in the gauge variation of the
covariantized action for the spin-$s$ field - and the Weyl tensor
cannot be cancelled by any variation of the gravitational part of
the action. More precisely, the argument proceeds as follows. To
introduce the coupling with gravity, it is necessary to introduce a
coupling with the spin-2 field by covariantizing derivatives,
$\partial\longrightarrow D=\partial+\Gamma$, both in the lagrangian
and in the gauge transformations. The lagrangian for the spin-$s$
field $\varphi$ contains the terms
$\frac{1}{2}(D\varphi)^{2}-\frac{s}{2}(D\cdot\varphi)^{2}$, and the
gauge transformation becomes
$\d\varphi_{\mu_{1}\mu_{2}...\mu_{s}}=D_{(\mu_{1}}\e_{\mu_{2}...\mu_{s})}$.
The variation of the lagrangian amounts to a commutator of covariant
derivatives, that is proportional to the Riemann tensor acting on
the gauge parameter,
\begin{equation}
\delta{\cal L} =\Re...(\epsilon...D\varphi...)\neq 0 \
.\label{varriem}
\end{equation}
For $s>2$ the \emph{full} Riemann tensor - both its trace (Ricci
tensor) and its traceless part (Weyl tensor) - contributes to this
variation, and this is what makes a consistent coupling
impossible, since it is only the trace part that can be
compensated by varying the gravitational action. It is interesting
to note that the case of spin 3/2  is the last one (together of
course with spin 2 self-couplings) in which a consistent coupling
is possible, and indeed gives rise to simple supergravity. Indeed,
the variation of the covariantized Rarita-Schwinger lagrangian
\bea I_{RS} \ = \ \bar{\psi}_\m\c^{\m\n\r}D_\n\psi_\r \eea
under $\d\psi^\a_\m=D_\m\e^\a$ is proportional to
\bea \bar{\psi}_\m\c^{\m\n\r}D_\n D_\r\e \ \sim
\bar{\psi}_\m\c^{\m\n\r}[D_\n D_\r]\e\  \ \sim \
\bar{\psi}_\m\c^{\m\n\r}\Re_{\n\r,\s\t}\c^{\s\t}\e \ \sim \
\bar{\psi}_\m (Ric)^{\m}{}_\n\c^\n\e\ ,\label{sugra}\eea
where we have used $\c$-matrices identities in the last step. Thanks
to the low rank of the gravitino, the only nontrivial part of the
variation that survives is proportional to the Ricci tensor, and can
be compensated by a corresponding supersymmetry transformation of
the metric,
\bea \d g_{\m\n} \ = \
\frac{i}{2}(\bar{\psi}_\m\c_\n\e+\bar{\psi}_\n\c_\m\e)\ ,\eea
in the spin-2 lagrangian, that is proportional to the Einstein field
equation, involving the Ricci tensor. Already for $s=5/2$ the eq.
\eq{sugra} does not hold anymore, and a Weyl contribution remains on
the right hand side.\\
At the beginning of the Eighties, however, the work of a few
different research groups showed that interactions of HS fields
among themselves are indeed possible, provided one relaxes certain
hypotheses - hidden or explicit - on which the no-go theorems and
arguments were crucially based. More in detail, in
\cite{Bengtsson:1983pd,Bengtsson:1983pg,Berends:wp,Berends:1984rq}
certain vertices among HS fields only in flat space were constructed
by explicitly evading one key assumption of the Coleman-Mandula
theorem and its generalizations, \emph{i.e.}, by dealing with an
\emph{infinite-dimensional} symmetry algebra. These were the first
works in which it was recognized that non-abelian HS gauge
transformations do not close on a finite set of generators, or,
equivalently, that as soon as one introduces a gauge field of spin
higher than 2 the closure of gauge transformations forces one to
introduce other fields of higher and higher spins, with no upper
bound. It was also noticed that, in general, vertices involving HS
fields should involve higher derivatives, as one would guess by
considering that a Lorentz -invariant cubic coupling involving
fields of spins 0-0-$s$ involves at least $s$ derivatives of the
scalar fields in order to saturate the indices of the spin-$s$ gauge
field.\\
A consistent coupling with gravity was obtained, a few years later,
by Fradkin and Vasiliev \cite{Fradkin:ks} reconsidering the
Aragone-Laroche problem in the framework of a perturbative expansion
around a \emph{nonflat} background - in particular, expanding the
metric around a constant curvature (A)dS space-time, that, being
Weyl-flat, has the virtue of allowing the free propagation of
massless HS fields without breaking the local HS symmetries, as
again can be understood from equation \eq{varriem}. Such a setting
evades, in the first place, the no-go theorems, since they assumed
$\miso(3,1)$ as space-time isometry algebra and, more generally,
since they are all S-matrix arguments, while there is no S-matrix in
(A)dS. Moreover, a nonvanishing cosmological constant $\L$ enables
to construct naturally higher-derivative cubic $s$-$s$-2 vertices
that restore the gauge invariance of the spin-$s$ lagrangian coupled
to gravity, at least to the first nontrivial order in interactions.
Indeed, the important fact is that, if $\L\neq 0$, it is possible to
expand in powers of the fluctuation $R$ of the Riemann tensor around
a nonvanishing background curvature $ R^{(0)}_{\m\n,\r\s} =
(\Lambda/3) \left( g^{(0)}_{\mu\rho} g^{(0)}_{\nu\sigma}-
g^{(0)}_{\nu\rho} g^{(0)}_{\mu\sigma} \right) $, and to write
nonminimal coupling terms that are, schematically, of the form
\begin{equation}
{\cal
L}^{int}=\sum_{A,B}\alpha(A,B)\Lambda^{-\left[\frac{(A+B)}{2}\right]}D^{A}\varphi
D^{B}\varphi R \ ,
\end{equation}
\emph{i.e.}, an appropriate combination of (A)dS-covariant
derivatives of the spin-$s$ field with certain coefficients
$\alpha(A,B)$, with $A$ and $B$ limited by the condition $A+B\leq
s$. The gauge variation of such terms again produces commutators of
two covariant derivative. However, the latter is now, to lowest
order, schematically,
\bea [D,D] \ \sim \ \L g^{(0)}g^{(0)}+\textrm{higher order terms}\
, \eea
and this in turn implies
\bea \delta {\cal L}^{int}\sim R\, D\varphi\,\epsilon\ ,  \eea
where the dependence on $\L$ has disappeared. A proper choice of the
coefficients $\alpha(A,B)$ is therefore enough to cancel
\eq{varriem} and to restore gauge invariance. Notice that, as
observed before, since in a fully interacting theory all spins $s>2$
must be included, the number of derivatives is not bounded from
above! In other words, there is strong evidence that interacting
HSGT are \emph{nonlocal}\footnote{Nonlocal theories do not
automatically suffer from the higher-derivative problem. For
instance, in some cases like String Field Theory, the problem is
somehow cured \cite{vt,Eliezer,nonloc} if the free theory is
well-behaved and if non-locality is treated perturbatively (see
\cite{higherder} for a comprehensive review on this point).}.
Moreover, HS vertices in general do not admit a flat limit
$\L\rightarrow 0$, \emph{i.e.}, a nonvanishing cosmological constant
is necessary for consistent HS interactions. The two facts are
connected, at least in a field theoretic context: indeed,
higher-derivative couplings need to be rescaled with negative powers
of a dimensionful parameter, and the only such parameter available
in HSGT is indeed the cosmological constant. However, the situation
is somewhat different in String Theory, where a natural dimensionful
parameter exists also in flat space, the string length $\a'$, or its
inverse, the string tension $T$. We will not comment further here on
such issue, but just recall that recent results
\cite{Boulanger:2006gr}, obtained with BRST techniques, show that
indeed, in field theory, the gravitational couplings obtained by
Fradkin and Vasiliev are the only ``minimal'' ones (where minimal
here means obtained by covariantizing derivatives with $\C$).\\
Thus, interestingly, a phase with unbroken HS gauge symmetries seems
to be related with a constant curvature gravitational background.
Amusingly, this is not in contradiction with the fact that we
observe, today, a very small - almost zero - positive cosmological
constant. Indeed, is HS symmetries play a role in Nature, they are
presently broken, and the spontaneous symmetry breaking mechanism
might be also responsible for a redefinition of $\L$ through the
vacuum values of the Higgs-like fields.

Still, the fact that interaction terms do not make sense in flat
space-time can be taken as an indication of the importance of not
forcing \emph{a priori} the cosmological constant to vanish. Similar
considerations motivated Fronsdal to undertake a series of works on
field theory in four-dimensional AdS space-time, starting from a
detailed study of the unitary irreducible representations (UIRs) of
its isometry group $SO(3,2)$. The greatest, striking difference with
respect to flat space-time was found by Flato and Fronsdal to be
that massless (and massive) particles in AdS are not fundamental
representations, but arise from the tensor product of two (or more)
ultra-short fundamental UIRs called \emph{singletons}
\cite{Flato:1978qz}. The latter had been first discovered in 1963 by
Dirac \cite{Dirac:1963ta}, and describe conformal particles (a
scalar and a spinor) living on the boundary of AdS. As such, they do
not admit a flat limit, although their tensor product - that
decomposes, under the adjoint action of the algebra $\mso(3,2)$,
into the direct sum of massless representations of all spins - does.
More in detail, the singleton representation becomes a trivial
representation under translations in the contraction $\L\rightarrow
0$ of the AdS isometry algebra to the Poincar\'e algebra ${\cal P}$.
The study of QFT in space-times with a nonvanishing cosmological
constant may therefore reveal some interesting subtlety, as indeed
has already been pointed out in several contexts.

Summarizing, the early analysis on interactions among massless HS
fields had shown that:

\begin{enumerate}
\item A consistent interacting HSGT requires the simultaneous
introduction of infinitely many gauge fields of all spins. We shall
be more precise on what ``all'' means, here. However, an
infinite-dimensional non-abelian HS algebra governing such
interactions can be postulated to underlie such a theory, and this
necessarily has to be spanned by generators of higher and higher
ranks;

\item The interaction with gravity is consistent with the HS gauge
symmetries only on a nonflat gravitational background, \emph{i.e.},
in the presence of a nonvanishing cosmological constant $\L$, since
interaction terms are nonanalytical in $\L$;

\item HS interaction vertices require higher derivatives of the
physical fields involved. This property is strictly connected with
the previous one, since, in order for the physical dimension of the
lagrangian to be preserved with more than two derivatives, a
dimensionful parameter must enter the vertices, and $\L$ is the only
natural candidate in a field theoretical context.

\end{enumerate}

Properties 1 and 3 together imply that a consistent interacting HSGT
is nonlocal. This fact, together with the consideration that fields
of all spins must be introduced, seems very reminiscent of how HS
fields appear in String Theory, together with low-spin fields as
vibration modes of extended fundamental objects, that are in general
$p$-dimensional ($p$-branes) \cite{Bergshoeff:1988jm}. Notice
moreover that the three properties listed above suggest that, in a
theory with unbroken HS symmetries that mix together fields of
different spin, higher and lower-derivative term must come on an
equal footing. In other words, there is no small parameter to set up
a low-energy effective action scheme, \emph{i.e.} an expansion in
derivatives.

For all these reasons, it should by now be clear that interacting
HSGT are a very challenging problem already at the classical level.
Indeed, by the mid Eighties, it was clear that the construction of a
full theory required a more systematic approach. This was developed
essentially by Vasiliev \cite{Vasiliev:sa, Vasiliev:1989yr,
Vasiliev:en, Vasiliev:vu, Vasiliev:1990bu, Vasiliev:1992av} (for
reviews, see \cite{Vasiliev:1995dn, Vasiliev:1999ba,
Vasiliev:2001ur, Bekaert:2005vh}), and it is at present available at
least for the class of totally symmetric fields.

This line of research began with a series of works by Vasiliev and
collaborators, where the free theory was cast in a frame-like
formalism generalizing the Einstein-Cartan approach to gravity. At
the same time, Fradkin and Vasiliev used this approach to construct
cubic interactions for HS fields, within the frame formalism, using
an action that was a HS generalization of the MacDowell-Mansouri
action for (super)gravity, \emph{i.e.}, a bilinear of type $\int
R\wedge R$ in suitable HS curvatures $R$ \cite{Fradkin:ks,
Fradkin:1986qy, Fradkin:ah}. This frame-like approach had the
advantage of giving hints on the possible structure of a non-abelian
HS algebra. Indeed, the reformulation of the free dynamics of a
spin-$s$ gauge field in terms of a set of one-form connections
$(e^{a_1\ldots a_{s-1}}, \o^{a_1\ldots
a_{s-1},b_1},...,\o^{a_1\ldots a_{s-1},b_1\ldots
b_{s-1}})$\footnote{As it will be explained in detail in Chapter
\ref{free}, the indices $a_i$ and $b_i$ are tangent-space
$\mso(D-1,1)$-vector indices, if the background manifold is
$(A)dS_D$, and in particular the one-form connections are valued in
all possible irreducible representations encoded in the Young
diagrams
\begin{picture}(60,13)(-3,2)
\multiframe(0,6.5)(13.5,0){1}(35,6){}\put(40,8){\tiny $s-1$}
\multiframe(0,0)(13.5,0){1}(20,6){}\put(23,0){\tiny $t$}
\end{picture},
where $0\leq t\leq s-1$.} instead of the the metric-like fields
$\varphi_{\m_1\ldots\m_s}$, gave a hint on the structure of the
generators of the algebra, much in the same way as the formulation
of gravity in terms of the frame field and of the Lorentz connection
$(e^a, \o^{a,b})$ bears a direct relation with the generators of the
Poincar\'e algebra $(P_a, M_{ab})$, via the correspondent
non-abelian Yang-Mills-like connection $\O=-i(e^a
P_a+\frac{1}{2}\o^{a,b}M_{ab})$. In other words, this approach
exploits the geometry of the group manifold, of which the generators
are a basis of tangent vectors and gauge fields - physical and
auxiliary ones - enter as the dual, cotangent basis of one-forms.
This framework was suitable for building a HSGT, since it provided a
systematic algorithm to gauge a Lie algebra \emph{\`a la}
Yang-Mills, and at the same time was powerful enough to treat in a
uniform way gravity and internal symmetries\footnote{Indeed,
diffeomorphisms are automatically included as field-dependent gauge
transformations, as we will see
later on.}. \\
The resulting candidate HS algebras should therefore have the same
index structure of the set of one-form connections and should be
constructed as appropriate infinite-dimensional extensions of the
AdS isometry algebra. Such requirements were essentially realized in
terms of a suitable quotient of the enveloping algebra of the
latter: the generators were therefore identified as certain
projections of all possible products of the generators $P_a$ and
$M_{ab}$ of the AdS isometry algebra\footnote{Similar considerations
extend to the case of other signatures, in particular to the dS
case. However, AdS is the suitable background for supersymmetric
extensions of HS algebras, that also have been constructed. The
study of the Vasiliev equations in other signatures will be one of
the subjects of this Thesis.}. Moreover, again in remarkable analogy
with String Theory, it was found that such algebras admit internal
extensions corresponding to Chan-Paton algebras \cite{Marcus:1982fr}
(i.e., only the classical $\msu(n)$, $\musp(n)$ and $\mso(n)$ are
admitted)\footnote{Interestingly, for this reason one would think
that Vasiliev equations might encode the dynamics of tensionless
\emph{open} strings, rather than that of closed strings. This does
not mean, however, that the spin-2 field that appears in the
equations cannot correspond to the graviton, since in the
tensionless limit there is also the possibility of a mixing between
open and closed string states \cite{Francia:2007qt}.}. Gauge fields
of all spins could therefore enter the equations as coefficients of
the expansion of an adjoint one-form, called \emph{master one-form},
over the generators of the HS algebra. However, in order to describe
the free dynamics in terms of the correct number of degrees of
freedom, the above mentioned action had to be supplemented by
certain torsion-like constraints that were not following from its
variation. Moreover, there was still no systematic way of building
interactions to all orders. Most importantly, it was found
\cite{Konstein:1988yg} that the physical spectrum of fields encoded
in the free equations could fit a unitary irreducible representation
of the HS algebra only if it contained a scalar.\\
It was therefore necessary to add a zero-form containing a scalar in
the game. It was then noted that, at the level of field equations,
the free dynamics admitted a useful reformulation in terms of the
master one-form and of a \emph{master zero-form}, valued in a
peculiar UIR of the HS algebra called \emph{twisted-adjoint}, that
included infinitely many auxiliary fields for every massless
physical spin-$s$ field. Constraints that were naturally included in
the system of equations relate such auxiliary fields to all on-shell
nontrivial combination of derivatives of the physical fields,
\emph{i.e.}, the scalar field, all spin-$s$ Weyl tensors and their
derivatives to all orders. Although there is no related action
principle, such a system of equations is a sort of covariant
first-order reformulation of the dynamics, called \emph{unfolded
formulation}, that presents several advantages:

\begin{itemize}

\item HS field equations are written in a manifestly HS-covariant
way, in terms of objects that have ``simple'' transformation
properties under the HS algebra, and without contracting space-time
indices with the (inverse) metric tensor. This latter property is of
great importance, as it enables to treat the spin-2 field on an
equal footing with all the other fields in a system of equations
that can encode nontrivial dynamics. As these equations only involve
differential forms, they are manifestly diffeomorphism invariant.

\item In such a scheme, the field equations are reformulated as
certain consistent zero-curvature and covariant constancy
conditions, defining a \emph{free differential algebra} (FDA)
\cite{D'Auria:nx} - \emph{i.e.}, some sort of generalization to
forms of arbitrary degree of the dual formulation of an algebra
through Maurer-Cartan one-forms. The problem of finding consistent
interactions is therefore reduced to finding consistent deformations
of the FDA. However, such systems are strongly constrained, which
makes it easy to control gauge symmetries, as they are a direct
consequence of the consistency of the field equations. Finally, this
setting enables the construction of gauge-invariant deformations in
an expansion in powers of the zero-form.

\item The twisted-adjoint zero-form guarantees a uniform treatment
of all higher-derivative interaction terms. Indeed, the latter are
expressed as gauge-invariant multi-linear combinations of fields
arising from multiple powers of the zero-form only or with one power
of the one-form. It is only on-shell, on the constraints that are
contained in the unfolded system, that the infinitely many auxiliary
fields sitting in every spin-$s$ sector of the zero-form are solved
in terms of derivatives of the physical fields.

\end{itemize}

In some sense, therefore, the unfolded formulation combines the
virtues of first order, Hamilton-like formulation of the dynamics
with jet-space methods. All relevant derivatives of the fields are
hidden in certain extra-variables, sitting in the twisted adjoint:
this implies, in particular, that the dynamical problem is
well-posed prior to specifying a background metric, and that to set
it up in this formulation it is in principle sufficient to specify
the values of all the zero-forms at a point $x_0$ in space-time.
Since the set of zero-forms is infinite-dimensional (for every fixed
spin), there is indeed room for nontrivial dynamics in the flatness
conditions of the unfolded system, because the fluctuating fields
can be reconstructed in an arbitrary neighborhood of $x_0$ via a
Taylor expansion, which is exactly what solving such first order
equation does. Locally, the dependence on space-time coordinates is
therefore purely auxiliary, and in this way this formulation can
achieve perfect background independence. We shall examine carefully
all these statements in the remainder of the Thesis.\\
Moreover, if the coordinate dependence is locally pure gauge, one
can always introduce additional dependence from some extra
coordinates without changing the physics, as long as one
correspondingly enlarges the original FDA with equations that
express the dependence on such extra-coordinates in terms of the
original physical degrees of freedom. This fact, combined with the
fact that consistency implies gauge-invariance, makes it possible to
``resum'' the perturbation series in powers of the master zero-form.
More precisely, one can enlarge the space-time manifold with a set
of auxiliary \emph{noncommutative} directions $Z$, and assign to all
the fields a dependence on such variables. The noncommutative nature
of such variables essentially ensures that the restriction of the
enlarged system to the physical subspace ${Z=0}$ is a nontrivial
deformation of the original system. It is then possible to write a
constraint that essentially solves the $Z$-dependence in terms of
the twisted adjoint zero-form (that contains the physical dof) as a
consistent equation - in other words, the new equations are an
enlargement of the original FDA. This ensures that the solutions of
such equations, \emph{i.e.}, the deformations of the original FDA,
enter as solutions of a consistent equation, and therefore, for the
properties of FDA, are automatically gauge-invariant. Notice that
relating the $Z$-dependence of adjoint fields with the twisted
adjoint zero-forms gives a nontrivial condition, and indeed
constrains the form of the commutation relations of the $Z$
variables. Therefore, the whole perturbative series in the
zero-forms is encoded in one shot as a solution of a consistent
flatness condition in this auxiliary set of noncommutative
directions $Z$. Solving this equation order by order in the twisted
adjoint zero-forms and substituting for the $Z$ dependence in the
remaining equations projected onto ${Z=0}$ gives
the long sought for interacting equations for massless HS fields.\\
Despite their relatively simple and elegant form, however, Vasiliev
equations encode the dynamics of a system of formidable complexity:
to begin with, they involve infinitely many physical fields, and the
proliferation of auxiliary fields makes it difficult to read
directly the interaction vertices among them. Furthermore, it is not
known, at present, how to derive the unfolded equations from a
conventional action principle, nor, consequently, how to quantize
such a theory.

However, the property of homotopy invariance of the equations,
\emph{i.e.}, the fact that all the local data are encoded in the
equations projected onto a point in space-time, can be read, in the
full system, as the fact that nontrivial dynamics can either be
encoded in the space-time equations (obtained after solving for the
$Z$-dependence) or, equivalently, in the $Z$-fiber over a fixed
point in space-time (after gauging away, locally, the dependence on
space-time coordinates). In other words, the unfolded system
involves some sort of duality between the space-time evolution and
the fiber evolution of the fields, that it is possible to use to
one's advantage. For example, as we shall see in Chapter
\ref{exactsol}, this observation leads to a general, very efficient
way of finding \emph{exact} solutions of the full equations! Indeed,
the fiber equations do not involve any space-time derivatives, and
are purely algebraic equations that are in principle more easily
solvable than the full space-time equations - that, moreover, are
only given as a perturbative expansion. The first example of a
nontrivial (\emph{i.e.}, other than the zeroth-order solution
representing the AdS background) solution was found for the
three-dimensional theory by Prokushkin and Vasiliev
\cite{Prokushkin:1998bq}. A few years later, solutions of this type
were elevated by Sezgin and Sundell to solutions of the
four-dimensional case and were shown to admit some interesting
cosmological interpretation \cite{Sezgin:2005pv}. They were found by
imposing symmetry requirements on the fiber-projection of the
master-fields, and were shown to describe a Lorentz-invariant
deformation of the vacuum consisting of a scalar field profile over
an asymptotically AdS metric. Recently,  a BTZ black hole solution
of the Vasiliev equations in three dimensions was also found
\cite{Didenko:2006zd}. Possibly, the algebraic methods developed so
far, can elevate it to the dynamically more interesting case of four
dimensions.

\vspace{0.5cm}

\begin{flushleft}
\textit{Original content of this Thesis}
\end{flushleft}

\vspace{0.2cm}

\noindent The research of new exact solutions of Vasiliev's
equations is of crucial importance for a better understanding of the
dynamics of higher-spin gauge fields, much in the same way as the
study of the Schwarzschild solution was of extreme importance to
uncover some peculiar feature of gravity. In the original part of
this Thesis, I shall describe new exact solutions that have been
found in collaboration with Ergin Sezgin and Per Sundell
\cite{Iazeolla:2007wt}. In this work, first the Vasiliev equations
and the correspondent symmetry algebras were generalized to
space-times with signature $(4-p,p)$ and nonvanishing cosmological
constant in $D=4$, and then certain families of exact solutions of
the equations have been found in the different resulting models.
Among them are chiral models in Euclidean $(4,0)$ and Kleinian
$(2,2)$ signatures involving half-flat gauge fields. Apart from the
maximally symmetric solutions, including de Sitter spacetime, we
find:

\begin{itemize}

\item $SO(4-p,p)$ invariant deformations, depending on a continuous
and infinitely many discrete parameters, including a degenerate
metric of rank one;

\item Non-maximally symmetric solutions with vanishing Weyl tensors
and higher spin gauge fields, that differ from the maximally
symmetric solutions in the auxiliary field sector;

\item Solutions of the chiral models furnishing higher-spin generalizations of Type
D gravitational instantons \cite{Lapedes:1980qw}, with an infinite
tower of Weyl tensors proportional to totally symmetric products of
two principal spinors. These are apparently the first exact 4D
solutions with nonvanishing massless HS fields.

\end{itemize}

We shall present the details of such solutions in Chapter
\ref{exactsol}. We shall also comment on the construction of certain
HS invariants that have been constructed in \cite{Sezgin:2005pv},
and that are crucial to distinguish gauge-inequivalent solutions and
to characterize them physically. Regrettably, however, at present no
``complete'' set of observables is known, and in particular the only
invariants that are available are built from the master zero-form
only. As we shall see, already certain vacuum solutions found in
\cite{Iazeolla:2007wt} cannot be distinguished from the trivial ones
that represent maximally symmetric space-times, although it seems
unlikely that gauge transformations can connect these two vacua.

Partly motivated by the study of Vasiliev equations in different
signatures, the construction of a precise map in $D$ dimension
between the $\mso(D;\Comp)$-covariant operators included in the
master fields of the Vasiliev system and the states in the complex
lowest (and highest) weight module representations of spin $s$ was
developed in \cite{us}, in collaboration with Per Sundell. Roughly
speaking, to each generator of the twisted adjoint representation
there corresponds a ``coherent'' superposition of infinitely many
states, and, viceversa, to every state in the lowest weight modules
there corresponds a nonpolynomial combinations of generators. The
map can be formulated at the level of complex representations, and
can then be restricted to different real forms corresponding to
models that admit $AdS_D$, $S^D$, $dS_D$, $H_D$ solutions. In the
first of such cases, that of the real form $\mso(D-1,2)$, the lowest
weight modules, that correspond to the massless representations
appearing in the tensor product of two singletons, are unitary
representations. As we shall illustrate, such state-operator
correspondence enables to read directly the on-shell content of the
adjoint one-form and of the twisted adjoint zero-form master-field
in terms of irreps of the background isometry algebra. Roughly
speaking, it exhibits the physical excitations that one would
discover from the unfolded system after solving the various
torsion-like constraints that express the auxiliary fields in terms
of the physical ones. Or again, in other words, it connects the
standard first-quantized description of localized fluctuations with
the master-fields entering the unfolded description. \\
Such a mapping provides some insight into various features of
Vasiliev equations. For example, it shows that, while the on-shell
content of the twisted adjoint zero-form can be analyzed in terms of
the tensor product of two singletons, that of the adjoint one-form
is related to the finite-dimensional $\mso(D+1;\Comp)$-modules that
arise from the tensor product of a singleton and his negative-energy
counterpart, called \emph{anti-singleton}. Another outcome is that
Vasiliev equations in a given signature can describe not only the
corresponding UIRs: indeed, for every spin $s$, a bigger
indecomposable module (containing a lowest-spin module, together
with the more familiar lowest and highest-energy modules) sits in
principle in the master-fields and all the states there contained
can, \emph{a priori}, take part in the dynamics. Finally, the
problem of potential local divergencies in HSGT, due to the
contribution of an arbitrary number of derivatives to some
interaction vertices (as, for example, in the scalar-field
corrections to the stress-energy tensor calculated in
\cite{Kristiansson:2003xx}), is mapped into the problem of divergent
products of nonpolynomial combinations of generators. This is
however a somewhat more transparent setting, and indeed we make a
proposal for an explicit regularization scheme.


\vspace{0.5cm}

\begin{flushleft}
\textit{Structure of this Thesis}
\end{flushleft}

\vspace{0.2cm}

\noindent The structure of this Thesis more or less follows the line
of this Introduction. \\
From Chapter \ref{free} to Chapter \ref{nonlin} we review in some
detail the main features of interacting HSGT. We begin by recalling
the main features of the free HS Field Theory in Chapter \ref{free}:
in particular, the free dynamics is reviewed both in its metric
formulation (concentrating our attention on the Fronsdal and the
Francia-Sagnotti local equations) and in the frame formulation. As a
preparation to the latter, we also describe the
MacDowell-Mansouri-Stelle-West formulation of gravity. In Chapter
\ref{absalg} we turn our attention to the structure of the HS
algebras that lie at the heart of the Vasiliev's system: in order to
be as general as possible, all its main properties will be examined
at the level of the complex abstract algebra, and only at a second
stage we introduce the real forms and the oscillator realizations
that will be crucial in the formulation of the interacting
equations. We also discuss in some detail the construction of the
representations that will be of interest in the remainder of the
Thesis. Chapter \ref{absalg} also presents some essential material
and notation for Chapter \ref{map}. Chapter \ref{unfolding} is then
devoted to the unfolded formulation of the free field equations for
arbitrary spin: the general scheme of \emph{free differential
algebras} is first described, and the features of the unfolded
systems are discussed. Then, some lower-spin examples are given, and
the unfolding procedure is analyzed in detail. Finally, the free
spin-$s$ unfolded system is presented. In Chapter \ref{nonlin},
finally, the nonlinear Vasiliev equations are reviewed, in their
four-dimensional realization that makes use of a simple oscillator
realization of the HS algebra, described earlier in Chapter
\ref{absalg}. The issue of uniqueness of interaction terms is also
discussed in some detail, as well as a perturbative expansion scheme
that makes contact with the free unfolded equations by linearizing
the full system around the $AdS$ background solution. \\
Chapters \ref{exactsol} and \ref{map} contain the original results
of this Thesis. In the first, we formulate the four-dimensional
Vasiliev equations in arbitrary signature, discuss some new features
that emerge in the various cases, and then find the new exact
solutions mentioned above. In the second, we elaborate further on
the representation theory underlying the Vasiliev system and
construct the above-mentioned \emph{reflection map} that will enable
us to connect the physical excitations to the basis of monomials of
the twisted adjoint zero-form. A number of tools and concepts that
are instrumental for such analysis is first introduced, and then
some outcome of this mapping is presented. Finally, we draw some
Conclusions. The Thesis also includes nine appendices, that provide
some background material or contain the detailed steps of some
calculations that are used in the main text. Some of the material
contained in Chapters \ref{free} and \ref{unfolding} is based on the
review paper \cite{Bekaert:2005vh}. The original results contained
in Chapter \ref{exactsol} were found in \cite{Iazeolla:2007wt},
while those of Chapter \ref{map} in \cite{us} and \cite{companion}.
However, we warn the reader that the paper \cite{us} was not yet
completed by the time this Thesis was typed, and therefore only a
subset of the results there included can actually be found in
Chapter \ref{map}.


\chapter{Free Fields}\label{free}

In this chapter, we shall review the formulation of free equations
for massless fields of arbitrary spin $s$, both in the metric and in
the frame formalism. We shall concentrate however only on bosonic
fields, and only on the aspects of the free theory that will be
directly of relevance to the following, therefore not mentioning
interesting features that have an importance of their own and that
are essential to the contemporary developments of the subject. We
therefore refer the interested reader to the original papers
\cite{Francia:2002aa, Francia:2002pt, Francia:2005bu,
Francia:2007qt, Bekaert:2002dt} and to some reviews
\cite{Bouatta:2004kk, Iazeolla:2004hj}.

\scs{Metric Formulation}

The Fronsdal formulation of linear HS gauge theories is somehow the
most straightforward generalization of the Maxwell and linearized
Einstein equations, in the metric formalism, to HS fields
represented by totally symmetric rank-s Lorentz tensors. The free
dynamics of a integer spin-s gauge field
$\varphi_{\m(s)}\equiv\varphi_{\m_1\ldots\m_s}$ in $D$-dimensional
Minkowski space-time is encoded in the equation
\begin{equation}\label{Fronsdals}
{\cal F}_{\m(s)} \ \equiv \ \Box
\varphi_{\m_1\ldots\m_s}-s\,\partial_{(\mu_{1}}\partial\cdot
\varphi_{\m_2\ldots\m_s)}+\frac{s(s-1)}{2}\,\partial_{(\mu_{1}}\partial_{\mu_{2}}\varphi^{\prime}_{\m_3\ldots\m_s)}
\ = \ 0 \ ,
\end{equation}
where the indices within parentheses are intended to be totally
symmetrized with unit strength and the prime over fields indicates
that a trace is being taken,
$\varphi^{\prime}_{\m_3\ldots\m_s)}\equiv
\eta^{\m_1\m_2}\varphi_{\m_1\m_2\ldots\m_s}$. It can be easily
checked that, under the local transformation
\begin{equation}\label{spinsgaugetr}
\delta\varphi_{\m_1\ldots\m_s}(x) \ =\
s\,\partial_{(\mu_{1}}\e_{\m_2\ldots\m_s)}(x) \ ,
\end{equation}
that is an natural generalization of linearized diffeomorphisms to a
totally symmetric rank-$(s-1)$ gauge parameter
$\e_{\m_1\ldots\m_{s-1}}$, the variation of \eq{Fronsdals} is
\begin{equation}\label{dopfronsd}
\delta{\cal F}_{\mu_{1}...\mu_{s}} \ = \
\frac{s(s-1)(s-2)}{2}\,\partial_{(\mu_{1}}\partial_{\mu_{2}}\partial_{\mu_{3}}\epsilon^{\prime}_{\mu_{4}...\mu_{s})}
\ .
\end{equation}
Thus, in order to achieve invariance of Fronsdal's kinetic operator
${\cal F}_{\mu_{1}...\mu_{s}}$, it is necessary to restrict the
gauge freedom to \emph{traceless} gauge parameters,
\be \e'_{\m(s-3)}(x) \ = \ 0 \ .\label{trace} \ee
Notice that, as announced, the Maxwell and linearized Einstein
equations are particular cases of \eq{Fronsdals} for $s=1$ and
$s=2$, respectively. Of course, there the gauge-invariance is fully
unconstrained, \emph{i.e.}, no algebraic constraints are imposed on
the gauge parameters, due to the low-rank of the latter. Indeed, it
is only from spin-3 onwards that the condition \eq{trace} is
nontrivial. \\
More precisely, the Frondal equations are similar to the free
Einstein's equations in vacuum $R_{\m\n}^{lin}=0$, both sharing the
feature of being non-lagrangian equations. The ``kinetic-operator''
that follows from the variation of the Einstein-Hilbert action is,
in fact, the divergenceless tensor
$\cG^{lin}_{\m\n}=R_{\m\n}^{lin}-\frac{1}{2}\eta_{\m\n}R^{lin}$. For
spin $s$ the situation is again more subtle: an action principle for
the Fronsdal equations indeed exists and is a generalization of the
Fierz-Pauli action (that is, the linearized Einstein-Hilbert
action),
\begin{eqnarray}
S^{(s)}_2[\varphi] &=&- {1\over 2} \int d^D x \,
\Big(\partial_\nu\varphi_{\mu_1...\mu_s}\partial^\nu
\varphi^{\mu_1...\mu_s}\nonumber
\\
&&\,\,\,\,- {s(s-1)\over 2}\, \partial_\nu
\varphi^{\lambda}{}_{\lambda\mu_3...\mu_s}\partial^\nu
\varphi_\rho{}^{\rho\mu_3...\mu_s} + s(s-1)\,\partial_\nu
\varphi^{\lambda}{}_{\lambda\mu_3...\mu_s}\partial_\rho \varphi^
{\nu\rho\mu_3...\mu_s}
\nonumber\\
&&\,\,\,\,- s\, \partial_\nu
\varphi^{\nu}{}_{\mu_2...\mu_s}\partial_\rho
\varphi^{\rho\mu_2...\mu_s} - {s(s-1)(s-2)\over 4}\,\partial_\nu
\varphi^{\nu\rho}{}_{\rho\mu_2...\mu_s}\partial_\lambda
\varphi_{\sigma}{}^{\lambda\sigma\mu_2...\mu_s} \Big) \
.\label{Fronsdalact}
\end{eqnarray}
It can be rewritten as
\bea S^{(s)}_2[\varphi] & \sim &
\int d^D x \, \varphi^{\m_1\ldots\m_s}\cG_{\m_1\ldots\m_s} \nn\\[5pt]
& = &
\varphi^{\m_1\ldots\m_s}\left(\cF_{\m_1\ldots\m_s}-\frac{s(s-1)}{4}\eta_{(\m_1\m_2}\cF'_{\m_3\ldots\m_s)}
\right)\ ,\eea
where $\cG_{\m_1\ldots\m_s}$ is a generalized linearized Einstein
tensor. Now, using \eq{spinsgaugetr}, integrating by parts in the
action, and taking the constraint \eq{trace} into account, it is
easy to find that its gauge invariance rests crucially on the
divergence-free nature of $\cG$. However, one gets
\bea \del^\n\cG_{\n\m_1\ldots\m_{s-1}} \ = \
-\frac{(s-1)(s-2)(s-3)}{4}\,\del_{(\m_1}\del_{\m_2}\del_{\m_3}\varphi''_{\m_4\ldots\m_{s-1})}
\ , \label{Bids} \eea
so that, in order to have a gauge-invariant spin-s free Fronsdal
Lagrangian, one has to supplement the theory with an additional
constraint on the fields, declaring them to be represented by
\emph{doubly traceless} tensors,
\bea \varphi''_{\m(s-4)} \ = \ 0 \ ,  \label{doubletr}\eea
which is a nontrivial condition for $s\geq 4$.

Indeed, the Fronsdal equations with their restricted
gauge-invariance propagate the correct number of degrees of freedom
associated to a massless field of spin $s$, in any dimension. To see
this, one can perform an analysis which is again a straightforward
generalization of the one that is proper of the low-spin cases.
First, we introduce a generalized de Donder gauge condition,
\begin{equation}\label{gendD}
{\cal D}_{\mu_{1}...\mu_{s-1}} \ = \
\partial\cdot\varphi_{\mu_{1}...\mu_{s-1}}-\frac{s-1}{2}\,\partial_{(\mu_{1}}\varphi^{\prime}_{\mu_{2}...\mu_{s-1})}
\ = \ 0 \ ,
\end{equation}
that reduces \eq{Fronsdals} to the usual wave equation
\begin{equation}
\Box\varphi_{\m(s)}\ = \ 0 \ .
\end{equation}
The gauge variation of (\ref{gendD}),
\begin{equation}\label{resgaufree}
\delta{\cal D}_{\mu(s-1)} \ = \ \Box\epsilon_{\mu(s-1)} \ ,
\end{equation}
allows for a residual gauge symmetry with parameters that satisfy
themselves a wave equation,
\begin{equation}
\Box\epsilon_{\m(s-1)} \ = \ 0 \label{res}\ .
\end{equation}
Notice also that, by virtue of (\ref{doubletr}), the de Donder
condition is traceless\footnote{As usual, eqs. \eq{resgaufree} and
\eq{res} are also the conditions that ensure that the de Donder
gauge \eq{gendD} is a good gauge, since it does not contain more
conditions than there are independent components of the gauge
parameter, and a parameter that enables to impose it is the solution
of a wave equation, that exists under very general conditions.},
\bea {\cal
D}'_{\mu{(s-3)}}=-\frac{s-3}{2}\,\partial_{(\mu_{1}}\varphi^{\prime\prime}_{\mu_{2}...\mu_{s-3})}\
=\ 0 \ .\eea
Being doubly traceless, the spin-$s$ field admits the
decomposition
\bea \varphi_{\m(s)} \ = \
\phi_{\m(s)}+\eta_{(\m_1\m_2}\s_{\m_3\ldots\m_s)} \ , \eea
where both $\phi_{\m(s)}$ and $\s_{\m(s-2)}$ are traceless
tensors. Correspondingly, the residual gauge transformation
\eq{res} splits into
\bea \d\phi_{\m(s)} \ = \ s\del_{\{\m_1}\e_{\m_2\ldots\m_s\}} \ ,
\eea
where $\{...\}$ denotes the symmetric traceless projection, and
\bea \d\s_{\m(s-2)} \ = \
\frac{s(s-1)}{2D+4(s-2)}\,\del\cdot\e_{\m(s-2)} \ .\eea
The divergence of the gauge parameter can be used to set
$\s_{\m(s-2)}$ to zero. Therefore, the number of independent degrees
of freedom propagated by the Fronsdal equations equals the number of
independent component of $\phi_{\m(s)}$ minus the number of
independent constraints (\ref{gendD}) minus the number of
independent components of the leftover divergenceless gauge
parameter. In $D=4$, for example, this number is\footnote{We recall
that a totally symmetric rank-$s$ tensor in $D$ dimensions, has
$\left({s+D-1 \atop D-1}\right)$ independent components (see, for
example, \cite{Bekaert:2006py}). The number of independent
components of a totally symmetric traceless rank-$s$ tensor in $D$
dimensions therefore follows immediately subtracting from these the
$\left({s+D-1 \atop D-1}\right)$ independent trace constraints. This
gives, in $D=4$, $\left({s+3 \atop 3}\right)-\left({s+1 \atop
3}\right)=(s+1)^2$.} $(s+1)^2-s^2-(s^2-(s-1)^2)=2$, as expected.

Although Fronsdal's formulation captures the fundamental features of
the massless HS free dynamics, it is desirable to overcome the need
for the algebraic constraints on the gauge field and parameter, for
a number of reasons: first, as already commented in the
Introduction, to establish a more direct contact with String Theory
and String Field Theory, whose vibration modes are represented by
unconstrained tensors; second, because, as our experience with
low-spin gauge theories suggests, a restricted gauge invariance is a
sign that the field equations are written in terms on non-fully
invariant objects. This expectation was indeed found to be true
first in \cite{Francia:2002aa}, where nonlocal unconstrained
equations were written in terms of the proper HS curvatures -
generalizations of the Maxwell field strength and the Riemann tensor
(a local equivalent version of the free unconstrained equations was
proposed in the first two references in \cite{Bekaert:2002dt}, see
the third reference therein for a review). The nonlocality was
moreover shown to be pure gauge, and Fronsdal's formulation was
recovered via a gauge fixing. An equivalent, \emph{local
unconstrained} formulation for the free dynamics of a massless
spin-$s$ field was first obtained in \cite{Francia:2002pt} and rests
on the fact that the variation \eq{dopfronsd} can be canceled by
introducing in the equations a spin-$(s-3)$ \emph{compensator} field
, $\a_{\m(s-3)}$, transforming as the trace of the parameter. The
Fronsdal equations are substituted with the system
\bea \cF_{\m(s)} & = &
\frac{s(s-1)(s-2)}{6}\del_{(\m_1}\del_{\m_2}\del_{\m_3}\a_{\m_4\ldots\m_s)}
\ ,\label{frcomp}\\[5pt]
\varphi''_{\m(s-4)} & = &
4\del\cdot\a_{\m(s-4)}+(s-4)\del_{(\m_1}\a'_{\m_2\ldots\m_{s-4})}
\ ,\label{dtrcomp}\eea
that is invariant under the \emph{unconstrained} gauge
transformations
\bea \d\varphi_{\m(s)} & = & s\del_{(\m_1}\e_{\m_2\ldots\m_s)}\ , \\[5pt]
 \d\a_{\m(s-3)} & = & 3\e'_{\m(s-3)} \ . \eea
Notice that \eq{dtrcomp} follows from \eq{frcomp} by using the
Bianchi identity \eq{Bids}, which ensures the compatibility of the
system. The latter equation characterizes $\a$ as a
Stueckelberg-like field, and therefore shows that it is
unphysical, since it can be gauged away by fixing the trace of the
gauge parameter to be $\e'_{\m(s-3)}=\frac{1}{3}\a_{\m(s-3)}$.
This reduces the system to the Fronsdal equations shown above.
Interestingly, the local unconstrained formulation can be shown to
follow from the variation of a minimal local lagrangian
\cite{Francia:2005bu} that only makes use of an additional,
spin-$(s-4)$ Lagrange multiplier to impose the constraint
\eq{frcomp}.


\scss{Interlude 1: Maximally Symmetric Space-times}\label{maxsym}

We have already commented in the Introduction on the importance of
the (A)dS background for building consistent HS interactions.
Indeed, the same reasoning can be extended to any maximally
symmetric background - which is a space-time whose metric has the
maximum number, $\frac{D(D+1)}{2}$, of isometries in $D$ dimensions
- with nonvanishing cosmological constant. Notable examples we shall
deal with later on are, together with $(A)dS_D$, their euclidean
versions: the hyperbolic space $H_D$, obtained from $AdS_D$ through
a ``Wick rotation ''of the time direction, and the sphere $S^D$,
obtained from $dS_D$ through a ``Wick rotation'' of the time
direction. In their turn, $AdS_D$ and $dS_D$ are connected by a
change in the sign of the curvature (\emph{i.e.}, of the
cosmological constant), and the same is true for $H_D$ and $S^D$. In
other words, all such spaces admit a unified description
characterized by two relevant parameters: the signature of their
tangent-space metric $\eta_{ab}$ and the sign of the cosmological
constant. The simplest one is given in terms of flat coordinates
that describe the embedding of any $D$-dimensional maximally
symmetric space-time in a flat, $(D+1)$-dimensional one via the
condition
\begin{equation}\label{gensph}
k\eta_{ab}x^{a}x^{b}+z^{2} \ = \ L^{2}\qquad a,b=0,1,...,D-1\ ,
\end{equation}
where for the moment we do not specify the signature of
$\eta_{\m\n}$, with the flat embedding space metric
\begin{equation}\label{ambsp}
ds^{2} \ = \ \eta_{ab}dx^{a}dx^{b}+\frac{1}{k}dz^{2}\ .
\end{equation}
Only the sign of the curvature constant $k$ will be of relevance,
since any rescaling with a positive factor can be absorbed into
the definition of the coordinates $x^\m$. Solving $z$ from
(\ref{gensph}), differentiating and substituting  $ dz^{2} $ in
(\ref{ambsp}) one gets
\begin{equation}
ds^{2} \ = \ \eta_{ab}dx^{a}dx^{b}+k\frac{\eta_{ac}
\eta_{bd}x^{c}x^{d}}{L^{2}-k\eta_{ab}x^{a}x^{b}}dx^{a}dx^{b} \ ,
\end{equation}
from which it follows that the metric for a maximally symmetric
space can be written as
\begin{equation}\label{max.sym}
g_{ab} \ = \
\eta_{ab}+k\frac{\eta_{ac}\eta_{bd}x^{c}x^{d}}{L^{2}-k\eta_{ab}x^{a}x^{b}}
\ ,
\end{equation}
that has the inverse
\begin{equation}
g^{ab} \ = \ \eta^{ab}-k\frac{x^{a}x^{b}}{L^{2}}\ .
\end{equation}
It is now simple to calculate the Christoffel connection,
\begin{equation}\label{chris}
\Gamma^{c}_{ab} \ = \
\frac{1}{2}g^{cd}(\partial_{b}g_{da}+\partial_{a}g_{db}-\partial_{d}g_{ab})\
= \ \frac{k}{L^{2}}x^{c}g_{ab} \ ,
\end{equation}
and the Riemann tensor
\begin{equation}
R^{c}_{dab} \ = \
\partial_{a}\Gamma^{c}_{db}-\partial_{b}\Gamma^{c}_{da}+\Gamma^{c}_{ea}\Gamma^{e}_{db}-\Gamma^{c}_{eb}\Gamma^{e}_{da}=\frac{k}{L^{2}}(\delta^{c}_{a}g_{db}-\delta^{c}_{b}g_{da})
\ ,
\end{equation}
so that
\begin{equation}
R_{cdab}\ = \ \frac{k}{L^{2}}(g_{ca}g_{db}-g_{cb}g_{da}) \ .
\end{equation}
The Ricci tensor is
\begin{equation}\label{ricci}
R_{ab}\equiv R^{c}_{acb}\ = \ \frac{k}{L^{2}}(D-1)g_{ab}\ ,
\end{equation}
and the curvature scalar
\begin{equation}\label{scal}
R\equiv R^{a}_{a} \ = \ \frac{k}{L^{2}}D(D-1)\ .
\end{equation}
Therefore, the Riemann tensor for a constant curvature space-time
is completely determined by the curvature scalar $R$, and
\begin{equation}
R_{cdab} \ = \ \frac{1}{D(D-1)}R(g_{ca}g_{db}-g_{cb}g_{da}) \ ,
\end{equation}
which implies that, for maximally symmetric space-times, the Weyl
tensor vanishes identically
\begin{eqnarray}\label{Weyl}
C_{cdab} & = &
R_{cdab}+\frac{1}{D-2}(g_{cb}R_{ad}-g_{ca}R_{bd}+g_{da}R_{bc}-g_{db}R_{ac}){}\nonumber\\
& & {} +\frac{1}{(D-1)(D-2)}R(g_{ca}g_{db}-g_{cb}g_{da}) \ = \ 0 \ .
\end{eqnarray}
Moreover, the curvature scalar is proportional to $k$, whose sign
therefore discriminates between the different types of such
space-times: $k=0$ represents a flat space-time with metric
$\eta_{\m\n}$ of arbitrary signature ; $k=+ 1$ ($-1$) represents a
positive (negative) curvature space-time with tangent space metric
$\eta_{\m\n}$. If the latter is fixed to be euclidean, then $k=1$
($-1$) corresponds to a $S^D$ ($H_D$) space, while for minkowskian
tangent space metric one $k=1$ ($-1$) corresponds to the $dS$
($AdS$) space-time.

All such space-times are solutions of Einstein equations in
absence of matter and in presence of a cosmological constant $\L$,
\begin{equation}\label{einstein}
R_{ab}-\frac{1}{2}g_{ab}R \ = \ -\Lambda g_{ab}\ ,
\end{equation}
that are extrema of the Einstein-Hilbert action
\begin{equation}\label{e-h}
S \ = \ \frac{1}{16\pi G_{D}}\int d^{D}x\sqrt{-g}(R-2\Lambda)\ .
\end{equation}
From (\ref{ricci}) and (\ref{scal}) it follows that
\begin{equation}
R_{ab}-\frac{1}{2}g_{ab}R \ = \ -k\frac{(D-1)(D-2)}{2L^{2}}g_{ab} \
,
\end{equation}
and by comparison with (\ref{einstein}) one has
\begin{equation}
\Lambda \ = \ k\frac{(D-1)(D-2)}{2L^{2}}\ ,
\end{equation}
from which one reads that the sign of $ \Lambda $ is related to that
of $ k $, i.e., of the curvature, for any\footnote{$ D=1,2 $ are
trivial cases, since in $D=1$ there is no curvature, and in $D=2$,
although a curvature can be defined, the Einstein-Hilbert action,
that encodes the dynamics of the gravitational field, is a
topological invariant, the Euler characteristic.} $D>2$. Thus, $S^D$
and $dS_D$ space-times have a positive cosmological constant, and
$H_D$ and $AdS_D$ a negative one.

A presentation of \eq{gensph} that puts the AdS case in greater
evidence is given by
\bea X^A X^B\y_{AB} \ \equiv \ \eta_{ab}x^a x^b -\l^2 z^2 \ = \
-\l^2L^2 \ ,\ \qquad A,B=0,1,...,D-1,0' \ , \label{gensph2}\eea
where $\l$ is a complex parameter and again of $\l^2=-1/k$ only the
sign matters (positive for AdS), and the ambient-space metric is
\bea ds^2 \ = dX^A dX^B\y_{AB}\ \equiv \ \eta_{ab}dx^a
dx^b-\l^2dz^2 \ \equiv \ -\t^2dt^2+\d_{rs}x^rx^s-\l^2dz^2\eea
Therefore, the embedding space-time has metric $\y_{AB}=(\y_{ab},
-\l^2)=(-\t^2,\d_{rs},-\l^2)$. Notice that the embedding direction
is a time for the AdS case, and this is the reason for denoting it
with $0'$ (we shall use this label for all signatures anyway, for
the sake of uniformity); similarly, the tangent-space metric is
denoted with an $\y_{ab}$ for all cases, including the euclidean
ones. Equation \eq{gensph2} clearly shows that the isometry algebra
of the different $D$-dimensional maximally symmetric space-times is
the one preserving the quadratic form at its left hand side, which,
for general tangent-space signature is $\mso(p',D+1-p')$ with
$p'\leq D+1$. In particular, the AdS isometry algebra corresponds to
the case $p'=D-1$, $\mso(D-1,2)$, while dS to $p'=D$. Moreover, one
can describe the different manifolds encoded in \eq{gensph2} as the
coset spaces
\bea AdS_D & = & \frac{\mso(D-1,2)}{\mso{(D-1,1)}} \qquad dS_D \
= \ \frac{\mso(D,1)}{\mso{(D-1,1)}} \nn\\[5pt]
H_D & = & \frac{\mso(D,1)}{\mso{(D)}} \qquad S^D \ = \
\frac{\mso(D)}{\mso{(D-1)}} \ ,\label{coset}\eea
where in all cases one factors out, from the isometry algebra of
the embedding metric $\y_{AB}$, the one of the tangent space
metric $\y_{ab}$.

The commutation relations that define the isometry algebra of
\eq{gensph2} are
\bea \left[M_{AB},M_{CD}\right] \ = \
i(\y_{AD}M_{BC}+\y_{BC}M_{AD}+\y_{AC}M_{BD}+\y_{BC}M_{AD}) \
.\label{genAdS}\eea
Splitting the indices as $A=(a,0')$ and defining the translation
generator to be $P_a=\frac{1}{\l L} M_{0'a}$ this can be rephrased
as
\bea [M_{ab},M_{cd}] &=&4i\eta_{[c|[b}M_{a]|d]}\ ,\qquad
[M_{ab},P_c] \ =\ 2i\eta_{c[b}P_{a]}\ ,\label{}\\[5pt]
[P_a,P_b] &=&\frac{i}{\l^2 L^2} M_{ab}\ ,\label{genAdSlor}\eea
which exhibits the difference with respect to the Poincar\'e
algebra, that can be obtained from the previous equations via an
In\"on\"u-Wigner contraction (\emph{i.e.}, in the limit
$\frac{\l}{L}\rightarrow 0$, or $L\rightarrow \infty$),
\bea [M_{ab},M_{cd}] &=&4i\eta_{[c|[b}M_{a]|d]}\ ,\qquad
[M_{ab},P_c] \ =\ 2i\eta_{c[b}P_{a]}\ ,\label{}\\[5pt]
[P_a,P_b] &=&0 \ .\label{genpoinc}\eea

In most of the review part of this Thesis we will mainly focus on
the AdS case, but for the bosonic HSGT everything can be rephrased
for $dS$ and for the euclidean signatures as well. One is mostly
interested in the $AdS$ case for the reason that it is more suitable
for supersymmetric extensions. Furthermore, a change in the
signature affects the Representation Theory: as it is well-known,
for example, $dS$ and $AdS$ have rather different unitary
representations (for $dS$ there are unitary irreducible
representations the energy of which is not bounded from below).
Nevertheless, in the original part of this Thesis we will explicitly
let the signature be arbitrary: examining HSGT in this more general
setting will also prove to be interesting in finding certain new
solutions to the Vasiliev equations, as announced already in the
Introduction.

\scss{Free Equations in $(A)dS$ Space-time}\label{frads}

The same reasoning (see, in particular, \eq{varriem}) that led to
recognize the importance of a nonflat background to build consistent
HS interactions, also leads to the conclusion that the free
propagation of HS fields in a gravitational background that solves
the Einstein equations is consistent with the HS gauge symmetries
only if the solution is Weyl-flat, \emph{i.e.}, if the Weyl tensor
calculated with the background metric is identically zero, which is
indeed the case for the (A)dS space-times - on which we focus here
for definiteness, keeping in mind that everything can be immediately
rephrased in the corresponding euclidean signatures. Therefore, it
is useful to look at the form of Fronsdal's equations in the
presence of a cosmological constant.

The interaction with the fixed gravitational background is
introduced, as usual, by covariantizing derivatives with respect to
the (A)dS Christoffel connection calculated in \eq{chris},
$\partial\rightarrow \nabla\equiv\partial+\Gamma$. Moreover,
\begin{equation}
\varphi'_{\mu_3...\mu_s}=g^{\mu_1\mu_2}\varphi_{\mu_1...\mu_s} \ ,
\end{equation}
where $g$ is the (A)dS metric tensor, and we are assuming
$\varphi''=0$ and $\epsilon'=0$. Now, in the (A)dS background,
these two conditions are no longer sufficient to ensure the
invariance under the covariantized spin-$s$ gauge tranformation
\bea \d\varphi_{\m(s)} \ = \ s\nabla_{(\m_1}\e_{\m_2\ldots\m_s)} \
,\label{trcov}\eea
since the covariant derivatives do not commute. Indeed,
\begin{equation}
[ \nabla_\mu , \nabla_\nu ] \, \varphi_{\rho_1...\rho_s} \ = \
\frac{s}{L^2} \left( g_{\nu(\rho_1|} \,
\varphi_{\mu|\rho_2...\rho_s)} \ - \
 g_{\mu(\rho_1|} \, \varphi_{\nu|\rho_2...\rho_s)} \right) \ , \label{noncommvect}
\end{equation}
for AdS ($\l^2=1$), while the analog for dS can be obtained from
(\ref{noncommvect}) changing sign to the curvature, \emph{i.e.}, by
continuing $\l$ to imaginary values ($\l^2=-1$). The direct
substitution of \eq{trcov} in the covariantized Fronsdal kinetic
operator,
\bea {\cal F}^{\textrm{cov}}_{\mu_1...\mu_s}(\varphi) \ = \
\Box\varphi_{\mu_1...\mu_s}-s\nabla_{(\mu_1}\nabla\cdot\varphi_{\mu_2...\mu_s)}+
\frac{s(s-1)}{2}\,\left\{\nabla_{(\mu_1},\nabla_{\mu_2}\right\}\,\varphi'_{\mu_3...\mu_s)}
\label{opcov}\eea
(where $\Box=g^{\mu\nu}\nabla_\mu\nabla_\nu$), produces terms such
as
\begin{equation}\label{stepinterm}
s\left[\Box,\nabla_{(\mu_1}\right]\,\epsilon_{\mu_2...\mu_s)}+\frac{1}{L^{2}}s(s-1)(D+s-3)\,\nabla_{(\mu_1}\,\epsilon_{\mu_2...\mu_s)}
\ .
\end{equation}
To eliminate these terms it is necessary to modify the kinetic
operator with appropriate terms of order $1/L^2$ that cancel the
variation of (\ref{opcov}) and vanish in the flat limit
$L\rightarrow\infty$. By explicitly calculating the commutator in
(\ref{stepinterm}) one can check that the invariant Fronsdal
equation in $AdS_D$ is
\begin{equation}\label{FropL}
{\cal F}^L_{\m(s)} \ \equiv \ {\cal F}^{\textrm{cov}}_{\m(s)} \ -
\ \frac{1}{L^2} \, \left\{ \left[ (3-D-s)(2-s) - s \right]\,
\varphi_{\m(s)} \ + \ \frac{s(s-1)}{4} \, g_{(\m_1\m_2} \;
\varphi'_{\m_3\ldots\m_s)} \right\} \ = \ 0\ .
\end{equation}
Notice that, although we deal with massless fields, requiring
invariance of the Fronsdal equations in a space-time with
nonvanishing cosmological constant results in the appearance of a
mass-like term, that in fact originates from the coupling with the
(constant) space-time curvature. One can repeat now for \eq{FropL}
the same considerations made above for the flat case. Again, the
Fronsdal equations are non-lagrangian, and one can define a
generalized Einstein tensor
\begin{equation}
{\cal G}^L_{\m(s)} \ = \ {\cal F}^L_{\m(s)} \ - \
\frac{s(s-1)}{4}\,g_{(\m_1\m_2}\,{\cal F}'^L_{\m_3\ldots\m_s)} \ ,
\end{equation}
in terms of which one can construct a Lagrangian from which
\eq{FropL} follows. Finally, a local unconstrained formulation in
(A)dS with the aid of a compensator field is also available
\cite{Francia:2002pt}, and again the presence of a cosmological
constant results in extra-terms of order $1/L^2$. In particular, if
the trace of the gauge parameter is not constrained to vanish,
\bea \d\cF^L_{\m(s)} \ = \
\frac{s(s-1)(s-2)}{2}\left(\nabla_{(\m_1}\nabla_{\m_2}\nabla_{\m_3}\e'_{\m_4\ldots\m_s)}-\frac{4}{L^2}
g_{(\m_1\m_2}\,\nabla_{\m_3}\e'_{\m_4\ldots\m_s)}\right) \eea
and therefore the compensator form of the equations should be
\bea \cF^L_{\m(s)} & = & \frac{s(s-1)(s-2)}{6}\left(\nabla_{(\m_1}
\nabla_{\m_2}\nabla_{\m_3}\a_{\m_4\ldots\m_s)}-\frac{4}{L^2}g_{(\m_1\m_2}\,\nabla_{\m_3}\a_{\m_4\ldots\m_s)}\right)
\ ,\label{frcompL}\\[5pt]
\varphi''_{\m(s-4)} & = &
4\nabla\cdot\a_{\m(s-4)}+(s-4)\nabla_{(\m_1}\a'_{\m_2\ldots\m_{s-4})}
\ ,\label{dtrcompL}\eea
with the gauge symmetries
\bea \d\varphi_{\m(s)} & = & s\nabla_{(\m_1}\e_{\m_2\ldots\m_s)}\ , \\[5pt]
 \d\a_{\m(s-3)} & = & 3\e'_{\m(s-3)} \  \eea
that are again consistent by virtue of the Bianchi identities.

\scs{Frame Formulation}\label{frame}

It is well-known that gravity admits a (first order) formulation in
terms of a frame field and a Lorentz connection, in which the
gauging of the Poincar\'e or (A)dS tangent-space isometry algebra is
manifest, and similar to the familiar Yang-Mills case. This fact
makes it interesting to examine whether the free massless HS theory
admits a reformulation in terms of one-form connections bearing a
direct relationship to the generators of an underlying symmetry
algebra. In other words, such a reformulation can give some hints
towards the construction of an appropriate HS symmetry algebra. If
this is the case, indeed, then the free equations could be
interpreted as the linearization of interacting equations that
involve such one-forms valued in a nonabelian algebra that admits
generators with the same index structure as that of the internal
(tangent-space) indices of the connections.

We shall first review how the frame formulation works for gravity,
by especially recalling the MacDowell-Mansouri and Stelle-West
formulations, that are of special interest for HS extensions. Then,
we shall extend our considerations to HSGT, describing the approach
to the free theory first developed in \cite{V80, LV, FVA}.

\scss{Interlude: Gravity \`a la MacDowell - Mansouri - Stelle -
West} \label{grav}

Einstein's theory of gravity is a non-abelian gauge theory of a
spin-two particle, in the same way (at least to a good extent, see
Appendix \ref{sptime} for more comments) as Yang-Mills theories are
non-abelian gauge theories of spin-one particles. Local symmetries
of Yang-Mills theories originate from the internal global
symmetries. Similarly, the gauge symmetries of Einstein gravity in
the vielbein formulation\footnote{See {\it e.g.} \cite{Zanelli} for
a pedagogical review on the gauge theory formulation of gravity and
some of its extensions, like supergravity.} originate from global
space-time symmetries of its most symmetric vacua. These symmetries
are manifest in the formulation of MacDowell, Mansouri, Stelle and
West \cite{MM,SW}.

This section is devoted to this formulation. First, the
Einstein-Cartan formulation of gravity is reviewed and the link with
the Einstein-Hilbert action without cosmological constant is
explained. Then, the same approach is extended to include a
cosmological constant. We review also an elegant action for gravity,
written by MacDowell and Mansouri, and its improved version
introduced by Stelle and West, where the covariance under all
symmetries is made manifest.

\vspace{0.5cm}

- \textit{Gravity as a Poincar\'e gauge theory}

\vspace{0.2cm}

The basic idea is as follows: instead of considering the metric
$g_{\m\n}$ as the dynamical field, two new dynamical fields are
introduced: the vielbein or frame field $e_{\m}^{a}$ and the Lorentz
connection $\o_\m^{L\hspace{.1cm} ab}$.

The relevant fields appear via the one-forms
   $e^a=e_{\mu}^{a}\, dx^{\mu}$ and
$\o^{{L}\,ab}=-\o^{{L}\,ba}=\o_{\mu}^{{L}\hspace{.1cm}
ab}\,dx^{\mu}\,$. The number of $1$-forms is equal to
$D+\frac{D(D-1)}{ 2}=\frac{(D+1)D}{ 2}$, which is the dimension of
the Poincar\'e group $ISO(D-1,1)$. So they can be collected into a
single $1$-form taking values in the Poincar\'e algebra as
$\o=-i(e^aP_a+\frac{1}{2}\,\o^{L\,ab}M_{ab})$, where $P_a$ and
$M_{ab}$ generate $\miso(D-1,1)$ (see \eq{genpoinc}). The
corresponding
curvature is the two-form: 
\bea R=d\o+\o^2\equiv -i(T^aP_a+\frac{1}{2}\,R^{L\,ab}M_{ab})\,,
\eea
where $T^a$ is the torsion, given by
\bea T^a = D^Le^a = d e^a + \o^{{L}\; a}_{\hspace{.5cm}b} e^b \
,\eea
and $R^{{L\;ab}}\,$ is the Lorentz curvature
\bea R^{{L}\; ab}= D^L\o^{{L}\; ab}= d \o^{{L}\; ab}+\o^{{L}\;
a}_{\hspace{.5cm}c} \o^{{L}\; cb}\,, \eea
as follows from the Poincar\'e algebra \eq{genpoinc}. Torsion and
Lorentz curvature are the invariant tensors under the symmetries of
the theory, i.e., diffeomorphisms and local Lorentz symmetry.

To make contact with the metric formulation of gravity, one must
assume that the frame $e^a_\m$ has maximal rank $d$ so that it gives
rise to the non-degenerate metric tensor $g_{\m\n}=\eta_{ab}e_\m^a
e_\n^b$. Moreover, the appearance of the extra local lorentz
symmetry, with gauge connection $\o^{L\hspace{.1cm} ab}$ and
antisymmetric parameter $\e^{ab}=-\e^{ba}$, is exactly what enables
to gauge away from $e^a_\m$ its antisymmetric part, therefore
reducing the field components to the $D(D+1)/2$ of the metric
tensor. One can also require the absence of torsion, $T_a=0$. Then
one solves this constraint and expresses the Lorentz connection in
terms of the frame field, $\o^L=\o^L(e,\partial e)$. It can be
checked that the tensor $R_{\m\n,\,\rho\s}=e^a_\m e^b_\n
R^L_{ab\;\rho\s}$ expressed solely in terms of the metric is the
familiar Riemann tensor.

The first order action of the frame formulation of gravity is due to
Weyl \cite{Weyl:1929}. In any dimension $D>1$ it can be written in
the form \bea S[\,e^a_\m,\o^{L\;ab}_\m]=\frac{1}{2\kappa^2}
\int_{{\cal M}^D} R^{{L}\; bc}e^{a_1} \ldots e^{a_{D-2}}
\epsilon_{a_1 \ldots a_{D-2}bc}\,\,, \label{mdma} \eea where
$\epsilon_{a_1 \ldots a_{D}}$ is the invariant tensor of the special
linear group $SL(D)$  and $\kappa^2$ is the gravitational constant,
so that $\kappa$ has dimension $(length)^{{D\over 2}-1}$. The
Euler-Lagrange equations of the Lorentz connection
\bea\frac{\d S}{\d \o^{{L}\,bc}}\propto \epsilon_{a_1 \ldots
a_{D-2}bc}\,e^{a_1} \ldots e^{a_{D-3}} T^{a_{D-2}} =
0\label{deltacomega} \eea
imply that the torsion vanishes. The Lorentz connection is then an
auxiliary field, which can be removed from the action by solving its
own (algebraic) equations of motion. The action
$S=S[\,e\,,\,\o^L(e,\partial e)\,]$ is now expressed only in terms
of the vielbein. Actually, only combinations of vielbeins
corresponding to the metric appear and the action $S= S[\,g_{\m\n}]$
coincides indeed with the second order Einstein-Hilbert action.

The Minkowski space-time solves $R^{L\;ab}= 0$ and $T^a= 0$. It is
the most symmetrical solution of the Euler-Lagrange equations, whose
global symmetries form the Poincar\'e group. The gauge symmetries of
the action (\ref{mdma}) are the diffeomorphisms and the local
Lorentz transformations. Together, these gauge symmetries correspond
to the gauging of the Poincar\'e group (see Appendix \ref{sptime}
for more comments).

\vspace{0.5cm}

- \textit{Gravity as an $\mso(D-1,2)$ gauge theory} \label{cosmo}

\vspace{0.2cm}

It is rather natural to reinterpret $P_a$ and $M_{ab}$ as the
generators of the $AdS_D$ isometry algebra $\mso(D-1,2)$.
The curvature $R=d\o+\o^2$ then decomposes as
$R=-i(T^aP_a+\frac{1}{2}\,R^{ab}M_{ab})$, where the Lorentz
curvature $R^{L\;ab}$ is deformed to \be R^{ab}\equiv R^{{L}\; ab}+
R^{cosm \; ab}\equiv R^{{L}\; ab}+\Lambda \, e^a  e^b\,,
\label{cosm}\ee since (\ref{genpoinc}) is deformed to
(\ref{genAdSlor}).

MacDowell and Mansouri proposed an action \cite{MM}, that can be
built from the product of two curvatures (\ref{cosm}) in $D=4$ \bea
S^{MM}[\,e,\o] =\frac{1}{4 \kappa^2 \Lambda}\int_{{\cal M}^4} R^{a_1
a_2} R^{a_3 a_4} \epsilon_{a_1 a_2 a_3 a_{4}}\,. \label{mmact} \eea
Expressing $R^{ab}$ in terms of $R^{{L}\;ab}$ and $R^{cosm\;ab }$ by
(\ref{cosm}), the Lagrangian is the sum of three terms: a term
$R^{{L}} R^{cosm}$, which is the previous Lagrangian (\ref{mdma})
without cosmological constant, a cosmological term $R^{cosm}
\,R^{cosm}$ and a Gauss-Bonnet term $R^{{L}} R^L$. The latter
contains higher-derivatives but does not contribute to the equations
of motion because it is a topological invariant.

The MacDowell-Mansouri action  admits a higher dimensional
generalization \cite{5d} \bea S^{MM}[\,e,\o] =\frac{1}{4 \kappa^2
\Lambda}\int_{{\cal M}^F} R^{a_1 a_2} R^{a_3 a_4} e^{a_5} \ldots
e^{a_{F}} \epsilon_{a_1 \ldots a_{F}}\,. \label{mmactd} \eea The
$AdS_D$ space-time is defined as the most symmetrical solution of
the Euler-Lagrange equations. As explained in more detail later on,
it is a solution of the system $R^{ab}= 0$, $T^a= 0$ such that
rank$(e^a_\m)=d$. The Gauss-Bonnet term \bea S^{GB}[\,e,\o]
=\frac{1}{4 \kappa^2 \Lambda}\int_{{\cal M}^d} R^{L\,a_1 a_2}
R^{L\,a_3 a_4} e^{a_5} \ldots  e^{a_{d}} \epsilon_{a_1 \ldots
a_{d}}\,.
\nonumber 
\eea is not topological beyond $D=4$, and therefore the field
equations resulting from the action (\ref{mmactd}) are different
from the Einstein equations in $D$ dimensions. However, the
difference involves nonlinear terms that do not contribute to the
free spin 2 equations \cite{5d}, apart from replacing the
cosmological constant $\Lambda$ by $ \frac{2(D-2)}{D}\Lambda$ (in
such a way that no correction appears in $D=4$, as expected). One
way to see this is by considering the action \bea S^{nonlin}[\,e,\o]
= S^{GB}[\,e,\o]+ \frac{D-4}{4 \kappa^2}\int_{{\cal M}^D} \Big
(\frac{2}{D-2}\, R^{L\,a_1 a_2} e^{a_3} \ldots  e^{a_{D}}
+\frac{\Lambda}{D} \,e^{a_1} \ldots e^{a_{D}}\Big ) \epsilon_{a_1
\ldots a_{D}}\,, \label{nonl} \eea which is the sum of the
Gauss-Bonnet term plus terms of the same type as the
Einstein-Hilbert and cosmological terms (note that the latter are
absent when $D=4$). The variation of (\ref{nonl}) is equal to \bea
\delta S^{nonlin}[\,e,\o] = \frac{1}{4 \kappa^2 \Lambda}\int_{{\cal
M}^D} R^{a_1 a_2} R^{a_3 a_4} \delta ( e^{a_5} \ldots  e^{a_{D}} )\,
\epsilon_{a_1 \ldots a_{D}}\,,\label{variation} \eea when the
torsion is required to be zero (i.e., applying the 1.5 order
formalism to see that the variation over the Lorentz connection does
not contribute). Indeed, the variation of the action (\ref{nonl})
vanishes when $D=4$, but when $D>4$ the variation (\ref{variation})
is bilinear in the $AdS_D$ field strength $R^{ab}$. Since the
$AdS_D$ field strength is zero in the vacuum $AdS$ solution, the
action $S^{nonlin}$  only contributes to corrections of the field
equations which are nonlinear in the fluctuations near the $AdS$
background, having no effect on the free spin 2 equations. As a
consequence, at the linearized level the Gauss-Bonnet term does not
affect the form of the free spin $2$ equations of motion, it merely
redefines  an overall factor in front of the action and the
cosmological constant via $\kappa^2 \to (\frac{D}{2}-1)\kappa^2$ and
$\Lambda \to \frac{2(D-2)}{D}\Lambda$, respectively (as can be seen
by substituting $S^{GB}$ in (\ref{mmactd}) with its expression in
terms of $S^{nonlin}$ from (\ref{nonl}) ). Beyond the free field
approximation the corrections to Einstein's field equations
resulting from the action (\ref{mmactd}) are nontrivial for $D>4$
and nonanalytic in $\Lambda$ (as can be seen from
(\ref{variation})), with no smooth flat limit. As will be shown
later, this is analogous to the structure of HS interactions which
also contain terms with higher derivatives and negative powers of
$\Lambda$. The important difference is that in the case of gravity
one can subtract the term (\ref{nonl}) without destroying the
symmetries of the model, while this is not possible in HS gauge
theories. The flat limit $\Lambda\rightarrow 0$ is perfectly smooth
at the level of the algebra ({\it e.g.} $\mso(D-1,2)\rightarrow
\miso(D-1,1)$ for gravity, see Section \ref{maxsym}) and at the
level of the free equations of motion, but it may be singular at the
level of the action and nonlinear field equations.

\vspace{0.5cm}

- \textit{MacDowell-Mansouri-Stelle-West gravity}

\vspace{0.2cm}

The gauge symmetries of the MacDowell-Mansouri action (\ref{mmact})
are diffeomorphisms and local Lorentz transformations. It is however
possible to make the $\mso(D-1,2)$ symmetry manifest by combining
the vielbein and the Lorentz connection  into a single field
$\o=-i\o_{\mu}^{\hspace{.2cm} AB}dx^{\mu}M_{AB}$. The fiber indices
$A,B$  now run from $0$ to $D$. They are raised and lowered by the
invariant mostly minus metric $\eta_{AB}$ of $\mso(D-1,2)\,$.

In order to promote local $\mso(D-1,2)$ transformations to gauge
symmetries, an additional field has to be introduced: the time-like
vector $V^A$ called {\it compensator}\footnote{This compensator
field compensates additional symmetries serving for them as a Higgs
field. It should \textit{not} be confused with the homonymous - but
unrelated - field $\a_{\m(s-3)}$ introduced in the previous
Subsection.}. The compensator vector is constrained to have a
constant norm $\r$, \bea V^A V^B \eta_{AB} = \rho^2\,. \label{norm}
\eea As we shall see, the constant $\r$ is related to the
cosmological constant according to  \be \label{rco}
\r^2=-\Lambda^{-1}\,. \ee

The MMSW action is (\cite{SW} for $D=4$ and \cite{5d} for arbitrary
$D$) \bea S^{MMSW}[\,\o^{AB},V^A]=-\frac{\r}{4\kappa^2}\int_{{\cal
M}^D}\epsilon_{A_1 \ldots A_{D+1}}R^{A_1 A_2} R^{A_3 A_4} E^{A_5}
\ldots E^{A_D} V^{A_{D+1}}\,, \label{mmswaction}\eea where the
curvature or field strength $R^{AB}$ is defined by $$ R^{AB}\equiv
d\o^{AB}+\o^{AC}\o_C^{~~B} $$ and the frame field $E^A$ by \be
\label{fram} E^A\equiv D V^A= d V^A+ \o^A_{~B} V^B \,. \nonumber \ee
Furthermore, in order to make contact with Einstein gravity, two
constraints are imposed: (i) the norm of $V^A$ is fixed, and (ii)
the frame field $E^A_\mu$ is assumed to have maximal rank equal to
$D$. As the norm of $V^A$ is constant, the frame field satisfies \be
E^AV_A=0\,. \label{EV} \ee If the condition (\ref{norm}) is relaxed,
the norm of $V^A$ corresponds to an additional dilaton-like field
\cite{SW}.

Let us now analyze the symmetries of the MMSW action. The action is
manifestly invariant under
\begin{itemize}
\item Local $\mso(D-1,2)$ transformations: \be\d \o^{AB} (x)= D
\epsilon^{AB}(x)\,,\qquad\d V^A (x)= -\epsilon^{AB}(x) V_B (x)
\label{localo}\,;\ee \item Diffeomorphisms: \be\d \o^{AB}_{\n}=
\del_{\n}(\xi^{\m})\o^{AB}_{\m}+\xi^{\m}\del_{\m}\o^{AB}_{\n}\,,
\qquad\d V^A=\xi^{\n}\del_{\n}V^A\,.\label{diffs}\ee
\end{itemize}
Let us define the covariantized diffeomorphism as the sum of a
diffeomorphism with parameter $\xi^\m$ and a $\mso(D-1,2)$ local
transformation with parameter
$\epsilon^{AB}(\xi^{\m})=-\xi^{\m}\o_{\m}^{AB}$. The effect of this
transformation is thus \be \d^{cov}
\o^{AB}_{\m}=\xi^{\n}R_{\n\m}^{AB}\,,\qquad\d^{cov} V^A=\xi^{\n}
E_{\n}^A \label{covd} \ee by (\ref{localo})-(\ref{diffs}).

The compensator vector is pure gauge. Indeed, by local $O(D-1,2)$
rotations one can gauge fix $V^A (x)$ to any values with $V^A(x) V_A
(x)=\r^2$. In particular, one can reach the standard gauge \be
\label{stgau} V^A=\r\,\d^A_{0'}\,. \ee Taking into account
(\ref{EV}), one observes that the covariantized diffeomorphism also
makes it possible to gauge fix fluctuations of the compensator
$V^A(x)$ near any fixed value. Since the full list of symmetries can
be represented  as a combination of covariantized diffeomorphism,
local Lorentz symmetry and diffeomorphisms, in the standard gauge
(\ref{stgau}) the algebra of gauge symmetries is broken to the local
$\mso(D-1,1)$ algebra and diffeomorphisms. In the standard gauge,
one therefore recovers the field content and the gauge symmetries of
the MacDowell-Mansouri action. Let us note that covariantized
diffeomorphisms (\ref{covd}) do not affect the connection
$\o^{AB}_{\m}$ if it is flat ({\it i.e.} has zero curvature
$R_{\n\m}^{AB}$). In particular covariantized diffeomorphisms do not
affect the background $AdS$ geometry. \vspace{.2cm}

To show the equivalence of the action (\ref{mmswaction}) with the
action(\ref{mmactd}), it is useful to define a Lorentz connection by
\be\o^{{L}\; AB}\equiv \o^{AB} -\r^{-2}
(E^AV^B-E^BV^A)\,.\label{lorentzcosm}\ee In the standard gauge, the
curvature can be expressed in terms of the vielbein $e^a\equiv E^a=
\rho \, \o^{a}_{~\hat d}$ and the non-vanishing components of the
Lorentz connection $\o^{{L}\; ab}= \o^{ab}$ as \bea
R^{ab}&=&d\o^{ab}+\o^{aC}  \o_C^{~~b}=d \o^{{L}\; ab}+\o^{{L}\;
a}_{~~~c}  \o^{{L}\; cb}-\r^{-2} e^a  e^b= R^{{L}\; ab}+ R^{cosm\;
ab}\,,\nonumber\\
R^{a0'}&=& \r^{-1}T^a\,.\nonumber \eea Inserting these gauge fixed
expressions into the MMSW action yields the action (\ref{mmactd}),
where $\Lambda =-\rho^{-2}$. The MMSW action thus reduces to
(\ref{mmactd}) after partially fixing the gauge symmetry.
\vspace{.2cm}

Let us now consider the vacuum equations $R^{AB}(\o_0)= 0$. They are
equivalent to $T^a= 0$ and $R^{ab}= 0$ and, under the condition that
rank$(E_{\n}^A)=d\,$, they uniquely define the local geometry of
$AdS_D$ with parameter $\r$, in a coordinate independent way. The
solution $\o_0$ obviously satisfies the equations of motion of the
MMSW action. To find the symmetries of the vacuum solution $\o_0$,
one first notes that vacuum solutions are sent to vacuum solutions
by diffeomorphisms and local $AdS$ transformations, because they
transform the curvature homogeneously. Since covariantized
diffeomorphisms do not affect $\o_0$, in order to find symmetries of
the chosen solution $\o_0$ it is enough to check its transformation
law under local $\mso(D-1,2)$ transformation. Indeed, adjusting an
appropriate covariantized diffemorphism it is always possible to
keep the compensator invariant.

The solution $\o_0$ is invariant under local $\mso(D-1,2)$
transformations if and only if the parameter $\epsilon^{AB}(x)$
satisfies \be 0= D_{0} \epsilon^{AB}(x)=d\epsilon^{AB}(x)+
\o_0{}^{A}{}_C(x)\,\epsilon^{CB}(x)-\o_0{}^{B}{}_C(x)\,\epsilon^{CA}(x)\,.
\label{AdSsol}\ee This equation fixes the derivatives $
\partial_\m\epsilon^{AB}(x)$ in terms of $\epsilon^{AB}(x)$ itself. In other
words, once $ \epsilon^{AB}(x_0)$ is chosen for some $x_0$,
$\epsilon^{AB}(x)$ can be reconstructed for all $x$ in a
neighborhood of $x_0$, since by consistency\footnote{The identity
$D_0^2=R_0=0$ ensures consistency of the system (\ref{AdSsol}),
which is overdetermined because it contains $\frac{D^2(D+1)}{2}$
equations for $\frac{D(D+1)}{2}$ unknowns. Consistency in turn
implies that higher space-time derivatives $\partial_{\nu_1} \ldots
\partial_{\nu_n} \epsilon^{AB}(x)$
obtained by hitting (\ref{AdSsol}) $n-1$ times with $D_{0\nu_k}$ are
guaranteed to be symmetric in the indices $\nu_1\ldots \nu_n$.}
   {\it all} derivatives of the parameter can be expressed in terms
of the parameter itself. The parameters $\epsilon^{AB}(x_0)$ remain
arbitrary, and are indeed parameters of the global symmetry
$\mso(D-1,2)$. This means that, as expected for $AdS$ space-time,
the symmetry of the vacuum solution $\o_0$ is the global
$\mso(D-1,2)$. \vspace{.2cm}

The lesson is that, to describe a gauge model that has a global
symmetry $h$, it is useful to reformulate it in terms of the gauge
connections $\o$ and curvatures $R$ of $h$ in such a way that the
zero curvature  condition $R=0$ solves the field equations and
provides a solution with $h$ as its global symmetry. If a symmetry
$h$ is not known, this observation can be used the other way around:
by reformulating dynamics \`a la MacDowell-Mansouri one might guess
the structure of an appropriate curvature $R$ and thereby the
non-abelian algebra $h$.

\scss{Frame-like Formulation of Free HS Dynamics}\label{frameHS}

It is possible to parallel the frame formulation of gravity for HS
fields. The doubly-traceless metric-like HS gauge field
$\varphi_{\m_1 \ldots \m_s}$ is replaced by a frame-like field $A_
{\m}^{~ a_1 \ldots a_{s-1}}$, a Lorentz-like connection $A_{\m}^{
~a_1 \ldots a_{s-1}, \, b}$
\cite{V80} and an extra set of connections\\
$A_{\m}^{ ~a_1 \ldots a_{s-1}, \, b_1 \ldots b_t}$, where
$t=2,\ldots, s-1$ \cite{Fort1,LV}. All fields are traceless in the
fiber indices $a,b$, which have the symmetry of the Young tableaux
\hspace{.1cm}
\begin{picture}(85,15)(0,2)
\multiframe(0,7.5)(13.5,0){1}(50,7){}\put(55,7.5){$s-1$}
\multiframe(0,0)(13.5,0){1}(35,7){}\put(40,0){$t \hspace{1.3cm},$}
\end{picture}
where $t=0$ for the frame-like field and $t=1$ for the Lorentz-like
connection. The metric-like field arises as the completely symmetric
part of the frame field \cite{V80}, $$ \varphi_{\m_1 \ldots
\m_s}=A_{\{\m_1 ,\, \m_2\ldots \m_s\}}\,,$$ where all fiber indices
have been lowered using the $AdS$ or flat frame field $e_0{}^a_\m$
defined in Section \ref{grav}. The fiber tracelessness of the frame
field implies automatically that the field $\varphi_{\m_1 \ldots
\m_s}$ is doubly traceless.

The frame-like field and other connections are then combined
   \cite{5d} into a connection one-form
$A^{A_1 \ldots A_{s-1},\, B_1 \ldots B_{s-1}}$ (where $A,B=
0,\ldots, d$) taking values in the irreducible $\mso(D-1,2)$-module
characterized by the two-row traceless rectangular Young tableau
\begin{picture}(60,15)(-5,2) \multiframe(0,7.5)(13.5,0){1}(50,7){}
\multiframe(0,0)(13.5,0){1}(50,7){}\put(53,0){}
\end{picture}
of length $s-1$, that is
\begin{eqnarray}
&A_\mu^{A_1\ldots A_{s-1},B_1\ldots B_{s-1}}=A_\mu^{\{A_1\ldots
A_{s-1}\},B_1\ldots
B_{s-1}}=A_\mu^{A_1\ldots A_{s-1},\{B_1\ldots B_{s-1}\}}\,,&\nonumber\\
&A_\mu^{\{A_1\ldots A_{s-1},A_s\}\,B_2\ldots B_{s-1}}=0\,,\quad
A_\mu^{A_1\ldots A_{s-3}\,C\,}{}_{C,}{}^{B_1\ldots
B_{s-1}}=0\,.&\label{adsy}
\end{eqnarray}

One also introduces a time-like vector $V^A$ of constant norm $\r$.
The component of the connection $A^{A_1 \ldots A_{s-1},\, B_1 \ldots
B_{s-1}}$ that is most parallel to $V^A$ is the frame-like field
$$A^{ A_1 \ldots A_{s-1}}=\o^{A_1 \ldots A_{s-1},\, B_{1} \ldots
B_{s-1}}V_{B_{1}} \ldots V_{B_{s-1}} \,,
$$
while the less $V$-longitudinal components are the other
connections. Note that the contraction of the connection with more
than $s-1$ compensators $V^A$ is zero by virtue of (\ref{adsy}). Let
us be more explicit in a specific gauge. As in the MMSW gravity
reformulation, one can show that $V^A$ is a pure gauge field and
that one can reach the  standard gauge $V^A=\d_{0'}^A \r$ (the
argument will not be repeated here). In the standard gauge, the
frame field and the connections are given by \bea A^{~ a_1 \ldots
a_{s-1}}&=& \r^{s-1} A^{~a_1 \ldots a_{s-1},\, 0'
\ldots 0'} \nonumber\\
A^{ ~a_1 \ldots a_{s-1}, \, b_1 \ldots b_t}&=&\r^{s-1-t} \Pi A^{~a_1
\ldots a_{s-1},\, b_1 \ldots b_t\, 0' \ldots 0'} \nonumber \eea
where the powers of $\rho$ originate from a corresponding number of
contractions with the compensator vector $V^A$ and $\Pi$ is a
projector to the Lorentz-traceless part of a Lorentz tensor, which
is needed for $t\geq 2$. These normalization factors are consistent
with the fact that the auxiliary fields $A_\mu^{ ~a_1 \ldots
a_{s-1}, \, b_1 \ldots b_t}$ will be found to be expressed via $t$
partial derivatives of the frame field $A_\mu^{~ a_1 \ldots
a_{s-1}}$ ($\rho$ is a length scale) at the linearized level.

The linearized field strength or curvature is defined as the
$\mso(D-1,2)$ covariant derivative of the connection $A^{A_1 \ldots
A_{s-1}, B_1 \ldots B_{s-1}}$, {\it {\it i.e.}} by \bea F_1^{A_1
\ldots A_{s-1}, B_1 \ldots B_{s-1}}&= &
D_0 A^{A_1 \ldots A_{s-1}, B_1 \ldots B_{s-1}}\nonumber \\
&=&d A^{A_1 \ldots A_{s-1}, B_1 \ldots B_{s-1}}+ \o_{0~~C}^{~A_1}
A^{C A_2 \ldots  A_{s-1}, \,B_1 \ldots B_{s-1}} + \ldots \nonumber \\
&& \hspace{3cm}+ \o_{0~~C}^{~B_1}  A^{A_1 \ldots  A_{s-1}, \,C B_2
\ldots B_{s-1}} + \ldots \;, \label{R1} \eea where the dots stand
for the terms needed to get an expression symmetric in $A_1 \ldots
A_{s-1}$ and $B_1 \ldots B_{s-1}$, and $\o_{0~B}^{~A} $ is the
$\mso(D-1,2)$ connection associated to the $AdS$ space solution, as
defined in Section \ref{grav}. The connection $A_\mu^{A_1 \ldots
A_{s-1},\, B_{1} \ldots B_{s-1}}$ has dimension $(length)^{-1}$ in
such a way that the field strength $F_{\m\n}^{A_1 \ldots A_{s-1},\,
B_{1} \ldots B_{s-1}}$ has proper dimension $(length)^{-2}$.

As $(D_0)^2 = R_0=0$, the linearized curvature $F_1$ is invariant
under Abelian gauge transformations of the form \bea \d A^{A_1
\ldots A_{s-1},\, B_1 \ldots B_{s-1}}&=&D_0 \epsilon^{A_1 \ldots
A_{s-1},\, B_1 \ldots B_{s-1}} \,  .\label{lingtransfo} \eea The
gauge parameter $\epsilon^{A_1 \ldots A_{s-1},\, B_1 \ldots
B_{s-1}}$ has the symmetry
\begin{picture}(60,15)(-5,2)
\multiframe(0,7.5)(13.5,0){1}(50,7){}
\multiframe(0,0)(13.5,0){1}(50,7){}
\end{picture} and is traceless.
\vspace{.2cm}

Before writing the action, let us analyze the frame field and its
gauge transformations, in the standard gauge. According to the usual
multiplication rule for $\mso(n)$-irreducible Young diagrams
\cite{Bekaert:2005vh, mishasalg, Bekaert:2006py}, the frame field
$A_{\m}^{~a_1 \ldots a_{s-1}}$ contains three irreducible
(traceless) Lorentz components characterized by the symmetry of
their indices:
\begin{picture}(40,12)(0,2)
\multiframe(0,0)(13.5,0){1}(35,7){}\put(38,1){\tiny{$ s$}}
\end{picture}\
\,,
\begin{picture}(55,15)(-5,2)
\multiframe(0,0)(13.5,0){1}(7,7){}\put(9,1){{\tiny $ 1$}}
\multiframe(0,7.5)(13.5,0){1}(35,7){}\put(38,9){\tiny{$ s-1$}}
\end{picture}
and
\begin{picture}(40,12)(0,2)
\multiframe(0,0)(13.5,0){1}(30,7){}\put(33,1){\tiny{$ s-2$}}
\end{picture}\,\,\,\,\,\,, where the last tableau describes the trace
component of the frame field $A_{\m}^{~a_1 \ldots a_{s-1}}$. Its
gauge transformations are given by (\ref{lingtransfo}) and read
$$\d A^{a_1 \ldots a_{s-1}}= D_0^L \epsilon^{a_1 \ldots a_{s-1}} -
e_{0\;c}\epsilon^{a_1 \ldots a_{s-1},\,c} \,.$$ The parameter
$\epsilon^{a_1 \ldots a_{s-1},\,c} $ is a generalized local Lorentz
parameter. It allows us to gauge away the traceless component
\begin{picture}(40,15)(0,2)
\multiframe(0,0)(13.5,0){1}(7,7){}
\multiframe(0,7.5)(13.5,0){1}(35,7){}
\end{picture}
of the frame field. The other two components of the latter just
correspond to a completely symmetric doubly-traceless Fronsdal field
$\varphi_{\mu_1\ldots \mu_s}$. The remaining invariance is then the
Fronsdal gauge invariance (\ref{Fronsdalg}) with a traceless
completely symmetric parameter $\epsilon^{a_1 \ldots a_{s-1}}$.

\vspace{0.5cm}

- \textit{Action for HS gauge fields}

\vspace{0.2cm}

For a given spin $s$, the most general $\mso(D-1,2)$-invariant
action that is quadratic in the linearized curvatures (\ref{R1})
and, for the rest, built only from the compensator $V^C$ and the
background frame field $E^B_0 = D_0 V^B$
   is \be S^{(s)}_2[\,A_\mu^{A_1 \ldots
A_{s-1},\, B_1 \ldots
B_{s-1}}\,,\o_0^{AB}\,,V^C\,]\,=\,\frac{1}{2}\,
\sum_{p=0}^{s-2}\,a(s,p)\, S^{(s,p)}[\,A_\mu^{A_1 \ldots A_{s-1},\,
B_1 \ldots B_{s-1}}\,,\o_0^{AB}\,,\,V^C\,]\label{HSaction} \ee where
$a (s,p)$ is the {\it a priori} arbitrary coefficient of the term
\bea S^{(s,p)}[\,\o,V\,]&=&\epsilon_{A_1 \ldots A_{D+1}}\int_{M^D}
E_0^{A_5}\ldots E_0^{A_{D}}V^{A_{D+1}} V_{C_1}\ldots
V_{C_{2(s-2-p)}}\times
\nonumber\\
& &  \quad\quad\quad\times F_1^{A_1 B_1 \ldots B_{s-2},\, A_2 C_1
\ldots C_{s-2-p} D_1\ldots D_p} F_1^{~A_3}{}_{B_1 \ldots
B_{s-2},}{}^{ A_4 C_{s-1-p} \ldots C_{2(s-2-p)}}{}_{D_1\ldots  D_p
}\,. \nonumber \eea This action is manifestly invariant under
diffeomorphisms, local $\mso(D-1,2)$ transformations (\ref{localo})
and Abelian HS gauge transformations (\ref{lingtransfo}) that leave
invariant the linearized HS curvatures (\ref{R1}). Having fixed the
$AdS_D$ background gravitational field $\o_0^{AB}$ and the
compensator $V^A$, diffeomorphisms and local $\mso(D-1,2)$
transformations break down to the $AdS_D$ global symmetry
$\mso(D-1,2)$.

The connections $A_{\m}^{~a_1 \ldots a_{s-1}, \, b_1 \ldots b_t}$
can be expressed via $t$ derivatives of the frame-like field,
according to HS analogues of the torsion constraint. Therefore the
coefficients $a (s,p) $ must be chosen in such a way that the
Euler-Lagrange derivatives are non-vanishing only for the frame
field and the first connection ($t=1$). All other fields, {\it i.e.}
the connections $A_{\m}^{~a_1 \ldots a_{s-1}, \, b_1 \ldots b_t}$
with $t>1$, appear only through total derivatives. They are called
extra fields\footnote{The extra fields show up in the non-linear
theory and are responsible for the higher-derivatives as well as for
the terms with negative powers of $\Lambda$ in the interaction
vertices.}. This requirement guarantees that higher-derivative terms
are absent in the free theory and fixes uniquely the spin-$s$ free
action up to an overall coefficient $b(s)$. More precisely, the
coefficient $a(s,p)$ is essentially a relative coefficient given by
\cite{5d}
$$a (s,p) = b (s) (-\Lambda)^{-(s-p-1)}
\frac{(D-5 +2 (s-p-2))!!\, (s-p-1)}{  \,(s-p-2)!} $$ where $b(s)$ is
the arbitrary spin-dependent factor.

The equations of motion  for $A_{\mu}^{~a_1 \ldots a_{s-1}, \, b}$
are equivalent to the ``zero-torsion condition'' $$ F_{1\,A_1 \ldots
A_{s-1} ,\,B_1\ldots B_{s-1}} V^{B_1} \ldots V^{B_{s-1}}=0\,.$$ They
imply that $A_{\m}^{~a_1 \ldots a_{s-1}, \, b}$ is an auxiliary
field that can be expressed in terms of the first derivative of the
frame field. Substituting the found expression for $A_{\m}^{~a_1
\ldots a_{s-1}, \, b}$ into the HS action yields an action only
expressed in terms of the frame field and its first derivative,
modulo total derivatives. As gauge symmetries told us, the action
actually depends only on the completely symmetric part of the frame
field, {\it i.e.} the Fronsdal field. Moreover, the action
(\ref{HSaction}) has the same gauge invariance as Fronsdal's one,
and hence it must be proportional to the Fronsdal action
(\ref{Fronsdalact}) because the latter is fixed up to an overall
factor by the requirements of being gauge invariant and of second
order in derivatives of the field \cite{curt}.\vspace{.2cm}

\chapter{HS Algebras and Representation Theory}\label{absalg}

In the previous section, the dynamics of free spin-$s$ gauge
fields has been expressed as a theory of one-forms, whose
$\mso(D-1,2)$ fiber indices have symmetries characterized by
two-row rectangular Young tableaux. This suggests that there
exists a non-Abelian HS algebra $h\supset \mso(D-1,2)$ that admits
a basis formed by a set of elements $\widehat T_{A_1\ldots A_{s-1}
,B_1\ldots B_{s-1}}$ in irreducible representations of
$\mso(D-1,2)$ characterized by such Young tableaux. More
precisely, the basis elements $\widehat T_{A_1\ldots A_{s-1}
,B_1\ldots B_{s-1}}$ satisfy the following properties $\widehat
T_{\{A_1 \ldots A_{s-1},A_s\} B_2\ldots B_{s-1} } =0$, $\widehat
T_{A_1 \ldots A_{s-3}C}{}_{C,}{}^{B_1\ldots B_{s-1} } =0$, and the
basis contains the $\mso(D-1,2)$ basis elements $\widehat
T_{A,B}=-\widehat T_{B,A}$ such that all generators transform as
$\mso(D-1,2)$ tensors \be [\widehat T_{C,D}\, ,\,\widehat T_{A_1
\ldots A_{s-1},B_1\ldots B_{s-1} }] = i\eta_{DA_1 } \widehat T_{C
A_2 \ldots A_{s-1}, B_1\ldots B_{s-1} }
+\ldots\,.\label{tensortransf} \ee

The question is whether a non-Abelian algebra $h$ with these
properties really exists. If it does, the Abelian curvatures $F_1$
can be understood as resulting from the linearization of the
non-Abelian field curvatures $F=dA+A^2$ of $h$ with the $h$ gauge
connection $A = \omega_0 +A^{(lin)}$, where $\omega_0$ is some
fixed flat ({\it i.e.} vanishing curvature) zero-order connection
of the subalgebra $\mso(D-1,2)\subset h$ and $A^{(lin)}$ is the
first-order dynamical part which describes massless fields of
various spins\footnote{ Notice that now we are extending the
notation previously used for HS one-form connections to the
fluctuational part of fields of all spins $s \geq 1$, maintaining
the notation $\o_0$ only for the background gravitational field
that is a solution of $R_0=d\o_0+\o_0^2=0$.}.

Summarizing, in a more general language: assuming that there
exists full equations with local HS symmetry $h$, and that such
equations admit some vacuum solution that breaks the local $h$
symmetry to a global one ($\mso(D-1,2)$ in the $AdS_D$ case); a
perturbative expansion around such vacuum yields the linearized
field equations seen in the previous Chapter, with massless HS
fields that possess local abelian gauge symmetry parameters in
$h$. In such a scheme, a candidate non-abelian HS algebra should
satisfy the following requirements:

\begin{itemize}

\item In order to be able to interpret the model in terms of
relativistic fields carrying some mass and spin, the vacuum
solutions has to be invariant under some space-time isometry
algebra (like $\mso(D-1,2)$) $g\subset h$.

\item $h$ must admit massless unitary representations that contain
all the gauge fields in the model and, possibly, some lower spin
fields with no associated gauge symmetries \emph{admissibility}
criterion).

\end{itemize}

The HS algebras with the above mentioned properties were originally
found for the case of $AdS_4$
\cite{FVA,V3,Fradkin:ah,Konstein:1988yg} in terms of spinor
algebras. Then this construction was extended to HS algebras in
$AdS_3$  \cite{bl,BBS,Aq} and to $4D$ conformal HS algebras
\cite{FLA,d4sym} equivalent to the $AdS_5$ algebras of \cite{SS5}.
$D=7$ HS algebras \cite{Sezgin/Sundell-7} were also built in
spinorial terms. Conformal HS conserved currents in any dimension,
generating HS symmetries with the parameters carrying
representations of the conformal algebra $\mso(D,2)$ described by
various rectangular two-row Young tableaux, were found in
\cite{KVZ}. The realization of the conformal HS algebra $h$ in any
dimension in terms of a quotient of the universal enveloping algebra
was given by Eastwood in \cite{Eastwood}.

Here we first illustrate the construction of an ``abstract'' HS
algebra, starting from the associative enveloping algebra of
$\mso(D-1,2)$, and factoring out an appropriate ideal. Later on,
in view of the presentation of the full Vasiliev equations, we
shall review the oscillator realization first given in
\cite{Vasiliev:2003ev} (see also \cite{Bekaert:2005vh}), which is
based on vector oscillator algebra (\emph{i.e.}, Weyl algebra).
Finally, we shall present the original four-dimensional spinor
oscillator realization of the four-dimensional algebra, which is
the simplest of all in that the above-mentioned ideal is
automatically factored out.

\scs{Complex HS Algebra}\label{absalg1}

As one of the requirements for a HS algebra in $D$ dimensions is
that of being an infinite-dimensional extension of the isometry
algebra of a maximally symmetric space-time (see Section
\ref{maxsym}), our starting point is the latter and its defining
commutation relations. To keep the discussion completely general
and valid for any signature, we will mostly work, unless
explicitly stated, at the level of \emph{complex} Lie algebras,
and only later specialize to the different real forms
(\emph{i.e.}, to the different signatures, or to the different
maximally symmetric backgrounds mentioned in Chapter 2) imposing
reality conditions on generators.

The complex Lie algebra $\mso(D+1;\Comp)$ has generators $M_{AB}$
obeying
\bea [M_{AB},M_{CD}]&=& 4i\eta_{[C|[B}M_{A]|D]}\ ,\label{soD+1h}\eea
where $A=(a,0')$,$a=(0,r)$, $r=1,\dots,D-1$ and
\bea \eta_{AB}&=& {\rm diag}(\eta_{ab}, -1)\ ,\qquad \eta_{ab}\ =\
{\rm diag}(-1,\d_{rs})\ .\label{signature}\eea
Although we work at the complex level, as outlined in Section
\ref{UIRs} the above choice of signature is convenient for
describing Harish-Chandra modules, that will be of relevance in
the following, and for examining the unitarity properties of the
representations for different real forms of the algebra.

The relevant infinite-dimensional extension of $\mso(D+1;\Comp)$
we are looking for is based on its universal enveloping algebra.
The universal enveloping algebra ${\cal U}$ of $\mso(D+1;\Comp)$
is the associative algebra with product $\star$ generated by the
unity $\1$ and monomials in $M_{AB}$ modulo the commutation rule
\eq{soD+1h}. As a consequence, a basis for ${\cal U}$ is given by
the unity and symmetrized products of $M_{AB}$. As we shall deal
all the time with this specific ordering prescription, it is
useful to adopt a convenient notation: we will henceforth denote
with $M_{AB}$ commuting variables that are symbols of the
corresponding operators, and implement the operator product on
them through a $\star$-product law. The latter is defined in such
a way that the $\star$-product of two symbols of operators is the
symbol of the product of the two operators. Therefore, the
commutation relation \eq{soD+1h} becomes now
\bea [M_{AB},M_{CD}]_\star&=& M_{AB}\star M_{CD}-M_{CD}\star M_{AB}\
=\ 4i\eta_{[C|[B}M_{A]|D]}\ ,\label{soD+1}\eea
and the totally symmetrized products of operators can simply be
denoted by juxtaposition of the commuting variables $M_{AB}$. With
this convention, the definition of ${\cal U}$ is
\bea {\cal U}&=&\bigoplus_{n=0}^\infty {\cal U}_n\ ,\qquad {\cal
U}_n\ =\ \left\{X_n\ =\ x^{A_1 B_1,\dots,A_n B_n}M_{A_1 B_1}\cdots
M_{A_n
B_n}\right\}\ ,\label{calU}\\[5pt] M_{A_1 B_1}\cdots M_{A_n
B_n}&\equiv&\left\{\ba{ll}{1\over n!}\sum_{\pi\in{\cal S}_n}
M_{A_{\pi(1)}B_{\pi(2)}}\star\cdots \star
M_{A_{\pi(n)}B_{\pi(n)}}&\mbox{for $n=1,2,...$.}\ ,\\[5pt] \1 &\mbox{for $n=0$}\ea
\right.\eea
where $x^{A_1 B_1,\dots,A_n B_n}$ are complex coefficients, and we
note that there is no separate symmetry on $A$ and $B$ indices.
The $\star$-product $X_m\star X_n$, which is computed by repeated
symmetrization using the commutation rule \eq{soD+1}, yields the
``classical'' product $X_m X_n$ together with terms of lower
order, since each commutation removes one generator. For example,
\bea M_{AB}\star M_{CD} \ = \
\frac{1}{2}\left\{M_{AB},M_{CD}\right\}_\star+\frac{1}{2}\left[M_{AB},M_{CD}\right]_\star
\ = \ M_{AB}M_{CD}+2i\eta_{[C|[B}M_{A]|D]}\ .\eea
The map
\bea \tau(X_n)&=&(-1)^n X_n\ ,\label{taumap}\eea
that is, $\tau(X(M_{AB}))=X(-M_{AB})$ where $X(M_{AB})$ is a
symmetrized function, is an (involutive) anti-automorphism of the
$\star$-product, \emph{i.e.}
\bea \tau(X\star Y)&=&\tau(Y)\star \tau(X)\ .\eea

However, the universal enveloping algebra of $\mso(D+1;\Comp)$ does
not satisfy our requirement for being a candidate HS algebra, since
the generators \eq{calU} do not really match the conditions we had
fixed at the beginning of this chapter. In general, indeed, a
symmetrized monomial of degree $n\geq 2$ is reducible under
$\mso(D+1;\Comp)$, as it contains both trace parts and irreps
labeled by Young diagrams with more than two rows. However, both of
them can be absorbed into the ideal
\bea {\cal I}[V]&=& \left\{ X\ =\ V\star X'\ \mbox{for $X'\in{\cal
U}$}\right\}\ ,\label{idealV}\eea
where $V=\l^{AB}V_{AB}+\l^{ABCD}V_{ABCD}$ with
$\l^{AB},\l^{ABCD}\in\Comp$, and
\bea V_{AB}&\equiv & \ft12 M_{(A}{}^C M_{B)C}-{1\over
2(D+1)}\eta_{AB}
M^{CD}\star M_{CD}\ ,\label{VAB}\\[5pt]
V_{ABCD}&\equiv & M_{[AB} M_{CD]}\ .\label{VABCD}\eea
The generator $V_{AB}$ absorbs the traces, while $V_{ABCD}$
absorbs the Young diagrams with with more than two rows (as it is
clear from \eq{3rows}) - \emph{i.e.}, the unwanted elements are
solved by such constraints in terms of allowed ones. So far, we
have a chain of proper ideals, namely ${\cal U}\supset {\cal
U}'\supset {\cal I}[V]$, where ${\cal U}'={\cal U}\setminus\1$.
Factoring out ${\cal I}[V]$ induces the infinite-dimensional
unital associative quotient algebra
\bea {\cal A}&\equiv & {{\cal U}\over {\cal I}[V]}\
,\label{calA1}\eea
and we shall use the notation
\bea X\simeq X'\qquad &\Leftrightarrow& \qquad X-X'\in {\cal
I}[V]\ .\eea
The constraints $V_{AB}\simeq 0$ and $V_{ABCD}\simeq 0$
\emph{together} fix the values of the Casimir operators\footnote{For
$D=5$, also the cubic Casimir operator plays a role.}
%
\bea C_{2n}[\mso(D+1;\Comp)]&\equiv &\ft12 M_{A_1}{}^{A_2}\star
M_{A_2}{}^{A_3}\star\cdots\star M_{A_{2n}}{}^{A_1}\ ,\qquad
n=1,2,\dots\ .\label{C2n}\eea
In what follows we shall denote the restriction of the Casimir
operators to a representation $\mR$ of $\mso(D+1;\Comp)$ by
$C_{2n}[\mso(D+1;\Comp)|\mR]$, or simply $C_{2n}[\mR]$ in case
there is no risk of confusion. To begin with, the higher-order
operators $C_{2n}$ with $n>1$ can be rewritten in terms of $C_2$
using $V_{AB}\simeq0$, which implies
\bea M_A{}^B\star M_{BC}&=&{i(D-1)\over 2}M_{AC}-{2\over
D+1}\eta_{AC}C_2\ ,\label{MABMBC}\eea
so that, for example,
\bea C_4[{\cal A}]&=&{2\over D+1}C_2[{\cal A}]^2+{(D-1)^2\over
4}C_2[{\cal A}]\ .\eea
The operator $C_2[{\cal A}]$ (and hence all $C_{2n}[{\cal A}]$) is
fixed by compatibility between $V_{AB}\simeq0$ and
$V_{ABCD}\simeq0$, which requires
\bea 0&\simeq& M_A{}^B\star V_{BCDE}\ \simeq\
\left(\mu^2-\e_0\right)\star\eta_{A[C}M_{DE]}\ ,\qquad \mu^2\
\equiv\ -{2C_2[{\cal A}]\over D+1}\ ,\label{mu2}\eea
where we have used \eq{MABMBC} and
\bea V_{ABCD}&=&M_{[AB}\star M_{CD]}\ =\ M_{[AB}\star
M_{C]D}-i\eta_{D[A}M_{BC]}\ ,\label{3rows}\eea
and we have also introduced the parameter
\bea \e_0&=&{D-3\over 2}\ .\label{epsilon0}\eea
Thus, one finds, for example,
\bea C_2[\mso(D+1;\Comp)|{\cal A}]&=& -\e_0(\e_0+2)\ ,\label{C2e0}\\[5pt] C_4[\mso(D+1;\Comp)|{\cal A}]&=&
-\e_0(\e_0+2)(\e_0^2+\e_0+1)\ ,\label{C4e0}\eea
and one can calculate higher-order Casimir operators as well, by
using \eq{MABMBC} recursively.

We can choose a canonical leveled basis for ${\cal A}$ as follows:
\bea {\cal A}&=& \bigoplus_{n=0}^\infty {\cal A}_n\ ,\qquad {\cal
A}_n\ \simeq \ \left\{X_n\ =\ x^{A(n),B(n)}\widehat
T_{A(n),B(n)}\right\}\ ,\label{calA}\eea
where $x^{A(n),B(n)}$ are traceless type $(n,n)$
tensors\footnote{Throughout this Thesis, tensors with the symmetry
of the Young diagram of height $\nu$ with $n_i$ cells in the $i$th
row ($i=1,\dots,\nu$) are referred to as type $(n_1,\dots,n_\nu)$
tensors (where thus $n_1\geq n_2\geq\cdots \geq n_\nu$ are positive
integers). We work with \emph{normalized and mostly symmetric Young
projections}
$$\mathbf P_{n_1,n_2,...,n_\nu}={1\over
\prod_{\mbox{cells}}(\mbox{hook-lengths})} \prod_{\mbox{rows $i$}}
\mathbf S_i \prod_{\mbox{columns $j$}}\mathbf A_j\ ,$$ where
$\mathbf S_i$ and $\mathbf A_j$ are symmetrizers and
anti-symmetrizers, respectively, acting on the indices of the
$i$th row ($i=1,\dots,\nu$) and $j$th column ($j=1,\dots,n_1$).
Thus, a type $(n_1,\dots,n_\nu)$ tensor
$T_{A^1(n_1),\dots,A^{\nu}(n_\nu)}$ has $\nu$ groups of
symmetrized indices $A^i(n_i)=A^i_1\dots A^1_{n_i}$, subject to
the over-symmetrization rule
$$T_{\cdots,(A^i_1\dots A^i_{n_i},A^{i+1}_1)A^{i+1}_2\dots A^{i+1}_{n_{i+1}},\cdots}\ =\ 0\ ,\qquad
i=1,\dots,\nu-1\ .$$ When $A=1,\dots,N$, such tensors are
$\msl(N;\Comp)$ irreps, and $\mso(N;\Comp)$ irreps when they are
traceless, of highest weight $(n_1,\dots,n_\nu)$.}, and
\bea &&\widehat T_{A(n),B(n)}\ =\ M_{\{ A_1B_1} \cdots
M_{A_nB_n\}}\ =\ M_{\{ A_1B_1}\star \cdots\star M_{A_nB_n\}}\
,\label{TAnBn}\eea
where we use the convention that curly brackets, $\{\cdots\}$,
enclose $\mso(D+1;\Comp)$ irreducible, \emph{i.e.} traceless and
Young-projected, groups of indices. More explicitly,
\bea \widehat T_{A(n),B(n)}&=&M_{\langle A_1B_1}\star \cdots\star
M_{A_nB_n\rangle }\nn\\[5pt]&&+\sum_{k=1}^{[n/2]}\k_{n;k} \eta_{\langle
A_1A_2}\eta_{B_1B_2}\cdots
\eta_{A_{2k-1}A_{2k}}\eta_{B_{2k-1}B_{2k}}M_{
A_{2k+1}B_{2k+1}}\star \cdots\star M_{A_{n}B_{n}\rangle}\
,\qquad\qquad\label{TAnBntraceparts}\eea
where we use the convention that hooked brackets,
$\langle\cdots\rangle$, enclose $\msl(D+1;\Comp)$ irreducible,
\emph{i.e.} Young projected, groups of indices, and the
coefficients $\k_{n;k}$ are fixed by
\bea \eta^{CD} \widehat T_{A(n),B(n-2)CD}&=&0 \
.\label{TAnBntraceless}\eea
We note that $V_{AB}\simeq 0$ implies that the trace parts in
\eq{TAnBntraceparts} only involve lower-order enveloping-algebra
monomials in rectangular Young projections. From \eq{TAnBn} and
onwards, we shall always use the convention that repeated indices
that are denoted by a single letter and distinguished by
subindices are always symmetrized, so that\footnote{We note that
prior to using this convention, the type $(n,n)$ Young projection
of $M_{A_1B_1}\cdots M_{A_nB_n}$ (no symmetry on $A$ and $B$
indices!) equals $\ft{2^n}{n+1} M_{A_1B_1}\cdots M_{A_nB_n}$
(symmetry on the $A$ and $B$ indices!).}
\bea M_{A_1}{}^{B_1} \cdots M_{A_n}{}^{B_n }&\equiv &
M_{(A_1}{}^{(B_1}\cdots M_{A_n)}{}^{B_n)}\ =\
M_{(A_1}{}^{(B_1}\star \cdots\star M_{A_n)}{}^{B_n)}\ ,\eea
that is,
\bea M_{A_1B_1} \cdots M_{A_nB_n }&= & M_{\langle A_1B_1}\cdots
M_{A_nB_n\rangle }\ =\ M_{\langle A_1B_1}\star \cdots\star
M_{A_nB_n\rangle }\ .\eea
For example, the simplest case is given by
\bea \widehat T_{A(2),B(2)}&=& M_{A_1 B_1}\star M_{A_2
B_2}-{2\over D(D+1)}(\eta_{A_1 A_2}\eta_{B_1 B_2}-\eta_{A_1
B_1}\eta_{A_2 B_2}) C_2[{\cal A}]\ .\eea
The $\mso(D+1;\Comp)$-transformations take the form
%
\bea \Ad_{M_{AB}}(T_{C(n),D(n)})&=& 2in
\eta_{[B|\{C_1}T_{|A]|C(n-1),D(n)\}}+2in\eta_{[B|\{D_1}T_{C(n),|A]|D(n-1)\}}\
,\label{OD+1transf}\eea
where the adjoint action of ${\cal A}$ on itself is defined by
\bea \Ad_X(Y)&=& [X,Y]_\star\ .\label{adj} \eea
Therefore, the generators $\widehat T_{A(n),B(n)}$ of $\cA$  indeed
have the correct index structure and transformation properties to be
candidate generators of the HS extension of $\mso(D+1;\Comp)$ we
were looking for. Let us now look at some other properties that will
be important for the following. We can also define the
anti-commutator action
\bea \Ac_X(Y)&=&\{X,Y\}_\star\ ,\eea
with the closure
\bea [\Ac_X,\Ac_Y]&=& \Ad_{[X,Y]_\star}\ .\label{Acclosure}\eea
From $V\simeq 0$ it follows that
\bea \Ac_{M_{AB}}(\widehat T_{C(n),D(n)}) &=& 2\D_n\widehat
T_{[A|\{C(n),|B]D(n)\}}+ 2\l_n
\eta_{[A\{C_1}\eta_{|B]\{D_1}\widehat T_{C(n-1),D(n-1)\}}\
,\label{AcMAB}\eea
with (suppressing the anti-symmetry on $AB$)
\bea &&\eta_{[A\{C_1}\eta_{|B]\{D_1}\widehat T_{C(n-1),D(n-1)\}} \
=\
\eta_{AC_1}\eta_{BD_1}\widehat T_{C(n-1),D(n-1)}\nn\\[5pt]&&+\beta_n\left(\eta_{C_1 C_2}\eta_{BD_1}\widehat
T_{AC(n-2),D(n-1)}+\eta_{AC_1}\eta_{D_1D_2}\widehat
T_{C(n-1),BD(n-2)}\right.\nn\\[5pt]&&+\left.\eta_{C_1 D_1}\eta_{C_1A}\widehat
T_{BC(n-2),D(n-1)}+\eta_{C_1D_1}\eta_{D_1B}\widehat
T_{C(n-1),AD(n-2)}\right)\nn\\[5pt]
&&+\a_n\left(\eta_{C_1 C_2}\eta_{D_1 D_2}\widehat
T_{AC(n-2),BD(n-2)}-\eta_{C_1D_1}\eta_{C_2D_2}\widehat
T_{AC(n-2),BD(n-2)}\right)\ ,\eea
where the coefficients 
\bea \D_n&=&2{n+1\over n+2}\ ,\nn\\[5pt] \a_n&=& {1\over
4}{(n-1)^2\over(n+\e_0-1)(n+\e_0-\ft12)}\ ,\qquad \b_n\ =\
-{1\over 2}{n-1\over n+\e_0-1} \label{Deltan}\ ,\eea
are fixed by traceless type $(n+1,n+1)$ Young projection (so that
$\mathbf P(M_{AB}\widehat T_{C(n),D(n)})=\D_n\mathbf P\widehat
T_{[A|\{C(n),|B]D(n)}=\D_n T_{AC(n),BD(n)}$ where $\mathbf P\equiv
\mathbf{P}_{AC(n),BD(n)}$), while the coefficient
\bea \l_n&=&-{1\over 2}{n(n+1)(n+\e_0-1)\over n+\e_0+\ft12}\
,\label{lambdan}\eea
can be computed either by solving the trace conditions on
\eq{AcMAB}, or by demanding closure under \eq{Acclosure} and
\eq{OD+1transf}, that is
\bea [\Ac_{M_{AB}},\Ac_{M_{CD}}](\widehat T_{E(n),F(n)})&=&
4i\eta_{BC}\Ad_{M_{AD}}\widehat T_{E(n),F(n)}\
,\label{MMclosure}\eea
(where the separate anti-symmetry on $AB$ and $CD$ has been
suppressed). In order to apply the first method, one first
substitutes the $\widehat T$-elements on the left-hand and
right-hand sides of \eq{AcMAB} by their trace expansions
\eq{TAnBntraceparts} up to ${\cal O}(\eta^2)$ and ${\cal
O}(\eta)$, respectively. One then contracts the equation by
$\eta^{BD_1}$ using $V\simeq0$ and \eq{TAnBntraceless},
respectively, to simplify the left-hand and right-hand sides. The
second method, on the other hand, relies entirely on the
$\mso(D+1;\Comp)$-covariance of the whole procedure of factoring
out ${\cal I}[V]$, and does not require any further use of
$V\simeq 0$. Instead, equation \eq{MMclosure} yields a recursive
relation between $\l_n$ and $\l_{n-1}$ that can be solved given
the initial datum $\l_0=0$. There is also a third method of
computing $\l_n$, namely to reduce \eq{AcMAB} under
$\mso(D+1;\Comp)\rightarrow \mso(D;\Comp)$, as we shall discuss in
the next Section.

As can be seen already by comparing \eq{C2e0} and \eq{C4e0} to
\eq{C2lhws} and \eq{C4lhws}, the values of the Casimir operators
in ${\cal A}$ are equal those assumed in the scalar-singleton
lowest-weight space $\mD_0\equiv \mD(\e_0;(0))$ described briefly
in Appendix \ref{App:Cas}, and that we shall look at more closely
later in this Chapter, \emph{i.e.}
\bea C_{2n}[\mso(D+1;\Comp)|{\cal A}]&=&
C_{2n}[\mso(D+1;\Comp)|\e_0;(0)]\ .\eea
In fact, the ideal ${\cal I}[V]$ is isomorphic to the
scalar-singleton annihilator in ${\cal U}$, \emph{i.e.}
\bea {\cal I}[V]\ \simeq\ {\cal I}[\mD_0]\ ,\label{annI}\eea
where ${\cal I}[\mD]$, for given lowest-weight space $\mD$, is the
ideal consisting of all elements in ${\cal U}$ that annihilate
\emph{all} states in $\mD$. To show this, one first derives the
lemma\footnote{The scalar singletons $\mD^\pm(\pm\e_0;(0))$ and
the 4D spinor singleton $\mD^\pm(\pm(\e_0+1/2);(1/2))$ are
annihilated by the ideal ${\cal I}[V]$, as we shall see in the
next Section.} that if $X$ belongs to a tensorial
$\mso(D+1;\Comp)$ irrep, then $X\in {\cal I}[\mD_0]$ iff
$X\ket{\e_0,(0)}=0$. Next one verifies that $V\ket{\e_0,(0)}=0$.

As found in \cite{Angelopoulos:1997ij, Angelopoulos:1999bz} (see
also \cite{Engquist:2007pr} for a more recent application in the
context of affine extensions of $\mso(D+1;\Comp)$), at the level
of lowest-weight spaces, the $V_{AB}$ constraint is by itself
sufficient to uniquely select the scalar singleton (and also the
spinor singleton in $D=4$), and the $V_{ABCD}$ constraint then
follows automatically. In the associative algebra ${\cal A}$, on
the other hand, which does not refer explicitly to lowest-weight
spaces, the values of $C_{2n}$ are instead fixed (to be those of
the singleton representation) by combining the $V_{AB}$ \emph{and}
$V_{ABCD}$ constraints. The enveloping-algebra construction thus
rests on a weaker set of assumptions than the lowest-weight
construction, and hence ${\cal A}$ has potentially an
algebraically richer structure than the space of operators on
$\mD_0$, as we shall explore in more detail in Chapter \ref{map}.


\scss{$\mso(D;\Comp)$-Covariant Form of the Quotient
Algebra}\label{Sec:OD}


Next, we turn to a $\mso(D;\Comp)$-covariant description of the
quotient algebra ${\cal A}$. We begin by splitting $M_{AB}$ into
$\mso(D;\Comp)$ generators $M_{ab}$ and translations
\bea P_a&=&M_{0'a}\ ,\label{defPa}\eea
obeying
\bea [M_{ab},M_{cd}]_\star&=&4i\eta_{[c|[b}M_{a]|d]}\ ,\qquad
[M_{ab},P_c]_\star\ =\ 2i\eta_{c[b}P_{a]}\ ,\label{}\\[5pt]
[P_a,P_b]_\star&=&iM_{ab}\eea
(comparing with the general notation of (\ref{genAdSlor}) we are now
choosing $L=1$ and $\l=1$). By definition, the translations are odd
under the automorphism $\pi$ of ${\cal A}$ (and ${\cal U}$),
\emph{viz.}
\bea \pi(P_a)\ =\ -P_a\ ,\qquad \pi(M_{ab})\ =\ M_{ab}\ ,\qquad
\pi(X\star Y)\ =\ \pi(X)\star \pi(Y)\ .\label{pimap}\eea
The constraints $V_{AB}\simeq0$ and $V_{ABCD}\simeq0$ then
decompose into
\bea V_{0'0'}&=&\ft12(P^a\star P_a -\mu^2)\ \simeq 0\ ,\label{v00}\\[5pt]
V_{0'a}&=& \ft14\{M_a{}^b,P_b\}_\star\ \simeq\ 0\ ,\label{v0a}\\[5pt]
V_{ab}&=& \ft12 (M_{(a}{}^c\star M_{b)c}-P_{(a}\star
P_{b)}+\mu^2\eta_{ab})\ \simeq 0\ ,\label{vab}\\[10pt]
V_{abcd}&=&M_{[ab}\star M_{cd]}\simeq 0\ ,\qquad V_{0'abc}\ =\
-P_{[a}\star M_{bc]}\ =\ 0\ ,\label{v0abc}\eea
where $\mu^2$ is defined in \eq{mu2}. As shown in Appendix
\ref{App:VAB}, the constraints \eq{v0a} and \eq{vab} follow from
\eq{v00}, and \eq{v0abc} is equivalent to $P_{[a}\star P_b\star
P_{c]}\simeq 0$. The value of $\mu^2$ is determined from
\bea P^a\star P_{[a}\star P_b\star P_{c]}&\simeq&
\ft{i}6(\mu^2-\e_0) M_{bc}\ .\label{detmu2}\eea
Thus, the ideal ${\cal I}[V]$ can be given the Lorentz covariant
presentation
\bea P^a\star P_a&\simeq &\e_0\ ,\qquad P_{[a}\star P_b\star
P_{c]}\ \simeq\ 0\ .\label{papa}\eea
and we note the auxiliary trace constraints:
\bea P^a\star M_{ab}&\simeq& M_{ba}\star P^a\simeq i(\e_0+1)P_b\
,\label{pamab}\\[5pt] M_{(a}{}^c\star M_{b)c}&\simeq & -P_{(a}\star
P_{b)}+\e_0\eta_{ab}\ .\label{macmcb}\eea
Correspondingly, the $\mso(D+1;\Comp)$-covariant expansion of the
quotient algebra ${\cal A}$ given in \eq{calA} reduces to the
following $\mso(D;\Comp)$-covariant expansion
\bea X&=&\sum_{n\geq m\geq 0} X^{a(n),b(m)} T_{a(n),b(m)}\
,\label{Xod}\eea
where $X^{a(n),b(m)}$ are traceless type $(n,m)$ tensors, and
\bea T_{a(n),b(m)}&\equiv & \widehat T_{\{ a(n),b(m)\} 0'(n-m)}\
=\ M_{\{ a_1b_1}\cdots M_{a_mb_m}P_{a_{m+1}}\cdots P_{a_n\}}\
,\label{Taskbs}\eea
where $\widehat T_{A(n),B(n)}$ are defined in \eq{TAnBn} and the
curly brackets indicate traceless type $(n,m)$ projection. We note
that $T_{a(n),b(m)}$ is a linear recombination of $\widehat
T_{a(n),b(m-2k) 0'(n-m+2k)}$, $k=0,1,\dots, [m/2]$, of the
form\footnote{In what follows, it is important that \eq{Taskbs},
or \eq{Tambn2}, is a strong equality, \emph{i.e.} it holds in
${\cal U}$ without the need to remove terms in the ideal ${\cal
I}[V]$.}
\bea \!\!\!\!\!\!\!\! T_{a(n),b(m)}=\widehat T_{\langle
a(n),b(m)\rangle 0'(n-m)}+\sum_{k=1}^{[m/2]}\k_{n,m;k} \widehat
T_{\langle a(n),b(m-2k) 0'(n-m+2k)}\eta_{ b_1b_2}\cdots
\eta_{b_{2k-1}b_{2k}\rangle}\ ,\label{Tambn2}\eea
where the $\langle\cdots \rangle$ indicate type $(n,m)$ Young
projection, and the coefficients $\k_{n,m;k}$ are fixed by the
requirement that $T_{a(n),b(m)}$ be traceless. For example, as
shown in Appendix \ref{App:VAB}, the simplest case is given by
\bea T_{a(n)}\,=\, P_{\{a_1}\cdots P_{a_n\}}\, \simeq \,
P_{(a_1}\star \cdots\star P_{a_n)}+ \k_{n,0;1}\eta_{(a_1
a_2}P_{a_3}\star\cdots \star P_{a_n)}+{\cal O}(\eta^2)\
,\label{lemmaApp}\eea
with
\bea \k_{n,0;1}&=& -{(n+1)n(n-1)(n+4\e_0-2)\over
48(n+\e_0-\ft12)}\ .\label{kn}\eea
The $O(D;\Comp)$-transformations are given by
\bea
\Ad_{M_{ab}}(T_{c(n),d(m)})&=&2in\eta_{[b|\{c_1}T_{|a]c(n-1),d(m)\}}+2im\eta_{[b|\{d_1}T_{c(n),|a]d(m-1)\}}\
.\eea
In Chapter \ref{map} we shall need the explicit form of the
anti-commutator
\bea \Ac_{P_c} (T_{a(s+k),b(s)})&=& 2\Delta_{s+k,s}T_{c\{
a(s+k),b(s)\}}+ 2\lambda_{k}^{(s)} \eta_{ c \{ a}
T_{a(s+k-1),b(s)\}}\ ,\label{TPa}\eea
with
\bea \eta_{ c \langle a} T_{a(s+k-1),b(s)\rangle}&=& \eta_{ c a}
T_{a(s+k-1),b(s)}+\a_{s+k,s}\eta_{a(2)}T_{a(s+k-2)c,b(s)}+\nn\\[5pt]
&&+\b_{s+k,s}\eta_{a(2)}T_{a(s+k-2)b,cb(s-1)}+\gamma_{s+k,s}
\eta_{ab}T_{a(s+k-1),cb(s-1)}\ ,\label{hooked}\eea
where the coefficients
\bea \D_{s+k,k}&=&{(k+2)(k+s+1)\over (k+1)(k+s+2)}\
,\label{Delta}\\[5pt]
\a_{s+k,s}&=&-{1\over 2}{s+k-1\over s+k+\e_0-\ft12}\
,\label{alpha}\\[5pt]
\b_{s+k,s}&=&{1\over 2}{(s+k-1)s\over
(s+k+\e_0-\ft12)(2s+k+2\e_0-1)}\
,\label{beta}\\[5pt] \c_{s+k,s}&=& -{s\over 2s+k+2\e_0-1}\ ,\label{gamma}\eea
are fixed by the traceless type $(s+k,s)$ Young projection (so
that $\mathbf P(P_{c} T_{a(s+k),b(s)})=\D_{s+k,s}\mathbf PT_{c\{
a(s+k),b(s)\}}=\D_{s+k,s} T_{c a(s+k),b(s)}$ where $\mathbf
P\equiv \mathbf{P}_{ca(s+k),b(s)}$), while the coefficient
\bea \lambda_k^{(s)}&=& {1\over 8}{k(k+s+1)(k+2s+2\e_0-1)\over
k+s+\e_0+\ft12}\ ,\label{lambda}\eea
can be computed in four ways: i) reducing \eq{AcMAB}; ii) solving
the trace conditions on \eq{TPa} (using \eq{papa}, \eq{pamab} and
\eq{macmcb}); iii) demanding closure under
\bea [\Ac_{P_a},\Ac_{P_b}](T_{c(s+k),d(s)})&=&
i\Ad_{M_{ab}}(T_{c(s+k),d(s)})\ ;\eea
or iv) solving the twisted-adjoint Casimir relation (\emph{i.e.}
the mass-formula for Weyl tensors) that we shall discuss in
Section \ref{Sec:adj}. The methods (ii) and (iii) are examined in
detail in the case of $s=0$ in Appendix \ref{App:VAB}.

In Chapter \ref{map} we shall also need the anti-commutator
\bea \Ac_{M_{ab}}(T_{c(n)})&=& 2
T_{c(n)[a,b]}+2\rho_n\eta_{[a|\{c_1}T_{c(n-1)\},|b]}\
,\label{MabT}\eea
with (suppressing the anti-symmetry on $ab$)
\bea \eta_{a\{c_1}T_{c(n-1)\},b}&=&\eta_{ac_1}T_{c(n-1),b}-{1\over
4}{n\over n+\e_0-\ft12} \eta_{c_1 c_2}T_{c(n-2)a,b}\ ,\eea
where the coefficient $\rho_n$ is
\bea \rho_n&=& -{(n-1)n(n+1)\over n+\e_0+\ft12}\ ,\eea
as shown in Appendix \ref{App:VAB}.


\scss{HS Algebras. Adjoint and Twisted-Adjoint Master
Fields}\label{Sec:adj}


We are finally ready to define a HS Lie algebra. The associative
algebra ${\cal A}$ plays a central role in Vasiliev's frame-like
formulation of higher-spin gauge theory: indeed, the vielbein
$e=dx^\mu e_\mu{}^a P_a$ and the Lorentz connection
$\omega=dx^\mu \omega_{\mu}{}^{ab} M_{ab}$ can be encoded, 
together with an infinite tower of higher-spin gauge fields, in a
\emph{master one-form} $A$ taking values in the \emph{adjoint
representation of an infinite-dimensional higher-spin Lie-algebra}
extension of $\mso(D+1;\Comp)$. The minimal extension, that is
unique in the sense that $\mso(D+1;\Comp)$ is its maximal
finite-dimensional Lie subalgebra, is given by
\bea \mho(D+1;\Comp)&=&\{Q\in{\cal A}:\t(X)\ =\ -X\}\
,\label{adj}\eea
where $\t$ is the anti-automorphism defined in \eq{taumap}, and with
Lie bracket induced by the associative $\star$-product, \emph{viz.}
\bea \Ad(Q)(Q')&=& [Q,Q']_\star\ =\ Q\star Q'-Q'\star Q\
.\label{adjrep}\eea
Decomposing $\mho(D+1;\Comp)$ under $\mso(D+1;\Comp)$, leads to an
expansion into \emph{finite-dimensional levels},
\bea \mho(D+1;\Comp)&=&\bigoplus_{\ell=0,1,2,\dots}^{\infty} {\cal
L}_{\ell}\ ,\label{Lell}\eea
where the $\ell$th level is spanned by monomials of degree
$2\ell+1$, \emph{i.e.}
\bea Q_{\ell}&=& Q^{A(2\ell+1),B(2\ell+1)} \widehat
T_{A(2\ell+1),b(2\ell+1)}\ =\ Q^{A(2\ell+1),B(2\ell+1)}
M_{A_1B_1}\star \cdots M_{A_n B_n}\ ,\eea
with $T_{A(2\ell+1),b(2\ell+1)}$ defined in \eq{TAnBn}, and where
$Q^{A(2\ell+1),B(2\ell+1)}$ are traceless type $(2\ell+1,2\ell+1)$
tensors. As expected, the Lie bracket mixes the levels as follows
(see \cite{Eastwood,mishasalg,Sagnotti:2005ns,Bekaert:2005vh} and
also \cite{Engquist:2007kz} for a more recent discussion)
\bea [Q_\ell,Q_{\ell'}]_\star&=&
\sum_{\ell''=|\ell-\ell'|}^{\ell+\ell'}Q_{\ell''}\
.\label{level1}\eea
Although, to the best of our knowledge, the explicit form of the
structure coefficients in the $\widehat
T_{A(2\ell+1),B(2\ell+1)}$-basis has not yet been worked out
explicitly, this formula shows that the HS algebra constructed
above recovers the feature that was long \cite{Berends:1984rq,
Berends:1985xx, Berends:wp} known to characterize HS interactions:
as soon as one massless field with spin $s\geq 3$ enters the game,
one must introduce infinitely many spins (at least all even
integer spins) as required by the closure of the gauge algebra.

The $\ell$th level decomposes further under $\mso(D;\Comp)$ into
traceless type $(s-1,t)$ tensors with $s=2\ell+2$ and $0\leq t\leq
2\ell+1$, so that the minimal adjoint one-form is
\bea A&=&\sum_{s=2,4,6,\dots} A_{(s)}\ ,\label{A}\\[5pt]
A_{(s)}&=& -i\sum_{t=0}^{s-1}dx^\mu A_{\mu,a(s-1),b(t)}(x)M^{a_1
b_1}\cdots M^{a_tb_t}P^{a_{t+1}}\cdots P^{a_{s-1}}\ .\label{As}\eea
where $A_{(2)}$ contains the vielbein and $\mso(D;\Comp)$
connection,
\bea A_{(2)}\ =\ e+\omega\ =\ -i(e^a P_a+\ft12 \omega^{ab}M_{ab})\
.\label{spin2sector}\eea
It is also convenient to collect the higher-spin gauge fields as
\bea W&=&\sum_{s=4,6,\dots} A_{(s)}\ .\label{W}\eea
Notice that, needless to say, for every spin-$s$ sector $A_{(s)}$
contains precisely the fields discussed in Section \ref{frame}. We
will later see that, starting from Vasiliev's equations, and
treating $A_{(2)}$ exactly (assuming $e_\mu{}^a$ to be invertible)
and $W$ perturbatively, one can derive a weak-field expansion in
which all component gauge fields are auxiliary except the metric
\bea g_{\mu\nu}&=& e_\mu{}^a e_{\nu a}\ ,\label{metric}\eea
and the symmetric rank-$s$ tensor gauge fields
\bea\phi_{a(s)}&=&(e^{-1})_{(a}{}^\mu W_{\mu,a(s-1))}\ ,\qquad
s=4,6,\dots\ .\label{tensorgaugefields}\eea

However, the master one-form \eq{A}, with its component field
\eq{As}, cannot be the only ingredient in a fully interacting HS
gauge theory. Indeed, we have not yet made sure that our candidate
HS symmetry algebra satisfies the second requirement, \emph{i.e.},
that it admits a unitary representation that contains all the
gauge fields that we have examined in \ref{frame}. As we will show
in Section \ref{UIRs}, massless UIRs of $\mso(D-1,2)$ and of its
infinite-dimensional HS extensions necessarily include a
\emph{scalar field}, which of course cannot sit in the master
one-form, but needs to be accommodated in a zero-form. Moreover,
as we shall discuss in detail in the next Chapter, the Vasiliev
equations have been written in a certain first-order form (called
\emph{unfolded} formulation) in which a zero-form, transforming in
a peculiar representation of the $\mso(D-1,2)$ algebra and its
infinite-dimensional HS extensions, plays a crucial role.
Therefore, aside from the adjoint master one-form $A$, the
admissibility criterion and the unfolded
formulation\footnote{Interestingly, in a recent paper
\cite{Engquist:2007kz} it has been proposed that full higher-spin
dynamics in even space-time dimensions can be induced by starting
from a Chern-Simons-like theory in one higher dimension based on
an adjoint one-form only.} require a \emph{master zero-form}
$\Phi$ taking values in a \emph{twisted-adjoint representation of
the higher-spin Lie algebra}. The minimal twisted-adjoint
representation is given by
\bea {\cal T}(D+1;\Comp)&=&\{S\in{\cal A}:\t(S)\ =\ \pi(S)\}\
,\label{twadj}\ ,\eea
where $\pi$ is the ${\cal A}$-automorphism defined in \eq{pimap},
and the higher-spin representation is defined by
\bea \widetilde{\Ad}_Q(S)&=&[Q,S]_\pi\ =\ Q\star S-S\star \pi(Q)\
.\label{twadjrep}\eea
The twisted-adjoint representation decomposes under
$\mso(D+1;\Comp)$ into \emph{infinite-dimensional levels}, 
\bea {\cal T}(D+1;\Comp)&=&\bigoplus_{\ell=-1,0,2,\dots} {\cal
T}_{\ell}\ ,\label{Tell}\eea
spanned by $O(D;\Comp)$-covariant elements, \emph{viz.}
\bea S_{\ell}&=& \bigoplus_{k=0}^\infty S^{a(s+k),b(s)}
T_{a(s+k),b(s)}\ ,\qquad s\equiv2\ell+2\ ,\label{levelell}\eea
where $S^{a(s+k),b(s)}\in\Comp$ and $T_{a(s+k),b(s)}$ is given by
\eq{Taskbs}. The twisted-adjoint transformations mixes the levels as
follows:
\bea
\widetilde{\Ad}_{Q_\ell}(S_{\ell'})&=&\sum_{\ell''=\max(-1,\ell'-\ell)}^{\ell+\ell'}S_{\ell''}\
,\label{level2}\eea
where the higher bound on $\ell''$ follows immediately, while the
the lower bound follows from the contraction rules
\eq{v00}-\eq{v0abc}.

The expansion of the minimal twisted-adjoint zero-form reads
\bea \Phi&=&\sum_{s=0,2,4,\dots} \Phi_{(s)}\ ,\label{Phi}\\[5pt]
\Phi_{(s)}&=& \sum_{k=0}^\infty{i^k\over k!}
\Phi^{a(s+k),b(s)}(x)M_{a_1b_1}\cdots M_{a_sb_s}P_{a_{s+1}}\cdots
P_{a_{s+k}}\ .\label{Phis}\eea
As we shall see, in the above-mentioned weak-field expansion of
the Vasiliev equations, the component fields $\Phi_{a(s),b(s)}$
become \emph{generalized spin-$s$ Weyl tensors} for $s=2,4,\dots$,
and a \emph{physical scalar} for $s=0$,
\bea \phi&=&\Phi\mid_{P_a=M_{ab}=0}\ ,\label{physicalscalar}\eea
while $\Phi_{a(s+k),b(s)}$ for $k=1,2,\dots$ become auxiliary
fields, given by the $k$th derivatives of $\Phi_{a(s),b(s)}$
on-shell.


There are many ways to extend the minimal model. In some sense,
the simplest extension is to add all odd spins $s=1,3,5,...$,
\emph{i.e.} half-integer levels $\ell=-1/2,1/2,3/2,\dots$, leading
to the (non-minimal) adjoint and twisted-adjoint
modules
\bea \mho_1(D+1;\Comp)&=& {\cal A}
,\label{adjprime}\\[5pt] {\cal T}_1(D+1;\Comp)&=&{\cal A}\ ,\eea
with representations given by \eq{adjrep} and \eq{twadjrep},
respectively. These modules decompose under $\mho(D+1;\Comp)$ into
\bea \mho_1(D+1;\Comp)&=&{\cal A}_- \oplus_s {\cal A}_{+}\ ,\qquad
{\cal A}_+\ \equiv\ \mho(D+1;\Comp)\
,\label{mhoprime}\\[5pt] {\cal T}_1(D+1;\Comp)&=& {\cal
T}_+(D+1;\Comp)\oplus_{\mho} {\cal T}_-(D+1;\Comp)\ ,\qquad {\cal
T}_+(D+1;\Comp)\ \equiv\ {\cal T}(D+1;\Comp)\ ,\eea
where $\oplus_s$ denotes the semi-direct sum, ${\cal A}_\pm=\{Q\in
{\cal A}:\tau(Q)=\mp Q\}$ and ${\cal T}_\pm(D+1;\Comp)=\{S\in{\cal
A}:\t(S)\ =\ \pm \pi(S)\}$. The spaces ${\cal A}_\pm$ and ${\cal
T}_\pm(D+1;\Comp)$ contain the integer $(+)$ and half-integer
$(-)$ adjoint and twisted-adjoint levels, respectively, associated
to gauge fields and Weyl tensors with even $(+)$ and odd $(-)$
spins, and the representations are given by $[{\cal A}_\pm,{\cal
A}_\sigma]_\star={\cal A}_{\pm\sigma}$ and $[{\cal
A}_\pm,T_\sigma(D+1;\Comp)]_\pi=T_{\pm\sigma}(D+1;\Comp)$ for
$\sigma=\pm$.


\scs{Lowest-Weight and Highest-Weight Representations of
$\mso(D+1;\Comp)$}\label{UIRs}


In this Section we recall the construction and the main features
of the representations of the $\mso(D+1;\Comp)$ algebra that will
be of interest in the following. In particular, we want to focus
on the massless irreducible representations of that algebra,
\emph{i.e.}, on those irreps that describe massless fields in
nonflat maximally symmetric space-times, as they will also be
irreps of the HS extension of their isometry algebras. All the
representations that will be of interest to us in this Chapter are
lowest-weight (or highest-weight) representations, where the
energy operator $E=P_0=M_{0'0}$ is bounded from below (or above)
and the energy levels consist of finitely many spins (\emph{i.e.}
tensorial or tensor-spinorial representations of the
$\mso(D-1;\Comp)$-subalgebra generated by $M_{rs}$). Among these
one finds finite-dimensional tensorial and tensor-spinorial
representations, which arise as invariant subspaces containing
both lowest-weight and highest-weight states, as well as
infinite-dimensional representations arising in harmonic analysis
of linearized field equations on maximally symmetric spaces (of
various signatures) with nonvanishing cosmological constant. As we
shall see in Chapter \ref{map}, the linearized fields also contain
lowest-spin modules, which contain neither highest-weight nor
lowest-weight states, although for fixed spin they consist of
finitely many energy levels.

We shall first characterize finite-dimensional and
infinite-dimensional highest and lowest representations of the
complex algebra $\mso(D+1;\Comp)$. The choices of real forms and
related unitarity issues will be discussed in Subsection \ref{real}.

\vspace{0.5cm}

In the standard basis, the $\mso(D+1;\Comp)$ commutation rules
read
\bea [M_{RS},M_{TU}]_\star&=& 4i\delta_{[T|[S}M_{R]|U]}\
,\label{soD+1}\eea
where $R=1,\dots,D+1$ and
\bea \delta_{RS}&=& {\rm diag}(+\cdots +)\ .\label{signature}\eea
Representations of $\mso(D+1;\Comp)$ can be described starting
from left modules consisting of eigenstates of the ``diagonal''
generators
\bea
(M_{D+3-2k,D+2-2k}-\lambda_k)\mid(\lambda_1,\dots,\lambda_\nu)\rangle&=&
0\ ,\qquad k=1,\dots,\nu=[(D+1)/2] \ ,\eea
on which the remaining generators act as suitable raising and
lowering operators. A particular class of representations are the
\emph{highest weight representations}. These arise assuming the
existence of a highest weight state $|(n_1,\dots,n_\nu)\rangle$
annihilated by all raising operators. From it, the lowering
operators generate a module $\mV(n_1,\dots,n_\nu)$, known as the
\emph{Verma module}. For generic values of $(n_1,\dots,n_\nu)$, it
is irreducible (and hence infinite-dimensional). However, for
special values, it contains at least one excited state, referred
to as a singular vector, that is annihilated by all the raising
operators. The singular vectors generate an invariant submodule
$\mN(n_1,\dots,n_\nu)$, and as a result the highest weight
representation is now defined as the quotient
\bea \mD(n_1,\dots,n_\nu)&=&{\mV(n_1,\dots,n_\nu)\over
\mN(n_1,\dots,n_\nu)}\ ,\eea
which is irreducible, and infinite-dimensional or
finite-dimensional depending on $(n_1,\dots,n_\nu)$. The
finite-dimensional \emph{tensorial representations} arise for
integer highest weights obeying
\bea n_1\geq n_2\geq\cdots\geq n_\nu\geq 0\ ,\eea
corresponding to an $\mso(D+1;\Comp)$ tensor with symmetry
properties given by the Young projection corresponding to the
diagram with $n_i$ cells in the $i$th row. In these
representations, the repeated action of any lowering operator on
any state sooner or later generates states in
$\mN(n_1,\dots,n_\nu)$. Thus, there exists a lowest weight state,
actually given by $\mid(-n_1,-n_2,\dots,-n_\nu)\rangle$, that is
annihilated by all lowering operators. Thus, the
finite-dimensional representations are \emph{highest and lowest
weight} spaces, and they are invariant under the mirror reflection
$\l_1\rightarrow -\l_1$.

Infinite-dimensional representations of interest for Field Theory
in a $D$-dimensional space-time arise in the case that
\bea {\bf s}_0&=& (n_2,\dots,n_\nu)\ ,\eea
remains an integer (or half-integer) highest weight, while
\bea e_0&=& -n_1\eea
becomes a sufficiently large positive number (for fixed ${\bf
s}_0$). Then $M_{D+1,D}$ becomes unbounded from below, \emph{i.e.}
$\mD(n_1,\dots,n_\nu)$ no longer contains any lowest weight state,
and thereby becomes infinite-dimensional. The mirror reflection
$\lambda_1\rightarrow -\lambda_1$ now sends the highest weight
space $\mD(n_1,\dots,n_\nu)$ to an infinite-dimensional lowest
weight space, denoted by $\mD^-(-n_1,\dots,n_\nu)$. Thus,
identifying
\bea E&=&-M_{D+1,D}\ ,\eea %
as the field theory Hamiltonian, and the generators
\bea M_{rs}\ ,\qquad r,s=1,\dots,D-1\ ,\eea
as the orbital plus internal angular momenta, the highest weight
$|(-e_0,n_2,\dots,n_\nu)\rangle$ becomes a ground state, or lowest
energy state, with energy $e_0$ and spin ${\bf s}_0$. It is
convenient to ``Wick-rotate'' the two spatial directions $D+1$ and
$D$ into two time-like directions, that we shall denote by $0$ and
$0'$, and describe the weight space starting from the commutation
rules of $\mso(D+1;\Comp)$ in the ``two-time'' basis \eq{soD+1}. The
energy operator therefore is
\bea E&=&M_{0'0}\ =\ P_0\ ,\label{energy}\eea
and the energy raising and lowering operators are identified with
\bea L^\pm_r&=& M_{0r}\mp iM_{0'r}\ =\ M_{or}\mp iP_r\
,\label{Lplusminus}\eea
leading to the following energy graded decomposition of the
commutation rules \eq{soD+1}:
\bea [L^-_r,L^+_s] & = & 2iM_{rs}+2\d_{rs}E \ , \label{ll}\\[5pt]
[E,L^{\pm}_r] & = & \pm L^{\pm}_r \ , \label{el}\\[5pt]
[M_{rs},M_{tu}]&=& 4i\d_{[t|[s}M_{r]|u]}\ ,\label{mm}\\[5pt] [M_{rs},L^\pm_t]&=&
2i\d_{t[s}L^\pm_{r]}\ .\label{ml}\eea
The highest weight state $|(n_1,\dots,n_\nu)\rangle$ of the
standard basis is a lowest weight state of the two-time basis, and
vice versa, and in order to avoid confusion we shall use the
notation
\bea |e_0;{\bf s}_0\rangle&\equiv &|(-e_0,n_2,\dots,n_\nu)\rangle\
.\eea
In order to accommodate also the negative energy states (resulting
from the reflections in weight space), one also needs to define
highest weight states. Thus, in general, we have \emph{highest
weight states} $(+)$ and \emph{lowest weight states} $(-)$,
obeying
\bea (E-e_0)\ket{e_0;{\bf s}_0}^\pm\ =\ 0\ ,\qquad
L^\mp_r\ket{e_0;{\bf s}_0}^\pm\ =\ 0\ .\eea
Moreover, since ${\bf s}_0$ is, by assumption, a positive integer
highest weight of $\mso(D-1;\Comp)$, one may, without loss of
generality, replace the original Verma module $\mV^\pm(-e_0,{\bf
s}_0)$ by the \emph{generalized Verma module} defined by
\bea \mC(e_0;{\bf s}_0)^\pm&=& {\mV^\pm(-e_0,{\bf s}_0) \over
\mN[\mso(D-1;\Comp)]}\ ,\eea
where $\mN[\mso(D-1;\Comp)]$ consists of all $\mso(D-1;\Comp)$
modules generated from states that are singular with respect to
$\mso(D-1;\Comp)$. In other words, the generalized Verma module is a
particular example of a \emph{Harish-Chandra module}\footnote{Given
a Lie algebra $\mg$, the definition of a Harish-Chandra module is a
more general one, which does not a priori involve any highest or
lowest-weight state, but only a certain slicing of an
infinite-dimensional irreducible $\mg$-module $\mR$. In particular,
the slicing $\mR|_{\mh}$ of $\mR$ under a subalgebra $\mh\subset
\mg$ is said to be \emph{admissible} if it contains only
finite-dimensional $\mh$ irreps, sometimes referred to as
$\mh$-types, with finite multiplicities, \emph{viz.}
\bea \mR|_{\mh}&=& \bigoplus_{\kappa} {\rm mult}(\kappa)\mR_\kappa\
,\qquad {\rm mult}(\kappa) \dim\mR_\kappa<\infty\
,\label{hcmodule}\eea
where $\k$ are referred to as the \emph{compact weights} of $\mh$.
If all $\mh$-types are generated by the universal enveloping algebra
${\cal U}[\mg]$ of $\mg$ starting from a finite number of
$\mh$-types then $\mR$ is referred to as a \emph{Harish-Chandra
module} (see \cite{us} and the general treatise \cite{Knapp} for
more details).}, with energies bounded from below by $e_0$ (+) or
above (-) by $-e_0$, consisting of all states that are generated by
the action of (only) the $L^\pm_r$ operators on $\ket{e_0;{\bf
s}_0}^\pm$, \emph{i.e.}
\bea \mC^\pm(e_0;{\bf s}_0)&=&\span{\Comp}\left\{L^\pm_{r_1}\cdots
L^\pm_{r_n}\ket{e_0;{\bf s}_0}^\pm\right\}_{n=0}^\infty\
.\label{HC}\eea
One refers to $\mso(2;\Comp)_E\oplus \mso(D-1;\Comp)_{M_{rs}}$ as
the \emph{compact subalgebra}, and to the corresponding basis of
$\mC^\pm(e_0;{\bf s}_0)$, consisting of states $\left\{\ket{e;{\bf
s}}^\pm\right\}$ labeled by energy eigenvalues $e$ and spins ${\bf
s}$, as the \emph{compact basis}. The various weights of a
Harish-Chandra module, that result from extracting the
$\mso(D-1;\Comp)$-irreducible parts from the states in \eq{HC},
can therefore be represented as discrete dots filling a wedge (and
its negative energy reflection) in the
($\mso(D-1;\Comp)$-)spin/energy plane. For example, concentrating
our attention on the lowest weight modules and starting, for
definiteness, from a lowest weight state
$\ket{e_0;(s_0,0,...,0)}_{t(s_0)}$, the first excited energy level
consists of
\bea L^+_r\ket{e_0;{\bf s}_0}^+_{t(s_0)} \ = \
\left\{\ba{c} \ket{e_0+1;(s_0+1,0,...,0)}^+_{rt(s_0)}=L^+_{\{r}\ket{e_0;(s_0,0,...,0)}^+_{t(s_0)\}}\ ,  \\
 \ket{e_0+1;(s_0,1,0,...,0)}^+_{t(s_0),r}=L^+_{\{ r}\ket{e_0;(s_0,0,...,0)}^+_{t(s_0)\}} \ , \\
 \ket{e_0+1;(s_0-1,0,...,0)}^+_{t(s_0-1)}=L^+_{r}\ket{e_0;{\bf s}_0}^+_{rt(s_0-1)}\ ,\ea\right.
\eea
where the brackets $\{...\}$ embrace the indices that are Young
projected and traceless according to the various projections
indicated in the middle terms of the equality above.

In general, the Harish-Chandra module $\mC^\pm(e_0;{\bf s}_0)$ is
not irreducible, as it may contain singular vectors, \emph{i.e.}
excited states $\ket{e^{\prime}_0;{\bf s}'_0}^\pm$ with
$e^{\prime}_0>e_0$ that are annihilated by $L^\mp_r$,
\bea L^\mp_r\ket{e^{\prime}_0;{\bf s}'_0}^\pm \ = \ 0 \
.\label{singvect}\eea
This can happen, for instance, when certain relations are imposed
between $e_0$ and ${\bf s}_0$. In a definite signature such a
constraint can arise from the requirement of unitarity, as we
shall see. It is clear from the commutation relations and the
definition of Harish-Chandra modules, however, that for $e_0\gg
s_0$ there cannot be singular vectors $\ket{e^{\prime}_0;{\bf
s}'_0}^\pm$: indeed, by construction excited states have
$e^{\prime}_0-{\bf s}'_0 \geq e_0 - s_0\gg 1$, and, schematically,
\bea L^-_r\ket{e^{\prime}_0;{\bf s}'_0}^+ \ = \
L^+...L^+(2iM+2E)L^+...L^+\ket{e_0;{\bf s}_0}^+ + ...\eea
where the commutation relation \eq{ll} has been used and the
ellipsis stand for other similar terms with all possible powers of
$L^+$ on the left and on the right of $2iM+2E$. But if $e_0\gg
s_0$, the action of $E$ and $M_{rs}$ can only extract strictly
positive eigenvalues, and there is no chance that \eq{singvect} be
verified. On the other hand, this argument shows that lowering
$e_0$ to some critical value
$e_{0,\textrm{crit}}=e_{0,\textrm{crit}}({\bf s}_0)$ a singular
vector may appear. Similarly (interchanging $-$ and $+$) for
highest-weight modules.

The singular vectors generate Harish-Chandra submodules
\bea \mC^\pm(e'_0;{\bf s}'_0) \ = \
\span{\Comp}\left\{L^\pm_{r_1}\cdots L^\pm_{r_n}\ket{e'_0;{\bf
s}'_0}^\pm\right\}_{n=0}^\infty\ ,\label{HC'}\eea
and $\mC^\pm(e_0;{\bf s}_0)$, that contains them, is therefore an
\emph{indecomposable} module: that is to say, the action of the
noncompact generators $L^\pm_r$ can bring from $\mC^\pm(e_0;{\bf
s}_0)$ to the inside of the singular submodule, but not back out
of the latter (because of \eq{singvect}). This implies that the
direct sum of such submodules is an ideal $\mI^\pm(e_0;{\bf
s}_0)\subset \mC^\pm(e_0;{\bf s}_0)$ that can be factored out
consistently, leaving the \emph{irreducible} lowest and highest
weight spaces
\bea \mD^\pm(e_0;{\bf s}_0)&=&{\mC^\pm(e_0;{\bf s}_0)\over
\mI^\pm(e_0;{\bf s}_0)}\ .\eea
The indecomposable structure can be also be presented by making use
of the semi-direct sum symbol $\oplus_s$, as $\mC^\pm(e_0;{\bf
s}_0)=\mI^\pm(e_0;{\bf s}_0)\oplus_s\mD^\pm(e_0;{\bf s}_0)$. The
lowest and highest weight spaces are isomorphic, namely
\bea \mD^\mp(e_0;{\bf s}_0)&=&\pi(\mD^\pm(-e_0;{\bf s}_0))\ ,\eea
where $\pi$ is defined by acting over $\mso(D+1;\Comp)$ as the
automorphism
\bea \pi(L^\pm_r)&=& L^\mp_r\ ,\qquad \pi(E)\ =\ -E\ ,\qquad
\pi(M_{rs})\ =\ M_{rs}\ ,\eea
and on lowest and highest weight states as the reflection
\bea \pi(\ket{e_0,{\bf s}_0}^\pm)&=& \ket{-e_0,s_0}^\mp\ .\eea
Of importance for our constructions is also the existence of the
linear anti-automorphism $\tau$, defined by
\bea \tau(M_{AB})&=&-M_{AB}\ ,\label{tau}\eea
and
\bea \tau(\ket{e_0,{\bf s}_0}^\pm)&=& \varphi(e_0,{\bf
s}_0)\,{}^\mp\bra{-e_0,{\bf s}_0}\ ,\label{tau2}\eea
where ${}^\pm\bra{e_0,{\bf s}_0}$ are the ground states of the
dual weight spaces $(\mD^\pm(e_0,{\bf s}_0))^\star$, and
$\varphi(e_0,{\bf s}_0)$ is a phase factor. Lowest weight states
and spaces will sometimes be written without the $+$ superscript,
and in that case the highest weight dittos will be denoted by a
tilde instead of the $-$ superscript.

After this general description, we now turn to examine some
important examples.

\vspace{0.5cm}

\noindent First, let us exemplify the lowest weight description of
the $\mso(D+1;\Comp)$ vector $\D(1,0)=\mD(-1;(0))$. The singular
vector is $L^+_{\{r}L^+_{s\}}\ket{-1;(0)}$, and $\mI(-1;(0))\simeq
\mC(1;(2))$ contains all states with energy $e\geq 2$, since
\bea L^-_r L^+_s L^+_t L^+_t\ket{-1;(0)}&=&
-6L^+_{\{r}L^+_{s\}}\ket{-1;(0)}\ ,\eea
leaving $\mD(-1,0)$ consisting of $\ket{-1;(0)}$,
$\ket{0;(1)}=L^+_r\ket{-1;(0)}$ and $\ket{1;(0)}=L^+_r
L^+_r\ket{-1;(0)}$. We note that $L^+_r\ket{1;(0)}\in \mI(-1;(0))$
so that $L^+_r\ket{1;(0)}\simeq 0$ in $\mD(-1;(0))$, reflecting
the fact that the finite-dimensional representations contain
lowest as well as highest weights.

\vspace{0.5cm}

\noindent Of the infinite-dimensional cases, those of main
interest to us are the singleton and massless representations,
\bea \mbox{scalar and spinor singletons}&:& e_0\ =\ s_0+\e_0\
,\qquad s_0\ =\ 0,\ft12\
,\label{singletons}\\[5pt] \mbox{massless particles}&:& e_0\ =\ s_0+2\e_0\ ,\qquad s_0\ =\
1,\ft32,2,\ft52,\dots\ ,\eea
where $\e_0=(D-3)/2$, and the singular vectors are
\bea \mbox{Singletons $s_0=0$}:&& \ket{\e_0+2;(0)}\ =\
L^+_r L^+_r|\e_0;(0)\rangle\ ,\label{singularsingleton}\\[5pt]s_0=\ft12:&&\ket{\e_0+\ft32;(\ft12)}_\a\ =\
(\c_r)_\a{}^\b L^+_r|\e_0+\ft12;(\ft12)\rangle_\b
\\[5pt]\mbox{Massless $s_0=
1,2,\dots$:}&& |s_0+2e_0+1;(s_0-1)\rangle_{r_1\dots r_{s_0-1}}\ =\
L^+_r|s_0+2e_0;(s_0)\rangle_{r r_1\dots r_{s_0-1}}\ ,\qquad \ ,\\[5pt]
s_0=\ft32,\ft52,\dots:&& |s_0+2e_0+1;(s_0-1)\rangle_{\a r_1\dots
r_{s_0-3/2}}\ =\ L^+_r|e_0;(s_0)\rangle_{\a r r_1\dots
r_{s_0-3/2}}\ ,\eea
with $\c_r$ given by $\mso(D-1)$ Dirac matrices. The
$\mso(D-1;\Comp)$ representations of the ground states are given
by
\bea M_{rs}\ket{e_0;(s_0)}^\pm_{t_1\dots t_{s_0}}&=& 2is_0\d_{s
\{t_1}\ket{e_0;(s_0)}^\pm_{t_2\dots t_{s_0}\}r}\ ,\qquad s_0\ =\ 0,1,2,\dots\ ,\\[5pt]
M_{rs}\ket{e_0;(s_0)}^\pm_{\a,t_1\dots t_{s_0-1/2}}&=&
2i(s_0-1/2)\d_{s \{t_1}\ket{e_0;(s_0)}^\pm_{t_2\dots
t_{s_0-1/2}\}r}\\[5pt]&-&\ft i2(\c_{rs})_\a{}^\b
\ket{e_0;(s_0)}^\pm_{\b,t_1\dots t_{s_0-1/2}}\ ,\quad s_0\ =\
\ft12,\ft32,\ft52,\dots\ ,\eea
where the curly brackets indicate the symmetric and traceless
projection on $t_1\dots t_{s_0}$, and $\c_{rs}=\c_{[r}\c_{s]}$.
The phase factors in \eq{tau2} are
\bea \tau(\ket{\pm\e_0;(0)}^\pm)&=& {}^\mp\bra{\mp\e_0;(0)}\ ,\\[5pt]
\tau(\ket{\pm (\e_0+\ft12);(1/2)}^\pm)&=& i~{}^\mp\bra{\mp (\e_0+\ft12);( 1/2)}\ ,\\[5pt]
\tau(\ket{\pm(s_0+2\e_0);(s_0)}^\pm)&=& (-)^{s_0}\
{}^\mp\bra{\mp(s_0+2\e_0);(s_0)}\
,\qquad s_0=1,2,...\ ,\\[5pt]
\tau(\ket{\pm(s_0+2\e_0);(s_0)}^\pm)&=& (-)^{s_0-1/2}i~\
{}^\mp\bra{\mp(s_0+2\e_0);(s_0)}\ ,\quad s_0=1/2,3/2,...\ ,\eea
and the action of $\tau$ on dual states is fixed by demanding that
\bea \tau^2&=&\left\{\ba{ll}\textrm{Id}&\mbox{integer
spin}\\-\textrm{Id}&\mbox{half-integer spin}\ea\right.\ .\eea
We now devote a more detailed discussion to the
infinite-dimensional representations that will be of importance in
the following.

\scss{Massless Irreducible Representations}

Even a proper definition of ``masslessness'' in maximally
symmetric spaces with nonvanishing cosmological constant is
nontrivial. All sensible definitions have in common the feature
that the massless representation should correspond, in the flat
limit $\L\rightarrow 0$, to a massless representation of the
Poincar\'e algebra, but this does not fix uniquely the notion of
masslessness, and additional conditions have to be introduced
(see, for example, \cite{Angelopoulos:1999bz} and references
therein). Stronger definitions of masslessness correspond to the
concepts of \emph{conformal masslessness} and \emph{composite
masslessness}: the first is related to the property of unique
extension from representations of $\mso(D+1;\Comp)$ to a
\emph{singleton} representation of the conformal group
\cite{Angelopoulos:1980wg}, while the second characterizes
massless particles as composites of two singletons
\cite{Flato:1978qz}, as we shall review in the next Subsection.
The two notions coincide only in $D=3,4$.

The latter definition is quite natural as it implies that the
appearance and factorization of the singular ideal
$\mI^\pm(e_0;{\bf s}_0)$ corresponds, for $s_0\geq 1$ to the
appearance of a gauge symmetry and elimination of gauge modes. Let
us show how this happens for the critical value
$e_{0,\textrm{crit}}=s_0+2\e_0$ mentioned above, henceforth
restricting our attention, for simplicity, to the case of totally
symmetric representations $\mD(e_0;(s_0,0,...,0))\equiv
\mD(e_0;s_0)$ and to integer spins only.

\begin{figure}[!h]
\begin{center}
\unitlength=.6mm
\begin{picture}(150,180)(0,-10)
\put(0,0){\vector(1,0){150}} \put(0,0){\vector(0,1){150}}
\put(150,-10){$s$} \put(-10,150){$E$}
\put(20,-10){1}\put(40,-10){2}\put(60,-10){3}\put(80,-10){4}\put(100,-10){5}\put(120,-10){6}
\put(-10,20){1}\put(-10,40){2}\put(-10,60){3}\put(-10,80){4}\put(-10,100){5}\put(-10,120){6}
\multiput(20,1)(20,0){6}{\!$|$} \multiput(1,20)(0,20){6}{\!$-$}
\multiput(20,40)(20,20){6}{\!$\bullet$}
\multiput(20,60)(20,20){5}{\!$\bullet$}
\multiput(20,80)(20,20){4}{\!$\bullet$}
\multiput(20,100)(20,20){3}{\!$\bullet$}
\multiput(20,120)(20,20){2}{\!$\bullet$}
\multiput(20,140)(20,20){1}{\!$\bullet$}
\end{picture}
\end{center}
\caption{{\small Weight diagram of the massless UIR of spin-1
$\mD(2,1)$ in four dimensions.}} \label{s1fig}
\end{figure}
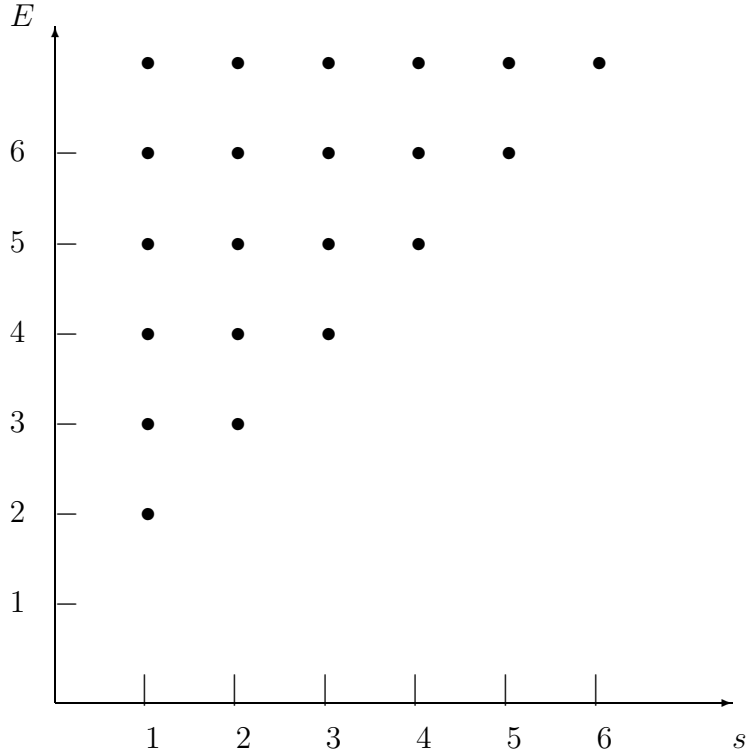

Indeed, for $e_{0,\textrm{crit}}=s_0+2\e_0$ and $s_0>1$ a singular
vector appears at the first excited level, with quantum numbers
\bea \ket{e'_0;s'_0}_{r(s'_0)} \ = \ \ket{e_0+1;s_0-1}_{r(s_0-1)}
\ = \ L^+_t\ket{e_0;s_0}_{tr(s_0-1)} \ . \eea
This simply follows from the assumption that $\ket{e_0;s_0}$ is a
lowest weight state and from the algebra,
\bea && L^-_q (L^+_t\ket{e_0;s_0}_{tr(s_0-1)}) \ = \
(2iM_{qt}+2\d_{qt}E)\ket{e_0;s_0}_{tr(s_0-1)} \nn\\[5pt]
& = & 2(e_0-s_0-2\e_0)\ket{e_0;s_0}_{tr(s_0-1)} \ = \ 0
\hspace{1cm} \textrm{for} \hspace{.3cm} e_0=s_0+2\e_0 \ ,\eea
as can be easily checked recalling also that
$\ket{e_0;s_0}_{ttr(s_0-2)}=0$. The whole Harish-Chandra module
$\mD(s_0+1;s_0-1)$ built on top of the singular vector, and that
encodes the gauge modes of the massless field of spin $s_0$
decouples completely from the rest of the representation, and can
therefore be consistently factored out. The leftover irreducible
weight space (illustrated in Figure \ref{s1fig} for the case of
the massless spin-1 field in four dimensions) is formed by states
that are ``divergence-free''
($L^+_p\ket{e;s}_{pr(s-1),q(t)}=L^+_p\ket{e;s}_{r(s),pq(t-1)}=0$)
and where each dot has multiplicity one\footnote{In the sense of
the Harish-Chandra module, of course, where every weight always
carries a finite-dimensional irrep of $\mso(D-1;\Comp)$.}
\cite{deWit:2002vz}.

The elements of the Harish-Chandra modules \eq{HC} can be
identified with the modes of a free quantum one-particle state (or
anti-particle state)\footnote{Strictly speaking, such a
terminology should be applied to the elements of \emph{unitary}
modules, that we shall examine later. With such a proviso, we
shall anyway extend it to every state of a Harish-Chandra module.}
on a maximally symmetric space with nonvanishing cosmological
constant. The corresponding field equations for general signature
can be obtained describing such space as the coset
$\frac{\mso(D+1;\Comp)}{\mso(D;\Comp)}$, that, in the various
signatures of interest, gives rise to the manifolds shown in
\eq{coset}. The D'Alembertian operator is related to the
difference of the quadratic Casimir operators
$C_2[\mso(D+1;\Comp)]$ and $C_2[\mso(D;\Comp)]$. For a general
signature $\S$ of $\y_{AB}$ and $\S'$ of $\y_{ab}$, and in the
notation of Section \ref{maxsym}, this amounts to
\bea \l^2L^2\nabla^a\nabla^b\y_{ab} \ = \
C_2[\mso(\S)]-C_2[\mso(\S')]  \ ,\eea
which characterizes the mass-shell condition as
\bea \l^2\t^2L^2m^2 \ = \ C_2[\mso(\S)]-C_2[\mso(\S')] \ , \eea
where $m$ is the mass of a spin-$s$ particle in the nonflat
maximally symmetric background at hand (that includes also the
mass-like term that originates from the coupling to the background
curvature) and we have temporarily reinstated, for clarity, the
factors of the radius of curvature $L^2$ elsewhere taken equal to
one. In the compact basis and in lowest or highest weight states
$\mD^\pm(e_0;{\bf s}_0)$, one can evaluate the Casimir operator of
$\mso(D+1;\Comp)$ as in \eq{C2lhws}, obtaining
\bea m^2 \ = \ e_0(e_0\mp (D-1))+C_2[\mso(D-1;\Comp)|({\bf s}_0)]-
C_2[\mso(D;\Comp)|({\bf s}_0)_D] \ ,\label{masscas}\eea
where $({\bf s}_0)_D$ is an $\mso(D;\Comp)$-irrep. One can check
from here that the value $m^2$ of the mass-like term, obtained in
Section \ref{frads} as the one preserving the gauge invariance of
the Fronsdal equations, corresponds to the one obtained from
\eq{masscas} at the critical energy $e_0=s_0+2\e_0$ of the
massless lowest weight representations $\mD(s_0+2\e_0;s_0)$.
Notice however that, in general, \eq{masscas} is a quadratic
equation for $e_0$, and admits therefore two roots. When there are
two real roots, they correspond to two different solutions of the
field equations with different boundary conditions at spatial
infinity\footnote{This is of course strictly true only in the
signature $(D-1,2)$, i.e., in the $AdS_D$ case. However, a notion
of boundary can be found also in other signatures: for example, in
the case of $S^D$, it is naturally associated with solutions of
the field equations that diverge at the poles, and that are
therefore well-behaved only if the latter are cut out, which
introduces a boundary.}. For fixed spin, such ``conjugate''
representations $\mD(E_0;s_0)$ are those with the same value of
the Casimir operator but different values of energy: from \eq{C2}
it is clear that these must have $E_0=D-1-e_0$ (here for lowest
weight representations). As we shall see, unitarity introduces
some other bound on the value of $e_0$ at fixed $s_0$, and in
general this rules out one of the two solutions as nonunitary (the
one with lower energy). One important exception that we shall
encounter is the scalar field in four dimension, that comes in two
varieties, $\mD(1,0)$ and $\mD(2,0)$, related to Neumann and
Dirichlet boundary conditions, respectively, and both unitary.

\begin{figure}[!h]
\begin{center}
\unitlength=.6mm
\begin{picture}(150,180)(0,-10)
\put(0,0){\vector(1,0){150}} \put(0,0){\vector(0,1){150}}
\put(150,-10){$s$} \put(-10,150){$E$}
\put(20,-10){1}\put(40,-10){2}\put(60,-10){3}\put(80,-10){4}\put(100,-10){5}\put(120,-10){6}
\put(-10,20){1}\put(-10,40){2}\put(-10,60){3}\put(-10,80){4}\put(-10,100){5}\put(-10,120){6}
\multiput(0,1)(20,0){6}{\!$|$} \multiput(1,20)(0,20){6}{\!$-$}
\multiput(0,20)(20,20){6}{\!$\bullet$}
\multiput(0,60)(20,20){4}{\!$\bullet$}
\multiput(0,100)(20,20){2}{\!$\bullet$}
\end{picture}
\end{center}
\caption{{\small Weight diagram of the massless scalar $\mD(1,0)$ in
four dimensions.}} \label{scalfig}
\end{figure}
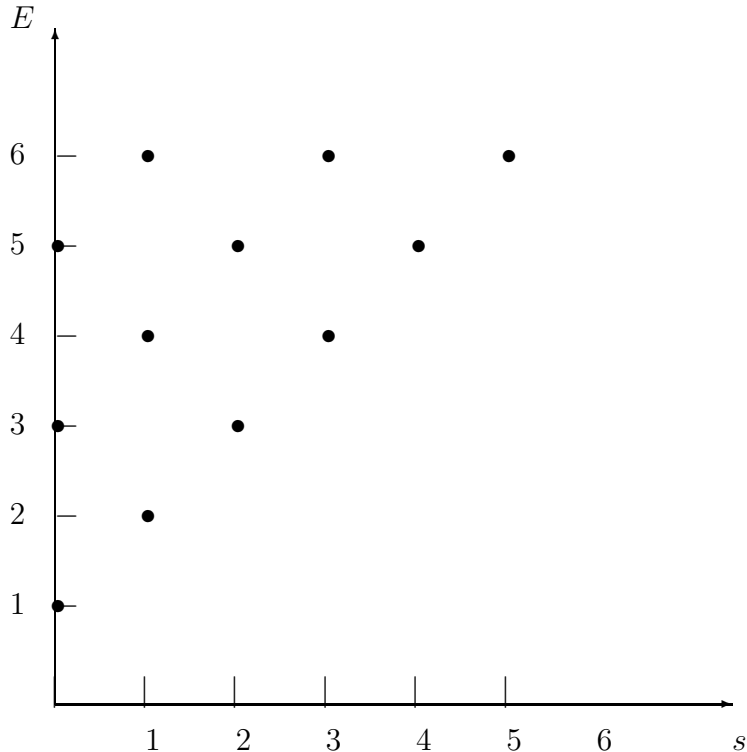

Not having any associated gauge symmetries, a massless scalar can
only be defined in accordance to the criteria of conformal or
composite masslessness. As we shall see, the latter leads, in
general, to a scalar representation $\mD(2\e_0;0)$, and possesses
a weight lattice that is half-filled (see Fig. \ref{scalfig}),
compared to that of the gauge fields, due to the fact that, as the
lowest weight state $\ket{2\e_0;0}$ has spin 0, there is no way of
forming a spin-0 combination at the first excited level
(\emph{i.e.}, with a single energy raising operator $L^+$).

In four dimensions, the composite massless scalars are of two
types, as we shall see, and coincide with the conformal massless
ones. Indeed, the conformal coupling gives a (fake) mass term of
the form
\bea m^2 \ = \ \frac{R}{6} \ = \ -\frac{12}{6} \ = -2 \ ,\eea
where eq. \eq{scal} has been used to determine the curvature
scalar $R$. Substituting in \eq{masscas} and solving for $e_0$
gives $\e_0=1,2$ (and notice that, in four dimensions $2\e_0=1$).
We shall soon examine the composite interpretation of these two
solutions.

\scss{Singleton Representations}

The most remarkable fact about massless representations in a
maximally symmetric background with nonvanishing cosmological
constant is that they are not fundamental! This is a dramatic
difference with respect to the case of a flat space-time, and relies
on the fact that the fundamental representation (unitary in the AdS
case) is a a very special one, an ultra-short representation that
admits no flat limit, called \emph{singleton}. Such representations
were first discovered by Dirac \cite{Dirac:1963ta}, who was quite
intrigued by their properties, subsequently studied by Fronsdal and
collaborators \cite{Fronsdal:1978vb, Flato:1978qz,
Angelopoulos:1980wg}, and then later found a very natural arena in
String Theory in the context of the AdS/CFT correspondence and of
recent studies on the tensionless limit of strings
\cite{Engquist:2005yt}.

Let us examine what happens to the scalar weight space in the case
that $e_0$ is lowered to the value $\e_0$: remarkably, also the
Harish-Chandra module of a scalar field becomes indecomposable,
and a singular vector appears at the second excited level,
\bea \ket{e'_0,s'_0} \ = \ \ket{\e_0+2,0} \ = \
L^+_rL^+_r\ket{\e_0,0} \ .\label{singforsing}\eea
Indeed,
\bea L^-_rL^+_sL^+_s\ket{e_0,0} & = &
(2iM_{rs}+2\d_{rs}E)L^+_s\ket{e_0,0} + 2L^+_r E\ket{e_0,0}
\nn\\[5pt]
& = & (4e_0-4\e_0)L^+_r\ket{\e_0,0} \ = \ 0\label{singproof}\eea
admits precisely the solution above, $e_0=\e_0$. The factorization
of the scalar singular submodule $\mD(\e_0+2;0)$ from the
Harish-Chandra module of a scalar representation leads to an
ultra-short irrep
\bea \mD_0 \ \equiv \ \mD(\e_0;0)\ = \
\span{\Comp}\left\{L^+_{\{r_1}L^+_{r_2}...L^+_{r_n\}}\ket{e_0,0}\right\}_{n=0}^\infty
\eea
that consists of a single line in the weight space (see Fig.
\ref{singfig}), hence the name singleton. The physical meaning of
this fact is that, for every excitation in such a representation,
the energy is always proportional to the $\mso(D-1;\Comp)$-spin:
this means that there are no radial excitations, \emph{i.e.}, that
such a representation only consists of boundary degrees of
freedom. In other words, the factorization of the singular
submodule does not correspond, as it was the case for $s_0\geq 1$,
to the elimination of gauge modes, but to the absence of bulk
degrees of freedom! Indeed, the singular vector \eq{singforsing}
can be shown to be related to $(D-1)$-dimensional equations of
motion of a conformal scalar field living at the boundary. This
property of being boundary objects gives a kinematical reason for
the unobservability of singletons. The defining property that, for
each energy level $\mD^{(n)}(\e_0;0)$,
\bea E\mD^{(n)}(\e_0;0) \ = \ (\e_0+n)\mD^{(n)}(\e_0;0) \ ,\eea
also gives a reason why such representations do not admit a flat
limit. Indeed, recall that for Poincar\'e irreps one has a
continuous tower of modes for every value of spin, which is
exactly what one gets from the flat limit of the massless irreps
found above\footnote{The discrete spectrum of energy for every
value of spin is related to the presence of a boundary, which is
in its turn related to a finite radius of curvature $L$ -
essentially, dealing with fields in a space-time of constant
curvature is analogous to analyzing waves in a box.}. However,
having a single energy eigenvalue for any given value of the spin
makes the excitations in the singleton representation peculiar,
and makes its flat limit result in a representation of the
Poincar\'e algebra that is trivial on the translations.

\begin{figure}[!h]
\begin{center}
\unitlength=.6mm
\begin{picture}(150,180)(0,-10)
\put(0,0){\vector(1,0){150}} \put(0,0){\vector(0,1){150}}
\put(150,-10){$s$} \put(-10,150){$E$}
\put(20,-10){1}\put(40,-10){2}\put(60,-10){3}\put(80,-10){4}\put(100,-10){5}\put(120,-10){6}
\put(-10,20){1}\put(-10,40){2}\put(-10,60){3}\put(-10,80){4}\put(-10,100){5}\put(-10,120){6}
\multiput(20,1)(20,0){6}{\!$|$} \multiput(1,20)(0,20){6}{\!$-$}
\multiput(0,10)(20,20){6}{\!$\bullet$}
\multiput(10,20)(20,20){5}{\!$\circ$}
\end{picture}
\end{center}
\caption{{\small Weight diagrams of the scalar singleton
$\mD(1/2,0)$ ($\bullet$) and of the spinor singleton $\mD(1,1/2)$
($\circ$) in four dimensions.}} \label{singfig}
\end{figure}

Nonetheless, as previously announced, the most remarkable
properties of the scalar singleton is that the tensor product of
two such representations (sometimes refereed to as
\emph{doubleton}) decomposes, under the action of
$\mso(D+1;\Comp)$, into the direct sum of all bosonic massless
representations, as first discovered in the case of $AdS_4$ by
Flato and Fronsdal \cite{Flato:1978qz} and later extended to $D$
dimensions \cite{Angelopoulos:1997ij, Angelopoulos:1999bz,
mishasalg, Engquist:2005yt},
\bea \mD_0\otimes \mD_0 \ = \ \bigoplus_{s_0=0,1,2,...}
\mD(s_0+2\e_0,(s_0))\ .\label{RacRac}\eea
The product above can be decomposed into the symmetric and
antisymmetric parts, that contain the even and odd massless spins,
respectively,
\bea [\mD_0\otimes \mD_0]_S & = & \bigoplus_{s_0=0,2,4,...}
\mD(s_0+2\e_0,(s_0))\\[5pt]
[\mD_0\otimes \mD_0]_A & = & \bigoplus_{s_0=1,3,5}
\mD(s_0+2\e_0,(s_0))\ .\eea
The composite massless lowest weight states can therefore be
written as a superposition of states in the doubleton,
\bea \ket{s+2\e_0;(s)}_{12;r(s)} &=&
f_{r(s)}(1,2)\ket{2\e_0;(0)}_{12}\ket{2\e_0;(0)}_{12}\
,\label{FFformula}\eea
where the composite operator\footnote{The coefficients $f_{s;k}$
are fixed by the condition
$(L^-_r(1)+L^-_r(2))\ket{s+2\e_0;(s)}_{12;r(s)} =0$, which is
equivalent to $a_k f_{s;k}+a_{s-k+1}f_{s;k-1}=0$, where
$a_k=2k(k+\e_0-1)$, with solution
$f_{s;k}=(-1)^sf_{s;s-k}=(-1)^k{a_{s-k+1}\cdots a_s\over a_k\cdots
a_1}f_{s;0}$, taking the form \eq{complw} for $f_{s;0}=1$.}
$f_{r(s)}(1,2)$ is given by
\bea f_{r(s)}(1,2)&=& (-1)^s f_{r(s)}(2,1)\ =\ \sum_{k=0}^s
f_{s;k}(L^+_{\{r_1}\cdots L^+_{r_{k}})(1)(L^+_{r_{k+1}}\cdots
L^+_{r_s\}})(2)\ ,\label{frs}\\[5pt] f_{s;k}&=&(-1)^s f_{s;s-k}\ =\ {s\choose k}{(1-s-\e_0)_k\over (\e_0)_k}\
.\label{complw}\eea

Eq. \eq{masscas} determines the mass of a free scalar singleton
field $\Psi$ to be $m^2_{\Psi}=-\frac{(D-3)(D+1)}{4}$, while its
field equation is
\bea \left(\Box -\frac{(D-3)(D+1)}{4}\right)\Psi \ = \ 0\ , \eea
and belongs to the ideal $\mD(\e_0+2,(0))$.

We note also that $\mso(D+1;\Comp)$ does not act transitively on
the singleton weight space. The smallest Lie algebra with this
property is the minimal bosonic HS algebra $\mho(D+1;\Comp)$
defined in \eq{adj}.

As announced earlier in this Chapter, the singleton representation
is also uniquely defined by the condition that it be annihilated
by every combination of generators of the enveloping algebra of
$\mso(D+1;\Comp)$ that is in the ideal $\cI[V]$ defined in
\eq{idealV}. Indeed, one can characterize $\cI[V]$ as the
annihilating (left and right) ideal of the singleton,
\bea \cI[V] \ = \ \span{\Comp}\left\{X\in \cU \ : \ X\ket{\psi}=0\
, \forall \ket{\psi}\in \mD_0\right\} \ , \label{annsing}\eea
and to prove this it is sufficient to show that the generating
elements $V_{AB}$ and $V_{ABCD}$ annihilate every element in the
singleton representation. This is easily done splitting the index
$A=(+,-,r)$, where $X_\pm\equiv X_0\pm iX_{0'}$, and noting that
$c_{--}=L^+_rL^+_r$, that exactly coincides with the combination
of energy-raising operators giving rise to the singular vector
\eq{singforsing} and is therefore consistently set to zero in the
singleton representation with all the submodule built on top of
it. Moreover, using $\mso(D+1;\Comp)$ rotations, this conclusion
can be extended to all the independent
$\mso(D-1;\Comp)$-components into which $c_{AB}$ is broken as a
consequence of the index splitting. Similarly, one can show that
also $V_{ABCD}$ gives zero acting on every state in the singleton.

Importantly, in $D=4$ the scalar singleton $\mD(1/2,0)$ is not the
only irrep satisfying \eq{annsing}: one can show indeed that the
latter condition admits another \emph{spinor singleton} irrep
$D_{1/2}=\mD(1,1/2)$, whose lowest weight state is a
$\mso(3;\Comp)$-spinor representation $\ket{1,1/2}_i$. Moreover, one
can also prove that
$C_2[\mso(5;\Comp);(1/2,0)]=C_2[\mso(5;\Comp);(1,1/2)]$. This
implies, among other things, that also the tensor product of two
spinor singletons decomposes into integer-spin massless
representations, and in particular
\bea \mD_{1/2}\otimes \mD_{1/2} \ = \
\mD(2,0)\oplus\bigoplus_{s_0=0,1,2,...}\mD(s_0+1,s_0) \ ,
\label{DiDi}\eea
while the tensor product of a scalar and a spinor singleton gives
rise to the half-integer-spin massless representations,
\bea \mD_{0}\otimes\mD_{1/2} \ = \
\bigoplus_{2s_0=1,3,...}\mD(s_0+1,s_0) \ . \eea
Notice that scalar and spinor doubletons admit the same
decomposition except for the scalar sector, where the former
contains the parity-invariant scalar $\mD(1,0)$ and the latter the
pseudo-scalar $\mD(2,0)$. We shall make use of this particular
feature of four dimensions in Chapter \ref{map}.

\scss{Real forms and unitary representation}\label{real}

So far, our analysis has been carried out at the complex level,
and with no notion of unitarity. We shall now examine in detail
the various different real forms of the $\mso(D+1;\Comp)$ algebra
with the different signatures $(p',D+1-p')$ that will be of
relevance in the following, and in which of these the
representations shown above are unitary. In particular, we shall
look at how the splitting

\be \mg \ = \ \mg_+\oplus\mg_0\oplus\mg_- \ ,\label{split}\ee

into the compact subalgebra $\mg_0=\mso(D-1;\Comp)\oplus
\mso(2;\Comp)$ and the noncompact parts $\mg_{\pm}=\span{\Comp}
\,\{L^{\pm}_r\}$ can be performed, and examine the effect of a
$\sigma$ automorphism that acts on the generators in such a way as
to always arrive at a positive definite
standard inner product. 

\begin{itemize}

\item $\mso(D-1,2)$ with tangent space signature $(D-1,1)$ ($AdS_D$)\
: \ Here we have $\y_{AB}=(-1,\d_{rs},-1)$.

We begin by assuming $M^\dagger_{AB}=M_{AB}$. One can realize the
splitting (\ref{split}) taking

\be E \ = \ M_{0'0} \ , \qquad M_{rs}\ee

as the generators of the compact subalgebra, and

\be L^{\pm}_r \ = \ M_{0r}\mp i M_{0'r}\ee

as the energy-raising and energy-lowering operators. Moreover, we
define $P_r=M_{0'r}$ as the spatial $AdS_D$ translations. One can
check that such definitions indeed satisfy the algebra

\bea \left[L^-_r,L^+_s\right] & = & 2iM_{rs}+2\d_{rs}E \ , \nn\\
\left[E,L^{\pm}_r\right] & = & \pm L^{\pm}_r \ , \label{alg}\eea

along with the reality conditions

\be E^\dagger \ = \ E \ , \qquad (L^{\pm}_r)^\dagger \ = \
L^{\mp}_r \ .\label{reality}\ee

The latter relation ensures that one can build a lowest weight
module $\{L^+_{r_1}...L^+_{r_n}\ket{e_0,s_0}\}$ of states with
positive norm. This, together with the hermitian nature of the
generators $M_{AB}$, implies the unitarity of such a representation,
at least for suitable values of $e_0, s_0$.

\item $\mso(D+1)$ with tangent space signature $(D,0)$ ($S^D$)\ : \
$\y_{AB}=\d_{AB}$.

Again we begin by assuming the hermiticity of the generators
$M_{AB}$. Here the splitting is completely arbitrary, due to the
compactness of the algebra. One way of realizing the algebra
(\ref{alg}) is to choose

\be L^{\pm}_r \ = \ iM_{0r}\mp M_{0'r} \ , \qquad E \ = \ M_{0'0} \
.\ee

However, here the reality conditions are such that

\be E^\dagger \ = \ E \ , \qquad (L^{\pm}_r)^\dagger \ = \
-L^{\mp}_r \ , \label{euclherm}\ee

and the latter relation prevents from constructing Fock space
states with positive norm. This is independent of the particular
chosen realization of the splitting (\ref{split}). The only way to
recover this property is to twist the reality conditions of the
generators, in such a way as to get rid of the minus sign in the
reality conditions of the energy-raising and energy-lowering
operators. This amounts to the requirement that

\be M_{AB}^\dagger \ = \ \s (M_{AB}) \ , \ee

where in this case

\be \s\ = 
 \ \left\{
\begin{array}{c}
+1\ , \ \textrm{on} \ M_{rs}, M_{0'0}=E \\
-1 \ , \ \textrm{on} \ M_{0r}, M_{0'r}=P_r \end{array} \right. \
.\ee

To reiterate, in the euclidean case one can have a positive definite
lowest weight module only by means of a ``Wick rotation'' of the
algebra that leaves hermitian only the generators $ M_{rs}, M_{0'0}
$, spanning the subalgebra $\mso(D-1)\oplus\mso(2)$. In any case, we
cannot have hermitian generators and states with positive norm at
the same time, so such lowest weight representations cannot be
unitary, for any value of $e_0, s_0$\footnote{Note that this
argument only holds for infinite-dimensional modules.
Finite-dimensional unitary representations, which are
lowest-and-highest-weight modules, are not ruled out. This is
essentially because, by $\pi$-invariance, their weight diagram is
symmetrical with respect to the $\{E=0\}$ axis, which in turn
implies that the lowest-weight state has a negative energy
eigenvalue that compensates for the minus sign appearing in the
hermiticity condition \eq{euclherm}, leading to positive norms.}.

\item $\mso(D-1,1)$ with tangent space signature $(D,0)$ ($H_D$)\
: \ $\y_{AB}=(\d_{ab},-)$.

The energy-raising and energy-lowering operators and the energy
operator can here be defined as

\be L^{\pm}_r \ = \ iM_{0r}\mp iM_{0'r} \ , \qquad E \ = \ iM_{0'0}
\ .\ee

We moreover define the space translations as
$P_r=M_{0'r}=\frac{1}{2i}(L^-_r-L^+_r)$. These satisfy the algebra
(\ref{alg}), but their reality conditions are

\be E^\dagger \ = \ -E \ , \qquad (L^{\pm}_r)^\dagger \ = \
-L^{\pm}_r \ ,\ee

as long as one insists in having hermitian $M_{AB}$ generators.
The twist which is needed here to have positive norms is

\be M_{AB}^\dagger \ = \ \s(M_{AB}) \ , \ee

where in this case

\be \s \ = 
 \ \left\{
\begin{array}{c}
+1\ , \ \textrm{on} \ M_{rs}, M_{0'r}=P_r \\
-1 \ , \ \textrm{on} \ M_{0r}, M_{0'0}\sim E\end{array} \right. \
.\ee

In this case, the $\s$-twist acts with an additional minus sign on
the energy and space translation generators, with respect to the
$\mso(D+1)$ case, so that they exchange their $\s$-eigenvalue. This
means that the leftover hermitian subalgebra changes, in this case,
and is indeed $\mso(D-1,1)$. As in the previous case, the
representation of the algebra on the lowest weight module defined
above cannot be made unitary.

\item $\mso(D,1)$ with tangent space signature $(D-1,1)$ ($dS_D$) \
: \ $\y_{AB}=(-,\d_{rs},+) $.

One can define

\be L^{\pm}_r \ = \ i M_{0'r}\pm i M_{0r} \ , \qquad E \ = \ i
M_{0'0} \ ,\ee

that satisfy the algebra (\ref{alg}), and $P_r=M_{0'r}$. The reality
conditions are, in this case,

\be E^\dagger \ = \ -E \ , \qquad (L^{\pm}_r)^\dagger \ = \
-L^{\pm}_r \ .\ee

Again, to have positive definite norms, one needs to twist the
reality conditions on generators, in the following way:

\be M_{AB}^\dagger \ = \ \s(M_{AB}) \ , \ee

where now

\be \s \ = \ \pi \ = \ \left\{
\begin{array}{c}
+1\ , \ \textrm{on} \ M_{rs}, M_{0r} \\
-1 \ , \ \textrm{on} \ M_{0'r}=P_r, M_{0'0}= E\end{array} \right. \
.\ee

Note that, in accordance with the action of the $\pi$ map
previously defined, here the whole space-time translation operator
$P_a=(E,P_r)$ becomes non hermitian as a consequence of the twist.
This amounts to say that the hermitian subalgebra is, in this
case, $\mso(D-1,1)$. The lowest weight realization of the $dS_D$
isometry algebra is then nonunitary.

\end{itemize}

It is therefore possible, in $AdS_D$, that the lowest weight
representations $\mD(e_0,(s_0))$ presented above are unitary, at
least for certain values of $e_0$ and $s_0$. One way to check this
is to check whether the norms of the various states of the
representation at hand are all positive, assuming that the norm of
the lowest weight state is, \emph{e.g.},
$\bra{e_0,(s_0)}e_0,(s_0)\rangle=1$. Equivalently, a unitarity
bound can be derived imposing that all states within a certain
lowest weight representation have the same value of the Casimir
operator $C_2$. Let us look, for example at a scalar
representation $\mD(e_0,(0))$, and in particular let us compare
the value of $C_2$ on the lowest weight state and on the scalar
excited state $\ket{e_0+2,(0)}$ obtained acting on the lowest
weight state with $L^+_rL^+_r$. We obtain
\bea e_0(e_0-D+1) \ = \ (e_0+2)(e_0-D+3)+|L^-_r\ket{e_0+2,(0)}|^2
\ , \eea
so that the requirement of unitarity implies the following lower
bound
\bea e_0\ \geq \ \e_0 . \eea
Note that this inequality is saturated for $e_0=\e_0$, \emph{i.e.},
for $L^-_r\ket{e_0+2,(0)}=0$, which tells us that the scalar
singleton is the representation that saturates the unitarity bound
for scalars.

A similar reasoning can be carried out for more complicated cases
\cite{Nicolai:1984hb, deWit:2002vz}. For a spin-(1/2)
representation, the bound is
\bea e_0\ \geq \ \e_0+\frac{1}{2}\ , \eea
which shows that the spinor singleton in $D=4$ ($\e_0=1/2$) is again
at the boundary of unitarity. For representations with $s_0\geq 1$,
on the other hand, one gets
\bea e_0\ \geq \ s_0+2\e_0\ , \eea
from which one sees that the massless representations
$\mD(s_0+2\e_0;(s_0))$ are unitary.

One can also notice that most of the ``conjugate'' representations
$\mD(D-1-e_0;(s_0))$ are nonunitary, with the exception of the
conjugate scalar fields $\mD(1,0)$ and $\mD(2,0)$ in four
dimensions. Given that the scalar singleton is a unitary
representation in any $D$, the massless lowest weight
representations into which the doubleton spectrum decomposes are
necessarily unitary in $AdS_D$ but nonunitary in other signatures.

\vspace{0.7cm}

However, the most important conclusion of this section is that the
spectrum of free equations we presented in Section \ref{frame}
contains indeed physical massless fields of every spin $s$, each
occurring once, and that, in order for it to fit a unitary module
of the infinite-dimensional extension $\mho(D-1,2)$ of the
$\mso(D-1,2)$ background isometry algebra it is necessary that
also a scalar enter the HS equations, since a scalar fields always
appears in the $\mso(D-1,2)$-decomposition of the tensor product
of the fundamental UIR of the HS algebra\footnote{Or in the
symmetric part of such tensor product if we compare with the
minimal HS algebra \eq{adj}.}, as it appears in \eq{RacRac}. This
gives a rationale for the introduction of a master zero-form
\eq{Phi}, that can contain such a scalar in a natural HS-covariant
``master field''. We shall examine further reasons for this choice
in the next Chapter.

\scs{Oscillator realizations}

The aim of this Section is to introduce two useful oscillator
realizations of the HS algebra, that are very much related to the
singleton representation. As we shall examine in Chapter
\ref{nonlin}, oscillator realizations are crucial for going to
full nonlinear level in the Vasiliev equations, essentially
because they allow to take ``square roots'' of the HS algebra
generators. The first oscillator realization we shall recall makes
use of ``vector'' oscillators $Y^A_i$, that have a vector index of
$\mso(D-1,2)$ and a doublet index of $\msp(2)$: they underlie a
$D$-dimensional formulation of Vasiliev equations
\cite{Vasiliev:2003ev, Bekaert:2005vh}, despite some subtlety due
to a redundancy that the $\msp(2)$ index brings in
\cite{mishasalg, Bekaert:2005vh, Sagnotti:2005ns}. The latter is
automatically absent in the four-dimensional spinor oscillator
realization, based on $\msl(2;\Comp)$-doublet oscillators, which
entered the first formulation of the Vasiliev equations in $D=4$
\cite{Vasiliev:en, Vasiliev:1992av, Vasiliev:vu, Vasiliev:1999ba,
Iazeolla:2004hj}, and that we shall review later on. For the
moment, we restrict our considerations to the $AdS$ case only.

\scss{Vector oscillator realization}\label{Y}

As stressed in \cite{Engquist:2005yt}, one can describe the
singleton as a massless particle living on the \emph{Dirac
hypercone} $X^A X_A=0$ in $\mathbb{R}^{D-1,2}$, where
$X^A=\sqrt{2}Y^A_1$ and $P^A=\sqrt{2}Y^A_2$ are its phase-space
coordinates. Upon quantization, one imposes the commutation
relations
\bea [Y^A_i,Y^A_j]_\star \ = \ 2i\e_{ij}\y^{AB} \ ,\label{Ycomm}\eea
where $\e_{ij}=-\e{ji}$ is the invariant tensor of $\msp(2)$ and we
use the conventions $\e^{ij}V_j=V^i$, $V^j\e{ji}=V_i$,
$\e^{ik}\e_{jk}=\d^i_j$. We are here, again, making use of a
$\star$-product defined on the oscillators that implements the
operator product on Weyl-ordered (i.e., totally symmetric)
combinations of oscillators, that we will simply denote by
juxtaposition,
\bea Y^A_{i_1}...Y^A_{i_n} \ = \
\frac{1}{n!}\sum_{\pi\in\cS_n}Y^{A_{\pi_1}}_{i_{\pi_1}}\star
...\star Y^{A_{\pi_n}}_{i_{\pi_n}}\ .\eea
The totally symmetric ordering of oscillators is preferred in that
it preserves $\msp(2)$-covariance, differently from other choices
such as the normal ordering of the creation/annihilation operators
$X^A\pm iP^a\sim a,a^\dagger$. More generally, for two Weyl-ordered
functions of oscillators,
\bea f(Y)\star g(Y) \ = \
f(Y)\exp\left(i\e^{ij}\y^{AB}\overleftarrow{\frac{\del}{\del
Y^{Ai}}}\overrightarrow{\frac{\del}{\del Y^{Bj}}}\right)g(Y) \ ,\eea
which admits the equivalent integral presentation
\bea  f(Y)\star g(Y) \ = \ \frac{1}{\pi^{2(D+1)}}\int dS dT\,
f(Y+S)\,g(Y+T)\exp(-iS_i^A T^i_A) \ .\eea

The constraints ensuring that this particle be massless, $P^2\approx
0$, and that it live on the hypercone $X^2\approx 0$ (i.e., in one
dimension less), together with their commutator, proportional to
${X_A,P_A}\approx 0$, which imposes symmetry under dilatations and
thus independence on the radial direction, actually characterize
such particle as a conformal massless particle in
$(D-1)$-dimensions, \emph{i.e.}, a singleton. One can see this also
from the fact that the constraints above generate $\msp(2)$, since
they correspond to the three independent components in
\bea K_{ij} \ = \ \frac{1}{2}Y^A_i Y_{Aj} \ , \label{sp2}\eea
with commutation relations
\bea [K_{ij},K_{kl}]_\star \ = \ 4i\e_{(i|(k}K_{l)|j)}\ ,\eea
which means that one can impose the constraints above by declaring
that the every state $\ket{\psi}$ in the Hilbert space of the
conformal particle is annihilated by $K_{ij}$. Notice also that,
with vector oscillators, the generators of $\mso(D-1,2)$ can be
realized as the bilinears\footnote{Note also that
$[M_{AB},K_{ij}]_\star=0$.}
\bea M_{AB} \ = \ \frac{1}{2}Y^{i}_A Y_{iB} \ ,\label{MY}\eea
and let us also introduce the combinations
\bea L_{ij,AB} \ \equiv \ \frac{1}{2}Y_{Ai}
Y_{Aj}-\frac{\y_{AB}}{D+1}K_{ij}\ , \eea
which are traceless in $A,B$. With the help of these constructs,
one can realize the generators $V_{AB}$ of the annihilating ideal
of the singleton $\cI[V]$ as
\bea V_{AB} \ = \ K^{ij}\star L_{ij,AB} \ = \ L_{ij,AB}\star K^{ij}
\ ,\eea
which means that imposing the $\msp(2)$-invariance of the states
amounts to describe the singleton weight space! Indeed, the other
constraint that generates $\cI[V]$ is trivially satisfied in this
setting, due to the fact that out of an $\msp(2)$-doublet such as
$Y^A_i$ one cannot form combination with more than two antisymmetric
indices $A,B$. Finally, the quadratic Casimir operator matches that
of the singleton, since
\bea C_2 \ = \ \frac{1}{2}M^{AB}\star M_{AB} \ = \
\frac{1}{2}K^{ij}\star K_{ij} -\e_0(\e_0+2) \ ,\eea
which gives the desired result on every state $\ket{\psi}$ such that
$K^{ij}\ket{\psi}=0$.

The realization of a HS algebra follows straightforwardly from the
realization of the generators of $\mso(D-1,2)$. It is sufficient to
consider arbitrary functions of the oscillators $f(Y)$ subject to
the conditions of being $\msp(2)$-singlets,
\bea [K_{ij},f(Y)]_\star \ = \ 0 \ , \eea
This restricts the generators of such an associative algebra, that
we shall denote with $\cS$, to be products of oscillators with
symmetry properties encoded into two-rows rectangular
$\mso(D-1,2)$-Young diagrams. However, the algebra formed by such
objects is still reducible, as it contains the left and right
ideal $\cI'$ spanned by all elements $g(Y)$ proportional to the
$\msp(2)$ generator, \emph{i.e.} of the form $g_{ij}\star K^{ij}=
K^{ij}\star g_{ij}$. Due to the definition of $K_{ij}$ \eq{sp2},
all traces of two-row Young diagrams are contained in $\cI'$.
After factoring it out, the resulting associative algebra $\cA =
\cS/\cI'$ contains only all traceless two-row rectangular Young
diagrams, and coincides with the associative algebra $\cA$ defined
in \eq{calA1}, generated by \eq{TAnBn}. One can then realize from
it the minimal bosonic HS algebra $\mho(D-1,2)$ \eq{adj} defining
the action of the antiautomorphism of the oscillator algebra
\eq{Ycomm} $\t$ on functions of the oscillators,
\bea \t(f(Y^A_i)) \ = \ f(iY^A_i) \ .\eea
Moreover, the reality conditions are
\bea (Y^A_i)^\dagger \ = \ Y^A_i \ .\label{Yreal}\eea
It is possible in particular to realize the $AdS$-translation
generator projecting one index onto the embedding direction, which
can be done covariantly making use of the compensator $V_A$
introduced in Section \ref{frame},
\bea P_a \ = \ \frac{1}{2}Y^i_A Y_{Bi}V^B \ ,\eea
which in the standard gauge becomes
\bea P_a \ = \ \frac{1}{2}Y^i_aY_{0'i} \ \equiv \ \frac{1}{2}Y^i_a
y_{i}\ .\label{PY}\eea
The action of the $\pi$ automorphism is
\bea \pi(f(Y^A_i)) \ = \ \pi (f(Y^a_i,y_i)) \ = \ f(Y^a_i,-y_i)\
,\eea
\emph{i.e.}, $\pi$ acts as a parity in the embedding direction.
The realization of the adjoint master one-form therefore follows
immediately putting \eq{MY} and \eq{PY} into \eq{A} and \eq{As},
and similarly for the twisted adjoint master zero-form \eq{Phi}
and \eq{Phis}. Notice that one can obtain real component fields
with the conditions \eq{Yreal} imposing
\bea A^\dagger \ = \ -A \ , \qquad \Phi^\dagger \ = \ \pi(\Phi) \
.\eea

As mentioned in the first Section of this Chapter, referring to
the singleton by factoring out its annihilating ideal is much more
convenient than doing it by working at the level of Hilbert
spaces, especially in view of formulating HS-covariant full field
equations that one would like to write in a manifestly background
covariant form. However, how to project out the traces,
concretely, at the level of the full field equations involves some
subtleties \cite{Bekaert:2005vh, Sagnotti:2005ns, mishasalg}. Such
factorization is however very important in the Vasiliev equations,
as it encodes standard second order dynamical equations in a set
of HS-covariant first order curvature constraints.

\scss{4D spinor oscillator realization}\label{y}

The factorization is automatic in the realization of
four-dimensional HS algebras in terms of commuting spinor
oscillators $y_\a,\yb_{\ad}$ of $\msl(2;\Comp)$ (first proposed in
\cite{V3, Fradkin:ah}) satisfying
\bea [y_\a,y_\b]_\star \ = \ 2i\e_{\a\b} \ , \qquad
[\yb_{\ad},\yb_{\bd}]_\star \ = \ 2i\e_{\ad\bd} \ ,\eea
that is to say
\bea y_\a\star y_\b \ = \ y_\a y_\b + i\e_{\a\b} \ ,& \qquad &
\yb_{\ad}\star\yb_{\bd} \ = \ \yb_{\ad}\yb_{\bd}+i\e_{\ad\bd} \ ,\\[5pt]
y_\a\star\yb_{\bd} & = & y_\a\yb_{\bd} \ , \eea
where $\e_{\a\b}$ is the invariant tensor of $\msl(2;\Comp)$ and our
spinor conventions are collected in Appendix \ref{App:F}. These are
particular cases of the most general $ \star$-product rule
\begin{equation}\label{defdiff}
f(y,\bar{y})\star g(y,\bar{y}) \ = \
f(y,\bar{y})e^{-i(\overleftarrow{\partial}^{\alpha}\overrightarrow{\partial}_{\alpha}+\overleftarrow{\partial}^{\dot{\alpha}}\overrightarrow{\partial}_{\dot{\alpha}})}g(y,\bar{y})
\ ,
\end{equation}
where $\del_\a\equiv \frac{\del}{\del y^\a}$, or, equivalently,
\begin{equation}\label{defint}
(f\star g)(y,\bar{y}) \ = \ \int \frac{d^{4}u d^{4}v}{(2\pi)^{4}}\,
f(y+u,\yb+\bar{u})\,g(y+v,\yb+\bar{v})\exp i(u^{\a}v_{\a}+
\bar{u}^{\ad}\bar{v}_{\ad})\ .
\end{equation}
The realization of the Lorentz and $AdS$ translation generators is
  \be
  M_{ab}\ =\ -\frac18 \left[~ (\s_{ab})^{\a\b}y_\a y_\b+
  (\sb_{ab})^{\ad\bd}\tilde y_{\ad}\yb_{\bd}~\right]\
,\qquad P_{a}\ =\
  \frac{1}{4} (\s_a)^{\a\bd}y_\a \yb_{\bd}\ .\label{mab2}
  \ee
To realize the HS algebra, here it is sufficient to consider the
associative algebra spanned by all possible monomials in
oscillators,
\bea T_{\a_1\ldots\a_n, {\ad}_1\ldots{\ad}_m} \ \equiv \
T_{\a(n),{\ad}(m)}\ ,\label{ymon}\eea
that have spin $s=\frac{n+m}{2}$, whose elements are therefore all
possible functions
\bea f(y,\yb) \ = \ \sum_{n,m}f^{\a(n),{\ad}(m)}T_{\a(n),{\ad}(m)} \
.\label{oscelement}\eea
Compared to the vector oscillator realization, however, this is a
simpler setting, due to the lack of the additional $\msp(2)$
redundancy. In particular, here traces are automatically factored
out, since for commuting spinors $y^\a y_\a =
\yb^{\ad}\yb_{\ad}=0$. In other words, the elements
\eq{oscelement} indeed span the four-dimensional associative
algebra $\cA$ \eq{calA1}. Here one can introduce the
antiautomorphism $\t$ as
\bea \t(f(y,\yb)) \ = \ f(iy,i\yb)\ ,\eea
and the automorphisms $\pi,\pb$, distinguishing Lorentz rotations
and $AdS$ translations, as
\bea \pi(f(y,\yb)) \ = \ f(-y,\yb)\ , \qquad \pb(f(y,\yb)) \ = \
f(y,-\yb)\ .\eea
Therefore, by imposing the $\t$-condition \eq{adj} on the elements
\eq{oscelement} one truncates the model to generators with
$m+n=2,6,10,...$, that exactly correspond to gauge fields with spin
$2,4,6,...$, \emph{i.e.}, one obtains the minimal bosonic HS algebra
$\mho(3,2)$. The adjoint master one-form can be written as
\bea A(x;y,\yb) \ = \ \frac{1}{2i}\sum_{\ell=0}^\infty A^{(\ell)}\
,\eea
where each level $\ell$ is expanded as
\begin{eqnarray}\label{master 1-form}
A^{(\ell)}(x;y,\bar{y}) \ = \
\sum_{n+m=4\ell+2}\frac{1}{n!m!}dx^{\mu}A_{\mu}^{(\ell)\,\alpha_{1}...\alpha_{n}\dot{\alpha}_{1}...\dot{\alpha}_{m}}(x)y_{\alpha_{1}}...y_{\alpha_{n}}\bar{y}_{\dot{\alpha}_{1}}...\bar{y}_{\dot{\alpha}_{m}}
\ ,
\end{eqnarray}
while imposing $\t(\Phi) = \pi(\Phi)$ one arrives at the
realization of the twisted adjoint master zero-form,
\bea \Phi(x;y,\yb) \ = \ \sum_{\ell=-1}^\infty \Phi^{(\ell)}\ ,\eea
where
\begin{equation}\label{master $0$-form}
\Phi^{(\ell)}(x|y,\bar{y}) \ = \
\sum_{|n-m|=4\ell+4}\frac{1}{n!m!}\Phi^{(\ell)\,\alpha_{1}...\alpha_{n}\dot{\alpha}_{1}...\dot{\alpha}_{m}}(x)
\,y_{\alpha_{1}}...y_{\alpha_{n}}\bar{y}_{\dot{\alpha}_{1}}...\bar{y}_{\dot{\alpha}_{n}}\
.
\end{equation}

We are also assuming the reality condition
\bea y_\a^\dagger \ = \ \yb_{\ad} \ .\eea
In Chapter \ref{exactsol} we will also examine more general spinor
oscillator realizations for the various signatures $\mso(p',5-p')$.

Another useful oscillator realization that we will make use of is
given in terms of linear combinations of the $y,\yb$ oscillators
that correspond to creation/annihilation operators building up the
Fock space of states of the scalar and spinor four-dimensional
singletons. In particular, we introduce the $\msu(2)$-doublet
$a_i,a^{\dagger\,i}$, with $a^{\dagger~i}=(a_i)^\dagger$ and $i=1,2$
as
\bea a_1 & = & \frac12 (y_1+i\yb_{\dot{2}}) \ , \qquad
a^{\dagger\,1}
\ = \ \frac12 (\yb_{\dot{1}}-iy_2) \ ,\\[5pt]
a_2 & = & \frac12 (-y_2+i\yb^{\dot{1}}) \ , \qquad a^{\dagger\,2} \
= \ \frac12 (-\yb_{\dot{2}}-iy_1) \ .\eea
They satisfy the following Heisenberg algebra
\bea [a_i,a^{\dagger\,j}]_{\star} \ = \ \d^j_i\ ,\label{acomm}\eea
and in terms of them the $\mso(3,2)$ generators can be expressed as
\bea E & = & \frac{1}{2}(a^{\dagger\,i}a_i+1)\ , \qquad M_{rs} \ = \
\frac{i}{2}(\sigma_{rs})_i{}^j a^{\dagger\,i}a_j \ , \\[5pt]
L^+_r & = & \frac{i}{2}(\sigma_r)_{ij}a^{\dagger\,i}a^{\dagger\,j} \
, \qquad L^-_r=\frac{i}{2}(\sigma_r)^{ij}a_{i}a_{j} \ .\eea
In terms of such oscillators one can build a Fock space on top of
the scalar singleton lowest weight state, which is declared to be
annihilated by the $a_i$, $a_i\ket{1/2,0}=0$,
\bea \cF \ = \
\textrm{span}\left\{a^{\dagger\,i_1}...a^{\dagger\,i_n}\ket{1/2,0}\right\}_{n=0}^\infty
\ ,\label{fock}\eea
that is reducible under the action of $\mso(3,2)$ generators:
indeed, being the latter bilinears in oscillators, their action
splits the Fock space into the two irreducible even and odd
subspaces,
\bea \cF \ = \ \cF_{even}\oplus\cF_{odd}\ ,\eea
that contain states with even or odd number of oscillators,
respectively. Notice that the lowest weight state of the latter is
\bea \ket{1,1/2}^i \ = \ a^{\dagger\,i}\ket{1/2,0} \ , \eea
\emph{i.e.}, the lowest weight state of the spinor singleton irrep
$\mD(1,1/2)$. The two singleton irreps are therefore connected very
naturally in the oscillator realization, in which a number of other
relations are manifest: for example, the lowest weights of the two
scalar $\mD(1,0)$ and $\mD(2,0)$, that are both composites and read
indeed (see \eq{RacRac} and \eq{DiDi})
\bea \ket{1,0} \ = \ \ket{1/2,0}_1\ket{1/2,0}_2 \ , \qquad \ket{2,0}
\ = \ \ket{1,1/2}^i_1\ket{1,1/2}_{2\,i} \ ,\eea
are also related by the oscillator combination
\bea \ket{2,0} \ = \ -y\ket{1,0}\ , \qquad y\ = \ a_i^\dagger(1)
a^{\dagger\,i}(2) \ .\eea
Similar relations will be useful in Chapter \ref{map}.

\chapter{Unfolded formulation}\label{unfolding}

As recalled in the Introduction, although certain gauge-invariant
vertices involving massless HS fields can be, and indeed were,
obtained in the early and mid-Eighties, addressing the full
``HS-interaction problem'' is much more demanding, for reasons that
we briefly repeat here for the reader's convenience:

\begin{enumerate}
\item A consistent interacting HSGT requires the simultaneous
introductions of infinitely many gauge fields of all spins;

\item The interaction with gravity is consistent with the HS gauge
symmetries only on a nonflat gravitational background, \emph{i.e.},
in presence of a nonvanishing cosmological constant $\L$, since
interaction terms are nonanalytical in $\L$;

\item HS interaction vertices require higher derivatives of the
physical fields involved. This property is strictly connected with
the previous one, since, in order for the physical dimension of the
lagrangian to be preserved with more than two derivatives, a
dimensionful parameter must enter the vertices, and $\L$ is the only
candidate in a field theoretical context.

\end{enumerate}

A step forward was described in the previous Chapter, where an
infinite-dimensional non-abelian HS algebra was constructed. As in
YM theories and in gravity, the invariance under the proper local
gauge transformations is the key to determine the form of the
interaction terms. However, the above-mentioned peculiar features
are important differences that make a ``traditional'' analysis
unyielding: first of all, a full HSGT will involve an infinite tower
of fields of different spin and, correspondingly, infinitely many
gauge symmetries, which makes impossible an order-by-order analysis
in terms of each different gauge fields. In order to overcome such a
problem, it is necessary to work with the proper variables,
\emph{i.e.}, with some proper ``superfields''\footnote{We are
borrowing this term from the well-known case of supersymmetric
theories, although, as it will be made clear in the rest of the
Chapter, here we do not mean that supercharges enter among the
symmetry generators of the theory (although they could, in
principle, since supersymmetric extensions of HS gauge theories have
indeed been constructed).} that have nice transformations properties
under the local HS symmetry, and that therefore enable to control at
once the whole tower of massless fields. We will see that this also
leads to a natural way out of the third problem, namely the
complication introduced by the fact that unbroken HS symmetry
implies no bound on derivatives in the vertices: indeed, a clever
first-order reformulation of the dynamics that involves crucially
the ``superfields'' as main variables will offer a way out of this
by enabling to ``hide'' higher derivatives in the component fields.
As the order of derivatives in the interaction vertices grows with
the spin, it is intuitively clear that the ``superfields'' involved
not only will have to encode infinitely many components to
accommodate all spins, but also, presumably, infinitely many
components for each spin-$s$ sector, since each spin can a priori
interact with any other one in couplings featuring higher and higher
derivatives of the lower spin field.

The end result will be that the required ``superfields'' will be the
adjoint master one-form and the twisted adjoint master zero-form
introduced in the previous chapter, and that the proper first-order
formulation of the dynamics is the so-called \emph{unfolded
formulation} (in which the twisted adjoint plays a crucial role),
that enables to write HS field equations in the form of
zero-curvature constraints. This is particularly appealing since
such form of the equations makes it easy to control
gauge-invariance, and this is especially important in view of the
search for consistent nonlinear deformations. Indeed, such a
formalism is today the only approach to full HS field equations,
although an action principle from which to derive them is still not
known. Moreover, \emph{unfolding} just means that every field enters
the field equations together with all its ``descendants'',
\emph{i.e.}, with all its derivatives: although this looks
unconventional, this will enable a ``canonical'', uniform treatment
of HS interactions. Finally, the introduction of infinitely many
derivatives will turn out not to be redundant, and the
zero-curvature equations will turn out to encode nontrivial dynamics
thanks to trace constraints on the component fields.

The unfolded formulation is a particular and extremely interesting
case of more general constructions known as \emph{free
differential algebras} (FDA) that we shall present first. As we
shall see, its peculiarity lies in the introduction of the
infinite-dimensional set of twisted-adjoint zero-forms within the
general FDA scheme. Indeed, as previously mentioned, this idea,
due to M. A. Vasiliev and first suggested in \cite{Vasiliev:sa},
is crucial for encoding nontrivial dynamics in a set of
zero-curvature conditions.

\scs{Free Differential Algebras}\label{defs}

FDA were first introduced in physics by D'Auria and Fr\`e
\cite{D'Auria:nx} (see \cite{Castellani:1992sv} for a review) as
 a way to formulate various supergravity theories
containing differential forms of higher degree through
zero-curvature equations. These generalize the Maurer-Cartan
equations that define an algebra through the dual cotangent basis
of one-forms of the corresponding Lie group manifold.

Let us consider an arbitrary set of differential $p-$forms
$W^\alpha\in\Omega^{p_\a}({\cal M}^D)$ with $p_\alpha\geq 0$
($0$-forms are included) and $\a$ is an index enumerating various
forms, which, in principle, may range in the infinite set $1\leq
\a<\infty$.

Let $R^\a \in\Omega^{p_\a+1}({\cal M}^D)$ be generalized
curvatures defined by the relations \be \label{uncur} R^\a= dW^\a
+G^\a(W^\b)\,, \ee where $G^\a(W^\b)$ are some power series in
$W^\b$ built with the aid of the exterior product of differential
forms (that is understood wherever is needed),
\bea G^\a(W^\b) \ = \ \sum_{n=1}^\infty
f^\a_{\b_1\ldots\b_n}W^{\b_1}...W^{\b_n} \ .\eea
The (anti)symmetry properties of the structure constants
$f^\a_{\b_1\ldots\b_n}$ are such that $f^\a_{\b_1\ldots\b_n}\neq
0$ for $p_a+1=\sum_{i=1}^n p_{\b_i}$, and the permutation of any
two indices $\b_i$ and $\b_j$ brings a factor of
$(-1)^{p_{\b_i}p_{\b_j}}$ (in the case of bosonic fields,
\emph{i.e.} with no extra Grassmann grading in addition to that of
the exterior algebra).

A function $G^\a(W^\b)$ satisfying the generalized Jacobi identity
\bea G^\b \frac{\delta^L G^\a }{\delta W^\b} \equiv 0\,
\label{prop}\eea (the derivative with respect to $W^\b$ is acting
from the left) defines a free differential algebra\footnote{We
remind the reader that a differential $d$ is a Grassmann odd
nilpotent derivation of degree one, {\it i.e.} it satisfies the
(graded) Leibnitz rule and $d^2=0$. A differential algebra is a
graded algebra endowed with a differential $d$. Actually, the
``free differential algebras" (in physicist terminology) are more
precisely christened ``graded commutative free differential
algebra" by mathematicians (this means that the algebra does not
obey algebraic relations apart from graded commutativity). In the
absence of $0$-forms, which however play a key r{\`o}le in the
unfolded dynamics construction, the structure of these algebras is
classified by Sullivan \cite{Sullivan}.}. We emphasize that the
property (\ref{prop}) is a condition on the function $G^\a (W)$ to
be satisfied identically for all $W^\b$. It is equivalent to the
following generalized Jacobi identity on the structure
coefficients
\bea \sum_{n=0}^m (n+1)f^\c_{[\b_1...\b_{m-n}}
f^\a_{\c\b_{m-n+1}...\b_m\}} \ = \ 0\ ,\eea
where the brackets $[ ... \}$ denote an appropriate
(anti)symmetrization of all indices $\b_i$. Strictly speaking, the
generalized Jacobi identities \eq{prop} have to be satisfied only
at $p_\a < D$ for the case of a D-dimensional manifold ${\cal
M}^D$ where any $(D + 1)$-form is zero. We shall call a free
differential algebra universal if the generalized Jacobi identity
holds for all values of the indices, \emph{i.e.}, independently of
a particular choice of space-time dimension. The HS free
differential algebras discussed in this paper belong to the
universal class.

The property (\ref{prop}) guarantees the generalized Bianchi
identity
$$ dR^\a = R^\b \frac{\delta^L G^\a }{\delta W^\b}\,, $$ which tells
us that the differential equations on $W^\b$ \be \label{eq} R^\a
=0 \ee are consistent with $d^2=0$ and supercommutativity.
Conversely, the property (\ref{prop}) is necessary for the
consistency of eq. (\ref{eq}).

One defines the gauge transformations as \be \label{delw} \delta
W^\a = d \varepsilon^\a -\varepsilon^\b \frac{\delta^L G^\a
}{\delta W^\b}\,, \ee where $\varepsilon^\a (x) $ has form degree
equal to $p_\a -1$ (so that $0$-forms $W^\a$ do not give rise to
any gauge parameter). With respect to these gauge transformations
the generalized curvatures transform as $$ \delta R^\a =-R^\c
\frac{\delta^L }{\delta W^\c} \left (\varepsilon^\b \frac{\delta^L
G^\a }{\delta W^\b} \right )\,, $$ due to the property
(\ref{prop}). This implies the gauge invariance of the equations
(\ref{eq}). Also, since the equations (\ref{eq}) are formulated
entirely in terms of differential forms, they are explicitly
general coordinate invariant. In fact, the diffeomorphisms are
incorporated in the gauge group, since the Lie derivative $\cL_\xi
W^\a\equiv \{d,i_\xi\}W^\a$, where $i_\xi$ is the inner derivative
with respect to a vector field $\xi=\xi^\m\del_\m$, is equivalent,
up to vanishing curvatures, to a field-dependent gauge
transformation with parameters $\e^\a=i_\xi W^\a$ (see Appendix
\ref{sptime} for an example).

\scss{Unfolding strategy}\label{unffda}

Unfolding means reformulation of the dynamics of one or another
system in the form (\ref{eq}) which, as we explain below, is always
possible by virtue of introducing enough auxiliary fields. Note
that, according to (\ref{uncur}), in this approach exterior
differential of all fields is expressed in terms of the fields
themselves, a feature that we had already encountered in Section
\ref{frame} in the context of the MMSW-reformulation of gravity.

The case of a FDA that only contains one-forms coincides with the
usual Maurer-Cartan dual formulation of an algebra. Indeed, let
$h$ be a Lie (super)algebra, a basis of which is the set
$\{T_\a\}$, and $\o=\o^\a T_\a$ be a 1-form taking values in $h$.
Choosing $G (\,\o)=\o^2\equiv \frac{1}{2} \o^\a \o^\b [T_\a , T_\b
]$, then eq. (\ref{eq}) with $W=\o$ is the zero-curvature equation
$d\o+\o^2=0$, and imposes the Maurer-Cartan equations on $\o^\a$,
\bea d\o^\a+\frac{1}{2}f^\a_{\b\c}\o^\b\o^\c \ = \ 0\ . \eea
The relation (\ref{prop}) then amounts to the usual Jacobi
identity for the Lie algebra $h$. In the same way, (\ref{delw}) is
the usual gauge transformation of the connection $\o$,
\bea \d\o^\a \ = \ D\e^a \ = \ d\e^\a+f^\a_{\b\c}\o^\b\e^\c \ .\eea
Note again that the whole philosophy of the MMSW reformulation of
gravity examined in Section \ref{frame} consisted in a
reformulation of gravity in AdS as a FDA with only one-forms, in
such a way that the background emerges in a coordinate-independent
way as a maximally symmetric solution $\o_0$ of the zero-curvature
field equation with its global stability algebra $h$ that solves
$\d\o_0^\a = 0$. Recall that also free HS fields were analyzed in
Section \ref{frameHS} as fluctuations around such a background.

If now the set $W^\a$ also contains some $p$-forms denoted by
$\cc^i$ ({\it e.g.} $0$-forms) and if the functions $G^i$ are linear
in $\o$ and $C$, \be \label{lin}
   G^i = \o^\a(T_\a)^i {}_j \cc^j\,,
\ee
   then the relation
(\ref{prop}) implies that the coefficients $(T_\a)^i {}_j$ define
some matrices $T_{\a}$ forming a representation $T$ of $h$, acting
in a module $V$ where the $\cc^i$ take their values. The
corresponding equation (\ref{eq}) is a covariant constancy
condition $D_\o \cc=0$, where $D_\o\equiv d+\o$ is the covariant
derivative in the $h$-module $V$, and admits the gauge symmetry
\bea \d\cc^i \ = \ \e^\a(T_\a)^i {}_j \cc^j \ .\eea
Suppose now that $\cc^i$ are zero-forms $C^i$. Notice that, if we
pursue the strategy of perturbatively expanding our FDA around a
vacuum solution $\o_0$, and we treat both the remaining one-forms
$\o$ and the zero-forms as small fluctuations, the equation for
the one-forms becomes
\bea d\o^\a+f^\a_{\b\c}\o_0^\b\o^\c+f^\a_{\b\c i}\o^\b_0\o^\c_0 C^i
\ = \ 0 \ . \label{deformcurv}\eea
We note that the vacuum equation $d\o_0+\o_0^2=0$ together with
\eq{deformcurv} and the zero-form equation $dC^i+G^i =0$ form a
consistent set of equations: indeed, the latter ensures
compatibility of the second with $d^2=0$, while the first is the
consistency condition of the last (and also its own consistency
condition). We shall see in the next Sections that free HS field
dynamics can be reformulated in this way - as, in fact, every
dynamical system, upon the addition of sufficiently many auxiliary
fields.

One may wonder how the set of equations
\bea d\o_0+\o_0^2 & = & 0 \ ,\\[5pt]
 D_{\o_0} C & = & 0 \label{fda0} \eea
could describe any dynamics, giving that it implies that (locally)
the connection $\o_0$ is pure gauge and $C$ is covariantly
constant, so that
\bea\o_0(x)&=&g^{-1}(x)\,dg(x)\,, \label{pureg1} \\
C(x)&=&g(x)\, \mathrm{C}\,,\label{pureg}\eea
where $g(x)$ is some function of the position $x$ taking values in
the Lie group associated with $h$ (by exponentiation), and
$\mathrm{C}$ is a constant vector of the $h$-module $T$. Since the
gauge parameter $g(x)$ does not carry any physical degrees of
freedom, all physical information is contained in the value
$C(x_0)=g(x_0)\mathrm{C}$ of the $0$-form $C(x)$ in a fixed point
$x_0$ of space-time. But as one shall see in the next Section, if
the $0$-form $C(x)$ somehow parametrizes all derivatives of the
original dynamical fields, then, supplemented with some algebraic
constraints (that, in turn, single out an appropriate $h$-module),
it can actually describe nontrivial dynamics. Indeed, the
restrictions imposed on the values of some $0$-forms at a fixed
point $x_0$ can lead to a nontrivial dynamics if the set of
$0$-forms is rich enough to describe all space-time derivatives of
the dynamical fields in a fixed point of space-time, provided that
the constraints just single out those values of the derivatives
which are compatible with the original dynamical equations.
Knowing a solution (\ref{pureg}) one knows all derivatives of the
dynamical fields compatible with the field equations, and can
therefore reconstruct these fields by analyticity in some
neighborhood of $x_0$.

The $p$-forms with $p>0$, that also satisfy a zero-curvature
condition, are still pure gauge in this setting. As will be clear
from the examples below, the meaning of the $0$-forms $C$
contained in $\cc$ is that they describe all gauge invariant
degrees of freedom ({\it e.g.} the spin-$0$ scalar field, the
spin-$1$ Maxwell field strength, the spin-$2$ Weyl tensor, etc.,
and all their on-mass-shell nontrivial derivatives). When the
gauge invariant $0$-forms are identified with derivatives of the
gauge fields which are $p>0$ forms, this is expressed by a
deformation of the equation of the latter,
 \be D_{\o_0} \cc= P(\o_0 ) \cc\,, \label{fdadel} \ee where $P(\o_0 ) $ is a linear operator
(depending on $\o_0$ at least quadratically) acting on $\cc$, as
seen explicitly in \eq{deformcurv}. If the deformation is trivial,
one can get rid of the terms on the right-hand-side of
(\ref{fdadel}) by a field redefinition. The interesting case
therefore is when the deformation is nontrivial. A useful criterium
for telling whether the deformation (\ref{fdadel}) is trivial or not
is given in terms of the $\sigma_-$ cohomology, discussed at length
in \cite{Bekaert:2005vh}.

Let us now stress some of the advantages of the unfolded formulation
(see \cite{Vasiliev:2007yc} for more comments) to understand why it
is a useful for gauge theories in general and, in particular, HS
gauge theories:

\begin{itemize}

\item As we have already stressed elsewhere, HS gauge
transformations mix fields of different spins: in particular, the
metric is not left invariant. This conflicts with the standard
implementation of general covariance in General Relativity, where
the inverse metric plays a crucial role. Therefore, the unfolded
formulation, where manifest gauge invariance and invariance under
diffeomorphisms (\emph{i.e.}, coordinate independence) is achieved
using the exterior algebra formalism and without any need \emph{a
priori} for singling out the metric, is perfectly suited for the
study of gauge invariant theories in the framework of gravity and,
in particular, HS gauge theories.

\item As seen above, in the topologically trivial situation, the
degrees of freedom are concentrated in the zero-forms at any point
$p$ in space-time. Indeed, the unfolded curvature constraints solve
such zero-forms in terms of all on-shell nontrivial derivatives of
the physical fields, that can be therefore reconstructed in a
neighborhood of $p$. This implies that, in order to describe a
system with an infinite number of degrees of freedom, it is
necessary to work with an infinite set of zero-forms that spans an
infinite-dimensional module of the space-time symmetry algebra
$g$\footnote{We will indeed construct, in Chapter \ref{map}, a
mapping between the operators contained in the twisted adjoint
zero-form at a fixed point in space-time and the states of the
doubleton spectrum, \emph{i.e.}, various massless irreducible
representation
$(\mD_0\otimes\mD_0)_S=\bigoplus_{s_0=0,2,4,...}\mD(s_0+2\e_0,s_0)$
(see also \cite{Shaynkman:2001ip} for related statements in the
context of conformal HS symmetries.).}. On the other hand, if the
set of zero-forms is finite, the corresponding unfolded system is
topological, describing at most a finite number of degrees of
freedom.

\item The unfolded formulation is thus an ultra-local approach to
the dynamical problem that is particularly appealing if one is
aiming at a background independent formulation of gauge theories:
indeed, thanks to the the introduction of an infinite set of
zero-forms, it is possible to achieve a generally covariant and
dynamically nontrivial formulation of gauge theories where the
metric is treated on an equal footing with the other fields. This is 
to be contrasted with what happens in Chern-Simons-like theories
that, although diffeomorphism invariant without involving
contractions of the indices by the inverse metric, are
topological\footnote{As pointed out in \cite{Banados:1995mq}
however, it is possible to encode local degrees of freedom in a
Chern-Simons (CS) theory by expanding it around a
\emph{non}-maximally symmetric solution. This is at the root of an
attempt \cite{Engquist:2007kz} towards the formulation of an
action principle for full HSGT starting from a CS-like action in
odd dimensions that does not make use of zero-forms.}. This in
particular means that in the full HS equations of motion the
inverse vielbein never appears, and therefore the theory also
naturally incorporates classical solutions with degenerate metrics
(as we shall see explicitly in Chapter \ref{exactsol}), that have
long been conjectured to be of importance in quantum gravity as
they can mediate space-time topology changes (see
\cite{Horowitz:1990qb} and references therein).

\item Equations \eq{pureg1} and \eq{pureg} show, in particular,
that the unfolded formulation based on universal FDAs makes the
dependence on space-time coordinates purely auxiliary. The dynamics
is entirely encoded in the functions $G^\a(W)$. This fact proves to
be extremely useful in the search for consistent HS interactions:
indeed, it means that one can search for them looking for
deformations of the $G^\a(W)$ that still respect a generalized
Jacobi identity, \emph{i.e.}, that preserve the consistency of the
system. As we shall see in the next Chapter, it also enables one to
express the whole infinite perturbative series of nonlinear
corrections to the free unfolded system as solution of some
additional equations of an enlarged unfolded system, where
differentials and differential forms live in a larger space: this
does not spoil consistency, nor alters the local dynamics, that is
still determined by the zero-forms at a point in space-time, as long
as the additional equations locally reconstruct the dependence on
the additional coordinates in terms of the original degrees of
freedom.

\item The unfolded formulation is in principle available for any
dynamical system, provided one introduces additional auxiliary
variables, since, as stressed in \cite{Vasiliev:2005zu}, it is
nothing but a generally covariant first-order formalism.

\end{itemize}

We now have to determine what is the appropriate
infinite-dimensional module of the space-time symmetry algebra in
which the zero-forms have to take values, in order to encode
nontrivial dynamics. This will be explained through the unfolding of
free lower-spin fields.

\scs{Unfolding of lower spins}

In this Section, we first show how one can indeed unfold an
arbitrary system \cite{Vasiliev:1992gr} and then apply such
technique for the free spin 0 and spin 2 system
\cite{Bekaert:2005vh}.

\scss{General procedure}

Let $\o_0=e_0^a\,P_a+\frac12 \o_0^{ab}M_{ab}$ be a vacuum
gravitational gauge field taking values in some space-time symmetry
algebra $s$. Let $C^{(0)}(x)$ be a given space-time field satisfying
some dynamical equations to be unfolded. Consider for simplicity the
case where $C^{(0)}(x)$ is a $0$-form. The general procedure of
unfolding free field equations goes schematically as follows:

For a start, one writes the equation \be D_0^L C^{(0)} \,=\,
e_0^a\,\,\, C_a^{(1)}\,,\label{unf1} \ee where $D_0^L$ is the
covariant Lorentz derivative and the field $C_a^{(1)}$ is auxiliary.
Next, one checks whether the original field equations for $C^{(0)}$
impose any restrictions on the first derivatives of $C^{(0)}$. More
precisely, some part of $\partial_\m C^{(0)}$ might vanish
on-mass-shell ({\it e.g.} for Dirac spinors). These restrictions in
turn impose some restrictions on the auxiliary fields $C_a^{(1)}$.
If these constraints are satisfied by $C_a^{(1)}$, then these fields
parametrize all on-mass-shell nontrivial components of first
derivatives.

Then, one writes for these first level auxiliary  fields an equation
similar to (\ref{unf1})
   \be D^L_0 C^{(1)}_a =  e_0^b\,\, C^{(2)}_{a,b}\,, \label{unf2}\ee
where the new fields $C^{(2)}_{a,b}$ parametrize the second
derivatives of $C^{(0)}$. Once again one checks (taking into account
the Bianchi identities) which components of  the second level fields
$C^{(2)}_{a,b}$ are non-vanishing provided that the original
equations of motion are satisfied.

This process continues indefinitely, leading to a chain of equations
having the form of some covariant constancy condition for the chain
of fields $C^{(m)}_{a_1,a_2,\ldots,a_m}$ ($m\in\mathbb N$)
parametrizing all  on-mass-shell nontrivial derivatives of the
original dynamical field. By construction, this leads to a
particular unfolded equation (\ref{eq}) with $G^i$ in (\ref{uncur})
given by (\ref{lin}). As explained in Section \ref{defs}, this means
that the set of fields realizes some module $T$ of the space-time
symmetry algebra $s$. In other words, the fields
$C^{(m)}_{a_1,a_2,\ldots,a_m}$ are the components of a single field
$C$ living in the infinite-dimensional $s$--module $T$. Then the
infinite chain of equations can be rewritten as a single covariant
constancy condition $D_0C=0$, where $D_0$ is the $s$-covariant
derivative in $T$.

\scss{The example of the scalar field}\label{scalar}

For simplicity, for the remaining of this Section, we will consider
a flat space-time background. The Minkowski solution can be written
as \be \o_0=dx^\m\d_\m^a P_a\label{omeg0}\ee {\it i.e.} the flat
frame is $(e_0)_\m^a=\d_\m^a$ and the Lorentz connection vanishes.
The equation (\ref{omeg0}) corresponds to the pure gauge solution
(\ref{pureg1}) with \be g(x)=\exp (x^\mu \,\delta_\mu ^a\,
P_a)\,,\label{gx} \ee where the space-time Lie algebra $s$ is
identified with the Poincar\'e algebra $iso(d-1,1)$.

As a preliminary to the gravity example considered in the next
subsection, the simplest field-theoretical case of unfolding is
reviewed, {\it i.e.} the unfolding of a massless scalar field
$\phi(x)$, which was first described in \cite{Vasiliev:1992gr}. The
``unfolding" of the massless Klein-Gordon equation \be \Box
\Phi(x)=0\label{KG} \ee is relatively easy to work out, so we give
directly the final result and we comment about how it is obtained
afterwards.

To describe dynamics of the spin zero massless field $\Phi (x)$, let
us introduce the infinite collection of $0$-forms $\Phi_{a_1\ldots
a_n}(x)$ ($n=0,1,2,\ldots$) which are completely symmetric traceless
tensors \be \label{tr} \Phi_{a_1\ldots a_n}=\Phi_{\{a_1\ldots
a_n\}}\ ,\quad \eta^{bc}\Phi_{bca_3\ldots a_n}=0\,. \ee The
``unfolded" version of the Klein-Gordon equation (\ref{KG}) has the
form of the following infinite chain of equations \be \label{un0} d
\Phi_{a_1\ldots a_n } =e_0^b \Phi_{a_1 \ldots a_n b}\quad
(n=0,1,\ldots)\,, \ee where we have used the opportunity to replace
the Lorentz covariant derivative $D_0^L$ by the ordinary exterior
derivative $d$. It is easy to see that this system is formally
consistent because applying $d$ on both sides of (\ref{un0}) does
not lead to any new condition,
$$ d^2 \Phi_{a_1\ldots a_n } =-
e_0^b d\Phi_{a_1 \ldots a_n b}= e_0^b e_0^c d\Phi_{a_1 \ldots a_n
bc}= 0 \quad (n=0,1,\ldots)$$ since $e_0^b e_0^c=-e_0^c e_0^b$
because $e_0^b$ is a 1-form.
   As we know from Section \ref{defs}, this property
implies that the space $T$ of $0$-forms $\Phi_{a_1 \ldots a_n}$
spans some representation of the Poincar\'e algebra $\miso(D-1,1)$.
In other words, $T$ is an infinite-dimensional
$\miso(D-1,1)$-module\footnote{ Strictly speaking, to apply the
general argument of Section \ref{defs} one has to check that the
equation remains consistent for any flat connection in
$\miso(D-1,1)$. It is not hard to see  that this is true indeed.}.

To show that this system of equations is indeed equivalent to the
free massless field equation (\ref{KG}), let us identify the scalar
field $\Phi (x)$ with the member of the family of $0$-forms
$\Phi_{a_1 \ldots a_n}(x)$ at $n=0$. Then the first two equations of
the system (\ref{un0}) read
$$\partial_\nu \Phi =\Phi_\nu \ ,$$
$$\partial_\nu \Phi_\mu= \Phi_{\mu\nu}\ ,$$
where we have identified the world and tangent indices via
$(e_0)_\mu^a=\delta_\mu^a$. The first of these equations just tells
us that $\Phi_\nu$ is the first derivative of $\Phi$. The second one
tells us that $\Phi_{\nu\mu}$ is the second derivative of $\Phi$.
However, because of the tracelessness condition (\ref{tr}) it
imposes the Klein-Gordon equation (\ref{KG}). It is easy to see that
all other equations in (\ref{un0}) express highest tensors in terms
of the higher-order derivatives \be \label{hder} \Phi_{\nu_1 \ldots
\nu_n}=
\partial_{\nu_1}\ldots\partial_{\nu_n}\Phi
\ee and impose no new conditions on $\Phi$. The tracelessness
conditions (\ref{tr}) are all satisfied once the Klein-Gordon
equation is true. {}From this formula it is clear that the meaning
of the 0-forms $\Phi_{\nu_1 \ldots \nu_n}$ is that they form a basis
in the space of all on-mass-shell nontrivial derivatives of the
dynamical field $\Phi(x)$ (including the derivative of order zero
which is the field $\Phi(x)$ itself).

Let us note that the system (\ref{un0}) without the constraints
(\ref{tr}), which was originally considered in \cite{Shaynk},
remains formally consistent but is dynamically empty just expressing
all highest tensors in terms of derivatives of $\Phi$ according to
(\ref{hder}). This simple example illustrates how algebraic
constraints like tracelessness of a tensor can be equivalent to
dynamical equations.

In a parallel fashion, one can also check that indeed the unfolded
dynamical problem is well-posed once one gives the values of all
the zero-forms at a point in space-time. In fact, one can show
that, again due to the trace constraints that the zero-forms
satisfy, this is equivalent to the Cauchy problem. To simplify
matters, let us apply the previous considerations to a
$2$-dimensional flat space-time, $a=0,1$. We suppose then that all
$\Phi_{a(n)}(p)$, where the space-time point $p$ has coordinates
$p=(t_0,x_0)$, are given. On the constraints that the unfolded
system imposes these are set equal to all the derivatives
\eq{hder}. However, because of trace constraints, there are fewer
independent derivatives, and they ``spread'', ``unfold'' the local
initial data onto the space-like line (hypersurface, in higher
dimensions) $\{t=t_0\}$. In particular: for $n=0$ one is given
$\Phi(p)$; $n=1$ fixes $\dot{\Phi}(p)$ and $\del \Phi(p)$ (where
we use the shorthand notations $\dot{\Phi}$ and $\del \Phi$ for
derivatives with respect to $t$ and $x$, respectively); the
independent local data for $n=2$ is
$\Phi_{00}(p)=\Phi_{11}(p)=\del^2\Phi(p)$ and
$\Phi_{01}(p)=\del\dot{\Phi}(p)$; for $n=3$ is
$\Phi_{000}(p)=\Phi_{110}(p)=\del^2\dot{\Phi}(p)$ and
$\Phi_{001}(p)=\Phi_{111}(p)=\del^3\Phi(p)$; and so on. In other
words, the independent local data fixes \emph{all} spatial
derivatives of $\Phi$ and $\dot{\Phi}$, \emph{i.e.}, it is
equivalent to the standard initial data of the Cauchy problem,
$\Phi$ and $\dot{\Phi}$ on the equal-time surface $\{t=t_0\}$.
Notice however that the unfolded formulation is more general,
since giving a set of zero-forms at a point in space-time can be
done prior to specifying a metric (the zero-form indices are fiber
indices only) and in a coordinate-independent way. Again, this is
because the first-order unfolded equations involve a trading of
space-time indices for fiber (tangent-space) indices, that makes
it possible to encode a nontrivial dynamics into an algebraic
constraint. This reasoning can be extended to arbitrary space-time
dimensions.

The above considerations can be simplified further by means of
introducing the auxiliary coordinate $u^a$ and the generating
function
\bea \Phi (x,u) \ = \ \sum_{n=0}^\infty \frac{1}{n\,!}\Phi_{a_1
\ldots a_n}(x) u^{a_1} \ldots u^{a_n} \label{genfun}\eea
with the convention that $$\Phi (x,0)=\Phi(x)\,. $$ This generating
function accounts for all tensors $\Phi_{a_1 \ldots a_n}$ once the
tracelessness condition is imposed, which in these terms implies
that \be \label{ubox} \Box_u \Phi (x,u)\equiv
\frac{\partial}{\partial u^a} \frac{\partial}{\partial u_a} \Phi
=0\,.\ee In other words, the $\miso(D-1,1)$-module $T$ is realized
as the space of harmonic formal power series in $u^a$. Eqns.
(\ref{un0}) then acquire the simple form \be \label{xu}
\frac{\partial}{\partial x^\mu} \Phi
(x,u)=\delta_\mu^a\,\frac{\partial}{\partial u^a} \Phi (x,u)\,. \ee
{}From this realization one concludes that the translation
generators in the infinite-dimensional module $T$ of the Poincar\'e
algebra are realized as translations in $u$--space, {\it i.e.}
$$P_a=-\frac{\partial}{\partial u^a}\,,$$ so that eqn.
(\ref{xu}) reads as a covariant constancy condition (\ref{fda0}) \be
d\Phi (x,u)+e_0^a P_a\Phi (x,u)=0\,.\label{xu2}\ee One can find a
general solution of eq. (\ref{xu2}) in the form $$ \Phi (x,u )=\Phi
(x+u,0) =\Phi (0,x+u )\, $$ from which it follows in particular that
\be \label{tay} \Phi (x)\equiv \Phi
(x,0)=\Phi(0,x)=\sum_{n=0}^\infty \frac{1}{n!}
\Phi_{\nu_1\ldots\nu_n}(0) x^{\nu_1} \ldots x^{\nu_n}\,. \ee {}From
(\ref{tr}) and (\ref{hder}) one can see that this is indeed the
Taylor expansion for any solution of the Klein-Gordon equation which
is analytic in $x_0=0$. Moreover one can recognize the equation
(\ref{tay}) as a particular realization of the pure gauge solution
(\ref{pureg}) with the gauge function $g(x)$ of the form (\ref{gx}).

The example of a free scalar field is so simple that one might think
that the unfolding procedure is always a trivial mapping of the
original equation, in this case (\ref{KG}), to the equivalent one,
here (\ref{ubox}), in terms of additional variables. This is not
true, however, for the less trivial cases of dynamical systems in
nontrivial backgrounds and, especially, for nonlinear systems. The
situation here is analogous to that in the Fedosov quantization
prescription \cite{fed} which reduces the nontrivial problem of
quantization in a curved background to the standard problem of
quantization of the flat phase space, that, of course, becomes an
identity when the ambient space itself is flat. It is worth to
mention that this parallelism is not occasional because, as one can
easily see, the Fedosov quantization prescription provides a
particular case of the general unfolding approach \cite{Vasfda} in
the dynamically empty situation ({\it i.e.}, with no dynamical
equations imposed).

The unfolded free scalar field can also be used as a prototype
example of how the ultra-local unfolded approach to the dynamical
problem can be mapped to the standard Cauchy problem, and viceversa.
The key point is the possibility of encoding a Taylor expansion into
an infinite-dimensional fiber at a point $p$: the corresponding
degrees of freedom can be unfolded in space-time via an expansion in
derivatives of the physical field, that is the content of the system
\eq{un0}. We shall encounter again a similar mechanism in Chapter
\ref{map}, when we shall see that space-time local fluctuation
fields, with different boundary conditions, are in correspondence
with certain well-defined nonpolynomial combinations of operators of
the HS algebra that are their fiber-( or tangent-space) duals, in
the same way as \eq{genfun} and \eq{tay} are dual via \eq{xu2}.

\scss{The example of gravity}\label{unfgrav}

The set of fields in the Einstein-Cartan's formulation of gravity
comprises the frame field $e_\m^a$ and the Lorentz connection
$\o_\m^{ab}$. One assumes that the torsion constraint $T_a=0$ is
satisfied, in order to express the Lorentz connection in terms of
the frame field. The Lorentz curvature can be expressed as
$R^{ab}=e_ce_d\,R^{[ab]\,;\,[cd]}$, where $R^{ab\,;\,cd}$ is a rank
four tensor with indices in the tangent space and which is
antisymmetric both in  $ab$ and in $cd$,  having the symmetries of
the tensor product \begin{picture}(60,16)(0,0)
\multiframe(0,5)(10.5,0){1}(10,10){$a$}
\multiframe(0,-5.5)(10.5,0){1}(10,10){$b$}\put(20,3){$\bigotimes$}
\multiframe(40,5)(10.5,0){1}(10,10){$c$}
\multiframe(40,-5.5)(10.5,0){1}(10,10){$d$}\end{picture}. The
algebraic Bianchi identity $e_b R^{ab}=0$, which follows from the
zero torsion constraint, forces the tensor $R^{ab\,;\,cd}$ to
possess the symmetries of the Riemann tensor, {\it i.e.}
$R^{[ab\,;\,c]d}=0$. More precisely, it carries an irreducible
representation of $GL(D)$ characterized by the Young tableau
\begin{picture}(30,16)(0,0)
\multiframe(1,4)(10.5,0){2}(10,10){$a$}{$c$}
\multiframe(1,-6.5)(10.5,0){2}(10,10){$b$}{$d$}
\end{picture}
in the antisymmetric basis. The vacuum Einstein equations state that
this tensor is traceless, so that it is actually irreducible under
the pseudo-orthogonal group $O(D-1,1)$ on-mass-shell. In other
words, the Riemann tensor is equal on-mass-shell to the Weyl tensor.

For HS generalization, it is more convenient to use the symmetric
basis. In this convention, the Einstein equations can be written as
\be T^a=0\,,\qquad
R^{ab}\,=\,e_c\,e_d\,\Phi^{ac,\,bd}\,,\label{Einst}\ee where the
$0$-form $\Phi^{ac,\,bd}$ is the Weyl tensor in the symmetric basis.
More precisely, the tensor $\Phi^{ac,bd}$ is symmetric in the pairs
$ac$ and $bd$ and it satisfies the algebraic identities
$$\Phi^{(ac,\,b)d}=0\,,\qquad \eta_{ac}\Phi^{ac,\,bd}=0\,.$$ Notice
that, while dynamically equivalent to the vanishing of the Ricci
tensor, this formulation of the vacuum Einstein equations is more
suitable for HS extensions, in that it is written purely in terms of
differential forms, and does not involve the inverse metric. In
other words, the metric is not given any special role, and this
makes this formulation of the spin-2 field equations manifestly
compatible with a symmetry that mixes different spins.

Let us now start the unfolding of linearized gravity around the
Minkowski background described by a frame 1-form $e^a_0$. The
linearization of the second equation of (\ref{Einst}) is
\be\label{HSpin2}
R_1^{ab}\,=\,e_{0\;c}\,e_{0\;d}\,\Phi^{ac,\,bd}\,,\ee where
$R_1^{ab}$ is the linearized Riemann tensor. This equation is a
particular case of eq. (\ref{fdadel}). What is lacking at this stage
is the additional set of equations containing the differential of
the Weyl $0$-form $\Phi^{ac,\,bd}$. Since we do not want to impose
any additional dynamical restrictions on the system, the only
restrictions on the derivatives of the Weyl $0$-form
$\Phi^{ac,\,bd}$ may result from the Bianchi identities for
(\ref{HSpin2}).

{\it A priori}, the first Lorentz covariant derivative of the Weyl
tensor is a rank-five tensor in the following representation
\begin{eqnarray}\footnotesize
\begin{picture}(38,15)(0,0)
\multiframe(-50,0)(10.5,0){1}(10,10){}\put(-30,2.5){$\bigotimes$}
\multiframe(-10,0)(10.5,0){2}(10,10){}{}
\multiframe(-10,-10.5)(10.5,0){2}(10,10){}{}\put(20,0){$=$}
\multiframe(40,0)(10.5,0){2}(10,10){}{}
\multiframe(40,-10.5)(10.5,0){2}(10,10){}{}
\multiframe(40,-21)(10.5,0){1}(10,10){}\put(70,0){$\bigoplus$}
\multiframe(90,0)(10.5,0){3}(10,10){}{}{}
\multiframe(90,-10.5)(10.5,0){2}(10,10){}{}
\end{picture}\normalsize
\label{Yougdec}\\\nonumber\end{eqnarray}decomposed according to
irreducible representations of $\msl(D)$. Since the Weyl tensor is
traceless, the right hand side of (\ref{Yougdec}) contains only one
nontrivial trace, that is for traceless tensors we have the
$\mso(D-1,1)$ Young decomposition by adding a three cell hook
tableau, {\it i.e.}
\begin{eqnarray}\footnotesize \nonumber
\begin{picture}(38,15)(0,0)
\multiframe(-50,0)(10.5,0){1}(10,10){}\put(-30,2.5){$\bigotimes$}
\multiframe(-10,0)(10.5,0){2}(10,10){}{}
\multiframe(-10,-10.5)(10.5,0){2}(10,10){}{}\put(20,0){$=$}
\multiframe(40,0)(10.5,0){2}(10,10){}{}
\multiframe(40,-10.5)(10.5,0){2}(10,10){}{}
\multiframe(40,-21)(10.5,0){1}(10,10){}\put(70,0){$\bigoplus$}
\multiframe(90,0)(10.5,0){3}(10,10){}{}{}
\multiframe(90,-10.5)(10.5,0){2}(10,10){}{} \put(135,0){$\bigoplus$}
\multiframe(155,0)(10.5,0){2}(10,10){}{}
\multiframe(155,-10.5)(10.5,0){1}(10,10){}
\end{picture}\normalsize
\\\nonumber\end{eqnarray}
The linearized Bianchi identity $dR_1^{ab}=0$ leads to \be
e_{0\;c}e_{0\;d}\,d\Phi^{ac,\,bd}=0\,.\label{linB1}\ee The
components of the left-hand-side, written in the basis $dx^\m dx^\nu
dx^\rho$, have the symmetry property corresponding to the tableau
$$\begin{picture}(38,15)(0,0)
\multiframe(10,9.5)(10.5,0){2}(10,10){$\m$}{$a$}
\multiframe(10,-1)(10.5,0){2}(10,10){$\n$}{$b$}
\multiframe(10,-11.5)(10.5,0){1}(10,10){$\r$}
\end{picture}\sim\partial_{[\r}\Phi^a{}_\m{}^{\,,b}{}_{\n]}\,,$$
which also contains the single trace part with the symmetry
properties of the three-cell hook tableau.

Therefore the consistency condition (\ref{linB1}) states that in the
decomposition (\ref{Yougdec}) of the Lorentz covariant derivative of
the Weyl tensor, the first term vanishes and the second term is
traceless but otherwise arbitrary. Let $\Phi^{abf,\,cd}$ be the
traceless tensor corresponding to the second term in the
decomposition (\ref{Yougdec}) of the Lorentz covariant derivative of
the Weyl tensor. This is equivalent to saying that
$$
d\Phi^{ac,\,bd}\,=\,{e_0}_f\,(2\Phi^{acf,\,\,bd}+\Phi^{acb,\,\,df}+\Phi^{acd,\,\,bf})\,,
$$
where the right hand side is fixed
   by the Young symmetry properties of the left hand side
modulo an overall normalization coefficient. This equation looks
like the first step (\ref{unf1}) of the unfolding procedure, with
$\Phi^{acf,\,\,bd}$ irreducible under $\mso(D-1,1)$.

One should now perform the second step of the general unfolding
scheme and write the analogue of (\ref{unf2}). This process goes on
indefinitely. To summarize the procedure, one can analyze the
decomposition of the $k$-th Lorentz covariant derivatives (with
respect to the  Minkowski vacuum background, so they commute) of the
Weyl tensor $\Phi^{ac,\,bd}$. Taking into account the Bianchi
identity, the decomposition goes as follows
\begin{eqnarray}\footnotesize
\begin{picture}(38,15)(0,0)
\multiframe(-100,0)(10.5,0){1}(50,10){}{} \put(-45,1){$k$}
\put(-30,0){$\bigotimes$} \multiframe(-10,0)(10.5,0){2}(10,10){}{}
\multiframe(-10,-10.5)(10.5,0){2}(10,10){}{}\put(20,0){$\cong$}
\multiframe(40,0)(10.5,0){2}(10,10){}{}{}\multiframe(51,0)(10.5,0){1}(50,10){}{}
\put(106,1){$k+2$} \multiframe(40,-10.5)(10.5,0){2}(10,10){}{}
   \end{picture}\normalsize
\label{Youngdecomp}\\\nonumber\end{eqnarray} As a result, one
obtains
\begin{eqnarray}
d\Phi^{a_1\ldots a_{k+2},\,b_1b_2}&=&
e_{0\;c}\,\Big(\,(k+2)\,\Phi^{a_1 \ldots a_{k+2}c,\, b_1
b_2}\,+\,\Phi^{a_1 \ldots a_{k+2} b_1,\,b_2c}\,+\,\Phi^{a_1 \ldots
a_{k+2}
b_2,\,b_1c}\,\Big)\,,\nn\\
&&\qquad\qquad(0\leq k\leq \infty)\,,\label{unfoldingggravity}
\end{eqnarray}
where the fields $\Phi^{a_1\ldots a_{k+2},\,b_1b_2}$ are in the
irreducible representation of $o(d-1,1)$ characterized by the
traceless two-row Young tableau on the right hand side of
(\ref{Youngdecomp}), {\it i.e.}
$$\Phi^{(a_1\ldots a_{k+2},\,b_1)b_2}=0\,,\quad
\eta_{a_1a_2}\Phi^{a_1a_2\ldots a_{k+2},\,b_1b_2}=0\,.$$ Note that,
as expected, the system (\ref{unfoldingggravity}) is consistent with
$d^2\Phi^{a_1\ldots a_{k+2},\,b_1b_2}=0$.

As in the spin-zero case, the meaning of the zero-forms
$\Phi^{a_1\ldots a_{k+2},\,b_1b_2}$ is that they form a basis in the
space of all on-mass-shell nontrivial gauge invariant combinations
of the derivatives of the spin-2 gauge field.

\scs{Free massless equations for any spin} \label{freemasslessequ}

In order to follow the strategy exposed in Subsection \ref{unffda}
and generalize the example of gravity treated along these lines in
Subsection \ref{unfgrav}, we shall begin by writing unfolded HS
field equations in terms of the linearized HS curvatures (\ref{R1}).
This result is called ``central on-mass-shell theorem".  It was
originally obtained in \cite{Fort1,Vasfda} for the case of $D=4$ and
then extended to any $D$ in \cite{LV,5d}. That these HS equations of
motion  indeed reproduce the correct physical degrees of freedom can
be shown via an elegant cohomological approach explained in
\cite{Bekaert:2005vh}.

The linearized curvatures $F_1^{A_1 \ldots A_{s-1},\, B_1 \ldots
B_{s-1}}$ were defined in (\ref{R1}). They decompose  into the
linearized curvatures with Lorentz ({\it i.e.,} $V^A$ transverse)
fibre indices which have the symmetry properties associated with the
two-row traceless Young tableau
\begin{picture}(71,15)(-2,2)
\multiframe(0,7.5)(13.5,0){1}(50,7){}\put(55,8.5){{\tiny $s-1$}}
\multiframe(0,0)(13.5,0){1}(35,7){}\put(40,0){{\tiny$t$}}
\end{picture}. It is convenient to use
   the standard gauge $V^A=\d^A_{0'}$ (we also
normalize $V$ to unity). In the Lorentz basis, the linearized HS
curvatures have the form \be F_1^{a_1 \ldots a_{s-1},\, b_1 \ldots
b_{t}}\,=\,D^L_0 A^{a_1 \ldots a_{s-1},\, b_1 \ldots
b_{t}}\,+\,e_{0\;c}\, A^{a_1 \ldots a_{s-1},\, b_1 \ldots b_{t}c}\,
+\,O(\Lambda)\,.\label{tauminus}\ee For simplicity, in this section
we discard the complicated $\Lambda$--dependent terms which do not
affect the general analysis, \emph{i.e.} we present explicitly the
flat-space part of the linearized HS curvatures. It is important to
note however that the $\Lambda$--dependent terms in (\ref{tauminus})
contain the field $A^{a_1 \ldots a_{s-1},\, b_1 \ldots b_{t-1}}$
which carries one index less than the linearized HS curvatures. The
explicit form of the $\Lambda$--dependent terms is given in
\cite{LV}.

For $t=0$, these curvatures generalize the torsion of gravity, while
for $t>0$ the curvature corresponds to the Riemann tensor. In
particular, as it is shown in \cite{Bekaert:2005vh}, the analogues
of the Ricci tensor and scalar curvature are contained in the
curvatures with $t=1$ while the HS analog of the Weyl tensor is
contained in the curvatures with $t=s-1$. (For the case of $s=2$
they combine into the level $t=1$ traceful Riemann tensor.)

The first on-mass-shell theorem states that the following free field
equations in Minkowski or $(A)dS$ space \be F_1^{a_1 \ldots
a_{s-1},\, b_1 \ldots
b_{t}}\,=\,\d_{t,\,s-1}\,\,\,e_{0\;c}\,e_{0\;d}\,\,\Phi^{a_1 \ldots
a_{s-1}c,\, b_1 \ldots b_{s-1}d}\,,\qquad (0\leq t\leq s-1)
\label{onmshell} \ee properly describe completely symmetric gauge
fields of generic spin $s\geq 2$. This means that they are
equivalent to the proper unfolded version of the Fronsdal equations
in any dimension, supplemented with certain algebraic constraints on
the auxiliary HS connections which express the latter via
derivatives of the dynamical HS fields. The zero-form $\Phi^{a_1
\ldots a_s,\, b_1 \ldots b_s}$ is the spin-$s$ Weyl-like tensor. It
is irreducible under $\mso(D-1,1)$ and is characterized by a
rectangular two-row Young tableau
\begin{picture}(60,15)(-5,2)
\multiframe(0,7.5)(13.5,0){1}(50,7){}\put(53,7.5){$s$}
\multiframe(0,0)(13.5,0){1}(50,7){}\put(53,0){$s$}
\end{picture} . The field
equations generalize (\ref{HSpin2}) of linearized gravity. The
equations of motion put to zero all curvatures  with $t\neq s-1$
and require $\Phi^{a_1 \ldots a_s,\, b_1 \ldots b_s}$ to be
traceless.

The analysis of the Bianchi identities of (\ref{onmshell}) works for
any spin $s\geq 2$ in a way analogous to gravity. The final result
is the following equation \cite{5d} which presents itself like a
covariant constancy condition \bea 0=D_0 \Phi^{a_1 \ldots a_{s+k},\,
b_1 \ldots b_s}&\equiv& D_0^L \Phi^{a_1 \ldots a_{s+k},\, b_1 \ldots
b_s}\nn\\&&-\,e_{0\;c}\Big((k+2)\Phi^{a_1 \ldots a_{s+k}c,\, b_1
\ldots b_s}\,+\,s\,\Phi^{a_1 \ldots a_{s+k} \{b_1,\,b_2 \ldots
b_s\}c}\Big)+O(\Lambda)\,,\nn\\&&(0\leq k\leq\infty
)\,,\label{unfHS} \eea where $\Phi^{a_1 \ldots a_{s+k},\, b_1 \ldots
b_s}$ are $\mso(D-1,1)$ irreducible ({\it i.e.}, traceless) tensors
    characterized by the Young tableaux
\begin{picture}(85,15)(0,2)
\multiframe(0,7.5)(13.5,0){1}(50,7){}\put(55,7.5){$s+k$}
\multiframe(0,0)(13.5,0){1}(35,7){}\put(40,0){$s$}
\end{picture}
. They describe on-mass-shell nontrivial $k$-th derivatives of the
spin-$s$ Weyl-like tensor, thus forming a basis in the space of
gauge invariant combinations of $(s+k)$-th derivatives of a spin $s$
HS gauge field. The system (\ref{unfHS}) is the generalization of
the spin-0 system (\ref{un0}) and  the spin-2 system
(\ref{unfoldingggravity}) to arbitrary spin and to $AdS$ background
(the explicit form of the $\Lambda$--dependent terms is given in
\cite{5d}). Let us stress that for $s \geq 2$ the infinite system of
equations (\ref{unfHS}) is a consequence  of (\ref{onmshell}) by the
Bianchi identity. For $s=0$ and $s=1$, the system (\ref{unfHS})
contains the dynamical Klein-Gordon and Maxwell equations,
respectively. Note that (\ref{onmshell}) makes no sense for $s=0$
because there is no spin-0 gauge potential while (\ref{unfHS}) with
$s=0$ reproduces the unfolded spin-0 equation (\ref{un0}) and its
$AdS$ generalization. For the spin-1 case, (\ref{onmshell}) only
defines the spin-1 Maxwell field strength $\Phi^{a,b} = -
\Phi^{b,a}$ in terms of the potential $A_\mu$. The dynamical
equations for spin 1, {\it i.e.} the Maxwell equations, are
contained in (\ref{unfHS}). The fields $\Phi^{a_1 \ldots a_{k+1},\,
b}$, characterized by the Lorentz irreducible ({\it i.e.} traceless)
two-row Young tableaux with one cell in the second row, form a basis
in the space of on-mass-shell nontrivial derivatives of the Maxwell
tensor $\Phi^{a,\,b}$.

Since $k$ goes from zero to infinity for any fixed $s$ in
(\ref{unfHS}), in agreement with the general arguments of Section
\ref{defs}, each irreducible spin-$s$ submodule of the twisted
adjoint representation is \emph{infinite-dimensional}. This means
that, in the unfolded formulation, the dynamics of any fixed
spin-$s$ field is described in terms of an infinite set of fields
related by the first-order unfolded equations. Of course, to make it
possible to describe a field-theoretical dynamical system with an
infinite number of degrees of freedom, the set of auxiliary
zero-forms associated with all gauge invariant combinations of
derivatives of dynamical fields should be infinite-dimensional.

It is clear that  the complete set of zero-forms $\Phi^{a_1 \ldots
a_{s+k},\, b_1 \ldots b_s}$ \begin{picture}(85,15)(0,2)
\put(0,5){$\sim$}
\multiframe(10,7.5)(13.5,0){1}(50,7){}\put(65,7.5){$s+k$}
\multiframe(10,0)(13.5,0){1}(35,7){}\put(50,0){$s$}
\end{picture}
$\,$ covers the set of all two-row Young tableaux. This suggests
that the Weyl-like zero-forms take values in the linear space of
$\mho$ that obviously forms an $\mso(D-1,1)$- ({\it i.e.} Lorentz)
module. Following the strategy sketched in Subsection \ref{unffda},
one can expect that they belong to an $\mso(D-1,2)$-module. But the
idea to use the adjoint representation of $\mho(D-1,2)$ defined in
\eq{adj} does not work because, according to the commutation
relation (\ref{tensortransf}), the commutator of the background
gravity connection $\o_0=\o_0^{AB}\widehat{T}_{AB}$ with a generator
of $\mho$ preserves the rank of the generator, while the covariant
derivative $D$ in (\ref{unfHS}) acts on the infinite set of Lorentz
tensors of infinitely increasing ranks. Fortunately, the appropriate
representation only requires a slight modification compared to the
adjoint representation. As it is clear by looking at
(\ref{twadj}-\ref{Phis}), the representation we are looking for is
nothing but the \emph{twisted adjoint} presented in Section
\ref{Sec:adj}, that naturally encodes the set of zero-forms that are
needed for the unfolding for every spin-$s$ sector and transforms
through the $\pi$-twisted adjoint action \eq{twadj}. Indeed, the
whole set of equations \eq{onmshell} and \eq{unfHS} can be written
in quite a compact form in terms of the adjoint master one-form
\eq{A} and of the twisted adjoint master zero-form \eq{Phi} as
\bea F_1 & \equiv & dA+\left\{\o_0,A\right\}_\star \ = \
e_0^a\,e_0^b\,\frac{\del\Phi}{\del
M^{ab}}\bigg|_{P^a=0} \ ,\label{lin1f}\\[5pt]
D_0\Phi & \equiv & d\Phi+\left[\o_0,\Phi\right]_\pi \ = \ 0 \ ,
\label{lin0f}\eea
where we recall that the anticommutator of the two one-forms in
\eq{lin1f} implements the adjoint action (\ref{tensortransf}). The
two linearized equations above admit the gauge symmetries
\bea \d^{(0)}A \ = \ D_0\e \ = \ d\e+[\o_0,\e] \ , \qquad
\d^{(0)}\Phi \ = \ 0 \ ,\label{lintransf}\eea
where $\e$ is an adjoint gauge parameter of $\mho(D-1,2)$, and
therefore contains all the linearized gauge transformations
$\{\e^{a(s-1),b(t)}\}$ for the set of one-form connections
$\{A^{a(s-1),b(t)}\}$ that span the spin-$s$ sector \eq{As} of the
master one-form, and that are needed for the frame-like description
of a spin-$s$ field. The equations above contain all the curvature
constraints of the frame formulation recalled in Section
\ref{frameHS}, supplemented with the unfolding of the Weyl ``source
term'' that follows from the Bianchi identities. As already stressed
before, the physical degrees of freedom contained in the system can
be conveniently analyzed through the so-called $\s_-$-cohomology
involving gauge parameters, gauge potentials, curvatures and Bianchi
identities. The details can be found in \cite{Bekaert:2005vh} (see
also \cite{Sezgin:1998gg, Sezgin:1998eh, Iazeolla:2004hj}), but the
end result is essentially that solving the generalized torsion
constraints (contained in the equations \eq{onmshell} for $t<s-1$)
expresses certain ($\mso(D-1,1)$-irreducible) components of the
auxiliary fields contained in each spin-$s$ sector in terms of the
generalized frame field $A_\m^{a(s-1)}$; all the remaining ones are
pure gauge, and can be eliminated by fixing the St\"{u}ckelberg-type
gauge parameters (all those with $t\geq 1$), except for the totally
symmetric component of the generalized frame field, \emph{i.e.}, the
Fronsdal doubly traceless field
$\varphi_{\m(s)}=(e_0^{-1})_{(\m_1}{}^{a_1}...(e_0^{-1})_{\m_{s-1}}{}^{a_{s-1}}A_{\m_s),a(s-1)}$.
The latter carries therefore the physical degrees of freedom, and is
accompanied by the remaining gauge symmetry $\e^{a(s-1)}$, which
coincides with the true (\emph{i.e.}, differential, and not shift
symmetry) Fronsdal traceless gauge parameter. Moreover, the
remaining components of the curvatures $F_{1\,\m\n}^{a(s-1),b(t))}$
are set to zero by the Bianchi identities except for the
$F_1^{a(s),b(s)}$ that is left free to fluctuate.

Eqns. (\ref{lin1f}), (\ref{lin0f}) provide the unfolded form of the
free equations of motion for completely symmetric massless fields of
all spins in any dimension. This fact is referred to as central
on-mass-shell theorem because it plays a distinguished role in
various respects. The idea of the proof is explained in
\cite{Bekaert:2005vh} (see references therein for the original
papers). Let us note that the right-hand-side of eq. (\ref{lin1f})
is a particular realization of the deformation terms
(\ref{deformcurv}) in free differential algebras.

\scss{A few Remarks}

As shown in the examples given in this Chapter, in the unfolded
formulation the trace constraints on the tangent-space indices of
the fields involved are crucial to ensure that the first-order
system of zero-curvature (or covariant constancy) conditions is not
dynamically empty. Indeed, they are responsible for encoding the
second-order physical field equation in the system. Once such
algebraic constraints are taken into account and one works with
traceless one-forms and zero-forms, however, the free unfolded
system can be shown to be equivalent to the Fronsdal equations, and
in this sense one can say that trace constraints put the system
on-shell - and, conversely, that the unfolded equations with
traceful fields describe an off-shell (or, better to say,
nondynamical) system.

However, it is important to stress at this point that, as discovered
first in \cite{Sagnotti:2005ns}, the free unfolded equations can
nicely recover also the local, unconstrained Francia-Sagnotti
equations \eq{frcomp}, and also their $(A)dS$ version \eq{frcompL}.
This can be achieved by declaring only the zero-forms to be
traceless in the internal indices, and letting the trace parts in
the one-forms free to adjust to their source terms in \eq{onmshell}.
Let us examine briefly how this can happen by looking at the example
of spin 3 in flat space-time. Let $\varphi_{\m ab}$ be the totally
symmetric $(3,0)$ component of the frame-like field. After solving
some generalized torsion constraints and using the shift symmetries,
one gets
\bea A_{\m,ab,cd}^{(2,2)} \ = \ \c_{2,2}\,\del_{\langle
c}\del_d\varphi_{ab\rangle\m}\ ,\eea
where $\c_{2,2}$ is a coefficient, unimportant for this discussion,
and we recall that the indices enclosed in the brackets $\langle
...\rangle$ are $\msl(D-1,1)$-projected. By tracing two indices this
becomes
\bea A_{\m,ab,c}^{(2,2)}{}^c \ = \ \frac{\c_{2,2}}{3}K_{\m,ab}\
,\eea
with
\bea K_{\m,ab} \ = \ \Box^2\varphi_{\m
ab}-2\del_{(a}\del\cdot\varphi_{b)\m}+\del_a\del_b\varphi'_\m \
.\eea
The curvature constraints involving such components of the gauge
potentials read
\bea \del_{[\m} A_{\n],ab,c}^{(2,2)}{}^c \ = \ 0 \ ,\eea
since the trace of the zero-form on the right hand side vanishes.
This implies that the curl of the tensor $K_{\m,ab}$ vanishes, i.e.,
that the latter is a pure gradient,
\bea K_{\m,ab} \ = \ \del_\mu\b_{ab} \ . \label{kbeta}\eea
This admits the two projections $\msl(D-1,1)$-irreducible
projections $(3,0)$ and $(2,1)$. The latter is a consistency
condition that admits the solution
\bea \b^{(2,0)}_{ab} \ = \
\del\cdot\varphi_{ab}-2\del_{(a}\varphi'_{b)}+\del_a\del_b\a \ ,\eea
where the last term is a homogeneous solution parametrized by an
unconstrained field $\a$. Finally, substituting back into the
solution \eq{kbeta} of the curvature constraint and taking the
$(3,0)$ projection one indeed finds the equations \eq{frcomp},
\bea \Box\varphi_{\m
ab}-3\del_{(a}\del\cdot\varphi_{b\m)}+3\del_{(a}\del_b\varphi'_{\m)}
\ = \ \del_{(a}\del_b\del_{\m)}\a \ .\eea
Moreover, in this case the leftover gauge parameter is, as the
adjoint fields, an $\msl(D-1,1)$-irreducible parameter. In other
words, the system has an unconstrained gauge symmetry. The
compensator therefore enters the unfolded constraints as a
cohomologically exact part \cite{DuboisViolette:2001jk} in the
solutions of a trace part of the curvature constraints.

This consideration points towards the fact that under many respects,
including a comparison with String Field Theory, it may be important
to have a formulation of the unfolded equations in terms of master
fields that contain traceful generators and coefficients. The
appropriate HS algebra corresponds to the enveloping algebra of
$\mso(D-1,2)$ (or $\miso(D-1,1)$ as well, as long as the free theory
is concerned) modulo the $V_{ABCD}$ element only, and not also
$V_{AB}$, leading to fields and generators represented by traceful
two-row Young diagrams. Notice moreover that this is exactly what
the vector oscillator realization, treated in Subsection \ref{Y}
enables one to do, prior to factoring out the $\msp(2)$ ideal. Then,
one can put the system on-shell, \emph{i.e.}, enforcing the
equations of motion on the physical field, by projecting out the
ideal in the zero-form sector only, if one wants to get the
unconstrained formulation; or, alternatively, both in the one-form
and the zero-form sector if one wants to obtain the Fronsdal
equations. On the other hand, the spinor oscillator formulation in
intrinsically on-shell, since trace parts are automatically factored
out, as explained in Subsection \ref{y}.

We are finally ready to see how interaction terms can be added. The
next step will be therefore to promote the linearized curvature
$F_1$ and the covariant derivative $D_0$ to full ones $F=dA+A\star
A$ and $D=d+[A,]_\pi$, involving the whole $\mho(D-1,2)$-connection
$A$, and to interpret \eq{lin1f} and \eq{lin0f} as the linearization
of a full system
\bea \cF&\equiv & F+{\cal J}(A,\Phi)\ =\ 0\ ,\\[5pt]
 \cD\Phi&\equiv & D\Phi+{\cal P}(A,\Phi)\ =\ 0\
,\label{generalizedconstraints}\eea
where ${\cal J}$ and ${\cal P}$ are non-linear zero-form
deformations obeying the compatibility condition \eq{prop} and
having the correct physical weak field limit \eq{lin1f} and
\eq{lin0f}, that excludes the trivial solution ${\cal J}={\cal
P}=0$. These functions are at least cubic and written using the
wedge product (and index contractions by the flat metric
$\eta_{ab}$ and possibly the anti-symmetric tensor $\e_{a_1\dots
a_D}$). Thus ${\cal J}$ is quadratic in $A$ and at least linear in
$\Phi$, while ${\cal P}$ is linear in $A$ and at least quadratic
in $\Phi$. Notice that such nonlinearities are required to
maintain consistency: indeed,
\begin{equation}
 D^{2}\Phi\sim F\Phi\sim \Phi^{2}+\textrm{higher order terms}
\ ,
\end{equation}
being $F$ at least of first-order in $\Phi$.

Before looking for the consistent nonlinear deformations, one can
make one more observation about the unfolding procedure. As already
stresses in the scalar field example, at the free level unfolding
may seem a cumbersome reformulation of the dynamics. The point is
that at the free level, and also at the interacting level for
lower-spin theories, higher-derivative interactions do not enter the
physical field equations: indeed, all the one-form connections
$A_\m^{a(s-1), b(t)}$ with $t\geq 2$, that are solved from torsion
constraints as \be \label{hder1} A^{~a_1 \ldots a_{s-1}, \, b_1
\ldots b_t} =\Pi \left ( \L^{-t/2} \frac{\partial}{\partial
x^{b_1}}\ldots \frac{\partial}{\partial x^{b_t}} A^{ ~a_1 \ldots
a_{s-1}}\right) +\mbox{lower derivative terms} \,, \ee and the
infinite chain of equations that constrains higher-rank zero-forms
$\Phi_{a(s+k),b(s)}$ have no effect on the dynamics. In particular,
$\Phi_{a(s+k),b(s)}$ with $k+s>1$ do not enter the contorsion or the
stress-energy tensor for matter-coupled gravity or supergravity, nor
the HS free theories, as we have seen. However, this is not the case
for interacting HS gauge theories, where higher-derivative
interaction vertices naturally enter the physical field equations
already at second order in the weak-field expansion: this was seen,
for example, in \cite{Fradkin:ks, Berends:1984rq, Berends:1985xx,
Berends:wp} among other works, and the quadratic scalar-field
contribution to $T_{ab}$ (\emph{i.e.} the scalar-scalar-graviton
vertex) in $D=4$ resulting from the full Vasiliev equations was
computed in \cite{Kristiansson:2003xx} (using a specific physical
gauge choice made at the level of the full Vasiliev equations). The
result is an infinite series of higher-derivative interactions given
by various contractions of $\nabla^m_{a(m)}\phi\nabla^n_{b(n)}\phi$
for arbitrarily high values of $m+n$. Higher-derivative interaction,
compensated by the inverse of the cut-off mass scale $M$, do arise
in lower-spin field theories, but they only yield small corrections
to classical solutions in which derivatives are small in units of
$M$. On the other hand, in a theory with local unbroken HS
symmetries, higher-derivative interactions are part of the minimally
coupled microscopic theory. Interacting HS can in principle give
rise to many complicated higher-derivative interaction terms in the
physical field equation: the unfolded formulation offers a way of
handling them systematically, in a first-order form: higher
derivatives are hidden in the auxiliary higher-rank zero-forms, that
enter the field equations through consistent deformations of the FDA
describing the free system. Therefore, the problem is reduced to
find such deformations, since gauge invariance is guaranteed by
consistency of the system (\emph{i.e.}, the deformed gauge symmetry
follows automatically through the relation \eq{delw}, and does not
need to be guessed).

\chapter{Nonlinear equations}\label{nonlin}

\scs{Preliminaries}\label{prel}

In the last Chapter, we have recalled the unfolded formulation of
the free theory. We still have to make use of the non-abelian HS
algebra constructed in Chapter \ref{absalg} to write nonlinear
corrections. We are now in a position similar to having the
linearized vacuum Einstein equations for the gravitational field
$h_{\m\n}$ (where of course $h_{\m\n}$ represents fluctuations over
a fixed background metric that can be chosen to be Minkowski,
$g_{\m\n}=\y_{\m\n}+h_{\m\n}$) and wanting to get the full nonlinear
Einstein equations without the possibility of resorting to
geometrical concepts. This program was indeed completed successfully
for gravity several years after the formulation of General
Relativity (see \cite{Feynman:1996kb, Deser:1969wk, Ortin:2004ms}
and references therein). The idea was to proceed by consistency: as
soon as one allows quadratic terms, the free equations of motion
become inconsistent, since while the linearized Einstein tensor is
divergenceless, this is not the case for its source, the
stress-energy tensor, as long as one does not introduce the
contribution of gravity to the latter, as calculated from the free
action. However, in order for the new equation to be derived from a
Lagrangian, one must add cubic terms to the free Lagrangian that
respect gauge invariance (up to $\cO(h^2)$). But the variation of
such cubic Lagrangian introduces new terms, that spoil consistency.
This process goes on indefinitely, and the infinite series of
nonlinear corrections sums up to the full Einstein equations, where
they appear through a crucial contribution of the inverse metric.

A similar program would still be hardly tractable for higher spins,
due to the presence of infinitely many fields and symmetries one
should control. As explained in the previous Chapter, the unfolded
formalism and the identification of the convenient master fields one
should work with bring a great simplification. Pursuing this
approach (first proposed in \cite{Vasiliev:sa}), the ``HS
interaction problem'' was solved, for totally symmetric massless
fields, by Vasiliev at the beginning of the Nineties
\cite{Vasiliev:en}. As shown above, indeed, in his formulation the
problem is reduced to find nonlinear corrections to \eq{lin1f} and
\eq{lin0f} that preserve consistency in the sense of \eq{prop}.

Let us begin our analysis by observing that, in the unfolded setup,
the natural expansion parameter is the zero-form $\Phi$: indeed,
containing all the (generalized) Weyl tensors and their derivatives,
it is the variable that measures the deviation from the background
solution; moreover, as already stressed below
\eq{generalizedconstraints}, in order to preserve the form-degree of
the curvature constraints, arbitrary nonlinearities can only involve
zero-forms contained in $\Phi$, and not $A$. Indeed, consistent
deformations of the free FDA at the lowest order in curvatures were
found in \cite{Vasiliev:sa, Vasiliev:1989yr}. However, to find the
whole series a more systematic approach was necessary, and that was
also developed exploiting the power of the unfolded formalism.

One could obtain all consistent nonlinear corrections in one shot if
some deformation of the HS algebra $\mho\rightarrow \widehat{\mho}$
existed such that the full nonlinear constraints
\eq{generalizedconstraints} could be seen as zero-curvature and
covariant constancy conditions for the deformed algebra
\cite{Vasiliev:2004cp}; \emph{i.e.}, if $\cF$ and $\cD$ could be
seen as curvature and covariant derivative with respect to the
connection $\hA$ of $\hmho$\footnote{In this chapter and in the next
one we denote with a hat the variables valued in the deformed
algebra that enter Vasiliev's nonlinear equations, and that should
not be confused with the hatted generators that appeared in Chapter
\ref{absalg}.}. In which case, the latter should contain all the
nonlinearities in the zero-forms contained in $\Phi$ with indices
contracted in various ways: $\hA=A+A\Phi+A\Phi\Phi+...$,
schematically. At the same time, the gauge symmetries get modified
correspondingly. How can one construct such a deformation?

The basic trick is to introduce an additional set of auxiliary,
noncommutative coordinates $Z$ and express the entire series of
nonlinear corrections as solution of a consistent equation with
respect to them. Such deformations will by this mechanism be
guaranteed to be consistent and, therefore, will preserve the gauge
invariance. Therefore, one enlarges the space-time with additional
variables $Z$, $x\longrightarrow (x,Z)$, and lets all the variables
and differentials acquire a dependence on them:
\bea d& \longrightarrow & \hd \ = \ d+d_Z \ ,\\[5pt]
A(x;M_{ab},P_a) & \longrightarrow & \hA(x;Z;M_{ab},P_a)\\[5pt] & = &
\!\!\!\!\-i\sum_s \sum_{t=0}^{s-1}dx^\mu
A_{\mu,a(s-1),b(t)}(x;Z)M^{a_1 b_1}\cdots
M^{a_tb_t}P^{a_{t+1}}\cdots P^{a_{s-1}} \ ,\\[5pt]
\Phi(x;M_{ab},P_a) & \longrightarrow & \hPhi(x;Z;M_{ab},P_a)\\[5pt] & = &
\!\!\!\!\sum_{s}  \sum_{k=0}^\infty{i^k\over k!}
\Phi^{a(s+k),b(s)}(x;Z)M_{a_1b_1}\cdots M_{a_sb_s}P_{a_{s+1}}\cdots
P_{a_{s+k}}\ ,\eea
where the coefficients are themselves power series in the $Z$
variables. The ordinary space-time corresponds to the subspace
$\{Z=0\}$. Notice, however, that due to the noncommutative nature of
the new coordinates the pull-back of the extended unfolded system to
$\{Z=0\}$ does not correspond to the original system,
\bea d\hA+\hA\star\hA \big|_{Z=0}\ \neq \ dA+A\star A \
,\label{nopb}\eea
since there will always be infinitely many contributions to the
identity coming from total contractions of the $Z$ variables. But
how is this connected to the nonlinear corrections in $\Phi$?

The crux of the matter is now to extend the FDA with a constraint (a
consistent one, by definition of FDA) that relates every contraction
of $Z$ to $\Phi$,
\bea \frac{\del}{\del Z}\hA \ \sim \ \Phi+... \ ,
\label{contraction}\eea
(where the ellipsis stands, possibly, for higher order terms)
thereby solving all the dependence on the extra variables in terms
of the physical degrees of freedom of the original system.
Therefore, the infinitely many terms that correct the pure
space-time curvature $dA+A\star A$ in \eq{nopb} are all expressed in
terms of $\Phi$ and build up the deformations ${\cal J}(A,\Phi)$ and
${\cal P}(A,\Phi)$ of equation \eq{generalizedconstraints}. In order
to write \eq{contraction} in a consistent way, it has to be cast in
the form of a curl in $Z$-space, a curvature with indices taking
values in the noncommutative subspace $Z$. The solutions of
\eq{contraction} will therefore be automatically consistent with the
gauge symmetries, that will turn out to be correspondingly deformed
but not broken down, which is essential in order to be able to
interpret the nonlinear theory in terms of the same degrees of
freedom present at the free level.

If such a $Z$-extension can be found, then the problem is solved,
since we have deformed the theory with infinitely many nonlinear
corrections while keeping gauge symmetries and diffeomorphism
invariance. Further constraints come from the fact that it must be
possible to relate, in this extension, the adjoint and twisted
adjoint representations, as the condition \eq{contraction} is
crucial. Note that in the linearized approximation \eq{lin1f} and
\eq{lin0f} this problem is absent, since the transformation of the
zero-forms produces terms of second order: in other words, because
\eq{lintransf} are valid, up to second order terms. However, at
the full level, in order to write a consistent equation like
\eq{contraction}, it is necessary that it involve some mapping
between adjoint and twisted adjoint representations. As we shall
examine in Chapter \ref{map} in greater detail, such a map has to
act on the ``square root'' of the $\mso(D-1,2)$ generators
that build up the expansion of $\Phi$. 

Moreover, other important constraints follow from the fact that
the appropriate noncommutative extension must be such that the
full equations reproduce the free equations \eq{lin1f} and
\eq{lin0f} once linearized around the $AdS$ vacuum solution and
pulled back on the space-time manifold. It turns out that this
requirement can be satisfied once one assumes that the $Z$
coordinates have a nontrivial contraction with the oscillators
that realize the $\mso(D-1,2)$ generators. This fact in its turn
implies that, in order to preserve covariance, the $Z$-coordinates
must have the same index structure of the oscillators,
\emph{i.e.}, the sought-after noncommutative extension must
correspond to a doubling of the oscillators. Associativity then
fixes the commutation relations of the $Z$ coordinates among
themselves. The final result is that they will satisfy an algebra
that is isomorphic to the Heisenberg algebra of the oscillators
(it only differs by a sign). A generalized $\star$-product can
then be defined, that controls contractions among oscillators and
$Z$ coordinates and therefore defines a proper composition rule
for the elements of an extended oscillator algebra $\hmho(D-1,2)$.

Indeed, an appropriate $Z$-extension has been constructed,
originally in the four-dimensional spinor oscillator realization and
later also in the $D$-dimensional vector oscillator formalism (see
\cite{Engquist:2005yt} for a geometric derivation of the $Z$
extension). In the following, we will present in detail only the
first realization, as some of the results of this Thesis, discussed
in Chapter \ref{exactsol}, only rest on them. The $D$-dimensional
realization of Vasiliev equations, first proposed in
\cite{Vasiliev:2003ev}, rests on the doubling of the vector
oscillators presented in Section \ref{Y}. Although many of the steps
performed in the following can be repeated in that setting, an
important subtlety arises in factoring out the traces (\emph{i.e.},
in putting the system on-shell). There are in principle two
different ways to do this: the first one (called \emph{weak
projection}, and proposed in \cite{mishasalg, Bekaert:2005vh})
consists in factoring out elements proportional to the (extended)
$\msp(2)$ generators, at the full level, both in the one-form and in
the zero-form sectors by multiplying the whole equations by a
projector $\hM$ such that $\widehat K_{ij}\star\hM=0$; the second
one (called \emph{strong projection}, and proposed in
\cite{Sagnotti:2005ns}), somehow in the spirit of the unfolding,
consists in factoring out traces only in the master zero-form by
multiplying it with the projector $\hM$. As mentioned in the
previous Chapter, at the free level the first formulation results in
the Fronsdal equations, while the second leads to the unconstrained
equations of Francia and Sagnotti. However, at the full level the
strong projection may suffer from potential divergencies arising in
the expansion in powers of the projected master zero-form
$M\star\Phi$, since $M\star M$ diverges. We will not comment further
on this issue, but send to the paper cited above for reference.

\scs{Z-Extension and Vasiliev Equations in $(3+1)$
Dimensions}\label{Z}

It is now time to examine more quantitatively how one can obtain the
nonlinear Vasiliev equations, working with the $SL(2;\Comp)$-doublet
spinor oscillator realization of Subsection \ref{y}. The free
equations one would like to make contact with, \emph{i.e.},
\eq{lin1f} and \eq{lin0f} read, in such realization,
\bea F_1 (x;y,\yb)& \equiv & dA+\left\{\O,A\right\}_\star \ = \
\frac{i}{4} \,e_0^{\a\ad}\,e_0^{\b}{}_{\ad}\,\frac{\del^2}{\del
y^\a \del y^\b}\,\Phi(y,0) - \textrm{h.c.}\ ,\label{lin1fosc}\\[5pt]
D_0\Phi(x;y,\yb) & \equiv & d\Phi+\left[\O,\Phi\right]_\pi \ = \ 0 \
, \label{lin0fosc}\eea
where the background connection is
\bea \O \ = \  \frac1{4i}
  dx^\mu\left[\omega_\mu^{\a\b}~y_\a y_\b
  +\overline{\omega}_\mu{}^{\dot\a\dot\b}~{\bar
y}_{\dot\a}{\bar y}_{\dot\b}
  + 2 e_\mu^{\a\dot\b}~y_\a {\bar y}_{\dot\b}\right]\ \eea
and our spinor conventions are listed in Appendix \ref{App:F}.

To formulate the nonlinear field equations we double the oscillators
by adding the $SL(2;\Comp)$-doublet spinors $z_\a$ their hermitian
conjugates $\zb_{\ad}$ to the theory. Together with the $y_\a$ they
generate an oscillator algebra with non-commutative and associative
$\star$-product defined by
\bea y_\a\star y_\b&=&y_\a y_\b+i\epsilon_{\a\b}\ ,\qquad
y_{\a}\star z_{\b}\ =\ y_{\a}z_{\b}-i\,\e_{\a\b}\ ,\label{osc1}\\[5pt] z_{\a}\star
y_{\b}&=& z_{\a}y_{\b}+i\,\e_{\a\b}\ , \qquad z_{\a}\star z_{\b}\ =\
z_{\a}z_{\b}-i\,\e_{\a\b} \ ,\label{osc2} \eea
and
\bea \bar y_{\dot\a}\star \bar y_{\dot\b}\ =\ \bar y_{\dot\a} \bar
y_{\dot\b}+i\epsilon_{\dot\a\dot\b}\ ,\qquad \bar z_{\dot\a}\star
\bar y_{\dot\b}\ =\ \bar z_{\dot \a} \bar y_{\dot\b}-
i\epsilon_{\dot\a\dot\b}\ ,\label{oscbar1}\\[5pt] \bar y_{\dot\a}\star \bar z_{\dot\b}\
=\ \bar y_{\dot\a} \bar z_{\dot\b}+i\epsilon_{\dot\a\dot\b}\ ,\qquad
\bar z_{\dot\a}\star \bar z_{\dot\b}\ =\ \bar z_{\dot\a} \bar
z_{\dot\b}-i\epsilon_{\dot\a\dot\b}\ ,\label{oscbar2}\eea
where, as usual, juxtaposition denotes symmetrized, or Weyl-ordered,
products. Equivalently, Weyl-ordered functions obey\footnote{The
integration measure is defined by $d^4\xi=d^2\x^1 d^2\x^2$, where
$d^2 z=idz\wedge d\bar z=2dx\wedge dy$ for $z=x+iy$. With this
normalization, $\mathbb{I}\star \widehat f=\widehat f$.}
 \bea
 &&\widehat f(y,\bar y,z,\bar z)~\star~ \widehat g(y,\bar
y,z,\bar z)\label{extdefint}\\[5pt]&=&\ \int \frac{d^4\xi d^4\eta}{(2\pi)^4}~ e^{i\eta^\a\xi_\a+
i\bar\eta^{\dot\a}\bar\x_{\dot\a}} ~\widehat f(y+\xi,\bar y+\bar
\xi,z+\xi,\bar z-\bar \xi)~\widehat g(y+\eta,\bar y+\bar
\eta,z-\eta,\bar z+\bar \eta)\ ,\nn
 \eea
which extends the definition \eq{defint} of the $\star$-product to
functions of all ($y,\yb,z,\zb$) oscillators, denoted with the hats,
while functions of only $y_\a$ and $\yb_{\ad}$ will be unhatted.
Notice the peculiar difference in sign in the contraction of two $z$
oscillators compared with that of two $y$ oscillators, and the
opposite sign of the contractions in $y_{\a}\star z_{\b}$ and
$z_{\a}\star y_{\b}$, so that the commutation relations are
\bea [z_\a,z_\b]_\star \ = \ -2i\e_{\a\b} \ , \qquad
[y_a,z_\b]_\star \ = \ 0 \ , \label{extcomm}\eea
together with their hermitian conjugates. Although $y$ and $z$
oscillators commute, it is crucial, as said in the beginning of this
Chapter, that they have a nontrivial contraction, as we shall stress
again later on.

The definitions of the master fields correspondingly extend to the
\emph{adjoint} one-form $\widehat A$ and the \emph{twisted-adjoint}
zero-form $\widehat \Phi$ defined by
\bea \widehat A&=&dx^\mu\widehat A_\mu(x;y,\bar y,z,\bar
z)+dz^\a\widehat A_\a(x;y,\bar y,z,\bar z)+d\bar z^{\dot\a}\widehat
A_{ \dot\a}(x;y,\bar y,z,\bar z)\ ,\label{ext1f}\qquad\\[5pt]
\widehat\Phi&=&\widehat \Phi(x;y,\bar y,z,\bar z)\
,\label{ext0f}\eea
where $x^\mu$ are coordinates on a commutative base manifold (which
can, but need not, be fixed to be four-dimensional space-time). One
also defines the total exterior derivative
\be d\ =\ dx^\m\partial_\m + dz^\a {\partial\over\partial z^{\a}}+
d\bar z^{\ad} {\partial\over\partial \bar z^{\ad}}\
,\label{spinder}\ee
with the property $d(\widehat f\wedge\star~\widehat g)=(d\widehat
f)\wedge \star~\widehat g+(-1)^{{\rm deg}\widehat f}\widehat
f\wedge\star ~d\widehat g$ for general differential forms. In what
follows we shall again suppress the $\wedge$ symbol as we did so
far. The master fields can be made subject to the following discrete
symmetry conditions\footnote{The exterior derivative obeys $\tau
d=d\tau$ and $\pi d=d\pi$, and the $\tau$ and $\pi$ maps do not act
on the commutative coordinates.}
\cite{Vasiliev:1995dn,Engquist:2002vr}
\bea \mbox{Minimal model ($s=0,2,4,...$)}&:& \tau(\widehat A)\ =\
-\widehat A\
,\qquad \tau(\widehat\Phi)\ =\ \bar\pi(\widehat \Phi)\ ,\label{minmod}\\[5pt]
\mbox{Non-minimal model ($s=0,1,2,3,...$)}&:& \pi\bar\pi(\widehat
A)\ =\ \widehat A\ ,\qquad \pi\bar\pi(\widehat \Phi)\ =\
\widehat\Phi\ ,\label{nonminmod}\eea
where $\tau$ is the $\star$-product algebra anti-automorphism
defined by
\be \tau(\widehat f(y,\bar y;z,\bar z))\ =\ \widehat f(iy,i\bar
y;-iz,-i\bar z)\ ,\label{tau}\ee
and $\pi$ and $\bar\pi$ are two involutive $\star$-product
automorphisms defined by
\bea \pi(\widehat f(y,\bar y;z,\bar z))&=&\widehat f(-y,\bar
y;-z,\bar z)\ ,\qquad \bar\pi(\widehat f(y,\bar y;z,\bar z))\ =\
\widehat f(y,-\bar y;z,-\bar z)\ .\eea
We note that
\bea \tau(\widehat f\star\widehat g)&=&(-1)^{{\rm deg}(\widehat
f){\rm
deg}(\widehat g)}\tau(\widehat g)\star\tau(\widehat f)\ ,\\[5pt]
\pi(\widehat f\star\widehat g)&=& \pi(\widehat f)\star\pi(\widehat
g)\ ,\\[5pt] \bar\pi(\widehat f\star\widehat g)&=& \bar\pi(\widehat
f)\star\bar\pi(\widehat g)\ ,\eea
and that $\tau^2=\pi\bar\pi$. As mentioned in the previous Section,
the crucial feature that selects the realization of the
noncommutative extension as a doubling of the oscillators, with the
specific $\star$-product rule \eq{extdefint}, is that the $\pi,\pb$
automorphisms, that distinguish adjoint and twisted adjoint, are
\emph{inner} and can be generated by conjugation with the functions
$\kappa$ and ${\bar\kappa}$ given by
 \be
\k\ =\ \exp(iy^\a z_\a)\ ,\quad\quad \bar{\k}\ =\ \exp(-i\bar
y^{\ad}\bar z_{\ad})\ ,
 \ee
such that
\be \kappa\star \widehat f(y,z)\ =\ \kappa \widehat f(z,y)\ ,\qquad
\widehat f(y,z)\star\kappa\ =\ \kappa \widehat f(-z,-y)\ ,\qquad
\kappa\star \widehat f\star\kappa\ =\ \pi(\widehat f)\ ,
\label{kappa}\ee \be \bar \kappa\star \widehat f(\bar y,\bar z)\ =\
\bar\kappa \widehat f(-\bar z,-\bar y)\ ,\qquad \widehat f(\bar
y,\bar z)\star\bar \kappa\ =\ \bar\kappa \widehat f(\bar z,\bar y)\
,\qquad \bar\kappa\star \widehat f\star\bar\kappa\ =\
\bar\pi(\widehat f)\ . \label{kappabar}\ee
In other words, $\kappa$ provides the map between adjoint and
twisted adjoint representation that is necessary to relate the
$Z$-dependence to the physical degrees of freedom of the theory,
and to make contact with the linearized equations \eq{lin1fosc}
and \eq{lin0fosc}. Notice also that, as one can check from
\eq{extdefint},
\bea\kappa\star\kappa \ = \ 1 \ , \qquad \bar\kappa\star\bar\kappa \
= \ 1 \ .\eea
Let us note that {\it a priori} the $\star$-product
(\ref{extdefint}) is well-defined for the algebra of polynomials
(which means that the $\star$-product of two polynomials is still a
polynomial). Thus the $\star$-product admits an ordinary
interpretation in terms of oscillators, as long as we deal with
polynomial functions. But $\k$ is not a polynomial because it
contains an infinite number of terms with higher and higher powers
of $ y^\a z_\a$. Thus, {\it a priori}, the $\star$-product with $\k$
may give rise to divergencies arising from the contraction of an
infinite number of terms (for example, an infinite contribution may
appear in the zeroth order like a sort of vacuum energy). What
singles out the particular $\star$-product (\ref{extdefint}) is that
this does not happen for the class of functions which extends the
space of polynomials to include $\k$ and similar functions. More
precisely, as was shown originally in \cite{Prokushkin:1998bq}, the
$\star$-product (\ref{extdefint}) is well-defined for the class of
functions, called ``regular", that can be expanded into a finite sum
of functions $f$ of the form \be \label{class} f(Z,Y) = P(Z,Y)
\int_{M^n}d^nt \ \rho (t)  \exp \Big(i\phi(t) y^\a z_\a \Big)\,, \ee
where the integration is over some compact domain $M^n \subset
{\mathbb R}^n$ parametrized by the coordinates $t_i$ $ (i=1,\ldots
,n )$, the functions $P(Z,Y)$ and $\phi(t)$ are arbitrary
polynomials of $(Z,Y)\equiv (z,\zb,y,\yb)$ and $t_i$, respectively,
while $\rho(t)$ is integrable in $M^n$. The key point of the proof
is that the $\star$-product (\ref{extdefint}) is such that the
exponential in the Ansatz (\ref{class}) never contributes to the
quadratic form in the integration variables simply because
$\xi^\a\xi_\a = \y^\a\y_\a = 0$. As a result, a $\star$-product of
two elements (\ref{class}) never develops an infinity and the class
(\ref{class}) turns out to be closed under $\star$-multiplication
pretty much as ordinary polynomials.

The full field equations are
\bea \widehat F&=& \frac{i}4 \left[dz^\a\wedge dz_\a \widehat
\Phi\star \kappa+ d\bar z^{\ad}\wedge d\bar z_{\ad}\widehat
\Phi\star\bar\kappa\right]\ ,\label{m1red}\\[5pt]\widehat D\widehat \Phi&=&0\
,\label{m2red}\eea
where the curvatures and gauge transformations are given by
\bea \widehat{F}&=& d\widehat{A}+\widehat{A}\star\widehat{A}\
,\qquad\ \delta_{\widehat \e}\widehat A\ =\ \widehat D\widehat
\e \label{extgaugeA}\\[5pt]\widehat{D}\widehat{\Phi}&=&d\widehat{\Phi}+[\widehat
A,\widehat \Phi]_\pi\ ,\qquad\delta_{\widehat \e}\widehat\Phi\ =\
-[\widehat \e,\widehat\Phi]_\pi\ ,\label{extgaugePhi}\eea
with
\bea [\widehat f,\widehat g]_\pi&=&\widehat f\star\widehat
g-(-1)^{{\rm deg}(\widehat f){\rm deg}(\widehat g)}\widehat
g\star\pi(\widehat f)\ .\eea
Notice that the \eq{m1red} and \eq{m2red} are indeed consistent: the
second guarantees the consistency of the first, since
\bea \hD\hF \ = \ 0 \ = \ \frac{i}4 \left[dz^\a\wedge dz_\a
\hD\widehat \Phi\star \kappa+ d\bar z^{\ad}\wedge d\bar
z_{\ad}\hD\widehat \Phi\star\bar\kappa\right]\ ,\label{cons1}\eea
where we have taken into account that
$\hD\kappa=dz^\a\frac{\del}{\del z^\a}\kappa$, and that the latter
term (and its hermitian conjugate) do not contribute to \eq{cons1}
because they involve a triple antisymmetrization $dz^\a \, dz^\b\,
dz^\c$, which vanishes identically; on the other hand, the first
ensures the consistency of the second, since
\bea \hD^2\hPhi \ = \ 0 \ = \ [\hF,\hPhi]_\pi \ = \ \frac{i}{4}\,
dz^\a\wedge dz_\a \left(\widehat \Phi\star
\kappa\star\hPhi-\hPhi\star\pi(\hPhi\star\kappa)\right)-\textrm{h.c.}\
,\eea
where the vanishing of the last two terms can be checked by using
the properties of $\kappa$ \eq{kappa}. The consistency of the
extended FDA \eq{m1red} and \eq{m2red} with respect to both $x$ and
$Z$ variables ensures that one can solve for $\hA$ in terms of
$\hPhi$ from the first, and then for $\hPhi$ in terms of the
``initial condition'' $\hPhi(Z=0)|_p=\Phi|_p$ (up to an extended
gauge transformation) from the second. This implies that the
extended unfolded system is equivalent to the nonlinear space-time
system \eq{generalizedconstraints}, since they both have the same
local data.

In components, the constraints read
\bea \widehat F_{\m\n}&=&0\ ,\qquad \widehat D_\mu\widehat\Phi\
\equiv\ \partial_\mu\widehat\Phi +[\widehat
A_\m,\widehat\Phi]_{\pi}\ =\ 0\ , \label{f1}\eea
\bea \widehat F_{\m\a}&=&0\ ,\qquad \widehat F_{\m\ad}\ =\ 0\
,\label{f2}\eea
\bea \widehat F_{\a\b}&=&-\ft{i}2\e_{\a\b}\widehat\Phi\star\kappa\
,\qquad \widehat F_{\ad\bd}\ =\
-\ft{i}2\e_{\ad\bd}\widehat\Phi\star\bar\kappa\ ,\label{f3}\eea
\bea \widehat F_{\a\ad}&=&0\ ,\label{f4}\eea
\bea \widehat D_\a\widehat\Phi\ \equiv\
\partial_\a\widehat\Phi+\widehat
A_\a\star\widehat\Phi+\widehat\Phi\star\pi(\widehat A_\a)&=&0\
,\label{s1}\eea
\bea \widehat D_{\ad}\widehat\Phi\ \equiv\
\partial_{\ad}\widehat\Phi+\widehat
A_{\ad}\star\widehat\Phi+\widehat\Phi\star\bar\pi(\widehat
A_{\ad})&=&0\ ,\label{s2}\eea
where \eq{s2} can be derived using $\pi\bar\pi(\widehat
A_{\ad})=-\widehat A_{\ad}$. Introducing \cite{Vasiliev:en}
\bea \widehat S_\a&=& z_\a-2i\widehat A_\a\ ,\qquad \widehat
S_{\ad}\ =\ \bar z_{\ad}-2i\widehat A_{\ad}\ ,\label{S}\eea
the component form of the equations carrying at least one spinor
index now take the form
\bea \partial_\m \widehat S_\a+[\widehat A_\mu,\widehat
S_\a]_\star&=&0\ ,\qquad
\partial_\m \widehat
S_{\ad}+[\widehat A_\mu,\widehat S_{\ad}]_\star\ =\ 0\
,\label{S1}\eea \bea [\widehat S_\a,\widehat S_\b]_\star&=&
-2i\e_{\a\b}(1-\widehat \Phi\star\kappa)\ ,\qquad[\widehat
S_{\ad},\widehat S_{\bd}]_\star\ =\  -2i\e_{\ad\bd}(1-\widehat
\Phi\star\bar\kappa)\ ,\label{S2}\eea\bea [\widehat S_\a,\widehat
S_{\bd}]_\star&=&0\ ,\label{S3}\eea \bea \widehat
S_\a\star\widehat\Phi+\widehat\Phi\star\pi(\widehat S_\a)&=&0\
,\label{S4}\eea\bea \widehat
S_{\ad}\star\widehat\Phi+\widehat\Phi\star\bar\pi(\widehat S_{\ad})\
=\ 0\ .\label{S5}\eea
%
%
%

\scss{Lorentz-Covariance and Uniqueness}

The introduction of the extra $(z_\a,\zb_{\ad})$-oscillators, which
are $\msl(2;\Comp)$-doublets, requires a corresponding modifications
of the Lorentz generators, that were realized at the free level as
in \eq{mab2}. Indeed, due to the commutativity of $y$ and $z$
oscillators \eq{extcomm}, it is impossible to rotate the content in
$z,\zb$ of the extended variables $\hA$ and $\hPhi$ if we do not
extend the realization of the Lorentz generators. This can be done
easily, however, by noticing that $z$-oscillators satisfy an algebra
which is identical to that of $y$-oscillators, up to a sign.This
means that the appropriate generalization of Lorentz rotations is
given by
\bea \hM_{\a\b} \ = \
 y_\a y_\b-z_\a z_\b \ ,\qquad
\hM_{\ad\bd} \ = \
 \yb_{\ad} \yb_{\bd}-\zb_{\ad} \zb_{\bd} \ .\label{1modM}\eea
In terms of such generators, indeed, one gets the usual Lorentz
transformation
\bea \d_{\he_0} y_\a \ \equiv \ [\frac{1}{4i}\L^{\b\c}\hM_{\b\c}  -
\textrm{h.c.}, y_\a]_\star & = & \L_\a{}^\b\, y_\b \ ,\\[5pt]
\d_{\he_0} z_\a \ \equiv \ [\frac{1}{4i}\L^{\b\c}\hM_{\b\c}  -
\textrm{h.c.}, z_\a]_\star & = & \L_\a{}^\b\, z_\b \ . \eea
But this only takes care of the internal $\msl(2;\Comp)$-doublet
indices of the master fields, while the extension also brought in
\emph{external} $\msl(2;\Comp)$-doublet indices associated with the
$Z$-space parts of the one-form connection $\hA$ introduced in
\eq{ext1f} and the corresponding $Z$-space ``covariant derivative''
$\hS_\a,\hS_{\ad}$ \eq{S}. Such indices are not rotated properly by
the modified Lorentz generators \eq{1modM}, since, as one can read
from \eq{extgaugeA}, the resulting transformation is not homogenous,
\bea  \d_{\he_0} \hA_\a \ = \ [\hA_\a,\he_0]_\star
+\frac{1}{2i}\L_\a{}^\b \, z_\b \ .\eea
It is important to be able to modify the local-Lorentz parameter
$\he_0$ in such a way as to get a standard homogeneous
transformation law under local Lorentz, since in order to make
contact with the free equations \eq{lin1f} and \eq{lin0f} a physical
gauge must be maintained (as we will see in the next Section) that
enables to solve the $Z$-space connection entirely in terms of
$\Phi$ in perturbation theory. To see how this can be done, let us
first rewrite the last equation in terms of $\hS_\a$,
\bea  \d_{\he_0} \hA_\a \ = \ \L_\a{}^\b \hA_\b+
[\hA_\a,\he_0]_\star +\frac{1}{2i}\L_\a{}^\b \, \hS_\b \
,\label{unw}\eea
where it is clear that the latter term is the one that must be
eliminated. Now, the crucial point is that a consistent modification
of the Lorentz generators that rotates properly the external spinor
indices fixes the form of the Vasiliev equations up to field
redefinitions. Indeed, notice that the only nontrivial equations
(that imply that the whole system is not pure gauge) \eq{S2} have
the form of a deformed Heisenberg algebra (see
\cite{Vasiliev:1999ba} and references therein)
\bea [\hat{y}_{\alpha},\hat{y}_{\beta}] \ = \
2i\varepsilon_{\alpha\beta}(1+\nu k)\ , \quad \{\hat{y}_{\alpha},k\}
\ = \ 0\ , \quad k^{2} \ = \ 1 \ ,\eea
(where $\n$ is a real number) which, together with \eq{S4} and
\eq{S5} that can be rewritten as
\bea \{\hS_{\alpha},\hPhi\star\k\}_\star \ = \ 0 \ = \
\{\hS_{\ad},\hPhi\star\bar\kappa\}_\star\ ,\eea
implies that the generator $\frac{1}{2}\{\hS_\a,\hS_\b\}_\star$ and
its hermitian conjugate satisfy the $\msl(2;\Comp)$ algebra and
rotate properly the external spinor indices
\bea \left[\frac{1}{2}\{\hS_\a,\hS_\b,\}_\star,\hS_\c\right]_\star \
= \ -4i\hS_{(\a}\e_{\b)\c}\ .\eea
Therefore, defining \cite{Sezgin:2002ru}
\bea \he_{\textrm{extra}} \ = \
\frac{1}{8i}\L^{\a\b}\{\hS_\a,\hS_\b,\}_\star \ ,\eea
one has
\bea [\hA_\a,\he_{\textrm{extra}}]_\star \ = \
-\frac{1}{2i}\L_\a{}^\b\,\hS_\b\ ,\eea
that removes the unwanted term in \eq{unw}. The necessary
modification of the Lorentz generator that implements
Lorentz-covariance in the full equations is therefore
\bea \he_L \equiv \he_0+\he_{\textrm{extra}} \ = \
\frac{1}{4i}\L^{\a\b}\left(\hM_{\a\b}+\frac12\{\hS_\a,\hS_\b,\}_\star\right)
- \textrm{h.c.} \ . \eea
What is important for the issue of uniqueness of the interaction
terms is that the need to maintain Lorentz-covariance in the full
equations only allows source terms of the type in \eq{m1red} and
no hermitian modifications\footnote{However, it would still
possible to substitute $\hPhi\star\k$ in \eq{m1red} by an
arbitrary function ${\cal V}(\hPhi\star \k)$, as long as such
function is \emph{odd} \cite{Vasiliev:1992av, Vasiliev:1999ba,
Sezgin:2002ru, Iazeolla:2004hj}, which ensures that \eq{S4} imply
$\hS_\a\star{\cal V}(\hPhi\star \k)+{\cal V}(\hPhi\star
\k)\star\hS_\a = 0 $, which is crucial for rotating properly the
external spinor indices, as we shall see. Notice, however, that
introducing higher odd powers of $\hPhi\star \k$ would only affect
higher order interaction terms, while leaving unaltered the free
equations \eq{lin1f} and \eq{lin0f}. The Vasiliev equations
correspond therefore to a ``minimal'' choice in this sense, that
reproduces the correct free field dynamics in the linearized
approximation.} such as $dz^\a\,d\zb^{\bd}
H_{\a\bd}(\hPhi\star\k)$. Finally, the local-Lorentz
transformations of the extended master fields are
\bea \d_{\he_L}\hPhi & = & -[\he_0,\hPhi]_\star \ , \\[5pt]
\d_{\he_L}\hA_\a & = & [\hA_\a,\he_0]_\star +\L_\a{}^\b\hA_\b \
,\\[5pt]
\d_{\he_L}\hA_\m & = & [\hA_\m,\he_0]_\star
+\frac{1}{4i}\del_\m\L^{\a\b}\left(\hM_{\a\b}+\frac12\{\hS_\a,\hS_\b,\}_\star\right)
 \ .\eea
The space-time constraints, that we will examine in the next
Section, will be left invariant by the pulled-back local-Lorentz
transformation
\bea \d_{\he_L}\Phi & = & -[\e_0,\Phi]_\star \ , \\[5pt]
\d_{\he_L}\hA_\m & = & [A_\m,\e_0]_\star
+\frac{1}{4i}\del_\m\L^{\a\b}\left[y_\a
y_\b-4\left(\hA_\a\star\hA_\b-{\partial\over\partial y^\a} \widehat
A_\b\right)_{Z=0}-\textrm{h.c.}\right]
 \ ,\eea
where $\e_0\equiv \frac{1}{2i}\l^{\a\b}y_\a y_\b-\textrm{h.c.}$. So
the transformations of the gauge fields have acquired a complicate
field dependent part. However, the latter can be reabsorbed into a
redefinition of the Lorentz connection $\o_\m^{\a\b}$ inside $A_\m$,
since the quantity $\o_\m+K_\m$, with
\bea \o_\m \ = \ \frac{1}{4i}\o_\m^{\a\b}y_\a y_\b-\textrm{h.c.}
\eea
and
\bea K_{\m}&=& {1\over 4i}\o_\m{}^{\a\b}\left.\widehat
S_\a\star\widehat S_\b\right|_{Z=0}+{1\over 4i}
\bar\omega_\m{}^{\ad\bd}\left.\widehat S_{\ad}\star\widehat
S_{\bd}\right|_{Z=0}\\[5pt]&=& i\o_{\m}{}^{\a\b} \left.(\widehat
A_\a\star\widehat A_\b-{\partial\over\partial y^\a} \widehat
A_\b)\right|_{Z=0}+ i\bar\o_{\m}{}^{\ad\bd} \left.(\widehat
A_{\ad}\star\widehat A_{\bd}-{\partial\over\partial \yb^{\ad}}
\widehat A_{\bd})\right|_{Z=0}\ ,\label{Kmu}\eea
transforms as
\bea \d_{\he_L}[\o_\m+K_\m] \ = \ [\o_\m+K_\m,\e_0]_\star
\frac{1}{4i}\del_\m\L^{\a\b}\left[y_\a
y_\b-4\left(\hA_\a\star\hA_\b-{\partial\over\partial y^\a} \widehat
A_\b\right)_{Z=0}-\textrm{h.c.}\right] \eea
and therefore
$\d_{\he_L}[A_\m-\o_\m-K_\m]=[A_\m-\o_\m-K_\m,\e_0]_\star$, which
means that every field inside $A_\m-\o_\m-K_\m$ is a Lorentz tensor.
The space-time nonlinear constraints will thus involve
\bea A_\m \ \equiv \hA_\m\big|_{Z=0} \ = \ e_\m+\o_\m+W_\m+K_\m \
,\eea
where we have singled out the vielbein part $e_\m$ and the
higher-spin ($s\geq 4,6,...$ for the minimal and $s\geq 3,4,...$ for
the nonminimal bosonic model) fields $W_\m$, and we conclude that
the fields, at the nonlinear level, transform as Lorentz tensors
under an \emph{undeformed} local-Lorentz symmetry $\e_0$,
implemented through a field-dependent Lorentz connection
$\o_\m+K_\m$.

\scs{Perturbative Expansion}

We now want to show that the Vasiliev equations admit a perturbative
expansion in the master zero-form $\Phi$ that, to the first order,
reproduces the free equations \eq{lin1f} and \eq{lin0f}. In
performing such an expansion we shall encounter and examine in
greater detail all the issues discussed qualitatively in Section
\ref{prel}. The strategy will be to solve the $Z$-dependence from
the components of the equations that have at least one spinor index,
in terms of the initial condition $\Phi=\hPhi|_{Z=0}$ and then plug
the solution into the pure space-time components \eq{f1}
\cite{Vasiliev:1992av, Sezgin:2002ru}.

At zeroth order in $\Phi$, a natural vacuum solution of the equation
is $AdS_4$ space-time, around which we shall expand the full
equations, in the end. Indeed, for $\Phi=0$ \eq{f3} (or \eq{S2})
implies
\bea \hS^{(0)} \ = \ dz^\a\,z_\a+d\zb^{\ad}\,\zb_{\ad} \ ,\eea
(where the superscript refers to the order zero in $\Phi$) since
$\hA^{(0)}_\a$ is then a flat connection, and one is choosing the
gauge condition
\bea \hA^{(0)}_\a \ = \ \hA^{(0)}_{\ad} \ = \ 0 \label{gauge}\eea
(which in its turn implies $\frac{\del}{\del y^\a}\hA_\b|_{Z=0}=0$,
that simplifies the expressions \eq{Kmu}). Plugging into \eq{S1} one
gets then $[\hA^{(0)}_\m,z_a]_\star=0$, which implies that the
space-time component of the master one-form, to zeroth order in
$\Phi$, is $Z$-independent. Therefore, one $Z$-independent solution
of $\hF_{\m\n}=0$ is the $AdS_4$ connection
\bea \hA^{(0)}_\m \ = \ \O_{\m} \ = \ \frac1{4i}
  \left[\omega_\mu^{\a\b}~y_\a y_\b
  +\overline{\omega}_\mu{}^{\dot\a\dot\b}~{\bar
y}_{\dot\a}{\bar y}_{\dot\b}
  + 2 e_\mu^{\a\dot\b}~y_\a {\bar y}_{\dot\b}\right]\ , \eea
which thus appears as a natural vacuum solution of the full
equations. Notice that the symmetry $\e^{gl}(Z,Y;x)$ of this vacuum
solution is just $\mho(3,2)$. Indeed, the vacuum symmetry parameters
$\epsilon^{gl}(Z,Y;x)$ must satisfy \be [S^{(0)} \,, \epsilon^{gl}
]_\star =0 \,,\qquad D_0 (\epsilon^{gl}) =0\,. \ee The first of
these conditions implies that $\epsilon^{gl}(Z,Y;x)$ is
$Z$--independent, {\it i.e.,}
$\epsilon^{gl}(Z,Y;x)=\epsilon^{gl}(Y;x)$
   while the second reconstructs the dependence
of $ \epsilon^{gl}(Y;x)$ on space-time coordinates $x$ in terms of
the values of the constant element $\epsilon^{gl}(Y;x_0)\in
\mho(3,2)$ at any fixed point $x_0$ of space-time.

Now we have to investigate whether the free equations on $AdS_4$
emerge from the full system as first order corrections to such
vacuum solution. Before doing that, however, we shall show that one
can also expand the Vasiliev equations in a HS-covariant curvature
expansion treating the gauge fields exactly. In general, we set up
an expansion scheme
\be \widehat{\Phi}=\sum_{n=1}^\infty \widehat{\Phi}^{(n)}\ ,\quad
\widehat{A}_{\a}=\sum_{n=0}^\infty \widehat{A}_\a^{(n)}\ ,\quad
\widehat{A}_{\mu}= \sum_{n=0}^\infty \widehat{A}_\mu^{(n)}\
,\la{ces}\ee
where $\widehat{\Phi}^{(n)}$ ($n=1,2,3,...$), $\widehat{A}^{(n)}_\a$
($n=0,1,2,...$) and $\widehat{A}_\m^{(n)}$ ($n=0,1,2,...$) are
functionals which are $n$th order in $\Phi$ and which obey the
initial conditions
\bea \widehat{\Phi}^{(n)}|_{Z=0}&=&\mx{\{}{ll}{\Phi\ ,&n=1\\0\
,&n=2,3,...}{.}\la{phiic} \w2 \widehat{A}_\m^{(n)}|_{Z=0} &=&
\mx{\{}{ll}{A_\m\ ,&n=0\\0\ ,&n=1,2,3,...}{.}\la{amic}\
.\la{aaic}\eea
Next, the constraints $\widehat{F}_{\a\bd}=0$,
$\widehat{F}_{\a\b}=-\ft{i}2\e_{\a\b}\hPhi\star\k$ and $\widehat
D_\a\hPhi=0$ can be solved in the $n$th order ($n\geq 1$) as
\bea \hPhi^{(1)}&=&\Phi(y,\yb)\ ,\\
\hA^{(1)}_\a&=& \partial_\a\widehat \x^{(1)}-{i \over 2} z_\a
\int_0^1 tdt~
 \F(-tz,\yb)\kappa(tz,y)\ ,\la{ah}\eea
and ($n\geq 2$):
\bea \hPhi^{(n)}&=& z^\a\sum_{j=1}^{n-1}\int_0^1 dt \Bigg(
\hPhi^{(j)}\star
\pb(\widehat{A}^{(n-j)}_\a)-\widehat{A}^{(n-j)}_\a\star
\hPhi^{(j)}\Bigg)_{z\ra tz,\zb\ra t\zb}\nn \w2 &&+~
\zb^{\ad}\sum_{j=1}^{n-1}\int_0^1 dt \Bigg( \hPhi^{(j)}\star
\pi(\widehat{A}^{(n-j)}_{\ad})-\widehat{A}^{(n-j)}_{\ad}\star
\hPhi^{(j)}\Bigg)_{z\ra tz,\zb\ra t\zb}\ ,\la{it1}\w2
\widehat{A}^{(n)}_\a&=&\partial_\a\widehat\x^{(n)}+z_\a \int_0^1
tdt\Bigg(-\ft{i}2 (\hPhi^{(n)}\star\kappa)+
\sum_{j=1}^{n-1}\widehat{A}^{(j)\b}\star
\widehat{A}^{(n-j)}_\b\Bigg)_{z\ra tz,\zb\ra t\zb} \nn\w2
&&+\zb^{\bd}\sum_{j=1}^{n-1}\int_0^1 tdt \left[
\widehat{A}^{(j)}_{\a},\widehat{A}^{(n-j)}_{\bd}\right]_{* (z\ra
tz,\zb\ra t\zb)}\ .\la{it2} \eea
We emphasize that in \eq{it1} and \eq{it2} the replacements
$(z,\zb)$ by $(tz,t\zb)$ are to be made {\it after} the
$\star$-products are carried out. The integration functions
$\widehat \x^{(n)}$ are gauge artifacts, which can be eliminated by
means of $\Phi$-dependent gauge transformations. We therefore impose
the gauge conditions \cite{Vasiliev:1992av, Sezgin:2002ru}

\be \widehat\x^{(n)}=0\ ,\qquad n=1,2,\dots\ \la{gauge1}\ee

The gauge conditions \eq{gauge} and \eq{gauge1} are left invariant
by $Z$-independent, and therefore $\mho(3,2)$-valued, gauge
transformations (which in general may be $\Phi$-dependent).

From the constraints $\widehat{F}_{\m\a}=0$ and
$\widehat{F}_{\m\ad}=0$ one can solve for the $Z$-dependence of
$\hA_\m$. It follows that
\begin{eqnarray}
\hA^{(0)}_\m & = & A_\m \\
\hA^{(n)}_\m(x;Y,Z) & = &
\frac{i}{2}\int_{0}^{1}dt\biggl\{z^{\alpha}\left(\sum_{j=0}^{n-1}\left[\hA^{(j)}_\m,\hA_{\alpha}^{(n-j)}\right]_{\star}\right)(x;Y,
Z) \nonumber \\
&&+\bar{z}^{\dot{\alpha}}\left(\sum_{j=0}^{n-1}\left[\hA^{(j)}_\m,\hA_{\dot{\alpha}}^{(n-j)}\right]_{\star}\right)(x;Y,
Z)\biggr\}_{z\ra tz,\zb\ra t\zb}\ ,\label{int3ordn}
\end{eqnarray}
where we note that the terms $z^\a d\hA_\a+\zb^{\ad} d\hA_{\ad}$ are
identically zero by virtue of \eq{it1} and \eq{it2} with the gauge
choice \eq{gauge1}, since $z^\a z_\a=\zb^{\ad}\zb_{\ad}=0$.

Finally, having solved the $Z$-space part of \eq{m1red} and
\eq{m2red}, the remaining constraints $\widehat{F}_{\m\n}=0$ and
$\widehat{D}_\m \widehat{\Phi}=0$ yield the following space-time
full nonlinear field equations\footnote{The integrability of
\eq{m1red} and \eq{m2red} implies that if
$\widehat{F}_{\m\n}|_{Z=0}=0$ and $\widehat{D}_\m
\widehat{\Phi}|_{Z=0}=0$ then $\widehat{F}_{\m\n}=0$ and
$\widehat{D}_\m \widehat{\Phi}=0$. Indeed, integrability ensures
that
\begin{equation}
[\hS,dx^\m dx^\n\hF_{\m\n}]_{\star}=0\ ,
\end{equation}
\begin{equation}
\hS\star dx^\m\hD_\m\hPhi+dx^\m\hD_\m\hPhi\star\bar{\pi}(\hS)=0 \ ,
\end{equation}
which imply that \eq{f1} are covariantly constant in $Z$ space.}:

\bea F_{\m\n} &=&-2\sum_{n=1}^\infty\sum_{j=0}^{n} \Bigg(
\widehat{A}_{[\m}^{(j)}\star
\widehat{A}_{\n]}^{(n-j)}\Bigg)_{|_{Z=0}}\ , \la{cn}\w2 D_\m \Phi
&=& \sum_{n=2}^\infty \sum_{j=1}^{n} \Bigg(\widehat{\Phi}^{(j)}\star
\bar \pi(\widehat{A}^{(n-j)}_\m)-\widehat{A}^{(n-j)}_\m\star
\widehat{\Phi}^{(j)} \Bigg)_{|_{Z=0}}\ , \la{bn} \eea

where

\be F=dA+A\star A\ ,\quad\quad D\Phi=d\Phi+A\star
\Phi-\Phi\star\bar{\pi}(A)\ .\ee

It is important that \eq{cn} and \eq{bn} are integrable equations.
As such they are invariant under gauge transformations whose form
follows uniquely from their functional variation of \eq{cn} and
\eq{bn}, according to the general scheme of FDAs. Equivalently,
these symmetries can be described as the residual $\mho(3,2)$-valued
gauge transformations discussed above.

Now, to make contact with the free unfolded equations in $AdS_4$,
one can further expand the one-form $A$ around the $AdS_4$ vacuum
connection,
\bea A \ = \ \O+A_1\ ,\eea
where $A_1$ contains all fluctuation fields (including spin 2) over
the $AdS_4$ background and is treated as a weak field.

To the first order in curvatures, $n=1$, \emph{and} in fluctuations
$A_1$, the space-time components of Vasiliev equations read
\bea F_{1\,\m\n} &=&- 2\left(\O_{[\m}\star
\widehat{\O}_{\n]}^{(1)}\right)_{Z=0}\ , \la{cn1}\\[5pt]
D_{0\,\m} \Phi &=&0\ , \la{bn1} \eea
with $F^{(1)}=d A_1+\{\O,A_1\}_\star$, and
\begin{eqnarray}
\widehat{\O}^{(1)}(x;Y,Z) & = &
-\frac{1}{2}\int_{0}^{1}dt'\int_{0}^{1}dt\,t
\biggl\{\left(itt'\,\omega_{0}^{\phantom{0}\alpha\beta}z_{\alpha}z_{\beta}+
e_{0}^{\phantom{0}\alpha\dot{\beta}}z_{\alpha}\bar{\partial}_{\dot{\beta}}\right)\Phi(x;-tt'
z,\bar{y})\kappa(tt'z,y) \nonumber \\
&&+\left(itt'\,\omega_{0}^{\phantom{0}\dot{\alpha}\dot{\beta}}
\bar{z}_{\dot{\alpha}}\bar{z}_{\dot{\beta}}-e_{0}^{\phantom{0}\alpha\dot{\beta}}
\bar{z}_{\dot{\beta}}\partial_{\alpha}\right)\Phi(x;y,
tt'z)\bar{\kappa}(tt'\bar{z},\bar{y})\biggr\} \ ,
\end{eqnarray}
where $\partial_{\alpha}=\frac{\partial}{\partial y^{\alpha}}$. We
can see from here that, had we not postulated the nontrivial
contraction rule \eq{osc2} and \eq{oscbar2}, there would be no
chance of obtaining a nontrivial right hand side in \eq{cn1}, since
$\O$ only contains $y,\yb$ oscillators. Equation \eq{bn1} is already
of the desired form \eq{lin0fosc}. Performing the $\star$-product in
\eq{cn1} and projecting onto $\{Z=0\}$ one finally obtains
\eq{lin1fosc} (where we note that what we called $A$ there is a
fluctuation filed over $AdS_4$, and therefore has to be identified
with $A_1$ in this context), which is what we wanted to prove.

Let us note that one could also expand $A_\m$ in terms of HS fields
only (as well as higher derivatives of all fields), \emph{i.e.}, by
treating exactly the whole vierbein and Lorentz connection while
assuming $W_\m$ to include weak fields. This procedure yields a
manifestly diffeomorphism and locally Lorentz invariant expansion.



\chapter{Exact Solutions}\label{exactsol}


\scs{Introduction}\label{sec:in}


In the last Chapter we have presented the four-dimensional Vasiliev
equations, that are naturally formulated in terms of $SL(2;\Comp)$
spinor oscillators in Lorentzian signature $(3,1)$, and we have
described an explicit perturbative expansion that makes contact with
the known free equations for massless fields of arbitrary spin. As
shown above, this procedure amounts to first solving for the
dependence from the auxiliary $Z$-variables and then plugging back
in the pure space-time components of the field equations, that can
be analyzed order by order in the interactions. In other words, one
solves first the evolution along the infinite dimensional $Z$-fiber
over each point of the space-time manifold, which generates
infinitely many nonlinear terms in $\Phi$, and then examines the
resulting very complicate equations in a subspace that can be taken
to be an ordinary four-dimensional space-time. Observe, indeed, that
according to the general discussion on unfolded systems of Section
\ref{unffda}, the FDA that corresponds to the Vasiliev equations
makes the dependence on space-time coordinates completely auxiliary,
and the dynamics is all encoded in the functions $G^\a(W)$
satisfying the generalized Jacobi identity (\ref{prop}). Moreover,
as it should be clear from the component form of the equations
(\ref{f1}-\ref{s2}), the only term that introduces a nontrivial
dynamics in the system sits in the $ZZ$ component of the curvature:
this implies that the equations are homotopy invariant, \emph{i.e.},
the dynamics is preserved by the restriction to a single point in
space-time. In other words, one may equivalently analyze the content
of the equations by first solving for the space-time dependence from
the equations that have at least one space-time index, and then for
the $Z$-evolution. This strategy has moreover the advantage that
solving the space-time equations is particularly easy, since they
are zero-curvature conditions. The $x$-dependence is thus all
encoded into gauge functions, and the local data (the zero-forms at
a fixed space-time point) is the only nontrivial information that
enters the $Z$-space constraints. Now, the latter are pure algebraic
equations, that one has a better chance of solving exactly! One may
always reconstruct the space-time dependence of the solution at a
later stage, by $\star$-multiplying with the gauge function.

This feature of the Vasiliev equations, namely the fact that their
projection to the fiber at a given space-time point preserves all
the dynamical information, is remarkable, and was indeed exploited
for finding the first nontrivial exact solution, other than the
$AdS$ background, in \cite{Sezgin:2005pv}. Locally, the solution
describes a scalar field in a FRW-like metric with a space-like
singularity, that can be resolved by the method of patches. The
solution is asymptotically AdS and periodic in time, so that one may
think of it as an ``instanton universe'' inside AdS
\cite{Sezgin:2005pv}.  More recently, the gauge function method has
been used to describe the BTZ black hole metric as a solution to
full three dimensional HS gauge theory \cite{Didenko:2006zd}.

This raises the question of how to Wick rotate solutions of the
Lorentzian theory into solutions of a Euclidean theory. The main
difficulty is to impose proper reality conditions given the
doubling of the spinor oscillators due to the Euclidean signature.
In this Chapter, we shall review the conclusions of the recent
paper \cite{Iazeolla:2007wt}, where the Vasiliev equations have
been formulated using spinor oscillators in Euclidean signature
$(4,0)$ and Kleinian signature $(2,2)$ as well, and new nontrivial
exact solutions with novel properties, such as the excitation of
all higher spin fields, have been found. The difficulty with the
Euclidean signatures is resolved taking the master fields to be
holomorphic functions of the left-handed and right-handed spinor
oscillators subject to pseudo-reality conditions, as we shall see.

In addition to the Euclidean signature, we shall consider the
Kleinian signature as well. While in all signatures there is the
possibility of a chiral asymmetry, in Euclidean and Kleinian
signatures, the extreme case of parity violation involving half-flat
gauge fields can also arise. We refer to the latter ones as
\emph{chiral models}. In HS gauge theory, the HS algebra valued
gauge-field curvatures can be made, say, self-dual, but the model
nonetheless contains the anti-self-dual gauge fields through the
master zero-form which contains the corresponding Weyl tensor
obeying the appropriate field equation. Although this is contrary to
what happens in ordinary Euclidean gravity, where the field
equations can contain only self-dual fields, it is not a surprise in
HS theory since the underlying higher spin algebra, which is an
extension of $SO(5)$, does not admit a chiral massless multiplet
\footnote{We shall leave a more detailed group-theoretical analysis
in all signatures to Chapter \ref{map}.}.

There are several reasons that make the investigation of HS theory
in Euclidean and Kleinian signatures worthwhile. To begin with, just
as the Euclidean version of gravity plays a significant role in the
path integral formulation of quantum gravity, it is reasonable to
expect that this may also be the case in the quantum formulation of
HS theory, despite the fact that an action formulation is yet to be
spelled out (see, however, \cite{Engquist:2007kz} for a recent
attempt). For reviews of Euclidean quantum gravity, see, for
example, \cite{GH} and \cite{Gibbons}.

Another well known  aspect of self-dual field theories is their
ability to unify a wide class of integrable systems in two and three
dimensions. It would be interesting to extend these mathematical
structures to self-dual HS gauge theories to find new integrable
systems.

The chiral HS theories in Kleinian signatures may also be of
considerable interest in closed $N=2$ string theory in which the
self-dual gravity in $(2,2)$ dimensions arises as the effective
target space theory \cite{Ooguri:1990ww}. However, there are some
subtleties in treating the picture-changing operators in the BRST
quantization which have raised the question of whether there are
more physical states \cite{Junemann:1997rx}, and in the case of open
$N=2$ theory an interpretation in terms of an infinite tower of
massless higher spin states has been proposed
\cite{Devchand:1997dq}. It would be very interesting to establish
whether these theories or their possible variants admit self-dual HS
theory in the target space. While the $N=2$ string theories may seem
to be highly unrealistic, it should not be ruled out that they may
be connected in subtle ways to all the other string theories which
are themselves connected by a web of dualities in M theory.

In this Chapter, we shall take the necessary first steps to start
the exploration of the Euclidean and Kleinian HS theories. We shall
start by determining the real forms of the complex HS algebra based
on an infinite dimensional extension of $SO(5;\Comp)$ and formulate
the corresponding higher-spin gauge theories in four-dimensional
spacetime with signature $(4-p,p)$. Maximally symmetric four-
dimensional constant curvature spacetimes, including de Sitter
spacetime, defined by the embedding into five-plane with signature
$(5-q,q)$ are readily exact solutions. Fluctuations about these
spaces arrange themselves into all the irreducible representations
of $SO(5-q,q)$ contained in the symmetric two-fold product of the
fundamental singleton representation of this group, each occurring
once. The details of this phenomenon will be provided in Chapter
\ref{map}.

We then devote the rest of the Chapter to finding a class of
nontrivial exact solutions of these models, including the Euclidean
and chiral cases. The key information about these solutions is
encoded in the master zero-form which we recall contains a real
ordinary scalar field, and the {\it Weyl\ tensors} $\Phi_{\a_1\cdots
\a_{2s}}$ and $\Phi_{\ad_1,\cdots \ad_{2s}}$ for spin $s=2,4,6,...$
in the minimal bosonic model and $s=1,2,3,4,...$ in a non-minimal
bosonic model \cite{Vasiliev:1995dn,Engquist:2002vr}. Our new exact
solutions are constructed by using the oscillators to build suitable
projectors, with slightly different properties in the minimal and
non-minimal models.

Our exact solutions fall into the following four classes:

\underline{Type 0}:

These are {\it maximally symmetric solutions} (see Table 1) with
\bea  \phi(x) &=& 0\ ,\qquad \Phi_{\a_1\cdots \a_{2s}}=0\ , \qquad
\Phi_{\ad_1\cdots\ad_{2s}}=0\ , \nn\w2
e_\mu^a &=& {4\delta_\mu^a\over (1-\lambda^2x^2)^2} \ ,
 \qquad W_\mu{}^{a_1\cdots a_{s-1}}=0\ , \label{type0}\eea
describing the symmetric spaces $S^4, H_4, AdS_4, dS_4,
H_{3,2}=SO(3,2)/SO(2,2)$, where $|\lambda|$ is the inverse radius of
the symmetric space, $x^2=x^a x^b\eta_{ab}$, and $\eta_{ab}$ is the
tangent space metric. In the above the zero-forms have spin
$s=2,4,6,...$ in the minimal model and $s=1,2,3,4,...$ in the
non-minimal model, while for $W_\mu{}^{a_1\cdots a_{s-1}}$,
$s=4,6,...$ in the minimal model, and $s=1,3,4,5,6,...$ in the
non-minimal model.

\underline{Type 1}:

These solutions, which arise in the minimal models (and therefore
are evidently solutions also to the non-minimal models with
vanishing odd spins), are {\it $SO(p,4-p)$ invariant deformations}
of the maximally symmetric solutions with
\bea  \phi(x) &=& \nu (1-\lambda^2 x^2)\ ,\qquad \Phi_{\a_1\cdots
\a_{2s}}=0\ , \qquad \Phi_{\ad_1\cdots\ad_{2s}}=0\ , \   \
(s=2,4,...) \nn\w2 e_\mu^a &=& f_1 \delta_\mu^a + \lambda^2 f_2
x_\mu x^a\ ,
 \qquad W_\mu{}^{a_1\cdots a_{s-1}}=0\ \ \ (s=4,6,...)\ , \eea
where $\nu$ is a continuous parameter and $f_1,f_2$ (see \eq{f12})
are highly complicated functions of $x^2$, $\nu$, and a set of
\emph{discrete} parameters corresponding to whether certain
projectors are switched on or off. The metric is Weyl-flat conformal
to the maximally symmetric solution with a complicated conformal
factor, and note that all the higher spin gauge fields vanish.
Interestingly, a particular choice of the discrete parameters yield,
in the $\nu \rightarrow 0$ limit, the {\it degenerate metric}:
\be g_{\mu\nu}= {1\over (1-\lambda^2 x^2)}{x_\mu x_\nu\over
\lambda^2 x^2}\ . \ee

Degenerate metrics are known to play a role in topology change in
space-time (see, for example, \cite{Horowitz:1990qb}, and references
therein). Interestingly, here they arise in a natural way by simply
taking a certain limit in the parameter space of our solution.

\underline{Type 2}:

These are solutions of the non-minimal model that are \emph{not}
solutions to the minimal model. The spacetime component fields are
identical to those of the maximally symmetric Type 0 solutions, but,
unlike in the Type 0 solution, the spinorial master one-form is
non-vanishing (see \eq{spinorform}). Even though all odd spin fields
are vanishing, the solution exists only for the non-minimal model
because the spinorial master field violates the kinematic conditions
of the minimal model. In particular, this means that this type of
solution cannot be a $\nu\rightarrow 0$ limit of the Type 1
solutions. Furthermore, the spinorial master field is parametrized
by {\it discrete} parameters, again associated with projectors.

\underline{Type 3}:

These are solutions of the {\it non-minimal chiral models} in
Euclidean and Kleinian signatures, in which {\it all gauge fields
are non-vanishing}. These solutions also depend on an infinite set
of {\it discrete parameters} and for simple choices of these
parameters we obtain  two such solutions, in both of which
\be \phi(x) \ =\  -1\ , \qquad \Phi_{\a_1\cdots \a_{2s}}\ =\ 0 \
,\qquad W_\mu{}^{a_1\cdots a_{s-1}}\ \neq\ 0\ . \ee
In one of the solutions the Weyl tensors and the vierbein take the
form
\bea && \Phi_{\ad_1\cdots\ad_{2s}}\ = \ -2^{2s+1}(2s-1)!!\left(
{h^2-1 \over \e h^2}\right)^s\, U_{(\ad_1}\cdots U_{\ad_s}\,
V_{\ad_{s+1}}\cdots V_{\ad_{2s})}\ ,\w2
&& e_\mu^a\ =\ {-2\over h^2(1+2g)}\left[g_3\delta_\mu^a +g_4
\lambda^2 x_\mu x^a + g_5\lambda^2 (Jx)_\mu (Jx)^a\right]\ , \eea
where $h,g,g_3,g_4,g_5$ are functions of $x^2$ defined in \eq{aad},
\eq{g345}, and the almost complex structure $J_{ab}$ and spinors
$(U,V)$ are defined in \eq{jj} and \eq{UV}, and $\e=\pm 1$ as
explained in Section 3.5.
For the other solution we have
\bea && \Phi_{\ad_1\cdots\ad_{2s}}\ = \ -2^{2s+1}(2s-1)!!\left(
{1\over \e h^2}\right)^s\, \bar\l_{(\ad_1}\cdots \bar\l_{\ad_s}\,
\bar\mu_{\ad_{s+1}}\cdots \bar\mu_{\ad_{2s})}\ ,\w2
&& e_\mu^a\ =\ {-2\over h^2(1+2{\tilde g})}\left[ \delta_\mu^a
+{\tilde g}_4 \lambda^2 x_\mu x^a + {\tilde g}_5\lambda^2 ({\tilde
J}x)_\mu ({\tilde J}x)^a\right]\ , \eea
where the functions ${\tilde g}, {\tilde g}_4, {\tilde g}_5$ are
defined in \eq{tildeg345}, and the almost complex structure ${\tilde
J}_{ab}$ is defined in \eq{jtilde}.

These are remarkable solutions in that they are, to our best
knowledge, the first exact solution of higher-spin gauge theory in
which higher-spin fields are non-vanishing. We also note that the
Weyl tensors in these solutions corresponds to higher-spin
generalization of the Type D Weyl tensor that takes the form
$\phi_{\ad\bd\cd\dd} \sim \lambda_{(\ad}\lambda_{\bd} \mu_{\cd}
\mu_{\dd)}$, up to a scale factor \cite{flaherty}. Type D
instanton solutions of Einstein's equation in Euclidean signature
with and without cosmological constant have been discussed in
\cite{Lapedes:1980qw}. Our solution provides their higher spin
generalization.

After we describe the full HS field equations in diverse signatures
in the next Section, repeating some of the steps of the previous
Chapter to show where the difference in signature plays a role, we
shall present the detailed construction of our solutions in Section
\ref{exact}. We shall comment further on these solutions and open
problems in the Conclusions to this Chapter.


\scs{The Bosonic 4D Models in Various Signatures}


We shall first describe the field equations without imposing reality
conditions on the master fields. These conditions will then be
discussed separately leading to five different models in
four-dimensional space-times with various signatures (see Table
\ref{Table1}).


\scss{The Complex Field Equations}


To formulate the complex field equations we use \emph{independent}
$SL(2;\Comp)_L$ doublet spinors $(y_\a,z_\a)$ and $SL(2;\Comp)_R$
doublet spinors $(\yb_{\ad},\zb_{\ad})$, generating an oscillator
algebra with non-commutative and associative product $\star$ defined
by
\bea y_\a\star y_\b&=&y_\a y_\b+i\epsilon_{\a\b}\ ,\qquad
y_{\a}\star z_{\b}\ =\ y_{\a}z_{\b}-i\,\e_{\a\b}\ ,\label{osc1}\\[5pt] z_{\a}\star
y_{\b}&=& z_{\a}y_{\b}+i\,\e_{\a\b}\ , \qquad z_{\a}\star z_{\b}\ =\
z_{\a}z_{\b}-i\,\e_{\a\b} \ ,\label{osc2} \eea
and
\bea \bar y_{\dot\a}\star \bar y_{\dot\b}\ =\ \bar y_{\dot\a} \bar
y_{\dot\b}+i\epsilon_{\dot\a\dot\b}\ ,\qquad \bar z_{\dot\a}\star
\bar y_{\dot\b}\ =\ \bar z_{\dot \a} \bar y_{\dot\b}-
i\epsilon_{\dot\a\dot\b}\ ,\label{oscbar1}\\[5pt] \bar y_{\dot\a}\star \bar z_{\dot\b}\
=\ \bar y_{\dot\a} \bar z_{\dot\b}+i\epsilon_{\dot\a\dot\b}\ ,\qquad
\bar z_{\dot\a}\star \bar z_{\dot\b}\ =\ \bar z_{\dot\a} \bar
z_{\dot\b}-i\epsilon_{\dot\a\dot\b}\ .\label{oscbar2}\eea
All the definitions given in Section \ref{Z} carry over, except that
the oscillators $y$ and $\yb$, $z$ and $\zb$ are \emph{not} related
by hermitian conjugation and that now the full field equation have
independent holomorphic and anti-holomorphic sources,
\bea \widehat F&=& \frac{i}4 \left[c_1 dz^\a\wedge dz_\a \widehat
\Phi\star \kappa+c_2 d\bar z^{\ad}\wedge d\bar z_{\ad}\widehat
\Phi\star\bar\kappa\right]\ ,\label{m1}\\[5pt]\widehat D\widehat \Phi&=&0\
,\label{m2}\eea
where $c_1$ and $c_2$ are complex constants. The curvatures and
gauge transformations are still given by
\bea \widehat{F}&=& d\widehat{A}+\widehat{A}\star\widehat{A}\
,\qquad\ \delta_{\widehat \e}\widehat A\ =\ \widehat D\widehat
\e\\[5pt]\widehat{D}\widehat{\Phi}&=&d\widehat{\Phi}+[\widehat
A,\widehat \Phi]_\pi\ ,\qquad\delta_{\widehat \e}\widehat\Phi\ =\
-[\widehat \e,\widehat\Phi]_\pi\ ,\eea
with
\bea [\widehat f,\widehat g]_\pi&=&\widehat f\star\widehat
g-(-1)^{{\rm deg}(\widehat f){\rm deg}(\widehat g)}\widehat
g\star\pi(\widehat f)\ .\eea
Since $\widehat \Phi$ is defined up to rescalings by complex
numbers, the model only depends on one complex parameter, that we
can take to be
\bea c&=& {c_2\over c_1}\ .\eea
In components, the constraints read
\bea \widehat F_{\m\n}&=&0\ ,\qquad \widehat D_\mu\widehat\Phi\
\equiv\ \partial_\mu\widehat\Phi +[\widehat
A_\m,\widehat\Phi]_{\pi}\ =\ 0\ , \label{f1c}\eea
\bea \widehat F_{\m\a}&=&0\ ,\qquad \widehat F_{\m\ad}\ =\ 0\
,\label{f2c}\eea
\bea \widehat
F_{\a\b}&=&-\ft{ic_1}2\e_{\a\b}\widehat\Phi\star\kappa\ ,\qquad
\widehat F_{\ad\bd}\ =\
-\ft{ic_2}2\e_{\ad\bd}\widehat\Phi\star\bar\kappa\ ,\label{f3c}\eea
\bea \widehat F_{\a\ad}&=&0\ ,\label{f4c}\eea
\bea \widehat D_\a\widehat\Phi\ \equiv\
\partial_\a\widehat\Phi+\widehat
A_\a\star\widehat\Phi+\widehat\Phi\star\pi(\widehat A_\a)&=&0\
,\label{s1c}\eea
\bea \widehat D_{\ad}\widehat\Phi\ \equiv\
\partial_{\ad}\widehat\Phi+\widehat
A_{\ad}\star\widehat\Phi+\widehat\Phi\star\bar\pi(\widehat
A_{\ad})&=&0\ ,\label{s2c}\eea
where \eq{s2c} can be derived using $\pi\bar\pi(\widehat
A_{\ad})=-\widehat A_{\ad}$. Introducing again
\bea \widehat S_\a&=& z_\a-2i\widehat A_\a\ ,\qquad \widehat
S_{\ad}\ =\ \bar z_{\ad}-2i\widehat A_{\ad}\ ,\label{Sc}\eea
the component form of the equations carrying at least one spinor
index now take the form
\bea \partial_\m \widehat S_\a+[\widehat A_\mu,\widehat
S_\a]_\star&=&0\ ,\qquad
\partial_\m \widehat
S_{\ad}+[\widehat A_\mu,\widehat S_{\ad}]_\star\ =\ 0\
,\label{S1c}\eea \bea [\widehat S_\a,\widehat S_\b]_\star&=&
-2i\e_{\a\b}(1-c_1\widehat \Phi\star\kappa)\ ,\qquad[\widehat
S_{\ad},\widehat S_{\bd}]_\star\ =\  -2i\e_{\ad\bd}(1-c_2\widehat
\Phi\star\bar\kappa)\ ,\label{S2c}\eea\bea [\widehat S_\a,\widehat
S_{\bd}]_\star&=&0\ ,\label{S3c}\eea \bea \widehat
S_\a\star\widehat\Phi+\widehat\Phi\star\pi(\widehat S_\a)&=&0\
,\label{S4c}\eea\bea \widehat
S_{\ad}\star\widehat\Phi+\widehat\Phi\star\bar\pi(\widehat S_{\ad})\
=\ 0\ .\label{S5c}\eea
This form of the equations makes the following $\integ_2\times
\integ_2$ symmetry manifest:
\bea \widehat S_\a&\rightarrow\pm \widehat S_\a\ ,\qquad \widehat
S_{\ad}\ \rightarrow\ \pm\widehat S_{\ad}\ ,\label{Z2c}\eea
(where the two transformations can be performed independently)
keeping $\widehat A_\m$ and $\widehat \Phi$ fixed. We note that
$\widehat S_\a\rightarrow -\widehat S_\a$ is equivalent to $\widehat
A_\a\rightarrow-\widehat A_\a-i z_\a$, \emph{idem} $\widehat
S_{\ad}$ and $\widehat A_{\ad}$.

All component fields are of course complex at this level. Next we
shall discuss various reality conditions on the (hatted) master
fields that will lead to models with real physical fields living in
space-times with different signatures.


\scss{Real Forms}


In order to define the real forms of the field equations one has to
impose reality conditions on both adjoint one-form and
twisted-adjoint zero-form, corresponding to suitable real forms of
the higher-spin algebra and signatures of spacetime. There are three
distinct real forms of the complex higher-spin algebra itself. In
two of these cases there are two distinct reality conditions that
can be imposed on the zero-form, leading to five distinct models in
total, as shown in Table \ref{Table1}. The reality conditions are
\be \widehat A^\dagger\ =\ -\sigma(\widehat A)\ ,\qquad
\widehat\Phi^{\dagger}\ =\ \sigma(\pi(\widehat \Phi))\
,\label{dagger}\ee
where the possible actions of the dagger
\footnote{The dagger acts as usual complex conjugation on component
fields; in this Chapter we shall denote the conjugate of a complex
number $x$ by $x^\ast$, while reserving the bar for denoting
quantities associated with the $R$-handed oscillators.} on the
spinor oscillators and consequential selections of real forms of
$SO(4;\Comp)\simeq SL(2;\Comp)\times SL(2;\Comp)$ are given by
\bea SU(2)_L\times SU(2)_R&:&\quad (y^\a)^\dagger\ =\
y^{\dagger}_{\a}\ ,\quad (z^\a)^\dagger\ =\ z^{\dagger}_{\a}\
,\label{su2}\\&&\quad  (\bar y^{\ad})^\dagger\ =\ \bar
y^{\dagger}_{\ad}\ ,\quad (\bar
z^{\ad})^\dagger\ =\ \bar z^{\dagger}_{\ad}\ ,\nn\\[5pt]
SL(2;\Comp)_{\rm diag}&:&\quad (y^\a)^\dagger\ =\ \bar y^{\ad}\
,\quad
(z^\a)^\dagger\ =\ \bar z^{\ad}\ ,\label{sl2}\\[5pt]
Sp(2;\Real)_L\times Sp(2;\Real)_R&:&\quad (y^\a)^\dagger\ =\ y^{\a}\
,\quad (z^\a)^\dagger\ =\ -z^\a\ ,\label{sp2}\\&&\quad (\bar
y^{\ad})^\dagger\ =\ \bar y^{\ad}\ ,\quad (\bar z^{\ad})^\dagger\ =\
-\bar z^{\ad} \ ,\nn
\eea
and the map $\sigma$ is given in Table \ref{Table1}, with the
isomorphism $\rho$ given by
\be \rho(\widehat f(y^\dagger_\a,\bar y^\dagger_\a,z^\dagger_\a,\bar
z^\dagger_\a))\ =\ \widehat f(y_\a,\bar y_\a,-z_\a,-\bar z_\a)
\label{iso}\ee
in the case of $(4,0)$ signature. Note that $\sigma$ is an
oscillator-algebra automorphism in signatures $(3,1)$ and $(2,2)$,
while it is an isomorphism in signature $(4,0)$. Here, the $SU(2)$
doublets are pseudo real in the sense that from
$(y_\a)^\dagger=-y^{\dagger\a}$ \emph{idem} $(z_\a)^\dagger$,
$(\bar y_{\ad})^\dagger$ and $(\bar z_{\ad})^\dagger$ it follows
that $(y_\a,\yb_{\ad};z_\a,\zb_{\ad})$ and
$(y^\dagger_\a,\yb^\dagger_\a;z^\dagger_\a,\zb^\dagger_{\ad})$
generate equivalent oscillator algebras with isomorphism $\rho$.
The reality property of the exterior derivative $d$, defined in
\eq{spinder}, takes the following form in different signatures:
\bea \mbox{Signature $(3,1)$ and $(2,2)$}&:& \quad d^\dagger\ =\ d\
,\\\mbox{Signature $(4,0)$}&:& \quad \rho \circ d^\dagger\ =\ d\circ
\rho\ .\eea
We note that the Euclidean case is consistent in the sense that
\bea \rho (dz^\a)^\dagger&=&\rho d^\dagger (z^\a)^\dagger\ =\ d\rho
(z^\dagger_\a)\ =\ -dz_\a\eea
is compatible with representing $d\widehat f$ using
$\partial\widehat f/\partial z^\a=\frac i2 [z_\a,\widehat f]_\star$,
which yields
\bea \rho\left(  \frac i2 dz^\a [z_\a,\widehat
f]_\star\right)^\dagger&=& \frac i2 dz_\a \rho\left([\widehat
f^\dagger,-z^{\dagger\a}]_\star\right)\ =\ \frac i2
dz_\a[\rho\widehat
 f^\dagger,z^{\a}]_\star\ =\ \frac i2 dz^\a [z_\a,\rho\widehat
 f^\dagger]_\star\ .\eea
Demanding compatibility between the reality conditions \eq{dagger}
and the master field equations \eq{m1} and \eq{m2}, and using
\bea \rho\left((\kappa)^\dagger\right)&=&\kappa\ ,\qquad
\rho\left((idz^\a\wedge dz_\a)^\dagger\right)\ =\ -idz^\a\wedge
dz_\a\ ,\eea
one finds the following reality conditions on the parameters
\bea \mbox{Signature $(3,1)$}&:& c_1^\ast\ =\ c_2\ ,\\
\mbox{Signature $(4,0)$ and $(2,2)$}&:& c_1^\ast\ =\ c_1\ ,\quad
c_2^\ast\ =\ c_2\ .\eea
As a result, the parameter $c$ is a phase factor in Lorentzian
signature and a real number in Euclidean and Kleinian signatures.
The parameters can be restricted further by requiring invariance
under the parity transformation
\bea P(y_\a)&=& \yb_{\ad}\ ,\qquad P\circ d\ =\ d\circ P\ ,\qquad
P^2\ =\ {\rm Id}\ .\eea
Taking $\widehat A$ to be invariant and assigning intrinsic parity
$\e=\pm 1$ to $\widehat\Phi$,
\bea P(\widehat A)&=&\widehat A\ ,\qquad P(\widehat\Phi)\ =\ \e
\widehat\Phi\ ,\eea
one finds that the master equations are parity invariant provided
that
\bea c&=&\e\ =\ \mx{\{}{ll}{1&\mbox{Type A model (scalar)}\\[5pt]
-1&\mbox{Type B model (pseudoscalar)}}{.}\eea
In Lorentzian signature, there is no loss of generality in choosing
$c_1=c_2=1$ in the Type A model and $c_1=-c_2=i$ in the Type B
model, while in Euclidean and Kleinian signatures one may always
take $c_1=c_2=1$ in the Type A model and $c_1=-c_2=1$ in the Type B
model. More generally, the parity transformation maps different
models into each other as follows,
\bea P(c_1)&=&\e c_2\ ,\qquad P(c_2)\ =\ \e c_1\ ,\qquad P(c)\ =\
{1\over c}\ ,\eea
leaving invariant the Type A and B models. The \emph{maximally
parity violating} cases are
\bea \mbox{Signature $(3,1)$}&:& c\ =\ \exp(i\pi/4)\ ,\\[5pt]
\mbox{Signature $(4,0)$ and $(2,2)$}&:& c\ =\ 0\ .\eea
The case with $c=0$ shall be referred to as the \emph{chiral model},
that we shall discuss in more detail below.

The HS equations in Lorentzian signature have the $\integ_2$
symmetry acting as $({\widehat S}_\a, {\widehat S}_{\ad})
\rightarrow (\epsilon {\widehat S}_\a, \epsilon {\widehat
S}_{\ad})$, and $\integ_2\times \integ_2$ symmetry in $(4,0)$ and
$(2,2)$ signatures acting as $({\widehat S}_\a, {\widehat S}_{\ad})
\rightarrow (\epsilon {\widehat S}_\a, \epsilon' {\widehat
S}_{\ad})$, where $\epsilon=\pm 1$ and $\e'=\pm 1$.


Finally, let us give the reality conditions at the level of the
$SO(5;\Comp)$ algebra and its minimal bosonic higher-spin extension.
The adjoint representation of the complex minimal bosonic
higher-spin Lie algebra is defined by
\footnote{A more detailed description of the complex higher-spin
algebra and its representations is given in Chapter \ref{absalg}.}
\bea \mho(5;\Comp) &=& \left\{Q(y,\bar y)\ :\quad \tau(Q)\ =\
-Q\right\} \ ,\label{chsa}\eea
and the corresponding minimal twisted-adjoint representation by
\bea T[\mho(5;\Comp)] &=& \left\{S(y,\yb)\ :\quad \quad \tau(S)\ =\
\pi(S)\right\}\ .\label{ctwadj}\eea
The real forms are defined by
\bea\mho(5-q,q) &=& \left\{Q(y,\bar y)\in \mho(5;\Comp)\ :\quad
Q^\dagger\ =\ -\sigma(Q)\right\} \ ,\label{chsareal}\\[5pt]
T[\mho(5-q,q)] &=& \left\{S(y,\yb)\in T[\mho(5-q,q)\ :\quad
S^\dagger\ =\ \sigma(\pi(S))\right\}\ .\label{ctwadj}\eea
The finite-dimensional $SO(5;\Comp)$ subalgebra is generated by
$M_{AB}$, that we split into Lorentz rotations and translations
$(M_{ab},P_a)$ defined by
\bea \pi(M_{ab})&=&M_{ab}\ ,\qquad \pi(P_a)\ =\ -P_a\ .\eea
For these generators, which by convention arise in the expansion of
the master fields together with a factor of $i$, the reality
condition \eq{dagger} implies
\bea (M_{AB})^\dagger&=&\sigma(M_{AB})\ .\eea
This condition is solved by
\be M_{ab}\ =\ -\frac18\left((\s_{ab})^{\a\b}y_\a
y_\b+(\bar\sigma_{ab})^{\ad\bd}\yb_{\ad}\yb_{\bd}\right),\quad P_a\
=\ \frac{\l}4(\sigma_a)^{\a\ad}y_\a\yb_{\ad}\ ,\ee
where the van der Waerden symbols are defined in Appendix
\ref{App:F} and $\l^2$ is proportional to the cosmological constant,
as shown in Table \ref{Table1}. The van der Waerden symbols encode
the space-time signature $\eta_{ab}$, and the commutation relations
among the $M_{AB}$ then fix the signature of the ambient space to be
\bea \eta_{AB}&=&(\eta_{ab};-\lambda^2)\ .\eea
%


\scss{The Chiral Model}


In the chiral model with $c=0$, the master field $\widehat\Phi$ can
be eliminated using \eq{f3c}, and expressed as
\bea \widehat \Phi&=&(1+\frac{i}2 \widehat S^\a\star\widehat
S_\a)\star\kappa\ ,\eea
where we have chosen $c_1=1$ and $\widehat S_\a$ is given by \eq{S}.
The remaining independent master-field equations now read
\bea \widehat F_{\m\n}&=&0\ ,\qquad \widehat D_\m\widehat S_\a\ =\
0\ ,\qquad \widehat D_\m\widehat S_{\ad}\ =\ 0\ ,\eea\bea[\widehat
S_\a,\widehat S_{\ad}]_\star&=&0\ ,\qquad [\widehat S_{\ad},\widehat
S_{\bd}]_\star\ =\ -2i\e_{\ad\bd}\ ,\eea\bea \widehat S_\a\star
\widehat S^\b\star\widehat S_\b+\widehat S^\b\star\widehat S_\b\star
\widehat S_\a&=&4i \widehat S_\a\ .\label{SplusScube}\eea
We note that \eq{S5c} holds identically in virtue of $\widehat
S_{\ad}\star\widehat\Phi+\widehat\Phi\star\bar\pi(\widehat
S_{\ad})=[\widehat S_{\ad},1+\frac i2 \widehat S^\a\star\widehat
S_\a]_\star\star\kappa=0$, where we used $\kappa\bar\kappa\star
\widehat S_{\ad}\star\kappa\bar\kappa=-\widehat S_{\ad}$ and
$[\widehat S_\a,\widehat S_{\ad}]_\star=0$. The chiral model can be
truncated further by imposing
\bea \widehat A_{\ad}&=&0\ ,\qquad {\partial\over \partial
z^{\ad}}\widehat A_\mu\ =\ 0\ ,\qquad {\partial\over \partial
z^{\ad}}\widehat A_\a\ =\ 0\ .\eea
In general, the chiral model also has interesting solutions with
non-vanishing $\widehat A_{\ad}$, since flat connections in
non-commutative geometry can be non-trivial.


\scss{Comments on Weak-Field Expansion and Spectrum}


The procedure, described in great detail in Chapter \ref{nonlin},
for obtaining the manifestly diffeomorphism and locally Lorentz
invariant weak-field expansion of the physical field equations can
be extended straightforwardly to arbitrary signature. The expansion
is in terms of spin-$s$ physical fields with $s\neq 2$ as well as
higher derivatives of all fields, while the vierbein and Lorentz
connection are treated exactly.

In this approach one first solves \eq{f2c}--\eq{s2c} subject to the
initial condition
\bea \Phi &=& \widehat\Phi|_{Z=0}\ ,\\[5pt]
A_\mu&=& \left.\widehat A_\m\right|_{Z=0}\ =\
e_\mu+\omega_\mu+W_\m+K_{\m}\ ,\label{Amu}\eea
where
\bea e_\m&=&{1\over 2i}e_\m{}^a P_a\ ,\qquad \o_\m \ =\ \ft1{4i}
\o_\m{}^{ab}M_{ab}\ ;\label{em}\eea
$W_\m$ contains the higher-spin gauge fields (and also the spin
$s=1$ gauge field in the non-minimal model); and we recall the field
redefinition
\bea K_{\m}&=& {1\over 4i}\o_\m{}^{\a\b}\left.\widehat
S_\a\star\widehat S_\b\right|_{Z=0}+{1\over 4i}
\bar\omega_\m{}^{\ad\bd}\left.\widehat S_{\ad}\star\widehat
S_{\bd}\right|_{Z=0}\\[5pt]&=& i\o_{\m}{}^{\a\b} \left.(\widehat
A_\a\star\widehat A_\b-{\partial\over\partial y^\a} \widehat
A_\b)\right|_{Z=0}+ i\bar\o_{\m}{}^{\ad\bd} \left.(\widehat
A_{\ad}\star\widehat A_{\bd}-{\partial\over\partial \yb^{\ad}}
\widehat A_{\bd})\right|_{Z=0}\ .\label{Kmu}\eea
One also imposes the gauge condition
\bea \widehat A^{(0)}_{\a}&=&0\ ,\qquad \widehat A^{(0)}_{\ad}\ =\
0\ ,\label{gauge}\eea
where we have defined the internal flat connection
\bea \widehat A^{(0)}_{\a}&=&\widehat A_\a|_{\Phi=0}\ ,\qquad
\widehat A^{(0)}_{\ad}\ =\ \widehat A_{\ad}|_{\Phi=0}\ .\eea
One then substitutes the resulting $\widehat\Phi$ and $\widehat
A_\mu$, which can be obtained explicitly in a perturbative expansion
in $\Phi$, into \eq{f1} and sets $Z=0$, which yields a manifestly
spin-2 covariant complex HS gauge theory on the base manifold. Up to
this point the local structure of the base-manifold, nor the
detailed structure of the gauge fields, have played any role. To
proceed, one may refer to an ordinary spacetime, take $e_\mu{}^a$ to
be an (invertible) vierbein, and treat $W_\mu$ as a weak field. This
allows one to eliminate a large number of auxiliary fields in $\Phi$
and $W_\mu$, leaving a model consisting of a physical scalar
$\phi=\Phi|_{y=\bar y=0}$, the vierbein $e_\mu{}^a$, and an infinite
tower of (doubly traceless) HS gauge fields $\phi_{a(s)}$ residing
in $W_\mu$.

The gauge choice \eq{gauge} is convenient since it implies
${\partial\over\partial y^\a} \widehat A_\b|_{Z=0}=0$ that
simplifies the expansion \cite{Sezgin:2002ru}. However, there are
also other gauges where $\widehat A_\a|_{\Phi=0}$ is a flat but
non-trivial internal connection, and indeed this will be the case
for the Type 1 and Type 2 solutions that we shall present in Section
\ref{exact}.

In the leading order in the weak fields, the two-form and one-form
constraints for the minimal model read
\bea \mbox{$s=2$} &:& \left\{ \ba{ll} {\cal R}_{\a\b,\c\d}\ =\
c_2\Phi_{\a\b\c\d}\ ,\qquad&{\cal R}_{\ad\bd,\c\d}\ =\ 0\ ,\w2 {\cal
R}_{\a\b,\c\dd}\ =\ 0\ ,&{\cal R}_{\ad\bd,\c\dd}\ =\ 0\ ,\w2 {\cal
R}_{\a\b,\cd\dd}\ =\ 0\ ,&{\cal R}_{\ad\bd,\cd\dd}\ =\
c_1\Phi_{\ad\bd\cd\dd}\ ,\ea \right. \label{s2b} \w4
\mbox{$s=4,6,...$}&:& \left\{ \ba{ll} F^{(1)}_{\a\b,\c_1\dots
\c_{2s-2}}\ =\ c_2\Phi_{\a\b\c_1\dots \c_{2s-2}}\
,&F^{(1)}_{\ad\bd,\cd_1\dots\cd_k\c_{k+1}\dots \c_{2s-2}}\ =\  0\
,\la{hse1}\w3 F^{(1)}_{\a\b,\c_1\dots\c_k\cd_{k+1}\dots \cd_{2s-2}}\
=\ 0\ ,&F^{(1)}_{\ad\bd,\cd_1\dots \cd_{2s-2}}\ =\
c_1\Phi_{\ad\bd\cd_1\dots \cd_{2s-2}}\ ,\ea\right. \label{hs} \w4
\mbox{0-forms}&:& \nabla_{\a}{}^{\ad}\Phi_{\b_1\dots\b_m}{}^{
\bd_1\dots\bd_n}\ =\
i\lambda\left(\Phi_{\a\b_1\dots\b_m}{}^{\ad\bd_1\dots\bd_n}-
 mn\e_{\a(\b_1}\e^{\ad(\bd_1}\Phi_{\b_2\dots
\b_m)}{}^{\bd_2\dots\bd_n)}\right)\ ,\phantom{aaaaa}
\la{linscalareq}\eea
where for higher spins $s=4,6,\dots$ and $k=0,\dots,2s-3$, and for
$0$-forms $|m-n|=0$ mod $4$. In all cases, the zero-form system
contains a physical scalar with field equation
\bea (\nabla^2+2\lambda^2)\phi&=&0\ .\eea
In the Lorentzian case, where both $c_1$ and $c_2=c_1^\ast$ are
non-zero, the spin-2 sector consists of gravity with cosmological
constant $-3\lambda^2$, and the spin-$s$ sectors with $s=4,6,\dots$
consist of higher-spin tensor gauge fields with critical masses
proportional to $\lambda^2$. The criticality in the masses, that
implies composite masslessness in the case of AdS, holds in the dS
case as well, where thus the physical spectrum is given by the
symmetric tensor product of two (non-unitary) $SO(4,1)$ singletons
(see Chapter \ref{absalg} for details).

In the Euclidean and Kleinian cases, the parameters $c_1$ and $c_2$
are real and independent. In case $c_1c_2\neq 0$, the Lorentzian
analysis carries over, leading to a composite massless spectrum
given by symmetric tensor products of suitable singletons. However,
unlike the Lorentzian case, the spin-$s$ sector of the twisted
adjoint representation can be decomposed into left-handed and
right-handed sub-sectors of real states, corresponding to
$\{\Phi_{\a_1\dots\a_m,\ad_1\dots\ad_n}\}$ with $m-n=\pm 2s$. These
sub-sectors mix under HS transformations.

In case either $c_1$ or $c_2$, but not both, vanishes, that we shall
refer to as the chiral models, the metric and the higher-spin gauge
fields become half-flat. For definiteness, let us consider the case
$c_2=0$. The components of the zero-form that drop out in the
two-form constraint, \emph{i.e.} $\Phi_{\a_1\dots\a_{2s}}$, now
become \emph{independent} physical fields, obeying field equations
following from \eq{linscalareq}.

{\footnotesize \tabcolsep=1mm \begin{table}[h!]
\begin{tabular}{|c|c|c|c|c|c|c|}\hline
& & & & & & \\
HSA & Signature & Spinors &  &
Reality & Symmetric & Hermitian \\
&  &  & $\mbox{\phantom{aaaaa}}$ &
  &  space&isometries  \\
 & \mbox{\footnotesize $\eta_{ab}$} &  &$\lambda^2$& $\sigma$ & &
 \\ \hline & & & & & & \\
\mbox{\footnotesize $\mho(5)$} & \mbox{\footnotesize $(4,0)$} &
\mbox{\footnotesize $SU(2)_L\times SU(2)_R$} & $-1$ & $\rho$ &
\mbox{\footnotesize $S^4$} & \mbox{\footnotesize $\mso(2) \otimes
\mso(3)$}
\\ \mbox{\footnotesize $\mho(4,1)$} & \mbox{\footnotesize $(4,0)$}
& \mbox{\footnotesize $SU(2)_L\times SU(2)_R$} & $+1$ & $\rho\pi$
& \mbox{\footnotesize $H_4$} & \mbox{\footnotesize $\mso(3,1)$} \\
\mbox{\footnotesize $\mho(4,1)$} & \mbox{\footnotesize $(3,1)$} &
\mbox{\footnotesize $SL(2,\Comp)_{\rm diag}$} & $-1$ & $\pi$ &
\mbox{\footnotesize $dS_4$} &
\mbox{\footnotesize $\mso(3,1)'$} \\
\mbox{\footnotesize $\mho(3,2)$} & \mbox{\footnotesize $(3,1)$} &
\mbox{\footnotesize $SL(2,\Comp)_{\rm diag}$} & $+1$ & id &
\mbox{\footnotesize $AdS_4$} &
\mbox{\footnotesize $\mso(3,2)$} \\
\mbox{\footnotesize $\mho(3,2)$} & \mbox{\footnotesize $(2,2)$} &
\mbox{\footnotesize $SL(2,\Real)_L\times SL(2,\Real)_R$} & $-1$ & id
& \mbox{\footnotesize $H_{3,2}$} & \mbox{\footnotesize $\mso(3,2)$}
\\ & & & & & & \\
\hline
\end{tabular}
\caption{{\footnotesize The minimal bosonic higher-spin algebras
$\mho(p',5-p')\supset \mso(5-p',p')$ in signature $(p,4-p)$ can be
realized with spinor oscillators transforming as doublets under the
groups listed in the third column. These realizations obey reality
conditions $(M_{AB})^\dagger=\sigma(M_{AB})$, with hermitian
subalgebras listed above. The symmetric spaces with unit radius have
cosmological constant $\Lambda=-3\lambda^2$.}} \label{Table1}
\end{table}}


\scs{Exact Solutions}\label{exact}


In this section we shall give four types of exact solutions to the
4D HS models given in the previous section. The salient features of
these are summarized in the Introduction to this Chapter. Here we
stress that (a) the Type 0 solutions are maximally symmetric spaces;
(b) the Type 1 solutions are $SO(4-p,p)$ invariant deformations of
Type 0; (c) the Type 2 solutions, which exist necessarily in the
non-minimal model, have vanishing spacetime component fields but
non-vanishing spinorial master one-form; (d) the Type 3 solutions,
which exist in the non-minimal chiral model only, have the
remarkable feature that all higher spin gauge fields are
non-vanishing in such a way that the Weyl zero-forms are covariantly
constant, in a certain sense that will be explained below. Before we
give these four types of solutions we shall describe briefly the
method for solving the master field equations using gauge functions.


\scss{The Gauge Function Ansatz}


In order to construct an interesting class of solutions we shall
use the $Z$-space approach \cite{Vasiliev:1992av, Bolotin:1999fa,
Sezgin:2005pv} in which the constraints carrying at least one
curved space-time index, \emph{viz.}
 \bea
 \widehat F_{\m\n}&=&0\ ,\qquad \widehat D_\m \widehat\Phi\ =\ 0\ ,\\[5pt]
 \widehat F_{\m\a}&=&0\ ,\qquad
 \widehat F_{\m\ad}\ =\ 0 \ ,\label{xsp2}
 \eea
are integrated in simply connected space-time regions given the
space-time zero-forms at a point $p$,
\be
 \widehat \Phi'\ =\ \widehat
\Phi|_{p}\ ,\qquad \widehat S'_\a\ =\  \widehat S_\a|_{p}\ ,\qquad
\widehat S'_{\ad}\ =\ \widehat S_{\ad}|_{p}\ , \label{phiprime}
 \ee
and expressed explicitly as
 \bea
\widehat A_\mu&=&\widehat L^{-1}\star \partial_\mu \widehat L\
,\qquad \widehat \Phi\ =\ {\widehat L}^{-1}\star \widehat\Phi'\star
\pi(\widehat L)\ ,\qquad\\[5pt]\widehat S_\a&=&\widehat L^{-1}\star
\widehat S'_\a\star \widehat L\ ,\qquad \widehat S_{\ad}\ =\
\widehat L^{-1}\star \widehat S'_{\ad}\star \widehat L\ ,
\label{Leq}
 \eea
where $\widehat L=\widehat L(x,z,\bar z;y,\yb)$ is a gauge function,
and
\be \widehat L|_{p}\ =\ 1\ ,\qquad
\partial_\m\widehat\Phi'\ =\ 0\ ,\qquad \partial_\m{\widehat S}'_\a\ =\
0\ ,\qquad \partial_\m{\widehat S}'_{\ad}\ =\ 0\ .\ee
The internal connections $\widehat A_\a$ and $\widehat A_{\ad}$ can
be reconstructed from $\widehat S_\a$ and $\widehat S_{\ad}$ using
\eq{S}. In particular note the relation
\be {\widehat A}_\a = {\widehat L}\star \partial_\a {\widehat L}
+{\widehat L}^{-1}\star {\widehat A}'_\a \star {\widehat L}\
,\label{Aalpha} \ee
from which it follows that
\be {\widehat S}_\a'= z_\a -2i {\widehat A}'_\a\ .\label{sprime} \ee
The remaining constraints in $Z$-space, \emph{viz.}
\bea [\widehat S'_\a,\widehat S'_\b]_\star&=&
-2i\e_{\a\b}(1-c_1\widehat \Phi'\star\kappa)\ ,\qquad[\widehat
S'_{\ad},\widehat S'_{\bd}]_\star\ =\  -2i\e_{\ad\bd}(1-c_2\widehat
\Phi'\star\bar\kappa)\ ,\label{z1}\eea\bea [\widehat S'_\a,\widehat
S'_{\bd}]_\star&=&0\ ,\label{z2}\eea \bea \widehat
S'_\a\star\widehat\Phi'+\widehat\Phi'\star\pi(\widehat S'_\a)&=&0\
,\label{z3}\eea\bea \widehat
S'_{\ad}\star\widehat\Phi'+\widehat\Phi'\star\bar\pi(\widehat
S'_{\ad})\ =\ 0\ ,\label{z4}\eea
are then to be solved with an initial condition
 \be
C'(y,\bar y)\ =\ \widehat\Phi'|_{Z=0}\ ,\label{C}
 \ee
and some assumption about the topology of the internal flat
connections
\bea
 \widehat S^{\prime(0)}_{\a}&=&\widehat S'_\a|_{C'=0}\ ,\qquad
 \widehat S^{\prime(0)}_{\ad}\ =\ \widehat S'_{\ad}|_{C'=0}\ .
 \label{physgauge}
\eea

In what follows, we shall restrict the class of solutions further by
assuming that
\be \widehat L\ =\ L(x;y,\yb)\ .\ee
The gauge fields can then be obtained from \eq{Amu}, \eq{Kmu} and
\eq{Leq}, \emph{viz.}
\be e_\m+\o_\m+W_\m\ =\ L^{-1}\partial_\m L-K_\m\ ,\label{gf}\ee
where
\bea K_\m&=&{1\over 4i}\left.L^{-1}\star \left(\o_{\m}{}^{\a\b}
\widehat S'_{\a}\star\widehat S'_{\b}+\bar\o_{\m}{}^{\ad\bd}
\widehat S'_{\ad}\star\widehat S'_{\bd}\right)\star L\right|_{Z=0}\
.\label{Kmu2}\eea
Hence, the gauge fields, including the metric, can be obtained
algebraically without having to solve any differential equations in
space-time.


\scss{Ordinary Maximally Symmetric Spaces (Type 0)}\label{Sec:symm}


The complex master-field equations are solved by
\be \widehat\Phi\ =\ 0\ ,\qquad \widehat S_\a\ =\ z_\a\ ,\qquad
\widehat S_{\ad}\ =\ \bar z_{\ad}\ ,\qquad \widehat A_\m\ =\
L^{-1}\star
\partial_\m L \ ,\ee
where the gauge function \cite{Bolotin:1999fa}
\bea L(x;y,\yb)&=& {2h\over 1+h} \exp\left[{i\lambda
x^{\a\dot\a}y_\a \bar y_{\dot\a}\over 1+h}\right]\ ,\label{wL} \eea
gives
\be ds^2_{(0)}\ =\ {4 dx^2\over (1-\lambda^2x^2)^2}\
,\label{vacmetric}\ee
which we identify as the metric of the symmetric spaces listed in
Table \ref{Table1} for the different real forms of the model, in
stereographic coordinates with inverse radius $|\lambda|$. This
metric is invariant under the inversion
\bea x^a\rightarrow -x^a/(\l^2 x^2)\ ,\eea
and $H_4$ is covered by a single coordinate chart, while the
remaining symmetric spaces require two charts, related by the
inversion. If we let $\tilde x^a= -x^a/(\l^2 x^2)$, the atlases are
given by
\bea S^4\ \quad(\l^2=-1)&:& \{x^\mu:0\leq -\lambda^2 x^2\leq 1\}\cup
\{\tilde x^\mu:0\leq
-\lambda^2 \tilde x^2\leq 1\}\ ,\label{atl1}\\[5pt]
H^4\quad\quad(\l^2=1)&:& \{x^\mu:0\leq \lambda^2 x^2<1\}\ ,\label{atl2}\\[5pt]
dS_4\quad(\l^2=-1)&:& \{x^\mu:-1< -\lambda^2 x^2\leq 1\}\cup
\{\tilde x^\mu:-1<
-\lambda^2 \tilde x^2\leq 1\}\ ,\label{atl3}\\[5pt]
AdS_4\ \quad(\l^2=1)&:& \{x^\mu:-1 \leq \lambda^2 x^2<1\}\cup
\{\tilde x^\mu:-1\leq
\lambda^2 \tilde x^2<1\}\ ,\label{atl4}\\[5pt]
H_{3,2}\quad(\l^2=-1)&:& \{x^\mu:-1< -\lambda^2 x^2\leq 1\}\cup
\{\tilde x^\mu:-1< -\lambda^2 \tilde x^2\leq 1\}\ ,\label{atl5}\eea
where the overlap between the charts is given by $\{x^\mu:\lambda^2
x^2=-1\}$ in the cases of $S^4$, $dS_4$, $AdS_4$ and $H_{3,2}$, and
the boundary is $\{x^\mu:\lambda^2 x^2=1\}$ in the case of $H_4$ and
$\{x^\mu:\lambda^2 x^2=1\}\cup \{\tilde x^\mu:\lambda^2\tilde
x^2=1\}$ in the cases of $dS_4$, $AdS_4$ and $H_{3,2}$. The
$H_{3,2}$ space can be described as the coset $SO(3,2)/SO(2,2)$.


\scss{$SO(4-p,p)$ Invariant Solutions to the Minimal Model (Type 1)}



\scsss{Internal Master Fields}


A particular class of $SO(4;\Comp)$-invariant solutions is given by
the ansatz
\be \widehat \Phi'\ =\ \nu\ ,\qquad \widehat S'_\alpha\ =\
z_\a~S(u)\ ,\qquad \widehat S'_{\ad}\ =\ \bar z_{\ad}~\bar S(\bar
u)\ee
where
\be u\ =\ y^\a z_\a\ ,\qquad \bar u\ =\ \bar y^{\dot \a} \bar
z_{\dot \a}\ .\ee
The above ansatz solves \eq{z2}-\eq{z4}. There remains to solve
\eq{z1}, which now takes the form
\be [\widehat S^{\prime\alpha},\widehat S'_\a]_\star\ =\ 4i(1-
c_1\nu e^{iu})\ ,\qquad [\widehat S^{\prime\ad},\widehat
S'_{\ad}]_\star\ =\ 4i(1- c_2\nu e^{-i\bar u})\label{SSnu}\ee
Following \cite{Prokushkin:1998bq}, we use the integral
representation
\bea S(u)&=&\int_{-1}^1 ds~ n(s) ~e^{\frac{i}2(1+s)u}\ ,\label{SSnu2}\\
\bar S(\bar u)&=& \int_{-1}^1 ds~ \bar n(s) ~e^{-\frac{i}2(1+s)\bar
u}\ .\label{Su}\eea
which reduces \eq{SSnu} to
\bea (n \circ n)(t)&=&\ \delta(t-1)-\frac{c_1\nu}2 (1-t)\ ,\\
(\bar n \circ \bar n)(t)&=& \delta(t-1)-\frac{c_2\nu}2 (1-t)\ .\eea
with $\circ$ defined by \cite{Prokushkin:1998bq}
 \be
 (f\circ g)(t)=\int_{-1}^1 ds \int_{-1}^1 ds'
 \delta(t-s s')~f(s)~g(s')\ .
 \ee
Even and odd functions, denoted by $f^\pm(t)$, are orthogonal with
respect to the $\circ$ product. Thus, one finds
\bea (n^{+}\circ n^{+})(t)&=&\iota^{+}_0(t)-\frac{c_1\nu}2\ ,\qquad
(n^{-}\circ n^{-})(t)\ =\ \iota^{-}_0(t)+\frac{c_1\nu}2 t\
,\label{mplus}\\(\bar n^{+}\circ \bar
n^{+})(t)&=&\iota^{+}_0(t)-\frac{c_2\nu}2\ ,\qquad (\bar n^{-}\circ
\bar n^{-})(t)\ =\ \iota^{-}_0(t)+\frac{c_2\nu}2 t\
,\label{mminus}\eea
where
\be \iota^{\pm}_0(t)\ =\
\frac12\left[\delta(1-t)\pm\delta(1+t)\right]\ .\ee
One proceeds \cite{Prokushkin:1998bq}, by writing
\bea n^\pm(t)&=& m^\pm(t)+\sum_{k=0}^\infty \l_k p^\pm_k\
,\label{mexp}\eea
where $m^\pm$ are expanded in terms of $\iota^{(\pm)}_0(t)$ and the
functions ($k\geq 1$)
 \bea
 \iota^{\s}_k(t)&=&\left[{\rm sign}(t)\right]^{\frac12(1-\s
 )}~\int_{-1}^1 ds_1 \cdots \int_{-1}^1 ds_k~\delta(t-s_1\cdots
 s_k)\nn\\[5pt]
 &=&\left[{\rm sign}(t)\right]^{\frac12(1-\s)}{\left(\log
 \frac1{t^2}\right)^{k-1}\over (k-1)!}\ ,
 \eea
obeying the algebra ($k,l\geq 0$)
 \be
 \iota^{\s}_k\circ \iota^{\s}_l\ =\ \iota^{\s}_{k+l}\ ,
 \label{ring}
 \ee
and $p^\s_k(t)$ ($k\geq 0$) are the $\circ$-product projectors
\bea p^{\s}_k(t)&=& {(-1)^k\over k!} \d^{(k)}(t)\ ,\qquad \s\ =\
(-1)^k\ ,\label{pk}\eea
obeying
\bea p^\s_k\circ f&=& L_k[f] p^\s_k\ ,\qquad L_k[f]\ =\ \int_{-1}^1
dt~ t^k f(t)\ .\label{proj1}\eea
In particular,
\bea p^{\s}_k\circ p^{\s}_l&=& \delta_{kl}p^{\s}_l\
.\label{proj2}\eea
Substituting the expansion \eq{mexp} into \eq{mplus} and
\eq{mminus}, one finds, in view of \eq{ring}, \eq{proj1} and
\eq{proj2}, manageable algebraic equations. Transforming back one
finds, after some algebra \cite{Sezgin:2005pv},
\bea m(t)&=& \d(1+t)+q(t)\ ,\\[5pt] q(t) &=&-{c_1\nu\over 4}\left({}_1\!
F_1\left[\frac12;2;{c_1\nu\over 2}\log \frac 1{t^2}\right]+t\,{}_1\!
F_1\left[\frac12;2;-{c_1\nu\over 2}\log \frac 1{t^2}\right]\right)\
,\label{mt}\eea
and
\bea \l_k&=&- 2\th_kL_k[m]\ ,\qquad \th_k\in\{0,1\}\
,\label{lambdak}\eea
where
\bea L_k[m]&=&(-1)^k+L_k[q]\ ,\label{Lkm}\\[5pt] L_k[q]&=& -{1+(-1)^k\over
2}\left(1-\sqrt{1-{c_1\nu\over 1+k}}\right)- {1-(-1)^k\over
2}\left(1-\sqrt{1+{c_1\nu\over 2+k}}\right)\ .\eea
The overall signs in $m^\pm$ have been fixed in \eq{mt} by requiring
that
\be S(u)=1\ \ {\rm for}\ \  \nu=0 \ \ {\rm and} \ \ \theta_k=0 \ .
\ee
Treating $\bar n$ the same way, one finds
\bea \bar m(t)&=& \d(1+t)+\bar q(t)\ ,\\[5pt] \bar q(t) &=&-{c_2\nu\over 4}\left({}_1\!
F_1\left[\frac12;2;{c_2\nu\over 2}\log \frac 1{t^2}\right]+t\,{}_1\!
F_1\left[\frac12;2;-{c_2\nu\over 2}\log \frac 1{t^2}\right]\right)\
,\label{mt2}\\[5pt]\bar \l_k&=&- 2\bar \th_kL_k[\bar m]\ ,\qquad \bar \th_k\in\{0,1\}\ ,\\[5pt]
L_k[\bar m]&=&(-1)^k+L_k[\bar q]\ ,\\[5pt] L_k[\bar q]&=& -{1+(-1)^k\over
2}\left(1-\sqrt{1-{c_2\nu\over 1+k}}\right)- {1-(-1)^k\over
2}\left(1-\sqrt{1+{c_2\nu\over 2+k}}\right)\ .\eea
Thus, the internal solution is given by
 \bea
 \widehat \Phi'&=&\nu\ ,\label{intsol1}\eea
together with ${\widehat S}'_\a$ and ${\widehat S}'_{\dot\a}$ as
given in \eq{sprime} with
 \bea \widehat A'_\a&=&
 \widehat A^{\prime(reg)}_\a+\widehat A^{\prime(proj)}_\a\ ,\qquad
 \qquad\qquad\qquad\!\! \widehat
 A'_{\ad}\ =\
 \widehat A^{\prime(reg)}_{\ad}+\widehat A^{\prime(proj)}_{\ad}\ ,\label{intconn}\\[5pt]
 \widehat A^{\prime(reg)}_\a&=&\frac{i}2 z_\a \int_{-1}^1 dt~ q(t)\,
 e^{\frac{i}2(1+t)u}\ ,\qquad\,\,\quad \widehat A^{\prime(reg)}_{\ad}
 \ =\ \frac{i}2 \bar z_{\ad} \int_{-1}^1 dt~ \bar q(t)\,
 e^{-\frac{i}2(1+t)\bar u}\ ,\label{part}\\[5pt]\widehat A^{\prime(proj)}_\a&=&
 -iz_\a \sum_{k=0}^\infty \theta_k (-1)^k L_k[m] P_k(u)\ ,\qquad \widehat
 A^{\prime(proj)}_{\ad}\ =\
 -i\bar z_{\ad}\sum_{k=0}^\infty \bar \theta_k (-1)^k L_k[\bar m] \bar P_k(\bar u)\
 ,\qquad \label{hom}\eea
where
\bea P_k(u)&=&\int_{-1}^1 ds ~e^{\ft{i}2 (1-s)u} p_k(s)\ =\ {1\over
k!}\left({-iu\over 2}\right)^ke^{\ft{iu}2}\ ,\label{Pk1}\\[5pt]
\bar P_k(\bar u)&=&\int_{-1}^1 ds ~e^{-\ft{i}2 (1-s)\bar u} p_k(s)\
=\ {1\over k!}\left({i\bar u\over 2}\right)^k e^{-\ft{i\bar
u}2}\label{Pk2}\eea
are projectors in the $\star$-product algebra given by functions of
$u$ and $\bar u$, \emph{viz.}
\bea P_k\star F&=& L_k[f]P_k\ ,\qquad P_k\star P_l\ =\
\delta_{kl}P_k\ ,\\[5pt] \bar P_k\star \bar F&=& L_k[\bar f]\bar P_k\ ,
\qquad \bar P_k\star \bar P_l\ =\ \delta_{kl}\bar P_k\ ,\eea
for $F(u)=\int_{-1}^1 ds e^{\ft{i}2(1-s)u}f(s)$ and $\bar F(\bar
u)=\int_{-1}^1 ds e^{-\ft{i}2(1-s)\bar u}\bar f(s)$ with $L_k[f]$
and $L_k[\bar f]$ given in \eq{proj1}. The projectors also obey
$(u-2ik)\star P_k=0$ and $y^\a\star P_k\star
z_\a=i(k+1)(P_{k-1}+P_{k+1})$ with $P_{-1}\equiv 0$. We note the
opposite signs in front of $s$ in the exponents of \eq{SSnu2},
\eq{Su} and \eq{Pk1}, \eq{Pk2}, resulting in the $(-1)^k$ in the
projector part \eq{hom} of the internal connection, which we can
thus write as
\bea \widehat
A^{\prime(proj)}_\a\!\!\!&=&\!\!\!-iz_\a\sum_{k=0}^\infty\left[\theta_k
P_k-\left(1-\sqrt{1-{c_1\nu\over
1+2k}}\right)\theta_{2k}P_{2k}+\left(1-\sqrt{1+{c_1\nu\over
3+2k}}\right)\theta_{2k+1}P_{2k+1}\right]\,,\qquad\qquad\\[5pt]
\widehat A^{\prime(proj)}_{\ad}\!\!\!&=&\!\!\!-i\bar
z_{\ad}\sum_{k=0}^\infty\left[\bar\theta_k \bar
P_k-\left(1-\sqrt{1-{c_2\nu\over 1+2k}}\right)\bar \theta_{2k}\bar
P_{2k}+\left(1-\sqrt{1+{c_2\nu\over 3+2k}}\right)\bar
\theta_{2k+1}\bar P_{2k+1}\right]\,,\qquad\qquad\eea
which are analytic functions of $\nu$ in a finite region around the
origin. For example, for $c_1=c_2=1$, they are real analytic for
$-3<{\rm Re}\nu<1$, where also the particular solution can be shown
to be real analytic \cite{Sezgin:2005pv}. The reality conditions on
the $\th_k$ and $\bar\th_k$ parameters are as follows:
\bea \mbox{$(4,0)$ and $(2,2)$ signature}&:& \th_k\ ,\quad
\bar\th_k\qquad \mbox{independent}\ ,\\[5pt]
\mbox{$(3,1)$ signature}&:& \th_k\ =\ \bar\th_k\ .\eea

Taking $\nu=0$ there remains only the projector part, leading to the
following ``vacuum'' solutions
\bea \widehat \Phi'&=&0\ ,\eea\bea \widehat
A'_\a&=&-iz_\a\sum_{k=0}^\infty\theta_k {1\over k!}\left({-iu\over
2}\right)^ke^{\ft{iu}2}\ ,\qquad  \widehat A'_{\ad}\ =\ -i\bar
z_{\ad}\sum_{k=0}^\infty\bar\theta_k {1\over k!}\left({i\bar u\over
2}\right)^k e^{-\ft{i\bar u}2}\ .\eea
The $\integ_2\times \integ_2$ symmetry \eq{Z2} acts by
\bea \th_k&\rightarrow& 1-\th_k\ ,\qquad \bar\th_k\ \rightarrow\
1-\bar\th_k\ .\eea
The maximally symmetric spaces discussed in Section 3.2 are
recovered by setting $\th_k=\th$ and $\bar \th_k=\bar\th$ for all
$k$. In Euclidean and Kleinian signatures, $\th$ and $\bar\th$ are
independent, leading to four solutions related by $\integ_2\times
\integ_2$ transformations. In Lorentzian signature, $\th=\bar\th$
leading to two solutions related by $\integ_2$ symmetry.


\scsss{Space-time Component Fields}


The calculation of the component fields follows the same steps as in
\cite{Sezgin:2005pv}. The spin $s\geq 1$ Weyl tensors vanish, while
the scalar field is given by
\bea\phi(x)&=& \nu h^2(x^2)\ =\ \nu(1-\lambda^2 x^2)\ .\eea
In order to compute the gauge fields, we first need to compute the
quantity $K_\m$ given in \eq{Kmu2}. This calculation is formally the
same as the one spelled out in the case of $\th_k=\bar\th_k=0$ in
\cite{Sezgin:2005pv}, and the result is
\bea K_\mu&=&{Q\over 4i}\omega_\mu^{\a\b}v_\a v_\b+{\bar Q\over
4i}\bar\omega_{\mu}^{\ad\bd}\bar v_{\ad}\bar v_{\bd}\ ,\eea
where
\bea Q&=& -{(1-a^2)^2\over 4}\int_{-1}^1ds \int_{-1}^1
ds'{(1+s)(1+s')n(s) n(s')\over (1-ss'a^2)^4}\ ,\label{Q}\\[5pt]
\bar Q&=& -{(1-a^2)^2\over 4}\int_{-1}^1ds \int_{-1}^1
ds'{(1+s)(1+s')\bar n(s) \bar n(s')\over (1-ss'a^2)^4}\
.\label{barQ}\eea
and
\bea v_\a&=&(1+a^2)y_\a+2(a\bar y)_\a\ ,\qquad \bar v_{\ad}\ =\
(1+a^2)\bar y_{\ad}+2(\bar a y)_{\ad}\ ,\eea
with ${\bar a}_{\ad \a}= a_{\a\ad}$ defined in \eq{aad}. We can
simplify $Q$ using $n(t)=\delta(1+t)+q(t)+\sum_k\lambda_k p_k(t)$,
with $p_k(t)$ given by \eq{pk} and $\lambda_k$ by \eq{lambdak} and
\eq{Lkm}. After some algebra we find
\bea Q(\nu;\{\th_k\})&=&Q^{(reg)}(\nu)+Q^{(proj)}(\nu;\{\th_k\})\ ,\\[5pt]
Q^{(reg)}&=& -{(1-a^2)^2\over 4}\int_{-1}^1ds \int_{-1}^1
ds'{(1+s)(1+s')q(s) q(s')\over (1-ss'a^2)^4}\ ,\\[5pt]
Q^{(proj)} &=& (1-a^2)^2\sum_{k=0}^\infty {4_k a^{2k}\over
k!}(\th_k-\th_{k+1})^2\times\nn\\[5pt]
& &
\hspace{2cm}\times\left((-1)^k+L_k(q)\right)\left((-1)^{k+1}+L_{k+1}(q)\right)\
,\eea
where we note that $Q$ depends on $\th_k$ only via
$\th_k-\th_{k+1}$. The same expression with $q\rightarrow\bar q$ and
$\th_k\rightarrow \bar\th_k$ holds for $\bar Q$. The regular part,
which was computed in \cite{Sezgin:2005pv}, is given by
\bea Q^{(reg)}&=& Q^{(reg)}_++Q^{(reg)}_-\ ,\\[5pt]
Q^{(reg)}_+&=&-{(1-a^2)^2\over 4}\sum_{p=0}^\infty {-4\choose
2p}a^{4p}\left(\sqrt{1-{c_1\nu\over 2p+1}}-\sqrt{1+{c_1\nu\over
2p+3}}\right)^2\label{Qpl}\\[5pt]
Q^{(reg)}_-&=&{(1-a^2)^2\over 4}\sum_{p=0}^\infty {-4\choose
2p+1}a^{4p+2}\left(\sqrt{1-{c_1\nu\over 2p+3}}-\sqrt{1+{c_1\nu\over
2p+3}}\right)^2\ ,\label{Omi}\eea
while a similar expression, obtained by replacing $c_1\rightarrow
c_2$, holds for $\bar Q$.

Since $K_\mu$ is bilinear in the $y_\a$ and $\yb_{\ad}$ oscillators,
it immediately follows that all higher spin fields vanish. Moreover,
after some algebra, we find that the vierbein and $\mso(4;\Comp)$
connection are given by
\bea e^a&=& f_1(x^2) dx^a+f_2(x^2) x^a dx^b x_b\ ,\\[5pt]
\omega_{\a\b}&=& f(x^2)\omega^{(0)}_{\a\b}\ ,\qquad
\bar\omega_{\ad\bd}\ =\ \bar f(x^2) \bar\omega^{(0)}_{\ad\bd}\ ,\eea
where
\bea f&=& {1+(1-a^2)^2\bar Q\over
\left[1+(1+a^2)^2Q\right]\left[1+(1+a^2)^2\bar
Q\right]-16a^4 Q\bar Q}\ ,\\[5pt] \bar f&=& {1+(1-a^2)^2 Q\over \left[1+(1+a^2)^2Q\right]\left[1+(1+a^2)^2\bar
Q\right]-16a^4 Q\bar Q}\ ,\eea
and
\bea f_1+\lambda^2 x^2 f_2&=& {2\over h^2}\ ,\qquad f_2\ =\
{2(1+a^2)^4\over (1-a^2)^2}(fQ+\bar f\bar Q)\ .\label{f12}\eea
By a change of coordinates, the metric can be written locally, in a
given coordinate chart, as a foliation
\bea ds^2&=& \e d\tau^2+R^2 d\O_3^2\ ,\qquad R^2(\tau)\ =\ \eta^2
|\sinh^2(\sqrt{\e}\tau)|\ ,\eea
where $x^2=\e \tan^2\ft{\tau}2$ with $\e=\pm 1$, and $d\Omega_3^2$
is a three-dimensional metric of constant curvature with suitable
signature, and \cite{Sezgin:2005pv}
\bea \eta&=&{f_1 h^2\over 2}\ .\label{etafactor}\eea
One has the following simplifications in specific models:
\bea \mbox{Type A model:}&& Q\ =\ \bar Q\ ,\qquad \eta\ =\
{1+(1-a^2)^2Q\over 1+(1+6a^2+a^4)Q}\ ,\\[5pt]
\mbox{Chiral model:}&& \bar Q\ =\ 0\ ,\qquad \eta\ =\
{1+(1-a^2)^2Q\over 1+(1+a^2)^2Q}\ .\eea
The metric may have conical singularities, namely zeroes
$R(\tau_0)=0$ for which $\partial_\tau R|_{\tau_0}\neq 1$ (we note
that $\eta|_{\tau=0}=1$, so that $\tau=0$ is not a conical
singularity). The scale factor depends heavily on $\nu$ as well as
on the choice of the infinitely many discrete parameters $\th_k$ and
$\bar\th_k$. This makes the analysis unyielding, and we shall
therefore limit ourselves to the case of vanishing discrete
parameters and $|\nu|\ll 1$. The resulting analysis was performed in
\cite{Sezgin:2005hf} in Lorentzian signature, and it generalizes
straightforwardly to Euclidean and Kleinian signatures. To this end,
one examines the integrals \eq{Q} and \eq{barQ} in the limits
$a^2\rightarrow\pm 1$, where there are potentially divergent
contributions from the region of the integration domain where $s$
and $s'$ approach $\pm 1$. These contributions are actually finite
for $a^2=1$, while they diverge as
\bea Q^{(reg)}&=&{c_1\nu\over 6}\log(1+a^2)+{\cal O}(\nu^2)\ ,\eea
when $a^2\rightarrow -1$ (and the $\cO(\n^2)$ contributions are
finite). Focusing on a single chart, as listed in
\eq{atl1}-\eq{atl5}, $a^2$ is bounded from below by
$(1-\sqrt{2})(1+\sqrt{2})^{-1}$, and hence, if $|\nu|\ll 1$, then
$|Q|\ll 1$, and consequently the factor $\eta$ defined in
\eq{etafactor} remains finite. Thus, for small enough $\nu$, there
are no conical singularities within the coordinate charts. However,
they may appear for some finite critical $\nu$.

While the $Q$ functions are complicated for $\nu\neq 0$, they
simplify drastically at $\nu=0$, where we find
\bea Q&=&-(1-a^2)^2\sum_{k=0}^\infty {4_k a^{2k}\over
k!}(\th_k-\th_{k+1})^2\ .\eea
An analogous expression can be found for $\bar Q$. Setting
$(\th_k-\th_{k+1})^2=1$, yields
\bea Q&=&-{1\over (1-a^2)^2}\ .\eea
If $Q=\bar Q=-(1-a^2)^{-2}$, which is necessarily the case in the
Lorentzian models, then the equation system for $\o_{\a\b}$ and
$\bar\o_{\ad\bd}$ becomes degenerate, and one finds
\bea \o_{\a\b}&=& -{(1-a^2)^2\over 8a^2} \omega^{(0)}_{\a\b}\ =\
{(\s^{ab})_{\a\b}dx_a x_b\over 2 x^2}\ ,\\[5pt]
\bar\o_{\ad\bd}&=& -{(1-a^2)^2\over 8a^2} \bar\omega^{(0)}_{\ad\bd}\
=\ {(\bar\s^{ab})_{\ad\bd}dx_a x_b\over 2 x^2}\ ,\eea
leading to the degenerate vierbein
\bea e_{\a\ad}&=&-{\l x_{\a\ad} x^a dx_a\over x^2h^2}\ ,\eea
and metric
\bea ds^2&=&{4(x^a dx_a)^2\over\l^2 x^2 h^2}\ .   \eea
%


\scss{Solutions of the Non-minimal Model (Type
2)}\label{sec:van0form}



\scsss{Internal Master Fields}


The non-minimal model admits the following solutions
\bea \widehat \Phi'&=&0\ ,\qquad \widehat S'_\a\ =\ z_\a\star
\C(y,\bar y)\ ,\qquad\widehat S'_{\ad}\ =\ \bar z_{\ad}\star \bar
\C(y,\bar y)\ ,\label{type2}\eea
provided that
\bea \C\star\C&=&\bar\C\star\bar\C\ =\ 1\ ,\qquad [\C,\bar
\C]_\star\ =\ 0\ ,\qquad \pi\bar\pi(\C)\ =\ \C\ ,\qquad
\pi\bar\pi(\bar\C)\ =\ \bar\C\ .\label{gprop}\eea
The elements $\C$ and $\bar\C$ can be written as
\bea \C&=&1-2P\ ,\qquad \bar \C\ =\ 1-2\bar P\ ,\eea
where $P(y,\bar y)$ and $\bar P(y,\bar y)$ are projectors obeying
\bea P\star P&=&P\ ,\qquad \bar P\star\bar P\ =\ \bar P\ ,\quad
[P,\bar P]_\star\ =\ 0\ ,\qquad \pi\bar\pi(P)\ =\ P\ ,\qquad
\pi\bar\pi(\bar P)\ =\ \bar P\ .\qquad\eea
A set of such projectors is described in Appendix \ref{AppProj},
where we also explain why the projectors can be subject to the
$\tau$-conditions of the non-minimal model, given in \eq{nonminmod},
but not to those of the minimal model, given in \eq{minmod}, unless
one develops some further formalism for handling certain divergent
$\star$-products.


\scsss{Space-time Component Fields}


Turning to the computation of the space components of the master
fields, since $z_\a$ star-commutes with $L$, it immediately follows
from \eq{gf}, \eq{Kmu} and \eq{type2} that
\bea K_\mu&=&0\ .\eea
From \eq{gf} this in turn implies that all HS gauge fields and the
spin-1 gauge field vanish, while the metric is that of maximally
symmetric spacetime. To that extent, the Type 1 solution looks like
the Type 0 solution, but it does differ in an important way, namely,
here the internal connection, \emph{i.e.} the spinor component
${\widehat A}_\a$ of the master $1$-form, is non-vanishing. Indeed,
\eq{type2}, \eq{gprop} and \eq{sprime} give the result
\bea{\widehat A}_\a&=& -iz_\alpha \star V(x;y,\yb)\ ,\qquad \widehat
A_{\ad}\ =\ -i\bar z_{\ad} \star \bar V(x;y,\yb)\
,\label{spinorform}\eea
where the quantities $V$ and $\bar V$, which shall be frequently
encountered in what follows, are defined by
\bea V&=& L^{-1}\star P\star L\ ,\qquad \bar V\ =\  L^{-1}\star \bar
P\star L \ .\eea
Their explicit evaluation is given in Appendix \ref{App2}, with the
result \eq{generalV}.

Whilst the internal connection does not turn on any spacetime
component fields, it does, however, affect the interactions as it
does not obey the physical gauge condition normally used in the
weak-field expansion \cite{Sezgin:2002ru}, namely that the internal
connection should vanish when the zero-form vanishes. In this sense,
the internal connection may be viewed as a non-trivial flat
connection in the non-commutative space.


\scss{Solutions of the Non-minimal Chiral Model (Type 3)}



\scsss{Internal Master Fields}


In the case of the non-minimal chiral model, defined in Section 2.3,
it is possible to use projectors $P(y,\bar y)$ to build solutions
with non-vanishing Weyl zero-form and higher spin fields. They are
\bea\widehat \Phi'&=& (1-P)\star\kappa\ ,\qquad \widehat S'_\a\ =\
z_\a\star P\ ,\qquad \widehat S'_{\ad}\ =\ \bar z_{\ad}\star \bar\C\
,\eea
where
\bea P\star P&=&P\ ,\qquad \bar\C\star\bar\C\ =\ 1\ ,\qquad
[P,\bar\C]_\star\ =\ 0\ ,\qquad \pi\bar\pi(P)\ =\ P\ ,\qquad
\pi\bar\pi(\bar\C)\ =\ \bar\C\ .\qquad\eea
These elements of the $\star$-product algebra can be constructed as
in Section \ref{sec:van0form} and Appendix \ref{AppProj}.

For the purpose of exhibiting explicitly the spacetime component
fields, we choose to work with the simplest possible projectors,
namely
\bea P_+(y)&=& 2e^{-2\e uv}\ =\ 2e^{\e yby}\ ,\label{Pplus}\\[5pt]
P_-(\yb)&=& 2e^{-2\e \bar u\bar v}\ =\ 2e^{\e \bar y\bar b \bar y}\
,\label{Pminus}\eea
where $\e=\pm 1$, and $u$, $v$, $\bar u$, $\bar v$, $b_{\a\b}$ and
$\bar b_{\ad\bd}$ are defined in Appendices \ref{AppDef} and
\ref{AppProj}.


\scsss{Space-time Component Fields}


The master gauge field and zero-form are given by
\bea e_\mu+\omega_\mu+W_\mu&=&
e^{(0)}_\mu+\omega^{(0)}_\mu+{\o_\mu^{\a\b}\over 4i} {\partial^2 V
\over\partial y^\a\partial y^\b}\ ,\label{Type3}\eea
and
\bea\Phi&=&\left[L^{-1}\star(1-P)\star\kappa\star
\pi(L)\right]|_{Z=0}\ =\ 1-V|_{y_\a=0}\ ,\eea
where $V$ is given by \eq{generalV} and we have used \eq{kappa}.
Remarkably, since there is no $y$-dependence in the Weyl zero-form
$\Phi$, it is covariantly constant in the sense that
$\Phi_{\a(m)\ad(n)}$ vanishes unless $m=0$. Moreover, using
\eq{generalV}, it is straightforward to compute the constant value
of the physical scalar field, with the result
\bea \phi(x)&=& 1-4\sum_{n_1,n_2\in
\integ+\ft12}(-1)^{n_1+n_2-\ft{\e_1+\e_2}2}\th_{n_1,n_2}\
,\label{VEV}\eea
where $\th_{n_1,n_2}$ are constrained as in \eq{constrth}. Summing
over all $n_2$, as explained in Appendix \ref{AppProj}, and using
\eq{generating} with $x=0$, \emph{i.e.} $\sum_{k=0}^\infty
(-1)^k=\ft12$, one finds that for the reduced projector
\eq{reduced}, the scalar field is given by
\bea \phi(x)&=& 1-2\sum_{n\in \integ+\ft12}(-1)^{n-\ft{\e}2}\th_{n}\
,\eea
where $\th_n$ obey the condition given below \eq{reduced}. Finally
setting $\theta_n=\th(\pm n)$, one ends up with $P=1$, \emph{i.e.}
in the Type 0 case, where indeed $\phi(x)=0$.

In the special cases of \eq{Pplus} and \eq{Pminus}, one finds
\bea V_+ &=& L^{-1}\star P_+\star L\ =\ 2 \exp \left(-\e{ [2\yb \bar
a -(1+a^2)y]\,b\,[2a\yb
+(1+a^2)y]\over (1-a^2)^2}\right)\ ,\label{vnonmin}\\[5pt]
V_-&=&L^{-1}\star P_-\star L\ =\ 2 \exp \left(-\e{ [2y a
-(1+a^2)\yb]\,\bar b\,[2\bar ay +(1+a^2)\yb]\over (1-a^2)^2}\right)
\label{vminus}\eea
where $a_{\a\ad}$ and $b_{\a\b}$ are defined in Appendix
\ref{AppDef} and $\e$ is defined in \eq{Pplus} and \eq{Pminus}. The
physical scalar is now given in both cases by
\bea\phi(x)&=&-1\ ,\eea
and the self-dual Weyl tensors in both cases by ($s=1,2,3,....$)
\bea \Phi_{\a(2s)}&=& 0\ ,\eea
while the anti-self-dual Weyl tensors take the form
\bea \Phi^+_{\ad_1\cdots\ad_{2s}}&=& -2^{2s+1}(2s-1)!!\left( {h^2-1
\over \e h^2}\right)^s\, U_{(\ad_1}\cdots U_{\ad_s}\,
V_{\ad_{s+1}}\cdots V_{\ad_{2s})}\ ,\label{phi+}\\[5pt]
\Phi^-_{\ad_1\cdots\ad_{2s}}&=&-2^{2s+1}(2s-1)!!\left( {1\over \e
h^2}\right)^s\, \bar\l_{(\ad_1}\cdots \bar\l_{\ad_s}\,
\bar\mu_{\ad_{s+1}}\cdots \bar\mu_{\ad_{2s})}\ ,\label{phi-}\eea
with spinors $(U, V)$ defined in \eq{UV}.

In the case of $\l^2=1$ in Euclidean signature, we only need to use
one coordinate chart, in which $0\leq h^2\leq 1$. The Weyl tensors
blow up in the limit $h^2\rightarrow 0$, preventing the solution
from approaching $H_4$ in this limit. In this sense the above
solution is a non-perturbative solution without weak-field limit in
any region of spacetime. Indeed, in the perturbative weak-field
expansion around the $H_4$ solution, the scalar field has
non-vanishing mass, preventing the linearized scalar field from
being a non-vanishing constant.

In the case of $\l^2=-1$ in Euclidean signature, the base manifold
consists of two charts, covered by the coordinates in \eq{atl1}.
Thus, in each chart we have $1\leq h^2<2$, and so the local
representatives \eq{phi+} and \eq{phi-} of the Weyl tensors are
well-defined throughout the base manifold.

Finally, in the case of $\l^2=-1$ in Kleinian signature, one also
needs two charts, with $0\leq h^2\leq 2$, and hence the Weyl tensors
blow up in the limit $h^2\rightarrow 0$, preventing the solution
from approaching $H_{3,2}$ in this limit.

From the Weyl tensors, which are not in themselves HS gauge
invariant quantities, one can construct an infinite set of invariant
(and thus closed) zero-forms \cite{Sezgin:2005pv}, namely
\bea {\cal C}^-_{2p}&=& \int {d^4y d^4z\over (2\pi)^4} [(\widehat
\Phi\star\pi(\widehat\Phi)]^{\star p}\star\kappa\bar\kappa\
.\label{inv}\eea
Remarkably, on our solution they all assume the same value, given by
the constant value of the scalar field, \emph{viz.}
\bea {\cal C}^-_{2p}&=&(1-V)^{\star 2p}|_{y=\bar y=0}\ =\
1-4\sum_{n_1,n_2\in
\integ+\ft12}(-1)^{n_1+n_2-\ft{\e_1+\e_2}2}\th_{n_1,n_2}\ .\eea

The calculation of the metric in the two models  proceeds in a
parallel fashion as follows:

%
{\bf The $P_+$ Solution:}
%
%

From \eq{Type3} and \eq{vnonmin} a straightforward computation
yields the result
\bea e_{\mu\ad\a} &=& e_{\mu\ad\a}^{(0)} +12(1+h)h^{-4}\,
b_{(\a\b}(ba)_{\gamma)\ad}\,\omega_\mu^{\beta\gamma}\
,\label{nm1}\w2
\omega_{\mu\a\b} &=& \omega_{\mu\a\b}^{(0)} +12
h^{-4}b_{(\a\b}b_{\gamma\delta)}\,\omega_\mu^{\gamma\delta}\
,\label{nm2}\w2
{\bar\omega}_{\mu\ad\bd}&=& {\bar\omega}_{\mu\ad\bd}^{(0)}
+4(1+h)^2h^{-4}\left[ -(\bar aba)_{\ad\bd} b_{\gamma\delta} +2 (\bar
ab)_{\ad\gamma} (\bar ab)_{\bd\delta}
\right]\omega_\mu^{\gamma\delta}\ .\label{nm3} \eea
First we solve for the spin connection from \eq{nm2} by inverting
the hyper-matrix that multiplies $\omega^{(0)}$, obtaining the
result
\be \omega_{\mu\a\b}= g_1\left[ \omega_{\mu\a\b}^{(0)} -8g
(b\omega_\mu^{(0)}b)_{\a\b}\right] + g_2 b_{\a\b} b^{\gamma\delta}
\omega_{\mu\gamma\delta}^{(0)}\ ,  \label{nm4}\ee
where
\be g_1={1\over 1-4g^2}\ ,\qquad g_2={4g\over (1-2g)(1-4g)}\ ,\qquad
g=h^{-4}\ . \ee
Substituting this result in \eq{nm1} then gives the vierbein
\be e_\mu^a= {-2\over h^2(1+2g)}\left[ g_3\delta_\mu^a +g_4
\lambda^2 x_\mu x^a + g_5\lambda^2 (Jx)_\mu (Jx)^a\right]\ , \ee
where
\be g_3=1+2h^{-2}\ , \qquad g_4 = 2g\, \qquad g_5= {6g\over 1-4g}\ ,
\label{g345}\ee
and the spin connections are given in \eq{nm4} and \eq{nm3}. Thus,
the metric $g_{\mu\nu}=e_\mu^a e_\nu^b\,\eta_{ab}$ takes the form
\be g_{\mu\nu}= {4\over h^4(1+2g)^2}\,\left[ g_3^2 \eta_{\mu\nu} +
g_4(\lambda^2 x^2 g_4+2g_3) x_\mu x_\nu +g_5(\lambda^2x^2
g_5+2g_3)(Jx)_\mu (Jx)_\nu\right]\ .\ee
The vierbein has thus potential singularities at $h^2=0$ and
$h^2=2$. The limit $h^2\rightarrow 0$ is a boundary in the case of
$\l^2=1$ in Euclidean signature and $\l^2=-1$ in Kleinian signature.
At these boundaries $e_\mu{}^a\sim h^{-2}x_\mu x^a$, \emph{i.e.} a
scale factor times a degenerate vierbein. In the limit
$h^2\rightarrow 2$ one approaches the boundary of a coordinate chart
in the case of $\l^2=1$ in Euclidean signature and $\l^2=-1$ in
Kleinian signature. Also in this limit, the vierbein becomes
degenerate, \emph{viz.} $e_\mu{}^a\sim h^{-2}(Jx)_\mu (Jx)^a$.

{\bf The $P_-$ Solution:}

A parallel computation that uses \eq{Type3} and \eq{vminus} yields
the result
\bea e_{\mu\ad\a} &=& e_{\mu\ad\a}^{(0)} +12\lambda^2 x^2 (1+h)
h^{-4}\,  {\tilde b}_{(\a\b}({\tilde
ba})_{\gamma)\ad}\,\omega_\mu^{\beta\gamma}\ ,\label{nnm1}\w2
\omega_{\mu\a\b} &=& \omega_{\mu\a\b}^{(0)} +12(\lambda^2x^2)^2
h^{-4}{\tilde b}_{(\a\b}{\tilde
b}_{\gamma\delta)}\,\omega_\mu^{\gamma\delta}\ ,\label{nnm2}\w2
{\bar\omega}_{\mu\ad\bd}&=& {\bar\omega}_{\mu\ad\bd}^{(0)}
+4(1+h)^2h^{-4}\left[ -(\bar a{\tilde b}a)_{\ad\bd} {\tilde
b}_{\gamma\delta} +2 ({\tilde b}a)_{\gamma\ad} ({\tilde
b}a)_{\delta\bd} \right]\omega_\mu^{\gamma\delta}\ ,\label{nnm3}
\eea
where ${\tilde b}_{\a\b}$ is defined in \eq{btilde}. As before,
solving for the spin connection from \eq{nnm2} by inverting the
hyper-matrix that multiplies $\omega^{(0)}$, we obtain
\be \omega_{\mu\a\b}= {\tilde g}_1\left[ \omega_{\mu\a\b}^{(0)}
-8{\tilde g} ({\tilde b}\omega_\mu^{(0)}{\tilde b})_{\a\b}\right] +
{\tilde g}_2 {\tilde b}_{\a\b} {\tilde b}^{\gamma\delta}
\omega_{\mu\gamma\delta}^{(0)}\ , \label{nnm4}\ee
where
\be {\tilde g}_1={1\over 1-4{\tilde g}}\ ,\qquad {\tilde
g}_2={4{\tilde g}\over (1-2{\tilde g})(1-4{\tilde g})}\ ,\qquad
{\tilde g}= (\lambda^2x^2)^2 h^{-4}\ . \ee
Substituting this result in \eq{nnm1} then gives the vierbein
\be e_\mu^a= {-2\over h^2\left(1+2{\tilde g} \right)}\left[
\delta_\mu^a +{\tilde g}_4 \lambda^2 x_\mu x^a + {\tilde
g}_5\lambda^2 ({\tilde J}x)_\mu ({\tilde J}x)^a\right]\ , \ee
where ${\tilde J}_{ab}$ is defined in \eq{jtilde}
\be {\tilde g}_4 = 2\lambda^2 x^2h^{-4}\, \qquad {\tilde g}_5=
{6\lambda^2 x^2h^{-4}\over 1-4{\tilde g}}\ , \label{tildeg345}\ee
and the spin connections are given in \eq{nnm4} and \eq{nnm3}. Thus,
the metric $g_{\mu\nu}=e_\mu^a e_\nu^b\,\eta_{ab}$ takes the form
\be g_{\mu\nu}= {4\over h^4\left[1+2{\tilde g}\right]^2}\,\left[
\eta_{\mu\nu} + {\tilde g}_4(\lambda^2 x^2 {\tilde g}_4+2) x_\mu
x_\nu +{\tilde g}_5(\lambda^2x^2 {\tilde g}_5+2)({\tilde J}x)_\mu
({\tilde J}x)_\nu\right]\ .\ee
The vierbein has potential singularities at $h^2=0$, $h^2=2$ and
$h^2=\ft23$. The singularities at $h^2=0$ and $h^2=2$ are related to
degenerate vierbeins exactly as for the $P^+$ solution. The
singularity at $h^2=\ft23$, which arises in the case of $\l^2=1$ in
Euclidean and Kleinian signature, also gives a degenerate vierbein.
This is an intriguing situation since the latter degeneration occurs
inside the coordinate charts.

\chapter{Singletons, Anti-Singletons and HS Master Fields}\label{map}

The aim of this Chapter is to elaborate further on the relevant
representations entering the Vasiliev equations. We have seen in
Chapter \ref{absalg} that the spectrum of physical fields that
emerges from the analysis of the linearized curvature constraints
actually fits the doubleton spectrum \eq{RacRac}
\cite{Konstein:1988yg, mishasalg}. In $AdS_D$ such states fill up a
unitary multiplet of the HS algebra, and are in correspondence with
the modes of normalizable fluctuation fields of all spins, with
finite Killing energy \cite{Breitenlohner:1982jf}. In Chapter
\ref{unfolding}, we have also remarked on the fact that, in an
unfolded system, all the local physical degrees of freedom are
contained in the zero-form at a point in space-time, that is to say,
in the basis monomials of the twisted adjoint zero-form $\Phi$ (see
eqs. \eq{levelell} and \eq{Phis}). Our purpose is now to have a
closer look to this basis of operators in order to establish a
direct correspondence between the master-field description and the
massless irreducible representations earlier examined. In other
words, we want to construct a map that exhibits the physical content
of the master fields, arranged into massless irreps of the
background isometry algebras, without solving the various
torsion-like constraints of the unfolded system. This aim will be
accomplished with the construction of a map that essentially
involves two steps:

\begin{enumerate}

\item Performing a nontrivial change of basis, that connects the
$\mso(D;\Comp)$-covariant tensors of the master fields with the
$\mso(2;\Comp)\oplus\mso(D-1;\Comp)$-covariant elements of the
massless modules;

\item Defining a \emph{reflection} map that essentially sends states to
operators and viceversa, preserving the representation properties.

\end{enumerate}

This mapping will be defined at the level of complex
representations, and later restricted to the various real forms of
the HS algebra. As we shall see, the outcome will be that to each
basis monomial of the twisted adjoint representation there
corresponds a ``coherent'' superposition of infinitely many states,
and, viceversa, to every state in the lowest weight modules there
corresponds a nonpolynomial combinations of basis monomials.

This analysis provides some insight into various features of
Vasiliev equations. For example, it shows that, while the on-shell
content of the twisted adjoint zero-form is related to the tensor
product of two singletons, that of the adjoint one-form is related
to the finite-dimensional $\mso(D+1;\Comp)$-modules that arise from
the tensor product of a singleton and its negative-energy
counterpart, called \emph{anti-singleton}. Thus, an intertwiner,
that transforms singletons in anti-singletons, connects the two
modules, in a way that is reminiscent of the role of the operator
$\kappa$ in the full Vasiliev equations.

Another outcome is that the twisted adjoint representation contains
not only the massless lowest-weight or highest-weight modules:
indeed, for every spin $s$, a bigger indecomposable module (that
contains also a lowest-spin module, rather than lowest energy
module) sits in principle in the zero-form master-field, and all its
elements can take part in the dynamics. The states in the
lowest-spin module do not admit an interpretation as composites of
singleton, and correspond to non-normalizable solutions of the
free-field equations, with divergent Killing energy. However, it is
important to know the full structure of the twisted adjoint module,
not only for analyzing the perturbative spectrum of physical
excitations in any signature, but also for including in the analysis
nonperturbative degrees of freedom, like the static linearized
solution $\Phi|_p=\frac{\sinh 4E}{4E}$ found in
\cite{Sezgin:2005pv}.

Finally, the problem of potential local divergencies in HSGT, due to
the contribution of an arbitrary number of derivatives to some
interaction vertices (as, for example, in the scalar-field
corrections to the stress-energy tensor calculated in
\cite{Kristiansson:2003xx}), is mapped to the problem of divergent
$\star$-products of nonpolynomial combinations of generators. This
is however a somewhat more transparent setting, and indeed we make a
proposal for an explicit regularization scheme.

This Chapter is organized as follows. We begin by defining some of
the key tools that will be useful for establishing the
correspondence: a nontrivial trace operation on the algebra $\cA$
(defined in \eq{calA1} and \eq{calA}), that will endow the latter
with a nondegenerate inner product, and certain \emph{reflector}
states, that possess a series of useful properties. Indeed, the
trace operation on $\cA$ can be defined via the expectation value
of its elements between such reflector states and their duals,
and, more generally, they lie at the heart of the state/operator
correspondence mentioned above. For example, they enable a
presentation of the master-fields as left bimodule constructs.
Then, we proceed to a general definition of the twisted adjoint
indecomposable module in compact basis, \emph{i.e.}, in the basis
of operators with definite $\mso(2;\Comp)\oplus\mso(D-1;\Comp)$
quantum numbers. As stated above, the latter are given by
nonpolynomial combinations of the twisted adjoint basis monomials.
We recover, in this fashion, the Flato-Fronsdal theorem and the
composite nature of massless lowest-weight modules (and their
negative energy counterparts, that are highest-weight modules).
Moreover, we also investigate the structure of the lowest-spin
module, containing the above mentioned nonperturbative degrees of
freedom: in particular, the static linearized solution
$\frac{\sinh 4E}{4E}$ will enter here as one of the two static
ground states from which the entire compact twisted adjoint module
can be generated. We also stress the appearance, in odd $D\geq 7$
dimensions, of scalar submodules that generalize the scalar
singleton representations, as their weight diagram consists of $p$
lines ($p=(D-5)/2$), which we shall refer to as
\emph{$p$-linetons}. In $D=3+2p$, $p=1,2,...$, we find similar
spin-1 $p$-linetons that generalize the 5D spin-1 singleton. Next,
we examine the inverse procedure of embedding
$\mso(D;\Comp)$-tensors into the compact basis as superpositions
of infinitely many states. In particular, we give a realization of
the (composite) reflector states as $\mso(D;\Comp)$-invariant
combinations of states in the massless scalar representations. We
also find an analog of the Flato-Fronsdal theorem for the tensor
product of the singleton with an anti-singleton: the decomposition
is given in terms of \emph{finite-dimensional} modules
(\emph{i.e.}, lowest \emph{and} highest-weight modules) of
$\mso(D+1;\Comp)$, in terms of which the content of the adjoint
master field can be analyzed. We also examine a relation between
adjoint and twisted-adjoint representation that provides a direct
explanation for the agreement of the eigenvalues of the Casimir
operators in the two representations. Finally, we comment on real
forms, and on the inner products on $\cA$ defined through the
trace operation. In $D=4$, for a composite reflector expanded over
states belonging to the scalar and spinor singleton
representations, the latter corresponds (modulo overall factors)
to the supertrace operation\footnote{For further details on the
relation between inner products, trace and supertrace operations
in the simpler context of one-dimensional Fock spaces we refer to
Appendix \ref{App:G}.} defined in \cite{V3} for the oscillator
realization. This setting furnishes a particularly simple example
of a composite reflector state with finite norm induced by the
standard inner product of the singleton representations. Finally,
we comment on the issue of divergent $\star$-products of
nonpolynomial elements and on a possible regularization scheme for
the weak-field expansion of the Vasiliev equations.

The results collected in this Chapter have been obtained in
\cite{us} and \cite{companion}.


\scs{Non-composite Trace and Reflector}\label{Sec:Tr}


The quotient algebra ${\cal A}$ can be equipped with a non-composite
trace operation $\Tr:{\cal A}\mapsto \Comp$ defined by the
projection\footnote{We note that if $\Tr$ is a cyclic trace
operation and $X$ is not a singlet then $\Tr[X]=0$, since $X$ can be
written as a commutator. Thus, up to normalization, any trace
operation on ${\cal A}$ is the projection onto the unity in some
basis, where the choice of basis is a crucial part of the definition
in the infinite-dimensional case.} on the $\mso(D+1;\Comp)$-singlet
$X_0$ in the basis \eq{calA}, or, equivalently (since \eq{Xod} and
\eq{Taskbs} are strong identities, \emph{i.e.} they hold in ${\cal
U}$, and $X_n$ contains $X^{(0,0)}$ if and only if $n=0$), on the
$\mso(D;\Comp)$-singlet $X^{(0,0)}$ in the basis \eq{Xod},
\emph{viz.}\footnote{We remark that at the level of the full
enveloping algebra ${\cal U}[\mg]$ of a general Lie algebra $\mg$,
\emph{i.e.} prior to factoring out any other ideal than the
commutation rules of $\mg$, the trace operation $\Tr[X]=X_{\1}$,
where $X_{\1}$ is the coefficient of $\1$, is trivial, that is
$\Tr[X\star Y]=(X\star Y)_{\1}=X_{\1}Y_{\1}$. At the level of
quotient algebras ${\cal U}[\mg]/{\cal I}[R]$, where ${\cal I}[R]$
is the annihilator of a representation $R$ of $\mg$, the trace $\Tr$
is equivalent to the composite trace over $R$, that is
$\Tr_R[X]=\sum_{n}\bra{n^\ast}X\ket{n}$ where $\ket{n}$ and
$\bra{n^\ast}$ are basis elements for $R$ and $R^\ast$ and
$\bra{m^\ast}n\rangle=\delta^m_n$, if $R$ is finite-dimensional. In
case $R$ is infinite-dimensional, more care is required. Thus, in
the case at hand, we shall distinguish between the non-composite
trace $\Tr$ on ${\cal A}$ (defined in the basis \eq{calA}) and the
composite trace $\Tr_{\mD_0}$ over the scalar-singleton
lowest-weight space $\mD_0$. We note that the composite trace can be
rewritten as $\Tr_R[X]={}_{12}\bra{\1^\ast}X(1)\ket{\1}_{12}$ where
the composite reflectors are defined by
$\ket{\1}_{12}=\sum_n\ket{n}_1\otimes \ket{n}_2$ and
${}_{12}\bra{\1^\ast}=\sum_n{}_1\bra{n^\ast}\otimes
{}_2\bra{n^\ast}$. Moreover, in case the ideal ${\cal I}[R]$ is
non-trivial, both types of traces are non-trivial, \emph{i.e.}
$\Tr[X\star Y]\neq (\Tr[X])(\Tr[Y])$ \emph{idem} $\Tr_R$. }
\bea \Tr[X]&=& X_0\ =\ X^{(0,0)}\ .\label{Trprime}\eea
This trace operation is well-defined on ${\cal A}$, since $X\simeq
X'$ iff $X$ and $X'$ have the same expansions in the bases
\eq{calA}, or, equivalently \eq{Xod}. The cyclicity follows from the
invariance of $\Tr$ under the anti-automorphism $\tau$ defined in
\eq{taumap}, that is
\bea \Tr[\t(X)]&=&\Tr[X]\ .\eea
Indeed, given two elements $X,Y\in\cA$, this implies
\bea \Tr[X\star Y] \ = \ \Tr[\t(X\star Y)] \ = \ \Tr[\t(Y)\star
\t(X))] \ = \ \Tr[Y\star X]\ ,\eea
as can be seen by splitting $X=X_++X_-$ with $\t(X_\pm)=\pm X_\pm$,
\emph{idem} $Y$, and noting that $\Tr(X_\pm\star Y_\mp)=0$. The
trace equips ${\cal A}$ with an non-degenerate bi-linear inner
product $(X,Y)\mapsto\Tr[X\star Y]$ that is invariant under the
adjoint action of ${\cal A}$ on itself, which includes not only
$\mso(D+1;\Comp)$ but also its higher-spin extension.

The basis elements $T_{A(n),B(n)}$ and $T_{a(n),b(m)}$ defined in
\eq{TAnBn} (we suppress the hat on the basis elements \eq{TAnBn}
in this Chapter, as that notation will be reserved here to
elements of the enlarged algebra \eq{enlA}) and \eq{Taskbs},
respectively, have inner products
\bea \Tr[T_{A(n),B(n)}\star T^{C(m),D(m)}]&=& \delta_{mn}
\delta_{\{A(n),B(n)\}}^{\{C(n),D(n)\}}{\cal
N}_{(n,n)_{D+1}}\ , \label{trtABtCD}\\[5pt]
\Tr[T_{a(n),b(m)}\star T^{c(n'),d(m')}]&=& \delta_{n,n'}
\delta_{m,m'}\delta_{\{a(n),b(m)\}}^{\{c(n),d(m)\}}{\cal
N}_{(n,m)_D}\ ,\label{trtabtcd}\eea
%
where the normalizations are given by
\bea {\cal N}_{(n,n)_{D+1}}&=& \l_n\l_{n-1}\cdots \l_1\ =\ (-2)^{-n}
{n!(n+1)!(\e_0)_n\over
(\e_0+\ft32)_n} \ ,\label{calNn}\\[5pt] {\cal N}_{(n)_D}&=& \l_n^{(0)}\l_{n-1}^{(0)}\cdots \l_1^{(0)}\ =\ 8^{-n}
{n!(n+1)!(2\e_0)_n\over (\e_0+\ft32)_n} \ ,\label{calNn0}\eea
as can be seen by repeated use of \eq{AcMAB} and \eq{TPa}, $(x)_n$
being the Pochhammer symbol, $(x)_n=\C(x+n)/\C(x)$. For example,
\bea &&\Tr[T_{a(n)}\star T_{b(n)}]\ =\ \Tr[P_{\{a_1}\star\cdots\star
P_{a_n\}}\star
T_{b(n)}]\nn\\[5pt]&&=\
\Tr[P_{\{a_1}\star\cdots\star P_{a_{n-1}}\star\left(
T_{n+1}+T_n+\l_n^{(0)}\eta_{a_n\}\{b_1}T_{b(n-1)\}}\right)]\nn\\[5pt]&&=\
\Tr[P_{\{a_1}\star\cdots\star P_{a_{n-1}}\star
\l_n^{(0)}\eta_{a_n\}\{b_1}T_{b(n-1)\}}]\ ,\eea
where $T_n$ denote traceless symmetric $\star$-products of $n$
translations, and we have used the fact that $\Tr[T_{n+1}\star
T_{n-1}]=\Tr[T_n\star T_{n-1}]=0$. In particular, using
\bea \d^{\{a(n)\}}_{\{b(n)\}}&=& \sum_{k=0}^{[n/2]}
t_{k}(\eta^{a(2)}\eta_{b(2)})^k\delta^{a(n-2k)}_{b(n-2k)}\ ,\qquad
t_k\ =\ {(-n)_{2k}\over 4^k k!(-n-\e_0+\ft12)_k}\ ,\eea
from which it follows that
\bea \d^{\{0(n)\}}_{\{0(n)\}}&=&\sum_{k=0}^{[n/2]} {(-n)_{2k}\over
4^k k!(-n-\e_0+\ft12)_k}\ =\ 2^{-n}{(2\e_0+1)_n\over
(\e_0+\ft12)_n}\ ,\eea
we obtain
\bea \Tr[T_{0(n)}\star T_{0(n)}]&=& (-1)^n
4^{-2n}{n!(n+1)!(2\e_0)_n(2\e_0+1)_n\over
(\e_0+\ft12)_n(\e_0+\ft32)_n}\ .\label{TrEnEn}\eea

The non-composite trace $\Tr$ can equivalently be realized as the
expectation value
\bea \Tr[X]&=& _{12}{}\bra{\1^\ast}X(1)\ket{\1}_{12}\
,\label{trvev}\eea
where $\ket{\1}_{12}\in {\cal B}$ and ${}_{12}\bra{\1^\ast}\in{\cal
B}^\ast$ are a \emph{non-composite reflector and non-composite dual
reflector} belonging to the ${\cal A}$ left and right
bimodules\footnote{In general, given a vector space $V$ that is a
left module of an algebra ${\cal A}$, the dual $V^\ast$ carries a
natural right action of ${\cal A}$, such that $v^\ast(Xw)=(v^\ast
X)(w)$ for $v^\ast\in V^\ast$, $w\in V$ and $X\in{\cal A}$. This
generalizes straightforwardly to bimodules. A an element $\widehat
X_{12}$ in a bimodule is referred to as composite in case it can be
represented as a sum of factorized elements of the form $u_1\otimes
v_2$. In finite-dimensional bimodules all elements are composite,
while this is not necessarily the case for infinite-dimensional
bimodules.}
\bea {\cal B}&=&\left\{\ket{\widehat X}_{12}:\ V(\xi)
\ket{\widehat X}_{12}=0\ ,\xi=1,2\right\}\ ,\label{defB}\\[5pt]
{\cal B}^\ast&=&\left\{{}_{12}\bra{\widehat X^\ast}:\
{}_{12}\bra{\widehat X^\ast}V(\xi)=0\ ,\xi=1,2\right\}\
,\label{defBast}\eea
(where V is the singleton annihilating ideal defined in
\eq{idealV}, \eq{VAB} and \eq{VABCD}) obeying the overlap
conditions
\bea (X(1)-(\tau\circ\pi)(X)(2))\ket{\1}_{12}&=&0\ ,\quad
{}_{12}\bra{\1^\ast}(X(1)-(\tau\circ\pi)(X)(2))\ =\ 0\
,\label{overlapX}\eea
for all $X\in{\cal A}$ and where $\tau\circ\pi$ is the
anti-automorphism composed by the anti-automorphism $\tau$ defined
in \eq{taumap} and the automorphism $\pi$ defined in \eq{pimap}, and
the pairing between the reflector and its dual is normalized to
\bea _{12}{}\bra{\1^\ast}\1\rangle_{12}&=& 1 \ .\eea
Eq. \eq{trvev} is a consequence of the cyclicity property
\bea _{12}{}\bra{\1^\ast}(X\star
Y)(1)\ket{\1}_{12}&=&_{12}{}\bra{\1^\ast}(Y\star X)(1)\ket{\1}_{12}\
,\eea
which follows from \eq{overlapX} and $(\pi\tau)^2={\rm Id}$ (the
cyclicity implies the $\mso(D+1;\Comp)$-invariance
$_{12}{}\bra{\1^\ast}[M_{AB}(1),X(1)]_\star\ket{\1}_{12}=0$ so that
if one expands $X$ as in the basis \eq{calA}, dropping the ideal
parts in view of the definitions \eq{defB} and \eq{defBast}, one
finds $_{12}{}\bra{\1^\ast} X(1)\ket{\1}_{12}=X_0=\Tr[X]$). We also
notice that the overlap conditions \eq{overlapX} are equivalent to
\bea (M_{ab}(1)+M_{ab}(2))\ket{\1}_{12}&=&0\ ,\qquad
(P_a(1)-P_a(2))\ket{\1}_{12}\ =\ 0\ ,\label{overlapM}\\[5pt]
{}_{12}\bra{\1^\ast}(M_{ab}(1)+M_{ab}(2))&=&0\ ,\qquad
{}_{12}\bra{\1^\ast}(P_a(1)-P_a(2))\ =\ 0\ ,\label{overlapP}\eea
which means that the reflectors are $\mso(D;\Comp)_{\rm
diag}$-invariant.

Next we observe that the map $\pi$ is an outer automorphism of
${\cal A}$. Upon defining the operator $k$ by
\bea k\star X&=& \pi(X)\star k\ ,\qquad k\star k\ =\ \1\ ,\qquad
\tau(k)\ =\ \pi(k)\ =\ k\ ,\eea
the map $\pi$ becomes an inner automorphism of the enlarged algebra
\bea \widehat \cA&=& \cA\oplus(\cA\star k)\ .\label{enlA}\eea
Since $\cA\star k$ does not contain the unity, we have
\bea \Tr[\widehat X]&=&\Tr[X]\ ,\eea
from which it in particular follows that $\Tr[k]=0$. Moreover, since
$\pi(V)=V$ it follows that if $\ket{\widehat X}_{12}\in {\cal B}$
and ${}_{12}\bra{\widehat X^\ast}\in{\cal B}^\ast$, then
$k(\xi)\ket{\widehat X}_{12}\in {\cal B}$ and ${}_{12}\bra{\widehat
X^\ast}k(\xi)\in{\cal B}^\ast$ for $\xi=1,2$. We can now establish a
sequence of reflection maps
\bea {\cal B}_{12}&\stackrel{R_2}{\longrightarrow}&
\widehat\cA_{12}\ \stackrel{R_1}{\longrightarrow}\ {\cal
B}^\ast_{12}\ ,\eea
by making the definitions
\bea _{23}\bra{\1^\ast}\star\ket{\1}_{13}&=& \1_{12}\
,\label{def1}\\[5pt] (k(1)-k(2))\ket{\1}_{12}&=&0\ ,\qquad
{}_{12}\bra{\1}(k(1)-k(2))\ =\ 0\ .\label{def2}\eea
Explicitly, the reflection maps are given by
\bea |\widehat X\rangle_{12}&=&\widehat X(1)\ket{\1}_{12}\ ,\qquad
{}_{12}\bra{\widehat X^\ast}\ =\ {}_{12}\bra{\1^\ast}\widehat X(1)\
,\label{doubleton}\\[5pt] \widehat X_{12}&=&
_{23}\bra{\1^\ast}\star\ket{\widehat X}_{13}\ =\
{}_{23}\bra{\widehat X^\ast}\star\ket{\1}_{13}\ ,\eea
with $\widehat X=X+Y\star k\in\widehat\cA$. We note that from
\eq{overlapM} and \eq{overlapP} it follows that the quantity ${\cal
O}_{12}=_{23}\bra{\1}\star \ket{\1}_{13}$ obeys $\widehat X\star
{\cal O}={\cal O}\star \widehat X$ for all $\widehat X\in \widehat
{\cal A}$, so that ${\cal O}=c\1$, and hence the definition
\eq{def1} amounts to setting $c=1$. We also define the \emph{twisted
non-composite reflectors}
\bea \ket{\j1}_{12}&=&k(1)\ket{\1}_{12}\ =\
k(2)\ket{\1}_{12}\ ,\label{twrefl}\\[5pt]
{}_{12}\bra{\j1^\ast}&=&{}_{12}\bra{\1^\ast}k(1)\ =\
{}_{12}\bra{\1^\ast}k(2)\ ,\eea
obeying the overlap conditions
\bea (X(1)-\t(X)(2))\ket{\j1}_{12} & = & 0 \ ,\qquad
{}_{12}\bra{\j1^\ast}(X(1)-\t(X)(2)) \ =\  0 \
,\label{twoverlap}\eea
which in particular imply that the twisted reflectors are
$\mso(D+1;\Comp)_{\rm diag}$-invariant, and having the
normalizations
\bea {}_{12}\langle{\j1^\ast}\ket{\j1}_{12}&=&1\ ,\qquad
{}_{12}\langle{\j1^\ast}\ket{\1}_{12}\ =\
{}_{12}\langle{\1^\ast}\ket{\j1}_{12}\ =\ 0\ ,\\[5pt] {}_{23}\bra{\j1^\ast}\star
\ket{\j1}_{12}&=&\1_{13}\ ,\qquad
{}_{23}\bra{\j1^\ast}\star\ket{\1}_{12}\ =\
{}_{23}\bra{\1^\ast}\star\ket{\j1}_{12}\ =\ k_{13}\ .\eea
The trace $\Tr$ can now be written as
\bea \Tr[X] \ = \ {}_{12}\bra{\j1^\ast}X(1)\ket{\j1}_{12}\ ,\eea
since ${}_{12}\bra{\j1^\ast}X(1)\ket{\j1}_{12}= {}_{12}\bra{\1^\ast}
k(2)X(1) k(2)\ket{\1}_{12}=
{}_{12}\bra{\1^\ast}X(1)\ket{\1}_{12}=\Tr[X]$.

The trace operation $\Tr$ induces higher-spin invariant bilinear
forms on the adjoint and twisted-adjoint representations:
\bea (Q,Q')_{\mho}&=&\Tr[Q\star Q']\ =\ _{12}\langle{Q}|
Q'\rangle_{12}\
,\label{hoinnerproduct}\\[5pt]
(S,S')_{\cal T}&=& \Tr[\pi(S)\star S']\ =\ _{12}\langle{\pi(S)}|
S'\rangle_{12}\ ,\label{Tinnerproduct}\eea
where \eq{doubleton} has been used to define
\bea \ket{Q_\pm}_{12}&=& Q(1)\ket{\1}_{12}\ =\ \ft12
(Q_\pm(1)\mp\pi(Q_\pm)(2))\ket{\1}_{12}\ ,\label{ketQ}\\[5pt]
\ket{S_\pm}_{12}&=& S_\pm(1)\ket{\1}_{12}\ =\ \ft12 (S_\pm(1)\pm
S_\pm(2))\ket{\1}_{12}\ ,\label{ketS}\eea
and
\bea {}_{12}\bra{Q_\pm}&=& {}_{12}\bra{\1^\ast}Q_\pm(1)\ =\ \ft12
{}_{12}\bra{\1^\ast}(Q_\pm(1)\mp\pi(Q_\pm)(2))\ ,\label{braQ}\\[5pt]
{}_{12}\bra{\pi(S_\pm)}&=& {}_{12}\bra{\1^\ast}\pi(S_\pm)(1)\ =\
\ft12{}_{12}\bra{\1^\ast} (\pi(S_\pm)(1)\pm \pi(S_\pm)(2))\
,\label{braS}\eea
carrying the higher-spin representations
\bea \ket{\Ad_Q Q'}_{12}&=& (Q(1)+\pi(Q)(2))\ket{Q'}_{12}\ \equiv \
\widetilde Q\ket{Q'}_{12}\ ,\label{ketAdQQ}\\[5pt]
\ket{\widetilde{\Ad}_Q S}_{12}&=& (Q(1)+Q(2))\ket{S}_{12}\ \equiv\
Q\ket{S}_{12}\ ,\label{ketAdQS}\eea
and
\bea {}_{12}\bra{\Ad_Q Q'}&=& -{}_{12}\bra{ Q'}\widetilde Q\ ,\qquad
{}_{12}\bra{\pi(\widetilde\Ad_Q S)}\ =\ -{}_{12}\bra{\pi(S)}Q\
.\label{braAdQ}\eea
Using instead the twisted reflector, the mappings read
\bea \ket{\widetilde Q}_{12}&=& Q(1)\ket{\j1}_{12}\ =\ \ft12
(Q(1)-Q(2))\ket{\j1}_{12}\ ,\label{ketQ2}\\[5pt] \ket{\widetilde S_\pm}_{12}&=&
S_\pm(1)\ket{\j1}_{12}\ =\ \ft12 (S_\pm(1)\pm
\pi(S_\pm)(2))\ket{\j1}_{12}\ ,\label{ketS2}\eea
carrying the higher-spin representations
\bea \ket{\Ad_Q \widetilde Q}_{12}&=& Q\ket{\widetilde Q'}_{12}\
,\qquad \ket{\widetilde\Ad_Q\widetilde S}_{12}\ =\ \widetilde
Q\ket{\widetilde S}_{12}\ ,\label{ketAdQ2}\\[5pt]
{}_{12}\bra{\Ad_Q \widetilde Q'}&=&-{}_{12}\bra{\widetilde
Q'}_{12}Q\ , \qquad {}_{12}\bra{\widetilde \Ad_Q\widetilde S}\ =\
-{}_{12}\bra{\widetilde S}\widetilde Q\ .\label{braAdQ2}\eea

The $\ell$th adjoint and twisted-adjoint levels ${\cal L}_{\ell}$
and ${\cal T}_\ell$ are finite-dimensional and infinite-dimensional
$\mso(D+1;\Comp)$ irreps, respectively. Nonetheless, as shown in
Appendix \ref{App:Cas}, the quadratic and quartic Casimir operators,
defined in \eq{C2} and \eq{C4}, assume the same values in ${\cal
L}_{\ell}$ and ${\cal T}_\ell$, namely ($s=2\ell+2$)
\bea C_2[O(D+1;\Comp)|\ell]&=& 2(s-1)(s+2\e_0)\ ,\label{c2ell}\\[5pt]
C_4[O(D+1;\Comp|\ell]&=&
2(s-1)(s+2\e_0)(s^2+(2\e_0-1)s+2\e_0^2-\e_0+1)\ .\label{c4ell}\eea
Moreover, as can be seen using \eq{C2lhws} and \eq{C4lhws}, these
values are also equal to those of the massless lowest-weight spaces
$\mD(s+2\e_0;(s))$, \emph{i.e.}
\bea C_2[O(D+1;\Comp)|\ell]&=& C_2[O(D+1;\Comp)|s+2\e_0;(s)]\
,\\[5pt]
C_4[O(D+1;\Comp|\ell]&=& C_4[O(D+1;\Comp|s+2\e_0;(s)]\ .\eea
These agreements follow direct relationships between ${\cal
L}_{\ell}$, ${\cal T}_\ell$ and $\mD(\pm(s+2\e_0);(s))$ that arise
upon going from the $\mso(D;\Comp)$-covariant bases of ${\cal
L}_{\ell}$ and ${\cal T}_\ell$ to \emph{compact bases} where
elements are labeled by quantum numbers of the $\mso(2)\oplus
\mso(D-1;\Comp)$ subalgebra (to be identified as the maximal compact
subalgebra in the case of two-time signature). In the case of ${\cal
T}_\ell $, the compact basis elements are series expansions in the
covariant basis elements $T_{a(s+k),b(s)}$ (related to Bessel
functions). This is a non-trivial change of basis, which corresponds
to the harmonic expansion of linearized Weyl tensors. To distinguish
between the $\mso(D;\Comp)$-covariant and the compact ``slicings''
of ${\cal T}_\ell $, we thus define the \emph{covariant
twisted-adjoint module}
\bea {\cal T}&=& \bigoplus_{\ell}{\cal T}_\ell\ =\
\bigoplus_{s}{\cal T}_{(s)}\ ,\label{covmod}\eea
consisting of polynomial elements of the form \eq{levelell}, and the
\emph{compact twisted-adjoint module}
\bea {\cal M}&=& \bigoplus_{\ell}{\cal M}_\ell\ =\
\bigoplus_{s}{\cal M}_{(s)}\ ,\label{compmod}\eea
whose elements are polynomial in the compact basis.

We next turn to a more careful analysis of the compact basis
elements, the harmonic expansions, and the relations between the
compact twisted-adjoint module, the massless weight spaces and the
adjoint modules.


\scs{Compact Weight-Space Description of the Master
Fields}\label{Sec:WS}


In this section we shall examine the properties of the adjoint and
twisted-adjoint representation spaces in compact bases. As a result,
we shall give the explicit embedding of the massless weight spaces
into the twisted-adjoint representation, and use this to describe
the harmonic expansion of the Weyl tensors. We shall also show how
to ``glue'' the adjoint and twisted-adjoint representations in
compact weight space.


\scss{Compact Twisted-Adjoint Modules}\label{Sec:LSM}


The \emph{compact twisted-adjoint $\mho_1(D+1;\Comp)$
module} is defined as 
\bea {\cal M}&=&\bigoplus_{s=0}^\infty {\cal M}_{(s)}\ ,\\[5pt]
{\cal M}_{(s)}&=& \bigoplus_{\ba{c}e\in \integ\\
s_1\geq s\geq s_2\geq 0\ea}\Comp\otimes
T^{(s)}_{e;(s_1,s_2)}\label{calMs}\eea
where ${\cal M}_{(s)}$ are $\mso(D+1;\Comp)$ submodules consisting
of components $T^{(s)}_{e;(s_1,s_2)}$ with spin $(s_1,s_2)$ and
energy $e$, that is
\bea T^{(s)}_{e;(s_1,s_2);r(s_1),t(s_2)}&=& \sum_{n=0}^\infty
f^{(s)}_{e;(s_1,s_2);n} T_{s;n;(s_1,s_2);r(s_1),t(s_2)}\
,\label{Ts}\eea
where
\bea T_{s;n;(s_1,s_2);r(s_1),t(s_2)}&=&T_{0(n)\{ r(s_1),t(s_2)\}
0(s-s_2)}\ ,\eea
with $T_{a(s_1+n),b(s)}$ defined in \eq{Taskbs} and $\{\cdots \}$
indicating $\mso(D-1)$-traceless type $(s_1,s_2)$-projection, and
$f^{(s)}_{e;(s_1,s_2);n}\in \Real$ are determined from
\bea \widetilde{\Ad}_E
(T^{(s)}_{e;(s_1,s_2)})&=&e~T^{(s)}_{e;(s_1,s_2)}\ .\label{TE}\eea
Using \eq{TPa}, which implies
\bea \widetilde{\Ad}_E
(T_{s;(s_1,s_2);n})&=&\l_{s;(s_1,s_2);n}T_{s;(s_1,s_2);n+1}+\l'_{s;(s_1,s_2);n}T_{s;(s_1,s_2);n-1}\
,\eea
where the coefficients $\l_{s;(s_1,s_2);n}$ and
$\l_{s;(s_1,s_2);n}'$ are non-vanishing for all $n$ except
$\l_{s;(s_1,s_2);0}'=0$, it follows that the generating functions
\bea f^{(s)}_{e;(s_1,s_2)}(z)&=& \sum_{n=0}^\infty
f^{(s)}_{e;(s_1,s_2);n} z^n\ ,\label{fses1s2}\eea
are \emph{analytical} at $z=0$, and determined \emph{uniquely} up to
an overall constant that can be fixed by the normalization condition
\bea f^{(s)}_{e;(s_1,s_2)}(0)&=& 1\ .\label{fnorm}\eea
Under the $\pi$-map
\bea \pi(T^{(s)}_{e;(s_1,s_2)})&=&
(-1)^{s_1-s}T^{(s)}_{-e;(s_1,s_2)}\ ,\qquad
f^{(s)}_{-e;(s_1,s_2)}(z)\ =\ f^{(s)}_{e;(s_1,s_2)}(-z)\
.\label{minuse}\eea
In what follows, we shall use the notation
\bea \widetilde Q S&=& \widetilde{\Ad}_{Q}(S)\ =\ Q\star S-S\star
\pi(Q)\ ,\eea
for non-minimal higher-spin generators $Q\in{\cal A}$. We note that
in ${\cal M}_{(s)}$, the condition $s_1\geq s\geq s_2$ and $s_3=0$
implies that
\bea \widetilde L^\pm_{r}
T^{(s)}_{e;(s,s_2);rt(s-1),u(s_2)}&=&0\qquad \mbox{for $s_1=s\geq 1$
and $s_2<s$}\ ,
\label{Mdiv}\\[5pt] \widetilde L^\pm_{[r_1}
T^{(s)}_{e;(s_1,s_2);r_2|t(s_1-1),|r_3]u(s_2-1)}&=&0\qquad \mbox{for
$s_2\geq 1$}\ .\label{Mcurl}\eea
Assuming that the twisted-adjoint action of the Casimir operator
$C_{2n}[\mso(D+1;\Comp)]$, defined in \eq{C2n}, commutes with the
summation in \eq{Ts}, that is
\bea
C_{2n}[\mso(D+1;\Comp)]\left(T^{(s)}_{e;(s_1,s_2)}\right)&=&\sum_{n=0}^\infty
f^{(s)}_{e;(s_1,s_2)} \ft12 \widetilde M_{A_1}{}^{A_2}\cdots
\widetilde M_{A_{2n}}{}^{A_1}T_{s;n;(s_1,s_2)}\ ,\eea
it follows that ($s=2\ell+2$)
\bea C_{2n}[{\cal M}_{(s)}]&=& C_{2n}[{\cal T}_{(s)}]\ =\
C_{2n}[\ell]\ ,\label{C2nMell}\eea
where $C_2[\ell]$ and $C_4[\ell]$ are given in \eq{c2ell} and
\eq{c4ell}.

The space ${\cal M}$ is a module also under \emph{separate} left and
right $\star$-multiplication by (polynomial) generators $Q\in{\cal
A}$, and as such it splits into \emph{even and odd submodules},
\bea {\cal M}&=&{\cal M}^+\oplus {\cal M}^-\ ,\eea
where
\bea {\cal M}^{\pm}&=& \bigoplus_{\ba{c}e;(s_1,s_2)\\
e+s_1+s_2=\ft12 (1\mp 1)\ \mbox{mod $2$}\ea}\Comp\otimes
T^{(s)}_{e;(s_1,s_2)}\ .\eea
As a consequence, also the $\mso(D+1;\Comp)$ submodules split into
even and odd parts,
\bea {\cal M}_{(s)}&=& {\cal M}_{(s)}^{+}\oplus {\cal M}_{(s)}^{-}\
.\eea
We propose that ${\cal M}_{(s)}^{\pm}$ are generated by
$\mso(D+1;\Comp)$ from the elements with $e=0$ and minimal
$s_1+s_2$, which we shall refer to as the \emph{static ground
states}, namely
\bea s=0&:&\qquad T^{(0)}_\pm\ =\ T^{(0)}_{0;(\sigma_\pm);t(\sigma_\pm)}\ ,\label{scalarstatic}\\[5pt]
s>0&:&\qquad T^{(s)}_\pm\ =\
T^{(s)}_{0;(s,\sigma_\pm);u(s),v(\sigma_\pm)}\ ,\label{hsstatic}\eea
where $\sigma_\pm=(1\mp 1)/2$. The scalar static ground states are
represented by 
\bea  f^{(0)}_{0;(0)}(z)& =& \sum_{n=0}^\infty
{4^{2n}(\e_0+\ft32)_{2n}\over (2)_{2n}(2\e_0+1)_{2n}}z^{2n}\nn \\[5pt]
& = &  {}_2 F_3({2\e_0+3\over 4},{2\e_0+5\over 4};
\ft32,\e_0+\ft12,\e_0+1;4z^2)\ ,\label{scalarf000plus}\\[5pt] f^{(0)}_{0;(1)}(z)& =&
\sum_{n=0}^\infty {(\e_0+\ft52)_{2n}\over
n!(2)_{n}(\e_0+1)_{n}(\e_0+2)_{n}}z^{2n}\nn \\[5pt]
& = &   {}_2 F_3({2\e_0+5\over 4},{2\e_0+7\over 4};
2,\e_0+1,\e_0+2;4z^2)\ .\label{scalarf000minus} \eea
In $D=4$ these functions take the following simple form:
\bea f^{(0)}_{0;(0)}(z)&=& {\sinh 4z\over 4z}\ ,\qquad
f^{(0)}_{0;(1)}(z)\ =\ {3\over 16 z^2}(\cosh 4z-{\sinh 4z\over 4z})\
, \eea
where we note that $f^{(0)}_{0;(0)}(E)$ was found in
\cite{Sezgin:2005pv}.

We propose that the static ground states $T^{(s)}_\pm$ with $s>0$
can be generated by $\mho_1(D+1;\Comp)$ starting from the scalar
static ground states $T^{(0)}_\pm$. Thus, in effect, we propose that
${\cal M}^\pm$ are generated by $\mho_1(D+1;\Comp)$ starting from
$T^{(0)}_\pm$, which therefore serves as ground states of ${\cal
M}^\pm$.

To generate an explicit basis one has to take into account
degeneracies of the form
\bea \widetilde L^\pm_t\widetilde L^\mp_t
T^{(s)}_{e;(s_1,s_2);r(s_1),t(s_2)}&=&{\cal
\mu}^{(s)}_{e;(s_1,s_2)}T^{(s)}_{e;(s_1,s_2);r(s_1),t(s_2)}\ ,\label{deg1}\\[5pt]
\widetilde x^+\widetilde x^-
T^{(s)}_{e;(s_1,s_2);r(s_1),t(s_2)}&=&{\cal
\mu}^{\prime(s)}_{e;(s_1,s_2)}T^{(s)}_{e;(s_1,s_2);r(s_1),t(s_2)}\
,\label{deg2}\\[5pt]
\widetilde x^\pm \widetilde L^\mp_{r_1}\widetilde L^\mp_{r_2}
T^{(s)}_{e;(s_1,s_2);s(s_1),t(s_2)}&=&{\cal
\mu}^{\prime\prime(s)}_{e;(s_1,s_2)}\widetilde L^\pm_{r_1}\widetilde
L^\mp_{r_2}T^{(s)}_{e;(s_1,s_2);s(s_1),t(s_2)}\ ,\label{deg3}\eea
for coefficients ${\cal \mu}^{(s)}_{e;(s_1,s_2)}$, ${\cal
\mu}^{\prime(s)}_{e;(s_1,s_2)}$ and ${\cal
\mu}^{\prime\prime(s)}_{e;(s_1,s_2)}$ (that may vanish), and where
we have defined
\bea  \widetilde x^\pm&=& \widetilde L^\pm_r\widetilde L^\pm_r\
.\label{tildeXpm}\eea
Thus, defining
\bea {\cal M}^+_{(s)}&=& {\cal M}^{+,\geq}_{(s)}\oplus {\cal
M}^{+,0}_{(s)}\oplus {\cal M}^{+,\leq}_{(s)}\ ,\eea
where
\bea {\cal M}^{+,\geq}_{(s)}&=& \{T^{(s)}_{e;(s_1,s_2)}\ :\ e\geq
s_1+s_2-s\}\ ,\\[5pt] {\cal M}^{+,0}_{(s)}&=& \{T^{(s)}_{e;(s_1,s_2)}\ :\
|e|\leq  s_1+s_2-s\}\ ,\\[5pt] {\cal M}^{+,\leq}_{(s)}&=& \{T^{(s)}_{e;(s_1,s_2)}\ :\
e\leq s_1+s_2-s\}\ ,\eea
and removing degeneracies of the types listed in
\eq{deg1}--\eq{deg3}, there should exist \emph{finite} coefficients
${\cal C}^{(s)}_{e;(s_1,s_2)}$ such that
\bea \hspace{-1cm}{\cal M}^{+,\geq}_{(s)}&:&
T^{(s)}_{e;(s_1,s_2);r(s_1),t(s_2)}\ =\ {\cal C}^{(s)}_{e;(s_1,s_2)}
(\widetilde x^+)^p \widetilde L^+_{\{r_1}\cdots \widetilde
L^+_{r_{s_1-s}}\widetilde L^+_{t_1}\cdots \widetilde L^+_{t_{s_2}}
T^{(s)}_{0;(s,0);r(s)\}}\ ,\label{stategeneration1}\eea
for $p=\ft12(e+s-s_1-s_2)$;
\bea \hspace{-1cm}{\cal M}^{+,\leq}_{(s)}&:&
T^{(s)}_{e;(s_1,s_2);r(s_1),t(s_2)}\ =\ {\cal C}^{(s)}_{e;(s_1,s_2)}
(\widetilde x^-)^p \widetilde L^-_{\{r_1}\cdots \widetilde
L^-_{r_{s_1-s}}\widetilde L^-_{t_1}\cdots \widetilde L^-_{t_{s_2}}
T^{(s)}_{0;(s,0);r(s)\}}\ ,\label{stategeneration1}\eea
for $p=\ft12(-e+s-s_1-s_2)$; and
\bea \hspace{-1cm}{\cal M}^{+,0}_{(s)}&:&
T^{(s)}_{e;(s_1,s_2);r(s_1),t(s_2)}\ =\ {\cal C}^{(s)}_{e;(s_1,s_2)}
((\widetilde L^+)^{m}(\widetilde L^-)^n)_{\{r_1\cdots r_{s_1-s}
t_1\cdots t_{s_2}} T^{(s)}_{0;(s,0);r(s)\}}\
,\label{stategeneration2}\eea
for $m=\ft12(s_1+s_2-s+e)$ and $n=\ft12(s_1+s_2-s-e)$, and where
$((\widetilde L^+)^{m}(\widetilde L^-)^n)_{r_1\cdots
r_{m+n}}=\widetilde L^+_{\{r_1}\cdots \widetilde L^-_{r_m}\widetilde
L^-_{r_{m+1}}\cdots \widetilde L^-_{r_{m+n}\}}$. We propose a
similar generation of the elements in ${\cal M}^-_{(s)}$.

Next, to generate $T^{(s)}_\pm$ with even spin $s=2p\geq 2$, one may
fix coefficients $\x_{2p;n}$ and $\x'_{2p;n}$ such that
\bea T^{(2p)}_{0;(2p);r(2p)}&=& \sum_{n=0}^{p-1} \x_{2p;n}
X^n_{r(2p)}\
,\label{gen1}\\[5pt]T^{(2p)}_{0;(2p,1);r(2p),s}&=& \sum_{n=0}^{p-1} \x'_{2p;n} X^n_{r(2p),s}\ ,\label{gen2}\eea
with
\bea X^n_{r(2p)}&=& \widetilde L^+_{\{r_1}\widetilde L^-_{r_2}\cdots
\widetilde L^+_{r_{2n-1}}\widetilde L^-_{r_{2n}}\widetilde
Q_{r(2p-2n)\}} T^{(0)}_{0;(0)}\ ,\\[5pt]X^n_{r(2p),s}&=& \widetilde L^+_{\{r_1}\widetilde L^-_{r_2}\cdots
\widetilde L^+_{r_{2n-1}}\widetilde L^-_{r_{2n}}\widetilde
Q_{r(2p-2n)}T^{(0)}_{0;(1);s\}}\ ,\eea
where the non-minimal higher-spin generator
\bea Q_{r(2n)}&=& L^+_{\{r_1}\star L^-_{r_2}\star\cdots \star
L^+_{r_{2n-1}}\star L^-_{r_{2n}\}}\ .\eea
The elements $T^{(2p)}_{0;(2p)}$ can also be generated by combining
minimal higher-spin transformations with $\mso(D+1;\Comp)$
transformations. Similarly, for odd spin $s=2p+1\geq 3$, there are
coefficients $\x_{2p+1;n}$ and $\x'_{2p+1;n}$ such that
\bea T^{(2p+1)}_{0;(2p+1);r(2p+1)}&=& \sum_{n=0}^{p-1} \x_{2p+1;n}
X^n_{r(2p+1),s}\
,\label{gen3}\\[5pt]T^{(2p+1)}_{0;(2p+1,1);r(2p+1),s}&=&
\sum_{n=0}^{p-1} \x'_{2p+1;n} X^n_{r(2p+1),s}\ ,\label{gen4}\eea
with
\bea X^n_{r(2p+1)}&=& \widetilde L^+_{\{r_1}\widetilde
L^-_{r_2}\cdots \widetilde L^+_{r_{2n-1}}\widetilde
L^-_{r_{2n}}\widetilde
Q_{r(2p-2n)} T^{(1)}_{0;(1);r_{2p+1}\}}\ ,\\[5pt]X^n_{r(2p+1),s}&=&
\widetilde L^+_{\{r_1}\widetilde L^-_{r_2}\cdots \widetilde
L^+_{r_{2n-1}}\widetilde L^-_{r_{2n}}\widetilde
Q_{r(2p-2n)}T^{(1)}_{0;(1,1);r_{2p+1},s\}}\eea
where the spin-$1$ static ground states, in their turn, can be
generated from the scalar static ground states. For example, to
generate $T^{(1)}_{0;(1)}$ from $T^{(0)}_{0;(1)}$, one may use
\bea \widetilde \Ad_{EM_{rs}}T^{(0)}_{0;(1);t}&=&
\{EM_{rs},T^{(0)}_{0;(1);t}\}\ =\ \d_{t[s}T^{(1)}_{0;(1);r]}\ ,\eea
as can be seen using $EM_{rs}=E\star M_{rs}=M_{rs}\star E$ and
\bea E\star T^{(0)}_{0;(1);r}&=&-T^{(0)}_{0;(1);r}\star E\ =\
\ft12\Ad_E T^{(0)}_{0;(1);r}\ =\ -\ft{i}2 T^{(1)}_{0;(1);r}\ .\eea
The generation of $T^{(1)}_{0;(1,1)}$ from $T^{(0)}_{0;(0)}$ is more
involved, since
\bea E\star T^{(0)}_{0;(0)}&=&\ft12\Ad_E T^{(0)}_{0;(0)}\ =\ 0\ .
\label{EstarT0}\eea
One may, for example, first use $\mso(D+1;\Comp)$ to transform
$T^{(0)}_{0;(0)}$ into $T^{(0)}_{0;(2)}$; then $EM_{rs}$ to go to
$T^{(1)}_{0;(2)}$; and finally $\mso(D+1;\Comp)$ to go down to
$T^{(1)}_{0;(1,1)}$. We note, however, that $T^{(1)}_{0;(1,1)}$ can
be generated immediately by separate $\star$-multiplication from the
left. For example,
\bea T^{(1)}_{0;(1,1)}&=&{(2\e_0+1)(2\e_0+2)\over 4\e_0} M_{rs}\star
T^{(0)}_{0;(0)}\ ,\label{MrsstarT0}\eea
as can be seen using \eq{MabT}, which implies
\bea M_{rs}\star T_{0(n)}&=&\ft12\Ac_{M_{rs}}T_{0(n)}\ =\
T_{0(n)[r,s]}-{(n-1)n^2(n+1)\over
16(n+\e_0-\ft12)(n+\e_0+\ft12)}T_{0(n-2)[r,s]}\ ,\eea
although one can check that \eq{MrsstarT0} is not a twisted-adjoint
$\mho_1(D+1;\Comp)$-transformation. We also notice that repeated
$\star$ commutation by $E$, \emph{i.e.} the maps $(\Ad_E)^n:{\cal
M}_{(s)}\rightarrow \bigoplus_{s'=|s-n|}^{s+n}{\cal M}_{(s')}$,
assume a relatively simple form, although also $\Ad_E$ is not a
twisted-adjoint $\mho_1(D+1;\Comp)$-transformation.

The compact twisted-adjoint modules described contain invariant
submodules generated by lowest-weight or highest-weight elements, to
which we now turn our attention.


\scs{Composite Lowest-Weight Spaces and Non-Composite Lowest-Spin
Spaces}\label{Sec:LWS}


The compact twisted-adjoint modules contain invariant lowest-weight
and highest-weight submodules. Suppose $T^{(s)}_{e;(s_1,s_2)}\in
{\cal M}^{(s)}$ is a lowest-weight state, \emph{i.e.}
\bea \widetilde L^-_r T^{(s)}_{e;(s_1,s_2);t(s_1),u(s_2)}&=&
L^-_r\star
T^{(s)}_{e;(s_1,s_2);t(s_1),u(s_2)}-T^{(s)}_{e;(s_1,s_2);t(s_1),u(s_2)}\star
L^+_r\ =\ 0\ .\label{candidatelws}\eea
Then the second and quartic Casimir operators are given by, on the
one hand, \eq{C2lhws} and \eq{C4lhws}, and, on the other hand,
\eq{C2nMell}, which leads to the necessary conditions
\bea x+y+z&=& C_2[\ell]\ ,\qquad x(x+\D)+y(y+\D')+z(z+\D'')\ =\
C_4[\ell]\ ,\label{c2c4conds}\eea
where we have defined
\bea x&=& e(e-D+1)\ ,\qquad y\ =\ s_1(s_1+D-3)\ ,\qquad z\ =\
s_2(s_2+D-5)\ ,\\[5pt]
\D & =&\ft12(D-1)(D-2)\ ,\qquad \D'\ =\ \ft12 (D-3)(D-4)-1\ ,\nn\\[5pt]
&&\D''\ =\ \ft12(D-5)(D-6)-2\ ,\eea
and
\bea C_2[\ell]&=& x_0+y_0\ ,\qquad C_4[\ell]\ =\
x_0(x_0+\D)+y_0(y_0+\D')\ ,\eea
where
\bea x_0&=& e_0(e_0-D+1)\ ,\qquad y_0\ =\ s(s+D-3)\ ,\qquad e_0\ =\
s+D-3\ .\eea
Moreover, combining \eq{candidatelws} with the identities \eq{Mdiv}
and \eq{Mcurl}, respectively, yields
\bea e&=&s+D-3-{s_2\over s}\qquad\mbox{for $s_1=s\geq 1$ and
$s_2<s$}\
.\label{div}\\[5pt] e&=& {s_1+s_2+2(D-4)\over D-3}\qquad\mbox{for $s_2\geq 1$}\
.\label{curl}\eea
To begin with, let us take $s_2=0$. Then $z=0$ and \eq{c2c4conds}
have two roots
\bea x&=&x_0\ ,\quad y\ =\ y_0\ ,\qquad \mbox{and}\quad x\ =\
y_0+2-D\ ,\quad y\ =\ x_0+D-2\ .\eea
The second root corresponds to $s_1=s-1$, which is ruled out for all
$s$, or $s_1=4-D-s$, which is ruled for all $s$ except $s=0$ in
$D=4$, where it coincides with the first solution (which is thus a
double root). The first root corresponds to $s_1=s$ and $e=s+D-3$ or
$e=2-s$. The latter energy level is ruled out for $s\geq 1$ due to
the condition \eq{div}. Thus, the admissible lowest weight states
with $s_2=0$ are
\bea s_1&=& s\ =\ 0\ ,\qquad e\ =\ \left\{\ba{l} 2\e_0\\[5pt] 2\ea\right.\ ,\\[10pt]
s_1&=& s\ \geq\ 1\ ,\qquad e\ =\ s+2\e_0\ ,\eea
where we note the degeneracy in case $s=0$ and $D=5$. For $s_2\geq
1$ we find the admissible root
\bea s_1\ =\ s_2\ =\ s\ =\ 1\ ,\qquad e\ =\ 2\ .\eea
%

In the scalar sector, the two admissible roots are indeed
lowest-weight states, given by
\bea f^{(0)}_{2\e_0;(0)}(z)&=& {}_1 F_1(\e_0+\ft32;2;-4z)\ ,
\label{scalarf020plus}\\[5pt] f^{(0)}_{2;(0)}(z)&=& {}_1 F_1(\e_0+\ft32;2\e_0;-4z)\
,\label{scalarf020minus}\eea
taking the following particularly simple form in $D=4$:
\bea f^{(0)}_{1;(0)}(z)&=& e^{-4z}\ ,\qquad f^{(0)}_{2;(1)}(z)\ =\
(1-4z)e^{-4z}\ . \label{lws4d}\eea
Here we note that the functions $f^{(0)}_{e;(0)}(z)$ are fixed
(uniquely) by the twisted-energy condition $(\widetilde E-e)
T^{(0)}_{e;(0)}=0$, so that $\widetilde L^-_r T^{(0)}_{e;(0)}=0$
(which can be worked out explicitly using \eq{TPa}) becomes an
algebraic equation for $e$ with roots $e=2$ and $e=2\e_0$.

\begin{figure}[!h]
\begin{center}
\unitlength=.6mm
\begin{picture}(150,180)(0,-10)
\put(0,0){\vector(1,0){150}} \put(0,0){\vector(0,1){150}}
\put(150,-10){$s$} \put(-10,150){$E$}
\put(20,-10){1}\put(40,-10){2}\put(60,-10){3}\put(80,-10){4}\put(100,-10){5}\put(120,-10){6}
\put(-10,20){1}\put(-10,40){2}\put(-10,60){3}\put(-10,80){4}\put(-10,100){5}\put(-10,120){6}
\multiput(0,1)(20,0){6}{\!$|$} \multiput(1,20)(0,20){6}{\!$-$}
\multiput(0,40)(20,20){6}{\!$\bullet$}
\multiput(0,80)(20,20){4}{\!$\bullet$}
\end{picture}
\end{center}
\caption{{\small Weight diagrams of the scalar 2-lineton in $D=9$.}}
\label{2lin}
\end{figure}
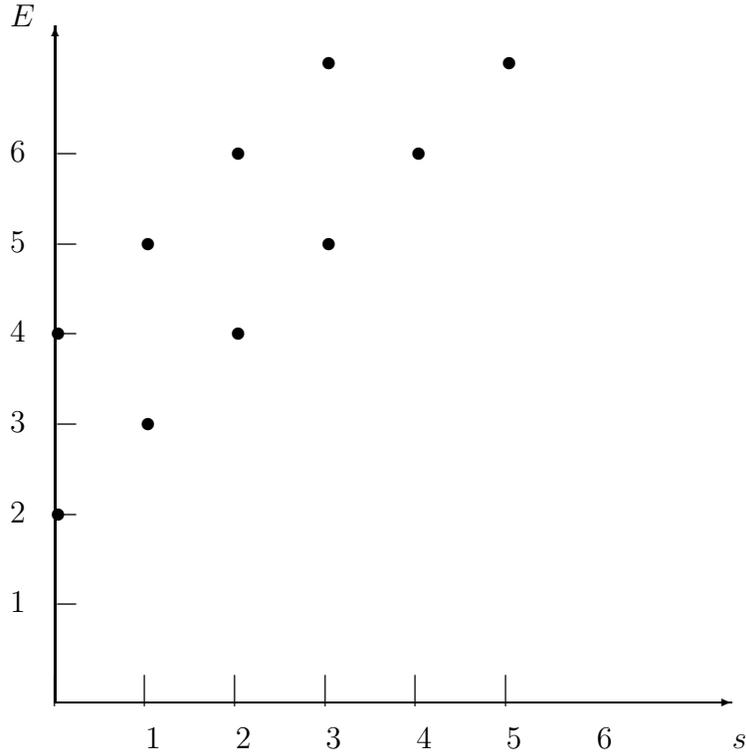

For $D=2p+5$ with $p=1,2,3,...$, the Harish-Chandra module
$\mC(2;(0))$ contains a singular vector at level $2p$, namely
$L^-_r (x^+)^p \ket{2;(0)}=0$ (where $x=L^+_r L^+_r$), that in its
turn generates $\mD(2\e_0;(0))$. The module $\mC(2;(0))$ is
isomorphic to the lowest-weight module inside ${\cal M}^+_{(0)}$
generated by the twisted-adjoint $\mso(D+1;\Comp)$ action on
$T^{(0)}_{2;(0)}$, where the singular vector obeys
\bea \widetilde L^-_r (\widetilde x^+)^p \,T^{(0)}_{2;(0)}&=&0\
,\qquad D=2p+5\ ,\quad p=1,2,\dots\ .\eea
The lowest-weight space $\mD(2;(0))$ hence occupies $p$ diagonal
lines in compact weight space, and we shall refer to it as a scalar
$p$-lineton\footnote{In $D$ dimensions, the $p$-lineton
$\mC(e_0;(0))$ with $e_0=\e_0-(p-1)$ has the singular vector
$x^p\mid e_0;(0)\rangle\simeq 0$, and therefore consists of $p$
lines in weight space,
$$\mD(e_0;(0))=\bigoplus_{k=0}^{p-1} \bigoplus_{n=0}^\infty |e_0+k+n;(n)\rangle\ ,$$
where $|e_0+k+n;(n)\rangle_{r(n)}=L^+_{r_1}\cdots L^+_{r_n}x^k|
e_0;(0)\rangle$. In particular, the $1$-lineton coincides with the
ordinary singleton. The scalar $p$-linetons can be thought of as
boundary particles satisfying a higher-derivative equation like
$\Box^p\phi=0$.} (see fig. \ref{2lin}). Thus, in summary, the scalar
compact twisted-adjoint modules contain the following invariant
subspaces (where we recall that the $\pm$ on ${\cal I}_{(0)}$ denote
$\s_\pm$-parity while on $\mD$ distinguish a module and its
negative-energy counterpart)
\bea D=4,6,\dots&:& {\cal I}^{+}_{(0)}\ =\ \mD^+(2;(0))\oplus
\mD^-(-2;(0))\\[5pt]&& {\cal I}^{-}_{(0)}\ =\ \mD^+(1;(0))\oplus
\mD^-(-1;(0))\ ,\\[10pt]
 D=5&:& {\cal I}^{+}_{(0)}\ =\ \mD^+(2;(0))\oplus
\mD^-(-2;(0))\\[5pt]&&
{\cal I}^{-}_{(0)}\ =\ 0\ ,\\[10pt]
D=7,9,\dots&:&{\cal I}^+_{(0)}\ = \ \left[\mD^+(2;(0))\oplus
\mD^-(-2;(0))\right]\\[5pt]
&& \hspace{1.5cm} \oplus_s\left[\mD^+(2\e_0;(0))\oplus
\mD^-(-2\e_0;(0)) \right]\ ,\\[5pt]&&
{\cal I}^-_{(0)}\ =\ 0\ ,\eea
where we recall that $\oplus_s$ denotes a semi-direct sum.

Turning to the case of $s>0$, we first use \eq{MabT} to show that
the scalar lowest-weight states actually obey slightly stronger
conditions that the lowest-weight condition, namely
\bea L^-_r\star T^{(0)}_{2\e_0;(0)}&=&L^-_r\star T^{(0)}_{2;(0)}\ =\
0\ ,\label{lminusleft}\\[5pt]
T^{(0)}_{2\e_0;(0)}\star L^+_r &=&T^{(0)}_{2;(0)}\star L^+_r \ =\ 0\
,\label{lplusright}\eea
and
\bea M_{rs}\star T^{(0)}_{2\e_0;(0)}&=&0\ ,\label{mrsleft}\eea
while $M_{rs}\star T^{(0)}_{2;(0)}$ is non-vanishing unless $D=5$.
Thus $L^+_{\{r_1}\star \cdots \star L^+_{r_s\}}\star
T^{(0)}_{2\e_0;(0)}={\cal C}^{(s)}T^{(s)}_{s+2\e_0;(s);r(s)}$ obeys
the lowest-weight condition, and we expect ${\cal C}^{(s)}$ to be
non-vanishing, so that $T^{(s)}_{s+2\e_0;(s)}$ is indeed a
lowest-weight element.

Alternatively, we notice that \eq{lminusleft}--\eq{mrsleft} imply
that $\ket{2\e_0;(0)}_{12}=T^{(0)}_{2\e_0;(0)}\ket{\1}_{12}$ obeys
\bea L^-_r(\xi)\ket{2\e_0;(0)}_{12}&=&
M_{rs}(\x)\ket{2\e_0;(0)}_{12} \ =\ 0\ ,\qquad \x=1,2\ ,\eea
so that one may formally reflect the \emph{Flato-Fronsdal formula}
\cite{Flato:1978qz}, which we recall here,
\bea \ket{s+2\e_0;(s)}_{12;r(s)} &=& f_{(s)}
f_{r(s)}(1,2)\ket{2\e_0;(0)}_{12}\ ,\label{FFformula}\eea
where the composite operator\footnote{The coefficients $f_{s;k}$ are
fixed by the condition
$(L^-_r(1)+L^-_r(2))\ket{s+2\e_0;(s)}_{12;r(s)} =0$, which is
equivalent to $a_k f_{s;k}+a_{s-k+1}f_{s;k-1}=0$, where
$a_k=2k(k+\e_0-1)$, with solution
$f_{s;k}=(-1)^sf_{s;s-k}=(-1)^k{a_{s-k+1}\cdots a_s\over a_k\cdots
a_1}f_{s;0}$, taking the form \eq{complw} for $f_{s;0}=1$.}
\bea f_{r(s)}(1,2)&=& (-1)^s f_{r(s)}(2,1)\ =\ \sum_{k=0}^s
f_{s;k}(L^+_{\{r_1}\cdots L^+_{r_{k}})(1)(L^+_{r_{k+1}}\cdots
L^+_{r_s\}})(2)\ ,\label{frs}\\[5pt] f_{s;k}&=&(-1)^s f_{s;s-k}\ =\ {s\choose k}{(1-s-\e_0)_k\over (\e_0)_k}\
,\label{complw}\eea
to obtain the following expression for the lowest-weight elements
\bea \hspace{-0.5cm}T^{(s)}_{s+2\e_0;(s);r(s)}&=& f_{(s)}
\sum_{k=0}^s (-1)^{s-k} f_{s;k}L^+_{\{r_1}\star\cdots \star
L^+_{r_k}\star T^{(0)}_{2\e_0;(0)}\star L^-_{r_{k+1}}\star \cdots
\star L^-_{r_s\}}\ ,\eea
where the (finite) renormalization $f_{(s)} $ is fixed by
\eq{fnorm}.

Turning to $M_{\{r_1 t_1}\star \cdots M_{r_s t_s\}}\star
T^{(0)}_{2;(0)}$, which are also lowest-weight elements, these must
vanish for $s\geq 2$, since $e=2$ and $s_1=s_2=s\geq 2$ are
inadmissible compact quantum numbers. On the other hand, for $s=1$
and $D\neq 5$, the element $M_{rs}\star T^{(0)}_{2;(0)}$ is
non-vanishing and proportional to $T^{(1)}_{2;(1,1);r,s}$, which is
thus a lowest-weight element for $D\neq 5$, and hence it is a
lowest-weight element also for $D=5$ (since $\widetilde L^-_r
T^{(1)}_{2;(1,1)}=0$ for $D\neq 5$ implies that $\widetilde L^-_r
T^{(1)}_{2;(1,1)}=0$ for $D=5$ by analytical continuation in $D$).
Thus, ${\cal M}^+_{(1)}$ contains the generalized Verma module
\bea \mC'(2;(1,1))&=& {\mC(2;(1,1))\over{\cal I}[V]}\ ,\eea
generated by the twisted-adjoint $\mso(D+1;\Comp)$ action on
$T^{(1)}_{2;(1,1);r,s}$ (modulo \eq{Mdiv} and \eq{Mcurl} which hold
modulo elements in ${\cal I}[V]$). For $D=3+2p$, $p=1,2,\dots$, the
lowest-weight state of $\mD(1+2\e_0;(1))=\mD(1+2p;(1))$ is a
singular vector in $\mC'(2;(1,1))$, \emph{viz.}\footnote{At the
level of the ordinary Harish-Chandra module $\mC(2;(1,1))$, $$L^-_t
x^p \widetilde L^+_s
\ket{2}_{s,u}=-4px^{p-1}L^+_uL^+_s\ket{2}_{s,t}+2p(2p+7-D)
x^{p-1}L^+_tL^+_s\ket{2}_{s,u}+(10-D)x^p\ket{2}_{t,u}\ ,$$ where
$\ket{2}_{s,u}=\ket{2;(1,1)}_{s,u}$. For $D=2p+5$, the
$(tu)$-projection vanishes, while the $[tu]$-projection is
proportional to $(D-5)L^+_{[u}\ket{2}_{s,t]}$ that vanishes for
$D\neq 5$ only at the level of $\mC'(2;(1,1))$. The 1-lineton in
$D=5$ coincides with the spin-1 singleton (which exists already in
$\mC(2;(1,1))$). }
\bea \widetilde L^-_t (\widetilde x^+)^p \widetilde L^+_s
T^{(1)}_{2;(1,1);u,s}&=&0\ ,\qquad D=3+2p\ ,\quad p=1,2,\dots\ .\eea
Factoring out the submodule $\mN'(2;(1,1))$ generated from the
singular vector, yields the irreducible lowest-weight space
\bea \mD'(2;(1,1))&=&{\mC'(2;(1,1))\over \mN'(2;(1,1))}\ ,\eea
occupying $p$ lines in compact weight space, and which we shall
therefore refer to as a spin-1 $p$-lineton. We note that the
$1$-lineton in $D=5$ is the spin-1 singleton, that is
$\mD'(2;(1,1))=\mD(2;(1,1))$.

In the case of $D=4$, we note that $\mD(2;(0))$ and
$\mD(2;(1,1))\simeq \mD(2;(1))$ initiate the sequence of composite
massless lowest-weight spaces $\mD(s+1;(s,1))\simeq \mD(s+1;(s))$
contained in the tensor product $\mD_{\ft12}\otimes \mD_{\ft12}$
of two spinor singletons. We expect that all these lowest-weight
spaces are realized in ${\cal M}_+$, \emph{i.e.} we expect the
following lowest-weight elements
\bea \hspace{-0.5cm}D=4:\, T^{(s)}_{s+1;(s);r(s)} =
f'_{(s)}\e_{tu\{r_{1} }\sum_{ k=1}^{s} f'_{s;k}
L^+_{r_2}\star\cdots\star L^+_{r_{k}}\star M_{tu}\star
T^{(0)}_{2;(0)}\star L^-_{r_{k+1}}\star\cdots\star L^-_{r_{s}\}}\
,\eea
where the coefficients $f'_{s;k}$ and (finite) renormalizations
$f'_{(s)}$, respectively, are fixed by \eq{candidatelws} and
\eq{fnorm}.

So far, for $s\geq 1$, we have found the invariant subspaces
\bea D=4&:&{\cal I}^+_{(s)}\ =\ {\cal I}^-_{(s)}\ \ =\
\mD^+(s+1;(s))\oplus
\mD^-(-(s+1);(s))\ ,\\[10pt]
D=5&:& {\cal I}^+_{(1)}\ =\ \left[\mD^+(2;(1,1))\oplus
\mD^-(-2;(1,1))\right]\nn\\[5pt]
&& \ \oplus_s \left[\mD^+(3;(1))\oplus \mD^-(-3;(1))\right]\ ,\\[5pt]
&&{\cal I}^+_{(s)}\ \supset\ \mD^+(s+2;(s))\oplus
\mD^-(-(s+2);(s))\qquad \mbox{for $s\geq 2$}
,\\[10pt]
D=7,9,\dots &:& {\cal I}^+_{(1)}\ =\
\left[\mD^{\prime+}(2;(1,1))\oplus
\mD^{\prime-}(-2;(1,1))\right] \nn\\[5pt]
&& \ \oplus_s \left[\mD^+(1+2\e_0;(1))\oplus \mD^-(-(1+2\e_0);(1))\right]\ ,\\[5pt]
&&{\cal I}^+_{(s)}\ \supset\ \mD^+(s+2\e_0;(s))\oplus
\mD^-(-(s+2\e_0);(s)\quad \mbox{for $s\geq 2$}\\[10pt]
D=6,8,\dots&:& {\cal I}^+_{(1)}\ =\ \mD^+(2;(1,1))\oplus
\mD^-(-2;(1,1))\ ,\\[5pt] && {\cal I}^-_{(s)}\ =\ \mD^+(s+2\e_0;(s))\oplus
\mD^-(-(s+2\e_0);(s))\quad\mbox{for $s\geq 1$}\ ,\eea
where we notice the ordinary spin-1 singleton $\mD(2;(1,1))$  in
$D=5$ and the spin-1 $p$-linetons $\mD'(2;(1,1))$ in $D=7,9,\dots$,
whose higher-spin ``completion'' we leave for future work (for
example, in $D=5$, the higher-spin singletons $\mD(s+1;(s,s))$ obey
\eq{div} and \eq{curl} but violate the conditions on $C_2$ and
$C_4$, and are hence not realized in ${\cal I}_{(s)}$).

In summary, taking into account the expected results, the compact
twisted-adjoint modules have the following indecomposable structures
\bea D=4,6,\dots&:& {\cal M}^+\ =\ {\cal W}^+ \oplus_s \mD'\ ,\label{indec1}\\[5pt]
&&{\cal M}^-\ =\ {\cal W}^- \oplus_s \mD\ ,\label{indec2}\\[10pt]
D=5,7,\dots&:& {\cal M}^-\ =\ {\cal W}^-\ ,\label{indec3}\\[5pt]
&&{\cal M}^+\ =\ {\cal W}^+ \oplus_s \mD'\oplus_s \mD\
,\label{indec4}\eea
where $\mD$ consists of the massless scalar-singleton composites,
\emph{i.e.}
\bea \mD&=& \bigoplus_s [\mD^+(s+2\e_0;(s))\oplus
\mD^-(-(s+2\e_0);(s))]\ ;\label{mD}\eea
$\mD'$ contains the higher-spin completion of $\mD^+(2;(0))\oplus
\mD^-(-2;(0))$\, \emph{i.e.}
\bea D=4&:& \mD'\ =\ \bigoplus_{s=0}^\infty [\mD^+(s+1;(s,1))\oplus
\mD^-(-(s+1);(s,1))]\ ,\label{mDprime4}\\[5pt] D=5&:& \mD'\ \supset\
\mD^+(2;(1,1))\oplus \mD^-(-2;(1,1))\ ,\label{mDprime5}\\[5pt]
D=6,7,\dots&:&\mD'\ \supset\ [\mD^+(2;(0))\oplus
\mD^-(-2;(0))]\nn\\[5pt]
&& \ \oplus [\mD^{\prime +}(2;(1,1))\oplus \mD^{\prime
-}(-2;(1,1))]\ ;\label{mDprime}\eea
and the remaining quotient spaces ${\cal W}^\pm$ are
\emph{non-composite lowest-spin modules}. These spaces contain all
the states in ${\cal M}^{\pm,0}$, \emph{i.e.} with energy
$|e|<s_1+s_2-s$, such as static states, and they may also contain
states in ${\cal M}^{\pm,\geq}$ and ${\cal M}^{\pm,\leq}$, but no
lowest-weight nor highest-weight states (we notice that if ${\cal
I}^\pm_{(0)}$ is non-trivial, then ${\cal W}^\pm_{(0)}$ contains a
finite number of energy levels for fixed $s_1$).

We stress that, in accordance with our proposal, as given in
\eq{stategeneration1} and \eq{stategeneration2}, the spaces ${\cal
M}^\pm$ are generated by $\mho_1(D+1;\Comp)$ from the scalar static
ground states $T^{(0)}_\pm$ defined in \eq{scalarstatic}. On the
other hand, by the Flato-Fronsdal construction, the space $\mD$ is
also generated by $\mho_1(D+1;\Comp)$ from the scalar lowest-weight
element $T^{(0)}_{2\e_0;(0)}$ and highest-weight element
$T^{(0)}_{-2\e_0;(0)}$. Similarly, in $D=4$, $\mD'$ is generated by
$\mho_1(D+1;\Comp)$ from $T^{(0)}_{2;(0)}$ and $T^{(0)}_{-2;(0)}$
(we expect an analogous generation of $\mD'$ in $D\geq 5$).
Correspondingly, in an abbreviated notation where $\mso(D-1;\Comp)$
vector indices are suppressed, elements $S_{{\cal W}^\pm}\in{\cal
W}^\pm$ and $S_{\mD}\in\mD$ can be expanded as
\bea S_{{\cal W}^\pm}&=& \sum_{s=0}^\infty \sum_{m,n=0}^\infty
S^{\pm(s)}_{m,n}(\widetilde
L^+)^m (\widetilde L^-)^n \widetilde Q_s T^{(0)}_\pm\ ,\label{wedgestate}\\[5pt]
S_{\mD}&=& \sum_{s=0}^\infty
\sum_{m,p=0}^\infty\left[S^{(s)}_{m,p}(\widetilde x^+)^p(\widetilde
L^+)^m \widetilde R_s T^{(0)}_{2\e_0;(0)}+\overline
S^{(s)}_{m,p}(\widetilde x^-)^p(\widetilde L^-)^m \pi(\widetilde R_s
T^{(0)}_{2\e_0;(0)})\right]\ ,\label{1pstate}\eea
for $Q_s,R_s\in\mho_1(D+1;\Comp)$ such that $\widetilde Q_s
T^{(0)}_\pm=T^{(s)}_{\pm}$ and $\widetilde R_s
T^{(0)}_{2\e_0;(0)}=T^{(s)}_{s+2\e_0;(s)}$, and where
$S^{(s)}_{m,n}$ and $(S^{(s)}_{m},\overline S^{(s)}_{m})$ are
complex coefficients.





\scs{Harmonic Expansion}\label{Sec:Harm}


Physically speaking, the map from ${\cal T}_\ell$ to ${\cal
M}_\ell$, is the \emph{harmonic expansion} of a spin-$(2\ell+2)$
Weyl tensor obeying linearized field equations in the maximally
symmetric $D$-dimensional geometry with cosmological constant (which
is also solutions of the higher-spin gauge theory). The
corresponding vielbein and spin-connection form the flat
$\mso(D+1;\Comp)$ connection
\bea \Omega\ =\ -i(e^a P_a+\ft12 \omega^{ab}M_{ab})\ =\ L^{-1}\star
dL\ ,\label{maxsymmspace}\eea
where the coset element $L\in SO(D+1;\Comp)$, or gauge function, is
given in stereographic coordinates by
\bea L&=&{2\over h}\exp{ix^\mu \d_\mu^a P_a\over h}\ ,\qquad h\ =\
\sqrt{1-\l^2 x^2}\ ,\qquad x^2\ =\ x^\mu
x^\nu\d_\mu^a\d_\nu^b\eta_{ab}\ ,\label{L}\eea
for which
\bea e_\mu{}^a&=& {2\delta_\mu^a\over h^2}\ ,\qquad
\omega_\mu{}^{ab}\ =\ {\delta_\mu^{[a}\delta_\nu^{b]}x^\nu\over
h^2}\ .\eea
The linearized adjoint one-form $A$ and twisted-adjoint zero-form
$\Phi$ obey the constraints
\bea DA&\equiv & dA+\{\Omega,A\}_\star\ \equiv \ \nabla
A-ie^a\{P_a,A\}_\star\nn\\[5pt] & =& -\ft{i}2\sum_{s=2,4,6,\dots}
e^a\wedge e^b
\Phi_{ac(s-1),bd(s-1)}M^{c_1 d_1}\cdots M^{c_{s-1}d_{s-1}}\ ,\label{linoneform}\\[5pt]
D\Phi&\equiv & d\Phi+[\Omega,\Phi]_\pi\ \equiv \
\nabla\Phi-ie^a\{P_a,\Phi\}_\star\ =\ 0\ ,\label{linzeroform}\eea
where $D^2=0$, and the consistency of the one-form constraint
\eq{linoneform} follows from $e^a\wedge e^b \wedge e^c \nabla_c
\Phi_{ad(s-1),be(s-1)}=0$ which is a consequence of the zero-form
constraint \eq{linzeroform} (which is in itself consistent). As
shown in Appendix \ref{App:T}, inserting the component expansion
of $\Phi$ into the zero-form constraint and using \eq{TPa} yields
\bea \nabla_c \Phi_{a(s+k),b(s)}-2k\D_{s+k-1,s} \eta_{c\{
a}\Phi_{a(s+k),b(s)\} }+ {2\l^{(s)}_{k+1}\over k+1} \Phi_{c\{
a(s+k),b(s)\}}&=&0\ .\label{DPhicomponents}\eea
These constraints decompose into the auxiliary field identifications
\bea \Phi_{a(s+k),b(s)}&=& -{k+1\over 2\l^{(s)}_{k+1}} \nabla_{\{
a}\Phi_{a(s+k-1),b(s)\}}\ ,\qquad k=1,2,\dots\
,\label{Phiauxiliary}\eea
where we have used $\Phi_{\langle c\langle
a(s+k),b(s)\rangle\rangle}=\Phi_{c a(s+k),b(s)}$; and the Bianchi
identities and mass-shell conditions
\bea \nabla_{[\mu} \Phi_{\nu |a(s+k-1),|\rho]b(s-1)}&=&0\ ,\qquad s\geq 1\ ,\label{BI}\\[5pt]
(\nabla^2-m^2_{s,k})\Phi_{a(s+k),b(s)} &=&0\ ,\label{massshell}\eea
with mass-terms
\bea m^2_{s,k}&=& -4\e_0-2s-(k+2s+2\e_0+1)k\ .\label{msk}\eea
These values are consistent with the Casimir relation
\bea \nabla^2 \Phi_{a(s+k),b(s)}&=&
\left(C_2[\mso(D+1;\Comp)|\ell]-C_2[\mso(D;\Comp)|(s+k,s)]\right)\Phi_{a(s+k),b(s)}\
,\eea
with $C_2[\mso(D+1;\Comp)|\ell]$ given by \eq{c2ell} and
$C_2[\mso(D;\Comp)|(s+k,s)]=(s+k)(s+k+D-2)+s(s+D-4)$. The Casimir
relation follows by re-writing
$C_2[\mso(D+1;\Comp)]=C_2[\mso(D;\Comp)]-P^a\star P_a$ in the
twisted-adjoint representation as
\bea -\Ac_{P^a} \Ac_{P_a}\Phi_{(s)}&=& \ft12
(\widetilde{\Ad}_{M_{AB}}\widetilde{\Ad}_{M^{AB}}-
\Ad_{M_{ab}}\Ad_{M^{ab}})\Phi_{(s)}\ ,\label{Casimirrelation}\eea
and using the fact that \eq{linzeroform} implies that
$\widetilde{\Ad}_{P_a}\Phi_{(s)}=-i\nabla_a\Phi_{(s)}$.

The zero-form constraint \eq{linzeroform} is solved explicitly by
\bea \Phi&=& L^{-1}\star S\star \pi(L)\ ,\label{linPhi}\eea
where $S$ is a constant twisted-adjoint element. The zero-form
$\Phi=\sum_{s=0}^\infty \Phi_{(s)}$ can be expanded either
covariantly or compactly, \emph{viz.}
\bea \Phi_{(s)}&=&\sum_{e;s_1\geq s\geq s_2\geq 0}
\Phi^{(s)}_{e;(s_1,s_2);r(s_1),t(s_2)} L^{-1}\star
T^{(s)}_{e;(s_1,s_2);r(s_1),t(s_2)}\star
\pi(L)\nn\\[5pt]&=&
\sum_{e;s_1\geq s\geq s_2\geq 0}\sum_{k=s_1-s}^\infty
T_{a(s+k),b(s)}S^{(s)}_{e;(s_1,s_2);r(s_1),t(s_2)}
D^{(s);a(s+k),b(s)}_{e;(s_1,s_2);r(s_1),t(s_2)}\
,\label{harmexp}\eea
where $S^{(s)}_{e;(s_1,s_2);r(s_1),t(s_2)}\in\Comp$ and the
generalized harmonic functions ($k\geq s_1-s$)
\bea \hspace{-1cm}
D^{(s);a(s+k),b(s)}_{e;(s_1,s_2);r(s_1),t(s_2)}(x)&=& {\cal
N}_{s,k}^{-1} ~\Tr[T^{a(s+k),b(s)}\star L^{-1}(x)\star
T^{(s)}_{e;(s_1,s_2);r(s_1),t(s_2)}\star \pi(L(x)]\ ,\eea
as can be seen using \eq{trtabtcd} and \eq{trtabtcd}. The harmonic
functions obey the Bianchi identity \eq{BI} and the mass-shell
condition \eq{massshell} (for fixed $(s);e;(s_1,s_2)$). Representing
the trace as the expectation value \eq{trvev}, and using the overlap
condition \eq{overlapX}, which implies
$\pi(L)(1)\ket{\1}_{12}=L^{-1}(2)\ket{\1}_{12}$, these harmonic
functions can be rewritten as
\bea \hspace{-1cm}
D^{(s);a(s+k),b(s)}_{e;(s_1,s_2);r(s_1),t(s_2)}(x)&=& {\cal
N}_{s,k}^{-1} {}_{12}\bra{T_{a(s+k),b(s)}}\star L^{-1}(x)\star
\ket{(s);e;(s_1,s_2)}_{12;r(s_1),t(s_2)}\ ,\eea
where $L$ is given by \eq{L} with $P_a=P_a(1)+P_a(2)$, and we have
used \eq{doubleton} to define
\bea \ket{(s);e;(s_1,s_2)}_{12;r(s_1),t(s_2)}&=&
T^{(s)}_{e;(s_1,s_2);r(s_1),t(s_2)}(1)\star\ket{\1}_{12}\ ,\\[5pt]
\ket{T_{a(s+k),b(s)}}&=& T_{a(s+k),b(s)}(1)\star\ket{\1}_{12}\
.\label{test}\eea
For $x=0$, the harmonic functions are the (finite) overlaps
\bea
\hspace{-1cm}D^{(s);a(s+k),b(s)}_{e;(s_1,s_2);r(s_1),t(s_2)}(0)&=&
\d^{\{a(s+k),b(s)\}_D}_{\{0(n)\{r(s_1),t(s_2)\}_{D-1}0(s-s_2)\}_D}
f^{(s)}_{e;(s_1,s_2);n}\ ,\quad n=s+k-s_1\ ,\eea
where $\{\cdots\}_{D}$ and $\{\cdots\}_{D-1}$, respectively,
denote $\mso(D;\Comp)$-traceless and $\mso(D-1;\Comp)$-traceless
Young projections.

The harmonic expansions include not only the composite massless
lowest-weight spaces $\mD^+(s+2\e_0;(s))$ and highest-weight spaces
$\mD^-(-(s+2\e_0);(s))$ but also the non-composite lowest-spin
spaces ${\cal W}_{(s)}$. In the context of standard second-quantized
free higher-spin field theory on $AdS_D$, the former are unitary
spaces containing the one-particle states, corresponding to
wave-functions that fall off relatively fast at time-like infinity,
while the latter are excluded from the perturbative spectrum since
they correspond to non-normalizable wave-functions. However, as we
shall see below, the situation is to some extent reversed in the
norm induced by the trace operation $\Tr$. Moreover, all elements of
${\cal M}$ play a role in the broader context of constructing the
space of classical solutions to the full non-linear Vasiliev
equations, \emph{i.e.} the covariant phase-space, which is the
starting point for the covariant phase-space quantization of
higher-spin gauge theory.

Next we shall examine the inverse procedure of expanding the
twisted-adjoint $\mso(D;\Comp)$ tensors $\ket{T_{a(s+k),b(s)}}$ and
the adjoint $\mso(D+1;\Comp)$ tensors $\ket{Q_\ell}$ in the compact
basis.


\scs{Expanding Covariant Tensors in Compact Basis}\label{Sec:Refl}


\scsss{Twisted-Adjoint $\mso(D;\Comp)$ Tensors as
Singleton-Singleton Composites}

According to the Flato-Fronsdal formula \eq{FFformula}, the massless
lowest-weight states belong to $\mD_0\otimes \mD_0\subset {\cal B}$.
Letting $\ket{\1_{\mD_0}}_{12}$ denote the restriction of the
reflector to $\mD_0\otimes \mD_0$, the expansion of the type
$(s+k,s)$ twisted-adjoint $\mso(D;\Comp)$ tensor
\bea
\ket{\Phi_{(s+k,s);\mD_0}}_{12}&=&\Phi^{a(s+k),b(s)}\ket{T_{a(s+k),b(s);\mD_0}}_{12}\
=\ \Phi^{a(s+k),b(s)}T_{a(s+k),b(s)}(1) \ket{\1_{\mD_0}}_{12}\ .\eea
 in the lowest-weight module $\mD(s+2\e_0;(s))$ is given by
\bea \ket{\Phi_{(s+k,s);\mD_0}}& = & \sum_{s+k\geq j_1\geq s\geq
j_2\geq 0} {\cal C}^{(s+k,s)}_{(j_1,j_2)} \Phi^{r(j_1),t(j_2)}
\ket{(s+k,s);(j_1,j_2)}_{r(j_1),t(j_2)}\ , \label{embedding}\eea
where we have made the following definitions: i) the coefficients
${\cal C}^{(s+k,s)}_{(j_1,j_2)} $ are overall normalizations; ii)
$\Phi^{r(j_1),t(j_2)}=\Phi^{0(s+k-j_1)\{r(j_1),t(j_2)\}_{D-1}0(s-j_2)}$
are the type $(j_1,j_2)$ $\mso(D-1;\Comp)$ \emph{polarization}
tensors contained in $\Phi^{a(s+k),b(s)}$; iii) the states
\bea \ket{(s+k,s);(j_1,j_2)}_{r(j_1),t(j_2)}&=&
\psi^{(s+k,s)}_{(j_1,j_2)}(x)
\ket{(s);2\e_0+j_1+j_2;(j_1,j_2)}_{r(j_1),t(j_2)}\
,\label{ketj1j2}\eea
where\footnote{In this section the $\star$ is suppressed in products
of $L^+$ operators. We also notice that $L^+_r=L^+_r(1)+L^+_r(2)$ so
that $x=L^+_r L^+_r=2L^+_r(1)L^+_r(2)$ acting on composites.}
\bea \ket{(s);j_1+j_2+2\e_0;(j_1,j_2)}_{r(j_1),t(j_2)}\ =\ L^+_{\{
r_1}\cdots L^+_{r_{j_1-s}} L^+_{t_1}\cdots L^+_{t_{j_2}}
\ket{2\e_0+s;(s)}_{r(s)\}}\ ,\label{ketj1j2mine}\eea
are normalized type $(j_1,j_2)$ states in $\mD(2\e_0+s,(s))$ of
\emph{minimal} energy; iv) the \emph{dressing} functions
\bea \psi^{(s+k,s)}_{(j_1,j_2)}(x)&=& \sum_{n=0}^\infty
\psi^{(s+k,s)}_{(j_1,j_2);n}x^n\ ,\qquad x\ =\ y^2\ =\ L^+_r L^+_r\
,\label{dressing}\eea
are determined by the normalization
\bea \psi^{(s+k,s)}_{(j_1,j_2)}(0)&=&1\ ,\label{psinorm}\eea
and by the embedding requirement that
$\{\ket{(s+k,s);(j_1,j_2)}\}_{s+k\geq j_1\geq s\geq j_2}$ furnishes
a decomposition of the original type $(s+k,s)$ $\mso(D;\Comp)$
tensor, \emph{i.e.} ($\ket{(j_1,j_2)}\equiv\ket{(s+k,s);(j_1,j_2)}$)
\bea M_{0r}\ket{(j_1,j_2)}&=&  {\cal
C}^{(s+k,s)}_{(j_1,j_2);(1,0)}\ket{(j_1+1,j_2)}+{\cal
C}^{(s+k,s)}_{(j_1,j_2);(-1,0)}\ket{(j_1-1,j_2)}\\[5pt]&&+
{\cal C}^{(s+k,s)}_{(j_1,j_2);(0,1)}\ket{(j_1,j_2+1)}+{\cal
C}^{(s+k,s)}_{(j_1,j_2);(0,-1)}\ket{(j_1,j_2-1)}\
,\label{ccoeff}\eea
with $M_{0r}=\ft12(L^+_r+L^-_r)$ and
\bea {\cal C}^{(s+k,s)}_{(s+k,j_2);(1,0)}&=&{\cal
C}^{(s+k,s)}_{(j_1,s);(0,1)}\ =\ 0\ ,\\[5pt] {\cal
C}^{(s+k,s)}_{(s,j_2);(-1,0)}&=&0\qquad\mbox{for $j_2<s$}\ ,\eea
that enforce the conditions on the ranges of $(j_1,j_2)$ given in
\eq{embedding}. We also notice that the Casimir constraint
\bea
&&\hspace{-1.7cm}\left(C_2[\mso(D;\Comp)]-C_2[\mso(D;\Comp)|(s+k,s)]\right)
\
\nn\\[5pt]
&&\hspace{3.5cm} \psi^{(s+k,s)}_{(j_1,j_2)}(x)
\ket{(s);2\e_0+j_1+j_2;(j_1,j_2)}_{r(j_1),t(j_2)} \ = \ 0\ .\eea
can be turned into a differential equation in $x$ for the dressing
functions. The embedding conditions are equivalent to the condition
that the ``top'' state $\ket{(s+k,s);(s);(s+k,0)}$ obeys
\bea M_{0\{r}\ket{(s+k,s);(s+k,0)}_{r(s+k)\}}&=&0\
,\label{Morcond}\eea
with solution
\bea \psi^{(s+k,s)}_{(s+k,0)}(x)&=& \Psi^{(0,0)}_{(0,0)}(x)\ =\
\C(\n+1)\left(\frac{y}{2}\right)^{-\n} J_{\n}(y)
 \ ,
\quad \n=\e_0-1\ .\label{psi00D}\eea
In $D=4$, where the index $\nu=-\ft12$, the rescaled Bessel
functions become trigonometric, and the reflector takes the simple
form
\bea D=4&:& \ket{\1_{\mD_0}}_{12} \ = \ \cos(y)\ket{1;(0)}_{12}\ .
\label{cos}\eea
Let us provide a few more detailed remarks on the basic structure of
\eq{embedding}:

\begin{itemize}

\item The ranges of $(j_1,j_2)$ follow from the decomposition formula
\bea \widehat{\begin{picture}(52,13)(-3,2)
\multiframe(0,6.5)(13.5,0){1}(35,6){}\put(40,8){\tiny $s+k$}
\multiframe(0,0)(13.5,0){1}(20,6){}\put(23,0){\tiny $s$}
\end{picture}}\ = \ \bigoplus_{\ba{c}j_1,j_2 \in\mathbb{N}\\
{\tiny s+k\geq j_1\geq s\geq j_2\geq
0}\ea}\begin{picture}(52,13)(-3,2)
\multiframe(0,6.5)(13.5,0){1}(35,6){}\put(40,8){\tiny $j_1$}
\multiframe(0,0)(13.5,0){1}(20,6){}\put(23,0){\tiny $j_2$}
\end{picture}  \ ,\label{decomposition}\eea
where the $\mso(D-1;\Comp)$ Young diagrams on the right-hand side
can be obtained by the $\mso(D;\Comp)$ Young diagram (denoted with
a hat) on the left-hand side by projecting its indices along the
$0$-direction in all possible ways compatible with the
irreducibility of Young diagrams. We note that for $D=4$ the type
$(j_1,j_2)$ tensors with $j_2\geq 2$ are trivial, \emph{i.e.}
zero-dimensional, as follows from King's rule, which implies that
the dimension of a traceless Young-projected
$\mso(N;\Comp)$-tensor is zero if the sum of the lengths of the
first two columns exceeds $N$ (see, for example,
\cite{Bekaert:2006py}).

\item Starting from a state in $\mD(2\e_0+s;(s))$ at excitation level
$l$, that is
\bea \ket{(s);l+s+2\e_0}_{r(l);t(s)}&=&L^+_{r_1}\cdots L^+_{r_l}
\ket{2\e_0+s;(s)}_{t(s)}\ ,\eea
and decomposing it under $\mso(D-1;\Comp)$ by extracting traces
into powers of $x=L^+_r L^+_r$ (recall that
$L^+_r\ket{2\e_0+s;(s)}_{rt(s-1)}=0$), one finds
\bea \hspace{-1.5cm}
\ket{(s);l+s+2\e_0}&=&\sum_{n=0}^{[l/2]}\sum_{p=0}^{\textrm{min}(s,l-2n)}
x^n\ket{(s);2\e_0+l-2n;(s+l-2n-p,p)}\ .\eea
Thus, a general state in $\mD(s+2\e_0;(s))$ of type $(j_1,j_2)$ is
of the form
\bea \ket{(s);(j_1,j_2)}&=&\sum_{n=0}^\infty \psi_{(j_1,j_2);n}
x^n\ket{(s);j_1+j_2+2\e_0;(j_1,j_2)}\ ,\eea
where $\psi_{(j_1,j_2);n}$ are arbitrary coefficients and
$\ket{(s);j_1+j_2+2\e_0;(j_1,j_2)}$ is the type $(j_1,j_2)$ state of
minimal energy given in \eq{ketj1j2mine}.

\item To show \eq{psi00D}, we use the lemma
\bea \hspace{-1.5cm} M_{0\{r} x^n\ket{2\e_0+p;(p)}_{r(p)\}}&=&
\ft12 (1+4n(n+\e_0-1))L^+_{\{r} x^{n-1}\ket{2\e_0+p;(p)}_{r(p)\}}\
,\eea
where we notice the independence of $p=s+k$, and we have defined
\bea \ket{2\e_0+p;(p)}_{r(p)}&=&L^+_{\{r_1}\cdots
L^+_{r_k}\ket{s+2\e_0;(s)}_{r(s)\}}\ ,\label{excstate}\eea
with the property $L^-_{\{r_1}\ket{2\e_0+p;(p)}_{r(p)\}}=0$. Hence,
from the embedding condition \eq{Morcond}, it follows that
\bea \left(4x{d^2\over dx^2}+4\e_0{d\over dx}+1
\right)\psi^{(s+k,s)}_{(s+k,0)}(x)&=&0\ ,\label{diffeqn}\eea
with solution \eq{psi00D}, as can be seen by the following rescaling
and change of variables,
\bea \psi^{(s+k,s)}_{(s+k,0)}(x)&=& y^{-\nu}J(y)\ ,\qquad x=y^2\
,\eea
which brings \eq{diffeqn} to Bessel's differential equation
\bea \left({d^2\over dx^2}+y{d\over dy}+(1-{\nu^2\over
y^2})\right)J(y)\ ,\qquad \nu=\e_0-1\ .\eea
Thus, the solution of \eq{diffeqn} that is analytical at $x=0$ is
given uniquely up to a normalization by \eq{psi00D}.

\item For $s=0$, the independence of $\psi^{(k)}_{(k)}(x)$
(where we have dropped the second highest-weight labels) on $k$, can
be shown directly using
$P_r=\frac{1}{2i}(L^-_r-L^+_r)=-iM_{0r}+iL^+_r$ and
$M_{0r}\ket{(0);(0)}=0$, which implies
\bea \ket{(k);(k)}&= &-i^k P_{\{r_1}\cdots
P_{r_k\}}\ket{(0);(0)}\\[5pt]&=&\sum_{p=0}^k{k\choose p}(-1)^{k-p} L^+_{\{r_1}\cdots L^+_{r_p}M_{0r_{p+1}}\cdots
M_{0r_{k}\}}\ket{(0);(0)} \\[5pt]&=&  L^+_{\{r_1}\cdots
L^+_{r_k\}}\ket{(0);(0)}\ .\eea
Thus, since $\ket{(k);(k)}=\psi^{(k)}_{(k)}(x)L^+_{\{r_1}\cdots
L^+_{r_k\}}\ket{2\e_0;(0)}$, it follows that
$\psi^{(k)}_{(k)}(x)=\psi^{(0)}_{(0)}(x)$.  The energy operator, on
the other hand, acts as
\bea E \,f(x)\ket{2\e_0+p;(p)}_{r(p)} & = &
\left(2x\frac{d}{dx}+2\e_0+p\right)f(x)\ket{2\e_0+p;(p)}_{r(p)} \ ,
\label{diffE}\eea
where $\ket{2\e_0+p;(p)}_{r(p)}$ is defined in \eq{excstate}, which
means that the functional form of $\psi^{(k)}_{(j_1)}(x)$ with
$j_1<k$ differs from that of $\psi^{(0)}_{(0)}(x)$. For example,
from $\ket{(k);(k-1)}\propto E\ket{(k);(k)}$, it follows that
\bea \psi^{(k)}_{(k-1)}&=& {1\over
2\e_0+k}\left(2x\frac{d}{dx}+2\e_0+k\right) \psi^{(0)}_{(0)}(x)\
.\eea

\item The embedding formula can easily be adapted to
highest-weight spaces by applying the $\pi$-map, using
$\pi(\ket{s+2\e_0;(s)})=\ket{-s-2\e_0,(s)}$ and
$\pi(L^+_r)=L^-_r$.

\end{itemize}

Let us demonstrate how we can arrive at the embedding formula and
\eq{psi00D} by using compositeness. To begin with, for $s=k=0$ the
embedding formula \eq{embedding}, where now $\Phi_{(0,0)}\in \Comp$,
amounts to 
\bea \ket{\1_{\mD_0}}_{12} &=& (\Phi_{(0,0)})^{-1}
\ket{\Phi_{(0,0)}}_{12}\ = \ {\cal
C}^{(0,0)}_{(0,0)}\psi^{(0,0)}_{(0,0)}(x) \ket{2\e_0;(0)}_{12}\
.\label{scalrefl}\eea
In other words, the $\mso(D;\Comp)$-singlet in $\mD(2\e_0;(0))$ can
be identified with the singleton reflector, which can equivalently
be described as the map
\bea R&:& \mD_0\mapsto \mD^\ast_0\ ,\eea
defined by
\bea R(\ket{\e_0;(0)})&=&\bra{\e_0;(0)} \ ,\eea
and
\bea R(X\ket{\e_0;(0)})&=& \bra{\e_0;(0)}(\tau\circ \pi)(X)\ ,\eea
for $X\in {\cal A}$. It follows that
\bea R(L^{\pm}_r)&=&-L^\mp_r \ ,\qquad R(E)\ =\ E\ ,\qquad R(M_{rs})\ =\ M_{rs}\ ,\\[5pt]
R(\ket{n}) &=&  (-1)^n\bra{n}\ ,\label{reflmap}\eea
where we have defined the following basis elements
\bea \ket{n}_{r(n)}&=& \ket{n+\e_0;(n)}_{r(n)}\ =\  L^+_{r_1}\cdots
L^+_{r_n}\ket{\e_0;(0)}\ ,\label{basisD0ket}\\[5pt]
\bra{n}_{r(n)}&=&\bra{n+\e_0;(n)}_{r(n)}\ =\
\bra{\e_0;(0)}L^-_{r_1}\cdots L^-_{r_n}\ ,\label{basisD0bra}\eea
which are traceless as a consequence of the singular vector
\eq{singularsingleton}, and with normalization
\bea \bra{\e_0;(0)}L^-_{r_1}\ldots L^-_{r_n}L^+_{s_1}\ldots
L^+_{s_n}\ket{\e_0;(0)} \ = \ {\cal
N}_n\d_{\{r_1...r_n\}}^{\{s_1...s_n\}} \ ,\qquad {\cal N}_n \ =\
2^n\,n!(\e_0)_n  \ ,\eea
as can be seen using
\bea L^-_r\ket{n}_{s_1...s_n} \ = \
2n(n+\e_0-1)\d_{r\{s_1}\ket{n-1}_{s_2...s_n\}} \
.\label{lminusketn}\eea
For general $s$ and $k$, we substitute the Flato-Fronsdal formula
\eq{FFformula} into the right-hand side of the embedding formula
\eq{embedding}, which then reads
\bea && \hspace{-2cm}\sum_{s+k\geq j_1\geq s\geq j_2\geq 0}{\cal
C}^{(s+k,s)}_{(j_1,j_2)}\,
\Phi^{r(j_1),t(j_2)}\,\psi^{(s+k,s)}_{j_1,j_2}(x) \nn\\[5pt]
&& \hspace{4cm}
 L^+_{\{
r_{1}}\cdots L^+_{r_{j_1-s}}L^+_{t_1}\cdots L^+_{t_{j_2}}
f_{r(s)\}}(1,2) \ket{2\e_0;(0)}_{12}\ , \label{embedding2}\eea
where $L^+_r=L^+_r(1)+L^+_r(2)$. Turning to the left-hand side, it
can be rewritten using \eq{test} and by decomposing
$T_{a(s+k),b(s)}$ under $\mso(D-1;\Comp)$. Schematically,
suppressing trace parts in
$T_{0(s+k-j_1)\{r(j_1),t(j_2)\}0(s-j_2)}$, the result reads
\bea \sum_{s+k\geq j_1\geq s\geq j_2\geq 0}\Phi^{r(j_1),t(j_2)}
\left((M_{rs})^{j_2} (M_{r0})^{s-j_2} (P_r)^{j_1-s}
E^{k-j_1+s}\right)(1)\ket{\1_{\mD_0}}_{12} \ ,\label{lhs}\eea
where the generators can be (anti)-symmetrized under the exchange of
the ``flavor' indices $1$ and $2$ using \eq{overlapX} together with
\eq{Lplusminus} and \eq{ll} with the following result:
\bea
P_r(1)\ket{\1_{\mD_0}}_{12}&=&\ft{i}2L^+_r\ket{\1_{\mD_0}}_{12}\,,\quad
M_{r0}(1)\ket{\1_{\mD_0}}_{12} =
-\ft{1}2(L^+_r(1)-L^+_r(2))\ket{\1_{\mD_0}}_{12}\
,\\[5pt]
M_{rs}(1)\ket{\1_{\mD_0}}_{12}&=&(L^+_{[r}(1)L^+_{s]}(2)+L^-_{[r}(1)L^-_{s]}(2))\ket{\1_{\mD_0}}_{12}\
,\\[5pt] E(1)\ket{\1_{\mD_0}}_{12}&=&
\ft{1}{2(D-1)}(L^+_r(1)L^+_r(2)-L^-_r(1)L^-_r(2))\ket{\1_{\mD_0}}_{12}\
.\label{lhs2}\eea
Proceeding by substituting
$\ket{\1_{\mD_0}}_{12}=\psi^{(0,0)}_{(0,0)}(x)\ket{2\e_0;(0)}_{12}$
into \eq{lhs}, and noting that
\bea
L^-_{[r}(1)L^-_{s]}(2)\psi^{(0,0)}_{(0,0)}(x)\ket{2\e_0;(0)}_{12}&=&
L^+_{[r}(1)L^+_{s]}(2)(D_M
\psi^{(0,0)}_{(0,0)})(x)\ket{2\e_0;(0)}_{12})\ ,\\[5pt]
L^-_{r}(1)L^-_{r}(2)\psi^{(0,0)}_{(0,0)}(x)\ket{2\e_0;(0)}_{12}&=&
(D_E \psi^{(0,0)}_{(0,0)})(x)\ket{2\e_0;(0)}_{12})\ ,\eea
where $D_M$ and $D_E$ are analytical differential operators in $x$
(\emph{i.e.}, with coefficients that are analytical functions of $x$
at $x=0$), one sees that the type-$(j_1,j_2)$ sector of the
left-hand side \eq{lhs} is of the form
\bea
\Phi^{r(j_1),t(j_2)}(D^{(s+k,s)}_{(j_1,j_2)}\psi^{(0,0)}_{(0,0)})(x)
M_{r(j_1),t(j_2)}(1,2)\ket{2\e_0;(0)}_{12}\ ,\eea
where $D^{(s+k,s)}_{(j_1,j_2)}(x)$ is an analytical differential
operator in $x$ and $M_{r(j_1),t(j_2)}(1,2)$ is a monomial in
$L^+_r(1)$ and $L^+_r(2)$ of degree $2j_2+s-j_2+j_1-s=j_1+j_2$ with
flavor-exchange symmetry $M_{r(j_1),t(j_2)}(1,2)=(-1)^s
M_{r(j_1),t(j_2)}(2,1)$. Since the right-hand side \eq{embedding2}
is of exactly the same form, we conclude that
\bea \left(D^{(s+k,s)}_{(j_1,j_2)}\psi^{(0,0)}_{(0,0)}\right)(x)&=&
{\cal C}^{(s+k,s)}_{(j_1,j_2)}\psi^{(s+k,s)}_{(j_1,j_2)}(x)\ .\eea
Specializing to $(j_1,j_2)=(s+k,0)$, there are no contributions to
\eq{lhs} from $M_{rs}$ and $E$, so that $D^{(s+k,s)}_{(s+k,0)}$
becomes a constant\footnote{For $(j_1,j_2)=(s+k,0)$, the schematic
expression \eq{lhs} is of the form
$\Phi^{r(s+k)}(L^+_r(1)+L^+_r(2))^k(L^+_r(1)-L^+_r(2))^s\ket{\1_{\mD_0}}_{12}$,
with a binomial expansion, to be compared with the ``dressed''
binomial expansion in the (precise) Flato-Fronsdal formula
\eq{FFformula}. There is a precise agreement, however, once the
trace corrections to \eq{lhs} are included. For example, for
$(s+k,s)=(2,0)=(j_1,j_2)$, the precise form of the left-hand side of
\eq{embedding} reads
$$\Phi^{r(2)}T_{r(2),0(2)}(1)\ =\
\ft43 \Phi^{r(2)}(M_{r_10}\star M_{r_20}+\ft1{2\e_0+1}P_{r_1}\star
P_{r_2})\ ,$$
as can be seen using \eq{papa} and \eq{macmcb} to expand the
quantity $T_{a(2),b(2)}=\widehat T_{\{a(2),b(2)\}_D}$ as follows
$$
\ft43 M_{a_1 b_1}\star M_{a_2 b_2}-\ft{4}{3(2\e_0+1)}\left(
\eta_{a(2)}P_{b_1}\star P_{b_2}-2\eta_{a_1b_1}P_{(a_2}\star
P_{b_2)}+\eta_{b(2)}P_{a_1}\star P_{a_2}\right)+{4\e_0\over
3(2\e_0+1)}(\eta_{a(2)}\eta_{b(2)}-\eta_{a_1 b_1}\eta_{a_2 b_2})\
.$$
Thus, substituting $M_{r0}=-\ft12(L^+_r+L^-_r)$ and
$P_r=\ft{i}2(L^+_r-L^-_r)$, and using
$(L^\pm_r(1)-L^\mp_r(2))\ket{\1_{\mD_0}}_{12}=0$ and
$\ket{\1_{\mD_0}}_{12}=\psi^{(0,0)}_{(0,0)}(x)\ket{2\e_0;(0)}_{12}$,
one arrives at
\begin{eqnarray*}&&\Phi^{r(2)}T_{r(2),0(2)}(1)\ket{\1_{\mD_0}}_{12}\nn\\[5pt]
& =& {2\e_0\over 3(2\e_0+1)}
\Phi^{r(2)}\psi^{(0,0)}_{(0,0)}(x)\left(L^+_{r_1}(1)L^+_{r_2}(1)-{2(\e_0+1)\over
\e_0}L^+_{r_1}(1)L^+_{r_2}(2)+L^+_{r_1}(2)L^+_{r_2}(2)\right)\ket{2\e_0;(0)}_{12}
\ ,\end{eqnarray*}
in agreement with \eq{FFformula}, and we also have ${\cal
C}^{(2,0)}_{(2,0)}=\ft{2\e_0}{3(2\e_0+1)}$.} and
\bea \psi^{(s+k,s)}_{(s+k,0)}&=&\psi^{(0,0)}_{(0,0)}(x)\ ,\qquad
{\cal C}^{(s+k,s)}_{(j_1,j_2)}\ =\  D^{(s+k,s)}_{(s+k,s)}.\eea
Finally, the precise functional form \eq{psi00D} follows by
expanding the unity $\1_{\mD_0}$ in the basis \eq{basisD0ket} and
\eq{basisD0bra} as
\bea \1_{\mD_0} & = & \sum_{n=0}^\infty\, \left[{\cal
N}_n\right]^{-1} \ket{n}\bra{n}\ =\ \sum_{n=0}^\infty {1\over 2^n
n!(\e_0)_n} L^+_{r_1}\cdots L^+_{r_n}\ket{\e_0;(0)}\bra{\e_0;(0)}
L^{-}_{r_1}\cdots L^-_{r_n} \ .\label{idD0}\eea
Letting $\cross f(L^-_r,L^+_r)\cross$ denote the standard
normal-ordering, and introducing the variable
\bea z &=& 2L^+_r L^-_r\ ,\eea
we can write
\bea \1_{\mD_0} & = & \sum_{n=0}^\infty {1\over 2^n n!(\e_0)_n}
\cross
(L^+_r L^-_r)^n \ket{\e_0;(0)}\bra{\e_0;(0)} \cross \\[5pt]&=&
\sum_{n=0}^\infty \frac{1}{4^n\,n!(\e_0)_n} \cross
(\sqrt{z})^{2n} \ket{\e_0,(0)}\bra{\e_0,(0)} \cross  \\[5pt]
& = & \C(\n+1)\left(\frac{\sqrt{z}}{2}\right)^{-\n}\cross
I_\n(\sqrt{z})\ket{\e_0,(0)}\bra{\e_0,(0)}\cross\ ,\eea
where $\nu=\e_0-1$ and $I_\n$ is the modified Bessel function,
related to $J_\n$ according to
\bea I_\n(w) \ = \ \left\{\ba{c}
e^{-\frac{i\pi\n}{2}}J_\n(e^{\frac{i\pi}{2}}w)\ ,
\quad\mbox{for $-\pi<\textrm{arg}\,w\leq\frac{\pi}{2}$} \\
 e^{\frac{2i\pi\n}{3}}J_\n(e^{-\frac{3i\pi}{2}}w)\ ,
\quad\mbox{for $-\frac{\pi}{2}<\textrm{arg}\,w\leq\pi$} \ea\right.
\eea
We note that in $D=4$,
\bea \1_{\mD_0} \ = \  \cross
\cosh{\sqrt{z}}\ket{\ft12;(0)}\bra{\ft12;(0)} \cross \
.\label{4Dsinglunity}\eea
Applying the inverse reflection $R^{-1}:\mD_0^\ast\mapsto \mD_0$ to
the dual states in the expansion of $\1_{\mD_0}$, and denoting the
resulting two copies of $\mD_0$ by $\mD_0(\x)$, $\x=1,2$, and using
\bea R^{-1}(z)&=&-2L^+_r(1)L^+_r(2)\ =\ -x\ ,\eea
which formally implies $R^{-1}(\sqrt{z})=iy$, we obtain
\bea R^{-1}\1_{\mD_0}&=&\sum_{n=0}^\infty
\frac{(-1)^n}{4^n\,n!(\e_0)_n} (\sqrt{x})^{2n}
\ket{\e_0;(0)}_1\otimes \ket{\e_0;(0)}_2\\[5pt]&=&
\C(\n+1)\left(\frac{\sqrt{x}}{2}\right)^{-\n}J_\n(\sqrt{x})\ket{2\e_0;(0)}_{12}\
=\ \ket{\1_{\mD_0}}_{12}\ .\label{leftidD0}\eea
%


\scss{Adjoint $\mso(D+1;\Comp)$ Tensors as Singleton-Anti-Singleton
Composites}


The $\ell$-th adjoint level $\L_\ell$ decomposes into a
finite-dimensional compact lowest and highest-weight space,
\emph{viz.} ($s=2\ell+2$)
\bea {\cal L}_\ell&=& \mD(-(s-1);(s-1))\ =\
\bigoplus_{(e;s)\in\Lambda_\ell} \Comp\otimes Q_{e;s}\
,\label{Lambdaell}\eea
where $Q_{e;s}$ are monomials of degree $2\ell+1$ built from
$L^\pm_r$, $M_{rs}$ and $E$. The lowest-weight element can be
written using \eq{ketQ2} as
\bea \ket{-(s-1);(s-1)}_{12;r(s-1)}&=& (L^-_{\{r}\cdots
L^-_{r_{s-1}\}})(1)\ket{\j1}_{12}\ ,\label{adjlws}\eea
where the $\mso(D+1;\Comp)$-invariant twisted reflector
$\ket{\j1}_{12}=k(1)\ket{\1}_{12}$ was defined in \eq{twrefl}. Thus,
the above lowest-weight elements are identified as the
singleton-singleton composites
\bea \hspace{-1cm}\ket{-(s-1);(s-1)}_{12;r(s-1)}&=&
\sum_{n=0}^\infty {(-1)^n\over 2^n n!(\e_0)_n}
\ket{n+s-1}^-_{1;r(s-1)t(n)}\otimes \ket{n}^+_{2;t(n)}\ ,\eea
where $\ket{n}^+_{r(n)}=\ket{n}_{r(n)}$ is the singleton basis
\eq{basisD0ket} and the corresponding anti-singleton basis is
defined by
\bea \ket{ n}^-_{r(n)}&=& k \ket{n}_{r(n)}\ =\ L^-_{\{r}\cdots
L^-_{r_{s-1}\}}\ket{-\e_0;(0)}^-\ ,\eea
such that $L^-_{r}\cdots
L^-_{r_{s-1}}\ket{n}^-_{s(n)}=\ket{n+s-1}^-_{r(s-1)t(n)}$. We notice
that \eq{adjlws} by construction obeys
\bea (L^-_r(1)+L^-_r(2))\ket{-(s-1);(s-1)}_{12;r(s-1)}&=&0\ ,\eea
which one can also check explicitly using \eq{lminusketn}. In other
words, the analog of the ordinary Flato-Fronsdal formula
\eq{FFformula}, which states that $\mD_0\otimes\mD_0$ can be
expanded in terms of infinite-dimensional massless lowest-weight
spaces, is the following \emph{twisted Flato-Fronsdal formula}:
\bea\widetilde{\mD}_0\otimes\mD_0&=& \bigoplus_{s=0}^\infty
\mD(-(s-1);(s-1))\ , \label{FFformulatwisted}\eea
which thus states that $\widetilde{\mD}_0\otimes\mD_0$ can be
expanded in terms of finite-dimensional lowest-weight spaces.


\scss{On Adjoining the Adjoint and Twisted-Adjoint
Representations}\label{Sec:adjoin}


The Flato-Fronsdal formulae \eq{FFformula} and its twisted version
\eq{FFformulatwisted} show, respectively, how the massless
lowest-weight spaces residing inside the compact twisted-adjoint
representation ${\cal M}(D+1;\Comp)$ and the adjoint
representation $\mho(D+1;\Comp)$ are mapped to singleton-singleton
and anti-singleton-singleton composites. As previously mentioned,
if $Q\in\mho(D+1;\Comp)$, then the element $k$ maps $Q$ to an
element $S=Q\star  k\in {\cal T}(D+1;\Comp)$.

In this section we shall propose another relation between the
adjoint and twisted-adjoint representations, that constitutes a
``fiber'' analog of the unfolding procedure in space-time, and
provides a direct explanation for the agreement between the Casimir
operators noted in \eq{c2ell} and \eq{c4ell}. The basic idea is to
adjoin the adjoint weight space ${\cal L}_\ell=\mD(-(s-1);(s-1))$,
given in \eq{Lambdaell}, to the massless lowest-weight space
$\mD(s+2\e_0;(s))$ via the intermediate conjugate massless
lowest-weight space $\mD(-s+2;(s))$, whose lowest-weight state has
the same quantum numbers of the singular vector
\bea \ket{-s+2;(s)}&=&L^+_{\{r_1}\ket{-s+1;(s-1)}_{r(s-1)\}}\in
\mI(-s+1;(s-1))\ .\label{singvector}\eea
We propose that ($x=L^+_r L^+_r$)
\bea D\geq 5&:& \ket{s+2\e_0;(s)}_{r(s)}\ =\
x^{\ft{D-5}2}L^+_{t_1}\cdots L^+_{t_s}\ket{2;(s,s)}_{r(s),t(s)}\
,\label{subsing1}\\[5pt]
D=4&:& \ket{s+1;(s)}_{r(s)}\nn\\[5pt]
&& \hspace{1.5cm}\ =\ \e_{r_1t_1u_1}\cdots \e_{r_s t_s
u_s}L^+_{u_1}\cdots L^+_{u_{s-1}}\,\ket{2;(s,1)}_{t(s),u_s}\
,\label{subsing2}\eea
where $\ket{2;(s,s)}$ and $\ket{2;(s,1)}_{t(s),u_s}$ are the
descendants of $\ket{-s+2;(s)}$ given by
\bea D\geq 5&:& \ket{2;(s,s)}_{r(s),t(s)}\ =\ L^+_{\{r_1}\cdots
L^+_{r_s}\ket{-s+2;(s,s)}_{t(s)\}}\ ,\label{wsweyl1}\\[5pt]
D=4&:& \ket{2;(s,1)}_{r(s),t} \nn\\[5pt]
& & = \ \e_{\{r_1|u_1v_1}\cdots
\e_{|r_{s-1}|u_{s-1}v_{s-1}}L^+_{u_1}\cdots
L^+_{u_{s-1}}L^+_{|r_s|}\ket{-s+2;(s,s)}_{v(s-1)|t\}}\ .\eea
We notice that \eq{subsing1} is a regular enveloping-algebra element
in $D=5,7,\dots$ while it is an irregular element (involving a
square root) in $D=6,8,\dots$, where we define the action of $L^-_r
x^n$ on lowest-weight states by analytical continuation of
\bea L^-_r x^n&=& x^nL^-_r+4nx^{n-1}\left(iL^+_sM_{rs}+ L^+_r
(E+n-\e_0-1)\right)\ \eea
in $n$. We have checked that \eq{subsing1} is a singular vector for
$s=1$ and for $s=2$ in $D=5$. We have also checked \eq{subsing2} for
$s=1$ and $s=2$. We notice that \eq{singvector} is a weight-space
analog of an abelian gauge transformation of a $(D-1)$-dimensional
gauge field, and that $\ket{2;(s,s)}$ and $\ket{2;(s,1)}$ are the
corresponding Weyl and Cotton tensors for $D\geq 5$ and $D=4$,
respectively, inducing the following short exact sequence
\bea 0\ \hookrightarrow\ \ket{-s+1;(s-1)}&\rightarrow
&\ket{-s+2;(s)}\ \rightarrow\ \ket{s+2\e_0;(s)}\ \rightarrow\ 0\
.\eea

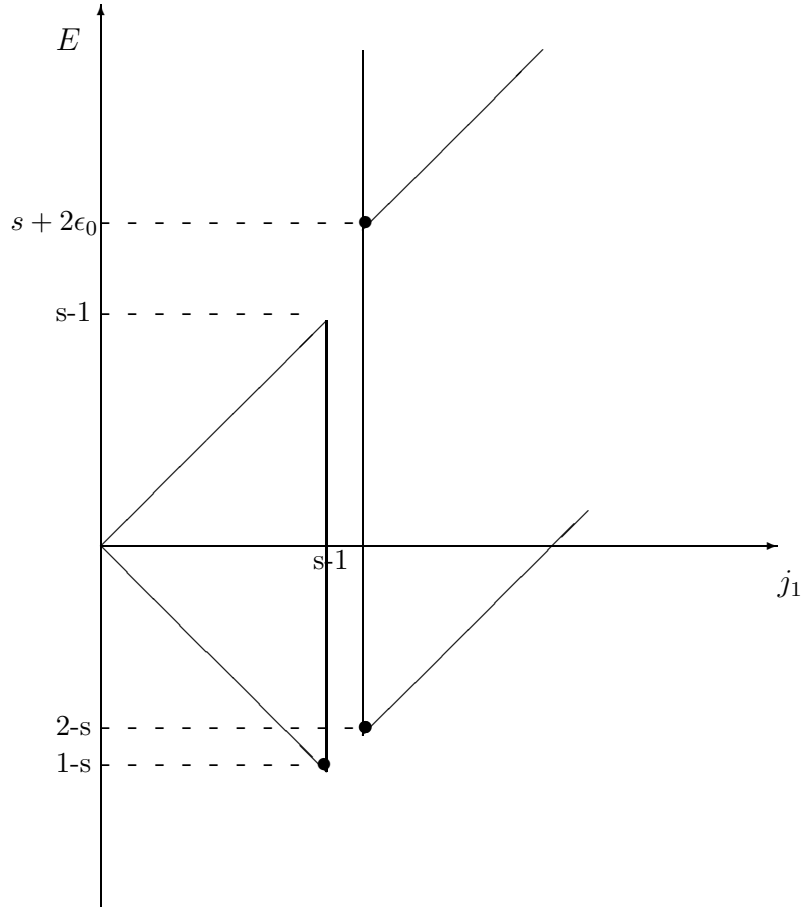
\begin{figure}[!h]
\begin{center}
\unitlength=.6mm
\begin{picture}(120,200)(0,-80)
\put(0,0){\vector(1,0){150}} \put(0,-80){\vector(0,1){200}}
\put(150,-10){$j_1$}
\put(-10,110){$E$}
\put(0,0){\line(1,1){50}}\put(0,0){\line(1,-1){50}}\put(50,-50){\!\!$\bullet$}\put(50,50){\line(0,-1){100}}
\put(47,-5){{\small s-1}}\put(-10,-50){{\small
1-s}}\put(-10,50){{\small s-1}}\multiput(0,50)(6,0){8}{-}
\multiput(0,-50)(6,0){8}{-} \put(-10,-42){{\small
2-s}}\put(58,-42){\line(0,1){130}}\put(58,-42){\!$\bullet$}\multiput(0,-42)(6,0){10}{-}
\put(58,-42){\line(1,1){50}}
\put(58,70){\line(0,1){40}}\put(58,-42){\!$\bullet$}\multiput(0,70)(6,0){10}{-}
\put(58,70){\line(1,1){40}}\put(58,70){\!$\bullet$}\put(-20,70){{\small
$s+2\epsilon_0$}}


\end{picture}
\end{center}
\caption{{\small The adjoint module $\mD(-(s-1);(s-1))$ is connected
via the conjugate massless module $\mD(-(s-2);(s))$ to the massless
module $\mD(s+2\e_0;(s))$.}} \label{wfig1}
\end{figure}

%


\scss{On Twisted-Adjoint $\mso(D;\Comp)$ Tensors in Non-Composite
Sectors}


The embedding formula \eq{embedding} describes the expansion of
enveloping-algebra operators $T_{a(s+k),b(s)}$ in the massless
lowest-weight (or highest-weight) spaces $\mD(s+2\e_0;(s))$, that
is, the expansion of the projected state
$T_{a(s+k),b(s)}(1)\ket{\1_{\mD_0}}_{12}$. Likewise, one may
consider their expansion around $T^{(0)}_{2;(0)}$ corresponding to
the lowest-weight state $\ket{2;(0)}_{12}$, obeying
$L^-_r(\xi)\ket{2;(0)}_{12}=0$ for $\x=1,2$ in view of
\eq{lminusleft} and \eq{lplusright} (although it has not simple
composite nature for $D\geq 6$). Denoting the corresponding dressing
functions by $\chi^{(s+k,s)}_{(j_1,j_2)}(x)$, the decoupling
condition \eq{Morcond} again implies
\bea \chi^{(s+k,s)}_{(s+k,0)}(x)&=&\chi^{(0,0)}_{(0,0)}(x)\ , \eea
where $\chi^{(0,0)}_{(0,0)}(x)$ now obeys \eq{diffeqn} with
$\e_0\rightarrow 2-\e_0$, that is
\be
\left(4x\frac{d^2}{dx^2}+4(2-\e_0)\frac{d}{dx}+1\right)\chi^{(0,0)}_{(0,0)}(x)
\ = \ 0 \ ,\ee
related to Bessel functions with index $\nu'=1-\e_0$. For $\n'\neq
-1,-2,\dots$, \emph{i.e.} in all even dimensions and also $D=5$, the
solution analytical at $x=0$ is the rescaled Bessel function
\be \chi^{(0,0)}_{(0,0)}(x) \ = \
\C(\n'+1)\left(\frac{\sqrt{x}}{2}\right)^{-\n'} J_{\n'}(\sqrt{x}) \
, \quad \n'=-\n=1-\e_0\ .\label{tpsi00D}\ee
In $D=4$, the dressing functions take the simple form
\bea \chi^{(0,0)}_{(0,0)}(x) &=& \frac{1}{\sqrt{x}}\sin(\sqrt{x})\
,\label{tpsi004}\eea
which one identifies with the spinor-singleton reflector
\bea D=4&:&\ket{\1_{\mD_{\ft12}}}_{12} \ =\  \chi^{(0,0)}_{(0,0)}(x)
\ket{2;(0)}_{12}\ ,\qquad \ket{2;(0)}_{12}\ =\
\ket{1;(\ft12)}^i_1\otimes \ket{1;(\ft12)}_{2,i}\
,\label{spinrefl}\eea
where $i=1,2$ is two-component spinor index.

For $\n'\neq -1,-2,\dots$, \emph{i.e.} $D=5+2p=7,9,\dots$,
$p=1,2,\dots$, the rescaled Bessel function develops a logarithmic
branch starting at order $x^p\log x$. Correspondingly, as examined
in Subsection \ref{Sec:LWS}, the module $\mD(2,(0)$ is actually a
$p$-lineton, so that $x^p\ket{2;(0)}$ is a singular vector. Thus,
the restricted reflector
\bea \ket{\1'}_{12}&=& \chi^{(0,0)}_{(0,0)}(x) \ket{2;(0)}_{12}\
,\eea
is a well-defined $\mso(D;\Comp)$-invariant state in $D=2p$ and
$D=5$, while for $D=5+2p$, $p=1,2,\dots$, it appears that some form
of logarithmic superpositions of states are required.

The compact twisted-adjoint module ${\cal M}$ also contains the
lowest-spin modules \eq{wedgestate}, whose elements can be reached
formally starting from the lowest-weight or highest-weight states by
considering dressing functions that are irregular elements of the
enveloping algebra. For example, starting from the scalar ground
states $\ket{2\e_0;(0)}_{12}$ and $\ket{2;(0)}$, the even static
ground state may be represented as either
$\ket{0;(0)}_{12}=x^{-\e_0}\ket{2\e_0;(0)}_{12}$ or
$\ket{0;(0)}_{12}=x^{-1}\ket{2;(0)}_{12}$, where $x=L^+_r
L^+_r=2L^+_r(1)L^+_r(2)$, although the precise meaning of these
expressions are not clear to us. Instead, we would like to stress
the fact that \emph{all} elements in ${\cal M}$ have regular
enveloping-algebra presentations.


\scs{Real Forms of the Master Fields}\label{Sec:RealForms}


The real forms of the master fields are defined by
\bea (A)^\dagger&=&-A\ ,\qquad (\Phi)^\dagger\ =\ \pi(\Phi)\
,\label{masterreal}\eea
where $\dagger$ acts as ordinary complex conjugation of the
component fields (on real spacetime with $(x^\m)^\ast=x^\m$) and as
hermitian conjugation on ${\cal A}$ defined by
\bea (M_{AB})^\dagger&=&\sigma(M_{AB})\ ,\eea
where $\sigma$ is an automorphism of ${\cal A}$ with the property
that $\sigma\circ\sigma={\rm Id}$, introduced in Chapter
\ref{absalg} and whose action we recall here. Decomposing
$M_{AB}\rightarrow (M_{ab};M_{0'a}=P_a)\rightarrow
(M_{rs},M_{r0};P_r,P_0)$, one may consider the following real forms
of $\mso(D+1;\Comp)\supset \mso(D;\Comp)$:
\bea \mso(D-1,2)\supset \mso(D-1,1)&:& \s(M_{AB})\ =\ M_{AB}\ ,\label{adsr}\\[10pt]
\mso(D,1)\supset \mso(D-1,1)&:& \s(M_{ab};P_a)\ =\ (M_{ab},-P_a)\ ,\label{dsr}\\[10pt]
\mso(D,1)\supset \mso(D)&:& \s(M_{rs},M_{r0};P_r,P_0)\ =\
(M_{rs},-M_{r0};P_r,-P_0)\ ,
\label{hr}\\[10pt]
\mso(D+1)\supset \mso(D)&:& \s(M_{rs},M_{r0};P_r,P_0)\ =\
(M_{rs},-M_{r0};-P_r,P_0)\ ,\label{sr}\eea
which we may summarize as
\bea \s(M_{rs},M_{r0};P_r,P_0)&=& (M_{rs},\s_{0}M_{r0};\s_{0'}
P_r,\s_{0'}\s_{0} P_0)\ ,\qquad \s_{0'}\ ,\s_0\ =\ \pm 1\ .\eea
Starting from \eq{adsr}, the three other reality conditions are
equivalent to using Wick rotations in either $0'$, or $0$, or both
$0'$ and $0$, respectively, to go to the \emph{real basis}
$M^{\Real}_{AB}$ obeying
\bea (M^{\Real}_{AB})^\dagger&=&M^{\Real}_{AB}\eea
and the commutation rules \eq{soD+1} with
$\eta_{AB}=(-\s_{0'};\eta_{ab})$ and
$\eta_{ab}=(-\s_{0},\delta_{rs})$, that is
\bea \mso(D-1,2)\supset \mso(D-1,1)&:& \eta_{AB}\ =\ (-;\eta_{ab})\
,\quad \eta_{ab}\ =\ (-,\delta_{rs})\ ,
\label{adseta}\\[10pt]
\mso(D,1)\supset \mso(D-1,1)&:& \eta_{AB}\ =\ (+;\eta_{ab})\ ,\quad
\eta_{ab}\ =\ (-,\delta_{rs})\ ,
\label{dseta}\\[10pt]
\mso(D,1)\supset \mso(D)&:&  \eta_{AB}\ =\ (-;\eta_{ab})\ ,\quad
\eta_{ab}\ =\ (+,\delta_{rs})\ ,\label{heta}\\[10pt]
\mso(D+1)\supset \mso(D)&:& \eta_{AB}\ =\ (+;\eta_{ab})\ ,\quad
\eta_{ab}\ =\ (+,\delta_{rs})\ .\label{seta}\eea
The corresponding component fields $A^{\Real}_{\mu,a(s-1),b(t)}$ and
$\Phi^{\Real}_{a(s+k),b(s)}$, defined by the real counterparts of
the covariant expansions \eq{As} and \eq{Phis}, \emph{i.e.}
\bea A&=&\sum_{s=2,4,6,\dots} A_{(s)}\ ,\label{A}\\[5pt]
A_{(s)}&=& -i\sum_{t=0}^{s-1}dx^\mu
A^{\Real}_{\mu,a(s-1),b(t)}(x)M^{a_1 b_1}_{\Real}\cdots
M^{a_tb_t}_{\Real}P^{a_{t+1}}_{\Real}\cdots P^{a_{s-1}}_{\Real}\
.\label{As}\eea 
\bea \Phi&=&\sum_{s=0,2,4,\dots} \Phi_{(s)}\ ,\label{Phi}\\[5pt]
\Phi_{(s)}&=& \sum_{k=0}^\infty{i^k\over k!}
\Phi^{\Real}_{a(s+k),b(s)}(x)M^{a_1b_1}_{\Real}\cdots
M^{a_sb_s}_{\Real}P^{a_{s+1}}_{\Real}\cdots P^{a_{s+k}}_{\Real}\
,\label{Phis}\eea
are real
\bea \left(A^{\Real}_{\mu,a(s-1),b(t)}(x^\mu)\right)^\ast&=&
A^{\Real}_{\mu,a(s-1),b(t)}(x^\m)\ ,\qquad
(\Phi^{\Real}_{a(s+k),b(s)}(x^\m))^\ast \ =\
\Phi^{\Real}_{a(s+k),b(s)}(x^\m)\ .\eea
For the compact basis elements $M_{rs}$, $E=M_{0'0}=P_0$ and
$L^\pm_r=M_{0r}\mp i M_{0'r}=M_{0r}\mp iP_r$, which obey the
commutation rules \eq{ll} and \eq{el} in all signatures, the reality
conditions read
\bea (M_{rs})^\dagger\ =\ M_{rs}\ ,\quad E^\dagger\ =\ \s_{0'0} E\
,\quad (L^\pm_r)^\dagger\ =\ \s_0 L^{\mp\s_{0'0}}_r\ ,\qquad
\s_{0'0}=\s_{0'}\s_0\ ,\eea
that is
\bea \mso(D-1,2)\supset \mso(D-1,1)&:& (L^\pm_r)^\dagger\ =\
L^\mp_r\ ,\quad (E)^\dagger\ =\ E\ ,\label{adsrc}\\[5pt]
\mso(D,1)\supset \mso(D-1,1)&:& (L^\pm_r)^\dagger\ =\
L^\pm_r\ ,\quad (E)^\dagger\ =\ -E\ ,\label{dsrc}\\[5pt]
\mso(D,1)\supset \mso(D)&:& (L^\pm_r)^\dagger\ =\ -L^\pm_r\ ,\quad (E)^\dagger\ =\ -E\ ,\label{hr}\\[5pt]
\mso(D+1)\supset \mso(D)&:& (L^\pm_r)^\dagger\ =\ -L^\mp_r\ ,\quad
(E)^\dagger\ =\ E\ ,\label{src}\ .\eea
Correspondingly, the compact basis elements obey
\bea (T^{(s)}_{e;(s_1,s_2)})^\dagger&=&(\s_{0'})^{s_1-s}
(\s_0)^{s-s_2} T^{(s)}_{\s_{0'0}e;(s_1,s_2)}\ ,\label{Tsdagger}\eea
as can be seen from \eq{Ts}, where $f^{(s)}_{e;(s_1,s_2);n}\in\Real$
and
\bea
(T_{s;n;(s_1,s_2);r(s_1),t(s_2)})^\dagger&=&(\s_{0'})^{s_1-s}(\s_0)^{s-s_2}(\s_{0'0})^n
T_{s;n;(s_1,s_2);r(s_1),t(s_2)}\ .\eea
Hence, using also \eq{minuse}, we see that $\Phi^\dagger=\pi(\Phi)$
implies the following conjugation rules for the harmonic-expansion
coefficients defined in \eq{harmexp}:
\bea
(\Phi^{(s)}_{e;(s_1,s_2)})^\ast&=&(-\s_{0'})^{s_1-s}(\s_0)^{s-s_2}\Phi^{(s)}_{-\s_{0'0}e;(s_1,s_2)}\
.\eea
Alternatively, using the ``ladder-operator'' bases \eq{wedgestate}
and \eq{1pstate} for ${\cal W}^\pm$ and $\mD$, we find the following
reality conditions in signatures $\mso(D-1,2)$ and $\mso(D+1)$:
\bea {\cal W}^\pm,\quad s=0&:& (\Phi^{\pm(0)}_{m,n})^\ast\ =\
(-\s_{0'})^{\ft{1\mp 1}2}
(-\s_{0})^{m+n}\Phi^{\pm(0)}_{n,m}\ ,\\[5pt]
{\cal W}^\pm,\quad s>0&:&(\Phi^{\pm(s)}_{m,n})^\ast\ =\
(\s_{0})^{s+\ft{1\mp 1}2}
(-\s_{0})^{m+n}\Phi^{\pm(s)}_{n,m}\ ,\\[10pt]\mD,\quad s\geq 0&:&
(\Phi^{(s)}_{m})^\ast\ =\ (\s_0)^s (-\s_{0})^{m}\overline
\Phi^{(s)}_m\ ,\eea
and in the signature $\mso(D,1)$:
\bea {\cal W}^\pm,\quad s=0&:& (\Phi^{\pm(0)}_{m,n})^\ast\ =\
(-\s_{0'})^{\ft{1\mp 1}2}
(-\s_{0})^{m+n}\Phi^{\pm(0)}_{m,n}\ ,\\[5pt]{\cal W}^\pm,\quad s>0&:&(\Phi^{\pm(s)}_{m,n})^\ast\ =\ (\s_{0})^{s+\ft{1\mp 1}2}
(-\s_{0})^{m+n}\Phi^{\pm(s)}_{m,n}\ ,\\[10pt]\mD,\quad s\geq 0&:&
(\Phi^{(s)}_{m})^\ast\ =\ (\s_0)^s (-\s_{0})^{m}\Phi^{(s)}_m\
,\qquad (\overline\Phi^{(s)}_m)^\ast\ =\ (\s_0)^s
(-\s_{0})^{m}\overline \Phi^{(s)}_m\ ,\eea
where we note the phase factors arising from
$\pi((T^{(0)}_\pm)^\dagger)= (-\s_{0'})^{\ft{1\mp 1}2}T^{(0)}_\pm$
and $\pi((T^{(s)}_\pm)^\dagger)= (\s_{0})^{s+\ft{1\mp
1}2}T^{(s)}_\pm$, and $\pi((T^{(s)}_{s+2\e_0;(s)})^\dagger)=
(\s_{0})^{s}T^{(0)}_{-\s_{0'0}(s+2\e_0);(s)}$. Hence, positive and
negative-energy modes are complex conjugated in signatures
$\mso(D-1,2)$ and unconjugated in the signature $\mso(D,1)$.


\scs{Inner Products and Unitarity in Two-Time
Signature}\label{Sec:Inner}


The bilinear inner product $(\cdot,\cdot)_{\cal
T}=Tr[\pi(\cdot)\star(\cdot)]$ on the covariant twisted-adjoint
module ${\cal T}$, consisting of the generalized polynomials defined
in \eq{covmod}, is no longer well-defined on ${\cal M}$, due to the
non-polynomiality of the compact basis. To define a bilinear form
$(\cdot,\cdot)_{\cal M}$ on ${\cal M}$, we thus declare the
twisted-adjoint $\mho_1(D+1;\Comp)$ action to be self-adjoint,
\emph{viz.} :
\bea ( \widetilde Q S,S')_{\cal M}&=&- (S, \widetilde Q S')_{\cal
M}\ ,\label{selfadjoint}\eea
for any $Q\in\mho_1(D+1;\Comp)$, and define $(S_\pm,S'_\mp)_{\cal
M}=0$ and
\bea (S_\pm,S'_\pm)_{\cal M}&=&\ft1{{\cal N}_\pm}\Tr[\pi(S_\pm)\star
S'_\pm]\ ,\label{s,s'}\eea
where ${\cal N}_\pm$ are the norms of the scalar static ground
states in $\cM^{\pm}$, \emph{viz.}
\bea {\cal N}_+&=& \Tr[T^{(0)}_{0;(0)}\star T^{(0)}_{0;(0)}]\
,\\[5pt]
{\cal N}_-&=& \ft1{D-1}\Tr[T^{(0)}_{0;(1);r}\star
T^{(0)}_{0;(1);r}]\ .\eea
Thus, to calculate $(S_\pm,S'_\pm)_{\cal M}$ one first expands
$S_\pm$ and $S'_\pm$ using \eq{gen1}--\eq{gen2} and
\eq{gen3}--\eq{gen4}, and then factors out ${\cal N}_\pm$ from
$Tr[\pi(S_\pm)\star S'_\pm]$ using \eq{selfadjoint}. By
construction, the resulting bilinear form is finite, symmetric
(which follows from $\Tr[\pi(X)]=\Tr[X]$) and
$\mho_1(D+1;\Comp)$-invariant. We note the twisted-adjoint energy
conservation law:
\bea (S_e,S'_{e'})_{\cal M}&=& \delta_{e+e',0}(S_e,S'_{e})_{\cal M}\
,\eea
provided $\widetilde E S_e =\{E,S_e\}_\star=e S_e$ \emph{idem}
$S'_{e'}$. We also note that $(\cdot,\cdot)_{\cal M^+}$ simplifies
in view of \eq{EstarT0}, which implies
\bea \Tr[T^{(0)}_{0;(0)}\star Q_{e}\star T^{(0)}_{0;(0)} \star
Q'_{e'}]\ =\ \delta_{e,0}\delta_{e',0}\Tr[T^{(0)}_{0;(0)}\star
Q_{0}\star T^{(0)}_{0;(0)} \star Q'_{0}]\ ,\label{T0lemma}\eea
provided $\Ad_E Q_e=[E,Q_e]_\star = eQ_e$ \emph{idem} $Q'_{e'}$.
Under the indecomposition \eq{indec1}--\eq{indec4}, the bilinear
form $(\cdot,\cdot)_{\cal M}$ splits into a non-degenerate inner
product on ${\cal W}^\pm$, and a completely degenerate bilinear form
on $\mD$ and $\mD'$. On $\mD$, we instead define the non-degenerate
inner product
\bea (S,S')_{\mD}&=& \ft1{{\cal N}_{2\e_0}}\Tr[\pi(S)\star S']\
,\eea
by declaring the twisted-adjoint $\mho_1(D+1;\Comp)$ action to be
self-adjoint, and using the state generation \eq{1pstate} to factor
out
\bea {\cal N}_{2\e_0}&=& \Tr[\pi(T^{(0)}_{-2\e_0;(0)})\star
T^{(0)}_{2\e_0;(0)}]\ =\ \Tr[T^{(0)}_{2\e_0;(0)}\star
T^{(0)}_{2\e_0;(0)}]\ .\eea

The normalizations ${\cal N}_\pm$ and ${\cal N}_{2\e_0}$, which thus
contain all non-polynomialities, can be computed using \eq{TrEnEn}
and \eq{TPa}, which implies
\bea &&\Tr[T_{r0(n)}\star T_{s0(n)}]\ =\ \ft12 \Tr[(\Ac_{P_r}
T_{0(n)})\star T_{s0(n)}]\ =\ \Tr[ T_{0(n)}\star
(\Ac_{P_r}T_{s0(n)})]\\[5pt]&=&
\l_{n+1}^{(0)}\delta_{r\{s}\Tr[T_{0(n)}\star T_{0(n)\}_D}]\ =\
{\l_{n+1}^{(0)}\over n+1}\,\d_{rs} \Tr[T_{0(n)}\star T_{0(n)}]\
,\eea
with $\l_{n+1}^{(0)}$ given in \eq{lambda}, and \eq{TrEnEn}. For the
scalar static ground states, whose generating functions are given in
\eq{scalarf000plus} and \eq{scalarf000minus}, we find the
non-oscillating series
\bea {\cal N}_+&=&{\e_0\over \e_0+\ft12}\sum_{n=0}^\infty
{n+\e_0+\ft12\over (n+\ft12)(n+2\e_0)} \ ,\eea
which is logarithmically divergent for $\e_0>0$, while
\bea {\cal N}_- &=&{4\e_0(\e_0+1)\over
(\e_0+\ft12)(\e_0+\ft32)} \ \times\nn\\[5pt]
&\times&\sum_{n=0}^\infty
{(n+\ft{\e_0}2+\ft34)(n+\ft{\e_0}2+\ft14)(n+\ft12)(n+1)\over
n+\e_0+1}{[(\e_0+\ft12)_n]^2[(\ft12)_n]^2\over
(\e_0+1)_n(\e_0+2)_n[(2)_n]^2}\ ,\qquad\eea
which is convergent for $\e_0>0$, as can be seen using
$(z)_n=\C(z+n)/\C(z)$ and
$\C(z)=\sqrt{2\pi}z^{z-\ft12}e^{-z}(1+{\cal O}(z^{-1}))$, which
implies that the summand goes like $n^{-2}(1+{\cal O}(n^{-1}))$ for
large $n$. For example, in $D=4$
\bea {\cal N}_+&=&{1\over 2} \sum_{n=0}^\infty {1\over n+\ft12} \
,\\[5pt] {\cal N}_-&=&{9\over 32}\sum_{n=0}^\infty
{1\over (n+\ft32)^2}\ .\eea
For the lowest-weight element, whose generating function is given in
\eq{scalarf020plus}, we find the oscillating series
\bea {\cal N}_{2\e_0}&=& \sum_{n=0}^\infty
(-1)^n{(\e_0+\ft32)_n(2\e_0)_n(2\e_0+1)_n\over
n!(2)_n(\e_0+\ft12)_n}\nn\\[5pt]
& =& {}_{3}F_{2}(\e_0+\ft32,2\e_0,2\e_0+1;\e_0+\ft12,2;-1)\ .\eea
To evaluate the hypergeometric function, we use rewrite it as
\bea &&{}_{3}F_{2}(\e_0+\ft32,2\e_0,2\e_0+1;\e_0+\ft12,2;x)\ =\
{1\over \e_0+\ft12} x^{\ft12-\e_0} {d\over
dx}\left(x^{\ft12+\e_0}{}_{2}F_{1}(2\e_0,2\e_0+1;2;x)\right)\nn\\[5pt]
&&=\ {1+x\over
(1-x)^{2\e_0+2}}{}_{2}F_{1}\left[1-\e_0,\e_0+\ft12;2;-{4x\over
(1-x)^2}\right]\nn\\[5pt]&&+{8(1-\e_0)\C(\ft12)\over
\C(\ft32-\e_0)\C(1+\e_0)}{x(1+x)\over
(1-x)^{2\e_0+4}}{}_{2}F_{1}\left[2-\e_0,\e_0+\ft32;\ft32;\left({1+x\over
1-x}\right)^2\right]\nn\\[5pt]
&&-{4\C(\ft12)\over \C(\e_0+\ft32)\C(1-\e_0)}{x\over
(1-x)^{2\e_0+3}}{}_{2}F_{1}\left[1+\e_0,\ft32-\e_0;\ft12;\left({1+x\over
1-x}\right)^2\right]\ ,\eea
with the result ($n=0,1,2,\dots$)
\bea {\cal N}_{2\e_0}&=&{\C(\ft12)\over
2^{2\e_0+1}\C(1-\e_0)\C(\e_0+\ft32)}\ =\ \left\{\ba{ll}
(-1)^n{(2n-1)!!\over 2^{3n+2}(n+1)!}&\mbox{for $\e_0=\ft12+n$}\\[5pt]0&\mbox{for $\e_0=1+n$}\ea\right.\
\eea
In particular, for $D=4$ we have
\bea {\cal N}_{1}&=&{1\over 4}\ .\label{4dnorm}\eea

The inner products between real twisted-adjoint elements, obeying
the reality condition \eq{masterreal}, are real,
\bea  (\Phi^\pm,\Phi^{\pm\prime})_{\cal M}\ =\ \ft1{{\cal
N}_\pm}\Tr[(\Phi^\pm)^\dagger\star \Phi^{\pm\prime}]\ =\ \ft1{{\cal
N}_\pm}\Tr[(\Phi^{\pm\prime})^\dagger\star \Phi^{\pm}]\ ,\\[5pt]
(\Phi,\Phi^{\prime})_{\mD}\ =\ \ft1{{\cal
N}_{2\e_0}}\Tr[\Phi^\dagger\star \Phi^{\prime}]\ =\ \ft1{{\cal
N}_{2\e_0}}\Tr[(\Phi^{\prime})^\dagger\star \Phi]\ .\eea
Expanding into components using the bases \eq{wedgestate} and
\eq{1pstate}, we find that in the signatures $\mso(D-1,2)$ and
$\mso(D+1)$,
\bea (\Phi^\pm,\Phi^{\pm\prime})_{\cal M}& = &
\sum_{m,n;m',n'=0}^\infty (\Phi^{\pm(0)}_{m,n})^\ast
(\s_0)^{m+n}(-\s_{0'})^{\ft{1\mp1}2}N^{\pm(0)}_{m,n;m',n'}\Phi^{\pm(s)\prime}_{m',n'}\nn\\[5pt]&+&
\sum_{s=1}^\infty \sum_{m,n;m',n'=0}^\infty
(\Phi^{\pm(s)}_{m,n})^\ast
(\s_0)^{m+n}(\s_0)^{s+\ft{1\mp1}2}N^{\pm(s)}_{m,n;m',n'}\Phi^{\pm(s)\prime}_{m',n'}\ ,\\[10pt]
(\Phi,\Phi')_{\mD}&=& \sum_{s=0}^\infty \sum_{m,m'=0}^\infty
\left((\Phi^{(s)}_m)^\ast (\s_0)^{s+m}M^{(s)}_{m,m'}
\Phi^{(s)\prime}_{m'}\right.\nn\\[5pt]
&+& \left.(\Phi^{(s)\prime}_{m'})^\ast (\s_0)^{s+m'} M^{(s)}_{m',m}
\Phi^{(s)}_m\right)\ , \label{mDAdS}\eea
and that in the signature $\mso(D,1)$,
\bea (\Phi^\pm,\Phi^{\pm\prime})_{\cal
M}&=&\sum_{m,n;m',n'=0}^\infty (\Phi^{\pm(0)}_{n,m})^\ast
(\s_0)^{m+n}(-\s_{0'})^{\ft{1\mp1}2}N^{\pm(0)}_{m,n;m',n'}\Phi^{\pm(s)\prime}_{m',n'}\nn\\[5pt]&+&
\sum_{s=1}^\infty \sum_{m,n;m',n'=0}^\infty
(\Phi^{\pm(s)}_{n,m})^\ast
(\s_0)^{m+n}(\s_0)^{s+\ft{1\mp1}2}N^{\pm(s)}_{m,n;m',n'}\Phi^{\pm(s)\prime}_{m',n'}\ ,\\[10pt]
(\Phi,\Phi')_{\mD}&=& \sum_{s=0}^\infty \sum_{m,m'=0}^\infty
\left(\overline \Phi^{(s)}_m (-1)^m M^{(s)}_{m,m'}
\Phi^{(s)\prime}_{m'}+\overline\Phi^{(s)\prime}_{m'} (-1)^{m'}
M^{(s)}_{m',m} \Phi^{(s)}_m\right)\ ,\label{mDdS}\eea
where the inner product matrices
\bea M^{(s)}_{m,m'}&=& \left(T^{(s)}_{-(s+2\e_0);(s)},(\widetilde
L^-)^m(\widetilde L^+)^{m'} T^{(s)}_{s+2\e_0;(s)}\right)_{\mD}\
,\eea
and
\bea N^{\pm(s)}_{m,n;m',n'}&=& \left(T^{(s)}_{\pm},(\widetilde
L^-)^m(\widetilde L^+)^n (\widetilde L^+)^{m'}(\widetilde
L^-)^{n'}T^{(s)}_{\pm}\right)_{\cal M}\ .\eea
More explicitly, including also the $\mso(D-1)$ vector indices, and
using \eq{stategeneration1} and \eq{stategeneration2}, the inner
product matrices $M^{(s)}$ and $N^{+(s)}$ read
\bea
M^{(s)}_{m,n}((s_1,s_2)|(s'_1,s'_2))&=&\d_{m,n}\d_{s_1,s'_1}\d_{s_2,s'_2}M^{(s)}(p;(s_1,s_2))\
,\label{innerprodM}\eea
where $p=\ft{m+s-s_1-s_2}2$ and
\bea &&
M^{(s)}(p;(s_1,s_2))_{r(s_1),t(s_2)}^{r'(s_1),t'(s_2)}\nn\\[5pt]&& \hspace{-1cm}=
\left(T^{(s)}_{-(s+2\e_0);(s);\{r(s)},(\widetilde
x^-)^{p}(\widetilde L^-_r)^{s_1-s} (\widetilde
L^-_{t\}})^{s_2}(\widetilde x^+)^{p}(\widetilde
L^+_{\{r'})^{s_1-s}(\widetilde
L^+_{t'})^{s_2}T^{(s)}_{s+2\e_0;(s);r'(s)\}}\right)_{\mD}\ ,\eea
and
\bea N^{+(s)}_{m,n;m',n'}((s_1,s_2)|(s_1',s_2'))&=&
\d_{m,m'}\d_{n,n'}\d_{m+n,s_1+s_2-s}\d_{s_1,s'_1}\d_{s_2,s'_2}N^{+(s)}
(p,q;(s_1,s_2))\ ,\eea
where $p=\ft{m-n+s-s_1-s_2}2$ and $q=\ft{n-m+s-s_1-s_2}2$, and we
note that $n=0$ for $p>0$ and $m=0$ for $q>0$, in order to avoid
redundancy, and
\bea &&
N^{+(s)}(p,q;(s_1,s_2))_{r(s_1),t(s_2)}^{r'(s_1),t'(s_2)}\nn\\[5pt]&=&
\left\{\ba{ll}\left(T^{(s)}_{\pm;\{r(s)},\left((\widetilde L^-)^{m}
(\widetilde L^+)^{n}\right)_{r(s_1-s),t(s_2)\}}\left((\widetilde
L^+)^{m} (\widetilde
L^-)^{n}\right)_{\{r'(s_1-s),t'(s_2)}T^{(s)}_{\pm;r'(s)\}}\right)_{\cal
M}&\mbox{for $p,q\leq 0$}\\[10pt]
\left(T^{(s)}_{\pm;\{r(s)},(\widetilde
L^-)^{m}_{r(s_1-s),t(s_2)\}}(\widetilde
L^+)^{m}_{\{r'(s_1-s),t'(s_2)}T^{(s)}_{\pm;r'(s)\}}\right)_{\cal
M}&\mbox{for $p>0$ }\\[10pt] \left(T^{(s)}_{\pm;\{r(s)},(\widetilde
L^+)^{n}_{r(s_1-s),t(s_2)\}}(\widetilde
L^-)^{n}_{\{r'(s_1-s),t'(s_2)}T^{(s)}_{\pm;r'(s)\}}\right)_{\cal
M}&\mbox{for $q>0$ }\ea\right. \ .\qquad\eea
the above inner product matrices do not depend on the signature. The
matrices $M^{(s)}$ are positive definite\footnote{The complex space
$\mD^+(e_0;(s))\oplus \mD^-(e_0;(s))$, where $e_0=2\e_0+s$, can be
equipped with the Hilbert-space inner product
\bea \widehat{M}(e;(s_1,s_2)|e';(s'_1,s'_2))&=&
\left(\ket{e;(s_1,s_2)}\right)^\dagger \ket{e';(s'_1,s'_2)}\ ,\eea
assuming the following forms in various signatures:
\bea \mso(D-1,2)\supset \mso(D-1,1)&:& \widehat M\ =\ \left[\ba{cc} M&0\\0&M\ea\right]\ ,\label{adsM}\\[5pt]
\mso(D,1)\supset \mso(D-1,1)&:&  \widehat M\ =\ \left[\ba{cc} 0&M\\M&0\ea\right]\ ,\label{dsM}\\[5pt]
\mso(D,1)\supset \mso(D)&:&  \widehat M\ =\ \left[\ba{cc} 0&(-1)^{-(e-e_0)}M\\(-1)^{e-e_0}M&0\ea\right]\ ,\label{hr}\\[5pt]
\mso(D+1)\supset \mso(D)&:&  \widehat M\ =\ \left[\ba{cc}
(-1)^{e-e_0}M&0\\0&(-1)^{-(e-e_0)}M\ea\right]\ ,\label{src}\ ,\eea
where $M$ is given in \eq{innerprodM}, and $\pm(e-e_0)$ is the total
number of $L^\pm_r$ operators applied to the ground states $\ket{\pm
e_0;(s)}^\pm$. We notice that $\widehat M$ couples states with the
same energy in signatures $\mso(D-1,2)$ and $\mso(D+1)$ (where
$\left(\ket{\pm e_0;(s)}^\pm\right)^\dagger={}^\pm\bra{\pm
e_0;(s)}$), and with the opposite energy in signature $\mso(D,1)$
(where $\left(\ket{\pm e_0;(s)}^\pm\right)^\dagger={}^\mp\bra{\mp
e_0;(s)}$), and that $\widehat M$ is positive definite only in
signature $\mso(D-1,2)$.}, and from \eq{mDAdS} and \eq{mDdS} it
follows that $(\cdot,\cdot)_{\mD}$ is positive definite (only) in
the signature $\mso(D-1,2)$. We conjecture that also the matrices
$N^{\pm(s)}$ are positive definite, so that also
$(\cdot,\cdot)_{\cal M}$ is positive definite on ${\cal W}^\pm$ in
the signature $\mso(D-1,2)$. For example, in the case of $N^{+(0)}$
in $D=4$, the elements in ${\cal W}^=_{(0)}$ have $s_2=0$ and
$p,q\leq 0$, and thus
\bea D=4&:& N^{+(0)}_{m,n;m',n'}((s_1)|(s_1'))\ =\
\d_{m,m'}\d_{n,n'}\d_{m+n,s_1}\d_{s_1,s'_1}N^{+(0)} (p,q;(m+n))\
,\eea
where the matrix elements
\bea N^{+(0)}(p,q;(m+n))_{r(m+n)}^{r'(m+n)}&=&
\left(T^{(0)}_+,\left((\widetilde L^-)^{m} (\widetilde
L^+)^{n}\right)_{\{r(m+n)\}}\left((\widetilde L^+)^{m} (\widetilde
L^-)^{n}\right)^{\{r'(m+n)\}}T^{(0)}_\pm\right)_{\cal
M}\nn\\[5pt]&=&
\d_{\{r(m+n)\}}^{\{r'(m+n)\}}N_{m,n}\ ,\eea
for coefficients
\bea N_{m,n}&=& {1\over \dim (m)}\left(T^{(0)}_+,\left((\widetilde
L^-)^{m} (\widetilde L^+)^{n}\right)_{\{r(m+n)\}}\left((\widetilde
L^+)^{m} (\widetilde
L^-)^{n}\right)^{\{r(m+n)\}}T^{(0)}_\pm\right)_{\cal M}\ ,\eea
with $\dim (m)=\d_{\{r(m)\}}^{\{r(m)\}}=2m+1$ being the dimension of
the type $(m)$ irrep of $\mso(3;\Comp)$. Positive definiteness
amounts to that $N_{m,n}$ are strictly positive for all $m$ and $n$.
Using \eq{T0lemma} and
\bea \widetilde L^+_r\widetilde L^-_r T^{(0)}_{0;(0)}&=&\widetilde
L^-_r\widetilde L^+_r T^{(0)}_{0;(0)} \ =\ 2 T^{(0)}_{0;(0)}\ ,\eea
we have found that this is indeed the case for the following
low-lying levels:
\bea N_{1,0}&=&{4\over 3}\ ,\qquad N_{2,0}\ =\ 8\ ,\qquad N_{1,1}\
=\ {32\over 15}\ ,\eea
although we have no proof to all levels.

\scs{Supersingleton and Oscillator Realization in
$D=4$}\label{supersing}

A particularly simple realization of a composite reflector can be
given, in four dimensions, in terms of the states belonging to the
scalar and spinor singletons. In other words, one can  make use of
the 4D oscillator realization \eq{acomm} defined in Subsection
\ref{y} and define the composite reflector state
$\ket{\1}_{\mD_0\oplus\mD_{1/2}}$ as the reflection of the identity
operator on the Fock space \eq{fock}, that contains both the scalar
and the spinor singleton.

We begin by defining the action of the reflection map on the vacuum
$\ket{0}\equiv\ket{1/2,0}$ and on the oscillators:
\bea R(\ket{0}) & = & \bra{0} \ , \qquad R(\ket{0}^-) \ = \
{}^-\bra{0} \ ,\\[5pt]
R(a^{\dagger\,i}) & = & ia^i \ , \qquad R(a^{i}) \ = ia^{\dagger\,i}
\ .\eea
One can check that $R$ is an antiautomorphism of the oscillator
algebra \eq{acomm}. On a state
$\ket{n}=a^{\dagger\,i_1}...a^{\dagger\,i_n}\ket{0}$ it acts as
\be R(\ket{n}) \ = \  i^n \bra{0}a^{i_1}...a^{i_n} \ ,\ee
and similarly for anti-singleton states
\be R(\ket{n}^-) \ =  i^n
{}^-\bra{0}a^{\dagger\,i_1}...a^{\dagger\,i_n} \ .\ee
We note that the variable $z$, defined in \eq{}, is
\bea z \ = \ 2L^+_rL^-_r \ = \
2\frac{i^2}{4}(\s_r)_{ij}(\s_r)^{kl}a^{\dagger\,i}a^{\dagger\,j}a_k
a_l \ = \  : \ (a^{\dagger\,i} a_i)^2 \ : \ .\eea
Thus, from the Fock-space point of view, eqs.
(\ref{idD0}-\ref{4Dsinglunity}) can be rewritten as
\bea \1_{\mD_0} \ = \ \1_{\cF_{even}} \ = \sum_{\mbox{n
even}}\ket{n}\bra{n} \ = \ \sum_{\mbox{n
even}}\frac{1}{n!}a^{\dagger\,i_1}\ldots
a^{\dagger\,i_n}\ket{0}\bra{0}a_{i_1}\ldots a_{i_n} \ ,\eea
or
\bea \1_{\cF_{even}} \ = \ : \ \sum_{\mbox{n
even}}\frac{1}{n!}N^n\ket{0}\bra{0} \ :  \ = \ : \ \cosh N
\ket{0}\bra{0}\ : \ ,  \label{cosh}\eea
where $N=a^{\dagger\,i}a_i=\sqrt{z}$. Analogously, the decomposition
of the reflector over spinor singleton states, that leads to
\eq{spinrefl}, can be expressed as
\bea \1_{\mD_{1/2}} \ = \ \1_{\cF_{odd}} \ = \ : \ \sum_{\mbox{n
odd}}\frac{1}{n!}N^n\ket{0}\bra{0} \ :  \ = \ : \ \sinh N
\ket{0}\bra{0}\ : \ .  \label{sinh}\eea
One can now combine these two into the identity over the entire Fock
module, obtaining
\bea \1_{\mD} & = & \cross \ e^{\sqrt{z}}\ket{1/2,0}\bra{1/2,0} \
\cross\ ,\label{idDexp} \eea
that can be reinterpreted simply as the completeness of the
Fock-space basis of states,
\bea  \1_{\mD} & = & \1_{\cF} \ = \ \sum_{n=0}^\infty \ket{n}\bra{n}
\ = \
 : \ \sum_{n=0}^\infty\,\frac{1}{n!}N^n\ket{0}\bra{0} \ : \ = \ : \ e^N\,\ket{0}\bra{0} \ : \ .\label{idF}\eea
The latter equations indeed simply amounts to the fact that the
Fock-space vacuum-to-vacuum projector $\ket{0}\bra{0}$ admits the
realization $:\, e^{-N} \,:\,$. On the other hand, \eq{cosh} and
\eq{sinh} give the definitions of the projectors onto $\cF_{even}$
and $\cF_{odd}$, respectively,
\bea \1_{\cF_{even}} \ = P_+ \ =  \ \frac{1}{2}(1+\C)  \ , &\qquad
&\1_{\cF_{odd}} \ = P_- \ = \ \frac{1}{2}(1-\C) \
,\\[5pt]
P_\pm \star P_{\pm} & = & P_\pm \ ,   \eea
where
\bea \C \ = \ : \ e^{-2N} : \ ,  \qquad \C\star\C \ = \ 1 \eea
(see Appendix \ref{App:G}). The reflection of \eq{idDexp} gives the
combination of scalar excitations in the doubleton basis
$\mD_0^{\otimes 2}\oplus\mD_{1/2}^{\otimes 2}$ onto which the
identity operator in the twisted adjoint representation can be
mapped,
\bea \ket{\1}_{} \ = \ e^{i y}\ket{1/2,0}_1\ket{1/2,0}_2 \
,\label{reflId} \eea
where we have defined
$y\equiv\sqrt{x}=a^{\dagger}_i(1)a^{\dagger\,i}(2)$.

The standard inner product on the Fock space induces the trace
operation $\textrm{Tr}$. The important feature of such realization
is that one can normalize the reflector appearing in the last
equation. As reviewed in Appendix \ref{App:G} for the simpler case
of a single oscillator, to any consistent inner-product law for the
Fock module is associated a corresponding trace operation in the
space of operators $f$ acting on that module. Now, there are two
possible trace operations which are consistent with the oscillator
algebra, namely
\bea \Tr_\pm(f) \ = \ \Tr_{\mD_0}(f)\pm\Tr_{\mD_{1/2}}(f) \eea
(note that these definitions are analogous to those given in
Appendix \ref{App:G}), where
\bea \Tr_{\mD_0}(f) \ = \ \sum_{k=0}^\infty 
\bra{2k}f\ket{2k} \ ,\eea
denotes the trace on the scalar singleton states, and
\bea \Tr_{\mD_{1/2}}(f) \ = \ \sum_{k=0}^\infty 
\bra{2k+1}f\ket{2k+1} \ ,\eea
the trace on the spinor singleton states. This means that the trace
of the identity operator amounts to a sum of the multiplicities of
each state $\ket{n}$ (that has spin $s=n/2$), \emph{i.e.}
\bea \Tr_{\mD_0}(\1) & = & \sum_{k=0}^\infty (2k+1) \ ,
\nn\\[5pt]
\Tr_{\mD_{1/2}}(\1) & = & \sum_{k=0}^\infty (2k+1+1) \ .\eea
This can be summarized as
\bea \Tr_{\pm}(\1) \ = \ \sum_{n=0}^\infty\,(\pm1)^n\,(n+1) \
= \ \left\{ \begin{array}{c} \infty \ , \ \mbox{for}  \ \Tr_+ \nn\\
 \frac{1}{4}\  \ \mbox{for}\  \Tr_- \end{array}\right. \ ,\eea
Notice that only $\Tr_-$ leads to a finite result, and for this to
happen is also necessary to compute $\Tr_-$ over the full Fock
space $\mD_0\oplus\mD_{1/2}$, while the traces on $\mD_0$ and
$\mD_{1/2}$ only are necessarily divergent. Moreover, as shown in
Appendix \ref{App:G}, for an arbitrary element of the oscillator
algebra $f=f(a_i, a^{\dagger\,i})$, $\Tr_-(f)$ actually coincides
with $\Tr_+((-1)^N_\star\star f)$, and the latter in its turn
coincides with the supertrace operation $\Str(f)=f(0,0)$ defined
in \cite{V3}, up to an overall factor. With two oscillators, the
precise relation is
\bea \Tr_-(f) \ = \frac{1}{4}\,\Str (f) \ = \ \frac{1}{4}\,f(0,0)\
.\eea

\scs{Divergencies in the Perturbative Expansion and a Proposal for
Their Regularization}

As we have already stressed, the map constructed above establishes a
correspondence between the master zero-form at a point in space-time
and the physical fluctuation fields with definite energy and
$\mso(3)$-spin. In doing so, we have mapped fluctuations in field
strengths and scalar fields to nonpolynomial combinations of twisted
adjoint elements. At this point, however, a potential subtlety
arises. As remarked in Section \ref{Z}, the $\star$-product of
nonpolynomial generators will in general give rise to divergencies.

Indeed, the component-field expansions of the master fields $A$ and
$\Phi$ are formal sums that are not subject to any convergence
criteria. Once the master fields are constrained on-shell (in the
context of the weak-field expansion) the component fields
$A_{\mu,a(s-1),b(t)}$ and $\Phi_{a(s+k),b(s)}$ become identified
with various \emph{non-linear and higher-derivative} constructs
built from the physical scalar $\phi$, metric $g_{\m\n}$ and
higher-spin gauge tensor gauge fields $\varphi_{\m_1\dots \m_s}$
($s=4,6,\dots$) defined in \eq{physicalscalar}, \eq{metric} and
\eq{tensorgaugefields}, respectively. There is nothing, in
principle, that prevents the resulting (full) master fields $A$ and
$\Phi$ from having potentially divergent $\star$-product
compositions\footnote{Divergencies were encountered recently in
\cite{Iazeolla:2007wt}, where exact solutions of the non-minimal
model based on $\mho_1(D+1;\Comp)$ were constructed using projectors
$P\in {\cal A}$ obeying $P\star P=P$. These solutions would admit a
consistent truncation to the minimal model provided $\tau(P)=P$.
However, the projectors used in \cite{Iazeolla:2007wt}, built using
Fock-space methods, are not $\tau$-invariant. The reason is that the
naive attempt to impose the $\t$-condition by replacing $P$ by
$P+\tau(P)$ leads to divergencies residing in $P\star\tau(P)$ that
remain unresolved at the level of the full Vasiliev equations
(although, mathematically speaking, they are related to the those
arising in the harmonic expansion).}. Such divergencies do indeed
arise in the composition $\pi(\Phi)\star\Phi$ of linearized
solutions (given by harmonic expansions), possibly jeopardizing the
weak-field expansion. However, due to the detailed nature of
Vasiliev's equations (having to do with doubling of oscillators) it
is possible that these divergencies do not affect the actual
higher-order corrections to the full master fields. In other words,
there is some evidence that the weak-field expansion of Vasiliev's
equations gives rise to well-defined classical interactions vertices
among states (one-particle as well as static states) with fixed
compact quantum numbers.

To understand why this can indeed be the case, we will use as a
prototype example the self-interaction of the lowest weight element
$T^{(0)}_{(2\e_0);(0)}$ in $D=4$, that has the simple form
\eq{lws4d}, \emph{i.e.} $\Phi'\equiv\Phi|_p=e^{-4E}$. At the second
order in perturbation expansion, the $Z$-dependence of the zero-form
master field is given by
\bea \del_\a\hPhi'^{(2)} \ = \ -\hA_\a^{\prime(1)}\star\Phi' +
\Phi'\star\pi(\hA_\a^{\prime(1)}) \equiv j^{(2)}_\a(z) \ .\eea
The source term is of second order in $\Phi'$, since
\bea \hA_\a^{\prime(1)} \ = \ z_\a \int_0^1
dt\,\Phi'(-zt,\yb)e^{ity^\a z_\a}\ ,\eea
and in particular, substituting $\Phi'(y,\yb)=\exp
((\s_0)^{\a\ad}y_\a \yb_{\ad})$ in the equation below, one finds
\bea \hA_\a^{\prime(1)}\star\Phi' & = &
\frac{\del}{\del\rho^\a}\int_0^1
dt\,\left(e^{ity^\a z_\a+\rho^\a z_\a}\Phi'(-zt,\yb)\star\Phi(y,\yb)\right)\nn\\[5pt]
& \sim & \sum_{n=0}^\infty (n+\frac12)t^n+... \ = \ \frac12
\frac{1+t}{(1-t)^2}+...\eea
where we only write the coefficient of the identity. This shows a
divergency in the upper bound of the integration domain, $t=1$.
However, being this an isolated singularity, one can apply the
method of regularization proposed in \cite{Sezgin:2005pv}, and
obtain a well-defined weak-field expansion by circumventing the
singularities that arise in the perturbative expansion using a
closed integration contour $\c$ as
\bea \hPhi'^{(n)} \ = \ \oint_\c \frac{dt}{2\pi
i}\ln\left(\frac{t}{1-t}\right)z^\a j^{(n)}_\a(zt) \ , \eea
where $\c$ encircles the branch cut from $t=0$ to $t=1$ (similarly
for $A^{\prime(n)}_\a$, see \cite{Sezgin:2005pv})\footnote{Notice
that the presence of the $t$-dependent exponential $e^{ity^\a z_\a}$
in the sources leads to an essential singularity at $t=\infty$,
which always prevents the closed-contour integrals from being
trivial.}. This more general presentation of the weak-field
expansion allows for singular initial data for the evolution along
$Z$. In the case of regular initial data, on the other hand, the
closed-contour can be collapsed onto the branch cut as to reproduce
the open-contour presentation.

An important subtlety however still needs to be resolved. Indeed, at
every order in the perturbative expansion one has to make sure that
the source terms $j^{(n)}_\a(z)$ are finite, and this request fixes
$\c$. But at higher orders, the isolated singularity is pushed
further and further away from the origin of the complex $t$-plane,
and the necessity of encircling it may lead to losing the
associativity of the $\star$-product composition, which is in its
turn crucial for the integrability of the equations. We hope this
issue admits a resolution, which we shall elaborate further on
elsewhere.

\chapter{Conclusions and Outlook}

In this Thesis, I have reviewed in some detail the main features of
interacting Higher-Spin Gauge Theories focusing on the Vasiliev
equations, and presented some original results concerning the
structure of the gauge algebra on which they are based and some new
exact solutions. As it can be appreciated from the review part of
this work, the study of HS gauge fields has made some very
significant progress in the last twenty years, and at the same time
the importance of a better understanding of their dynamics has grown
considerably also due to the developments in different research
fields, and most notably in String Theory. As a result, HSGT as we
know them today involve a number of physical and formal tools and
concepts that are not only fascinating and promising in themselves,
but also of potential interest to other fields at the forefront of
research in High-Energy Theoretical Physics.

The Vasiliev equations are arguably the most important achievement
in HSGT obtained so far, and are in fact the only known consistent
set of equations encoding the full dynamics of massless HS fields
(at least in four dimensions, while for higher dimensions this
statement is so far limited to totally symmetric tensors). As we
have seen, they encode a very complicate dynamics into a few elegant
curvature constraints according to the unfolded formulation. The
latter has been the key to overcome the main obstacles to the
formulation of a consistent nonlinear theory of HS fields, and also
enables a uniform treatment of higher-derivative couplings together
with, importantly, a background-independent description. On the
other hand, a conventional action principle from which the Vasiliev
equation descend is not known, at present, and in the unfolded
scheme it might not be easy to recover certain results that are
instead more readily within reach in the known low-spin Lagrangian
Field Theory. Moreover, making contact with the known lower-spin
gauge theories is also not easy, at present: this is due to the fact
that no consistent truncation of the equations down to the
lower-spin sector is known, as lower-spin fields serve as sources
for higher-spin fields, and some mechanism of spontaneous breaking
of the full HS symmetry must be known before one can turn off the
couplings to HS fields consistently. However, as we hope we have
made clear in this Thesis, the Vasiliev system offers other
important windows on the peculiar features of HS dynamics, and we
believe one should exploit such possibilities as much as possible.

For example, although the full equations in space-time are a system
of formidable complexity, and indeed are not even known in closed
form at present, we can nonetheless find exact solutions by
exploiting their relatively simple form in the extended space
$(x,Z)$. Indeed, the most important feature of the unfolded approach
is that, roughly speaking, it enables a trading of the space-time
evolution for the fiber evolution, and this makes it possible not
only to write the field equations in a background-independent way,
but also to solve them by means of purely algebraic methods. For
example, as we have explicitly shown in Chapter \ref{exactsol},
imposing symmetries on the zero-form, may simplify a lot the form of
the the fiber equations. Moreover, a large class of solutions to the
latter can be found by using projectors of the gauge algebra. The
methods developed for the construction of the exact solutions found
so far \cite{Sezgin:2005pv, Iazeolla:2007wt, Didenko:2006zd} are
likely to be useful to find new ones, and in general the homotopy
invariance of Vasiliev's equations gives some hope in this sense. Of
particular interest is the research of a spherically symmetric
solution, as well as of black hole solutions. For example, the exact
solution that describes the embedding of a BTZ black hole in the
three-dimensional HSGT (in which, however, higher-spin fields do not
carry local degrees of freedom), found in \cite{Didenko:2006zd},
might be elevated to the dynamically more interesting case of
$D=4$.\\
The importance of such issues actually goes beyond the
realm of higher spins. Since any system admits an unfolded
reformulation, one may even speculate that such algebraic methods
can be of use also in ordinary gravity, and that solutions could be
obtained \emph{algebraically} starting from the knowledge of the
Weyl zero-form at a point in space-time.

In this Thesis, starting from HS gauge theories in four dimensions
based on infinite dimensional extensions of $SO(5;\Comp)$, we have
determined their real forms in spacetimes with Euclidean $(4,0)$ and
Kleinian $(2,2)$ signature, in addition to the usual Lorentzian
$(3,1)$ signature. We have then found three new types of solutions
in addition to the maximally symmetric ones.\\
Type 1 solutions, which are invariant under an infinite dimensional
extension of $SO(4-p,p)$, give us a nontrivial deformation of the
maximally symmetric solutions, and depend on a continuous real
parameter as well as on an infinite set of discrete parameters.
Interestingly, a particular choice of the discrete parameters, in
the limit of vanishing continuous parameter, gives rise to a
degenerate, indeed rank one, metric. Given that degenerate metrics
are known to play an important role in topology change in quantum
gravity \cite{Horowitz:1990qb}, it is remarkable that such metrics
emerge naturally in HS gauge theory.\\
Type 2 solutions, which provide another kind of deformation of the
maximally symmetric solutions, have a non-vanishing spinorial master
one-form. \\
Type 3 solutions are particularly remarkable because all the higher
spin fields are non-vanishing, and the corresponding Weyl tensors
furnish a higher spin generalization of Type D gravitational
instantons. It would be interesting to apply the framework we have
used in this paper to finding pp-wave, black hole and domain wall
solutions with non-vanishing HS fields.\\
We stress that our models in Euclidean and Kleinian signatures are
formulated using the 4D spinor-oscillator formulation. It would be
interesting to compare these models to the vector-oscillator
formulation \cite{Vasiliev:2003ev, Bekaert:2005vh}, which exists in
any dimension and signature, and relies on the gauging of an
internal $\msp(2)$ gauge symmetry (as briefly discussed in Chapter
\ref{absalg}). At the full level, the vector-oscillator master field
equations, in any dimension and signature, are formulated using a
\emph{single} $\msp(2)$-doublet $Z$-oscillator, leaving, apparently,
no room for parity violating interactions. The precise relation
between the spinor and vector-oscillator formulations in $D=4$
therefore deserve further study.\\
In the context of supersymmetric field theories, including
supergravity, the non-Lorentzian signature typically presents
obstacles, since the spinor properties are sensitive to the
spacetime signature. Here, however, we have considered bosonic HS
gauge theories in which the spinor oscillators play an auxiliary
role, and we have formulated the non-Lorentzian signature theories
with suitable definition of the spinors without having to face such
obstacles. Remarkably, non-supersymmetric 4D theories in Kleinian
signature describing self-dual gravity arise in worldsheet $N=2$
supersymmetric string theories, known as $N=2$ strings. For reasons
mentioned in the Introduction to this Chapter, it is an interesting
open problem to find a niche for Kleinian HS gauge theory in a
variant of an $N=2$ string.

There are several other open problems that deserve investigation. To
begin with, we have not determined the symmetries of Type 2 and Type
3 solutions. moreover, as we have seen, some of the solutions found
so far admit an interesting cosmological interpretation. However,
the latter is not straightforward and, at present, cannot be carried
out in a HS covariant fashion. Indeed, there is no HS-invariant line
element to describe the geodesic motion of test particles and,
consequently, to define the notions of horizons and singularities in
a sensible way. In order to do this, and to be able to characterize
the solutions physically, it would be of extreme importance to
extend the set of invariants under the infinitely many symmetries of
HSGT. To date, a partial set of invariants \eq{inv} has been
constructed \cite{Sezgin:2005pv} only in terms of the master
zero-form, while no observable that involves the master one-form has
been found (although it is clear how to to extend the construction
to the case of one-forms). Moreover, while it may be useful in its
own right to determine whether our Type 3 solutions support a
complex, possibly K\"ahler, structure up to a conformal scaling,
such results may be limited in shedding light to the geometry
associated with infinitely many gauge fields present in HS gauge
theory. A proper formulation of the HS geometry would also provide a
framework for constructing the above-mentioned invariants that could
distinguish the gauge inequivalent classes of exact solutions.\\
It would also be interesting to study the fluctuations about our
exact solutions, and explore their potential application in quantum
gravity and cosmology. Similarities between the frameworks for
studying instanton and soliton solutions of the noncommutative field
theories (see, for example, \cite{Schaposnik:2003vr}), and in
particular open string field theory, are also worth investigating.

As we have stressed, there are many reasons why it would be
desirable to gain a better understanding of the unfolded formulation
and of the structure of the HS algebras. With such motivations in
mind, in Chapter \ref{map} we have elaborated further on them, and
on the representations that are of relevance to the Vasiliev
equations. In particular, an analysis of the physical content of the
Vasiliev system was developed, which is valid for arbitrary
signature, and is somehow in the spirit of the unfolded formulation.
Indeed, as the study of the chiral model in Chapter \ref{exactsol}
puts in greater evidence, the field-strengths are the natural place
where to look for the physical degrees of freedom, and this raises
the issue of examining carefully how the local data is encoded in
the zero-form master-field. This is investigated in Chapter
\ref{map} in some detail and in a general way, and normalizable and
non-normalizable fluctuation fields carrying definite energy and
spin quantum numbers are shown to emerge, \emph{a priori}, from the
unfolding of the local data. In particular, some nonperturbative
solutions to the linearized equations originally found in
\cite{Sezgin:2005pv} are nicely seen to come out from this analysis.
Although such states are nonunitary in the standard inner product in
$AdS$ \cite{Breitenlohner:1982jf}, they appear to be unitary in the
inner product $(.,.)_\cM$ based on the trace operation defined in
Chapter \ref{map}. It is however too early to understand what this
means in a quantum theory.

Moreover, the mapping developed in Chapter \ref{map} might admit
simple extensions to the case of other interesting representations
that we have not treated here, namely massive and partially massless
fields in maximally symmetric space-times with nonvanishing
cosmological constant. A lagrangian formulation is known for the
free massive case of arbitrary spin, and massive fields are known to
be related to the tensor product of three or more singletons
(although a complete classification is not available, at present).
An appropriate generalization of the map in \cite{us} to
multipletons might give an indication on the relevant master-fields
that would enter an unfolded formulation of massive fields.
Moreover, in the limit in which the mass of fields with spin $s\geq
2$ becomes proportional to the cosmological constant, with some
precise real or imaginary factors, the theory acquires a partial
gauge invariance under transformation that have a lower rank
parameter, compared to the massless case, and at least two
derivatives. Such fields are called ``partially massless'', and are
peculiar to space-times with nonvanishing cosmological constant.
Although both a lagrangian and an unfolded description of such
fields are available (see, for example, \cite{Skvortsov:2006at} and
references therein), to the best of our knowledge a
group-theoretical one is still unclear. The study carried on in
Chapter \ref{map} might shed some light on the corresponding
representations of the background isometry algebra and on the way
they can be related to its irreducible representations and their
negative-energy counterparts. For example, it might be the case that
partially massless representations arise from the decomposition of
the tensor product of the $p$-linetons presented in Chapter
\ref{map}, through some analog of the Flato-Fronsdal formula. We
plan to study such issue in a future work.

Finally, let us mention that some other partial results, obtained in
collaboration with A. Sagnotti and P. Sundell \cite{action}, seem to
point towards the possible existence of a non-standard action
principle for the Vasiliev equations.  Although of BF type (an
action of this form was first considered in \cite{Vasiliev:sa}), it
seems that it can encode the correct local degrees of freedom by
virtue of the properties of unfolded systems. Moreover, the first
attempts at a quantization of a simple field theoretical model
formulated in this way have given some positive results. Although
many related issues are unclear to us at present, a partial analysis
of the features of this ``unfolded action'' is encouraging, and
shows once more that a deeper understanding of the unfolded
formulation would be of relevance also for lower-spin gauge
theories.



\newpage


\begin{appendix}


\chapter{Gauging space-time symmetries}\label{sptime}

The usual Einstein-Hilbert action $S[g]$ is invariant under
diffeomorphisms. The same is true for $S[e,\o]$, defined by
(\ref{mdma}), since everything is written in terms of differential
forms. The action (\ref{mdma}) is also manifestly invariant under
local Lorentz transformations $\d \o=d\epsilon+[\o,\epsilon]$ with
gauge parameter $\epsilon=\epsilon^{ab}M_{ab}$, because
$\epsilon_{a_1 \ldots a_{d}}$ is an invariant tensor of
$SO(D-1,1)$. The gauge formulation of gravity shares many common
features with a Yang-Mills theory formulated in terms of a
connection $\o$ taking values in the Poincar\'e algebra.

However, gravity is actually \textit{not} a Yang-Mills theory with
Poincar\'e as (internal) gauge group. The aim of this section is to
express precisely the distinction between internal and space-time
gauge symmetries.

To warm up, let us mention several obvious differences between
Einstein-Cartan's gravity and Yang-Mills theory. First of all, the
Poincar\'e algebra $\miso(D-1,1)$ is not semisimple (since it is
not a \textit{direct} sum of simple Lie algebras, containing a
nontrivial ideal spanned by translations). Secondly, the action
(\ref{mdma}) cannot be written in a Yang-Mills form $\int Tr[F{}^*
F]$. Thirdly, the action (\ref{mdma}) is not invariant under the
gauge transformations $\d \o=d\epsilon+[\o,\epsilon]$ generated by
{\it all} Poincar\'e algebra generators, {\it i.e.} with gauge
parameter $\epsilon (x)=\epsilon^a (x) P_a+\epsilon^{ab} (x)
M_{ab}$.  For $D>3$, the action (\ref{mdma}) is invariant only
when $\epsilon^a= 0$. (For $D=3$, the action (\ref{mdma})
describes a genuine Chern-Simons theory with local $ISO(2,1)$
symmetries.)

This latter fact is not in contradiction with the fact that one
actually gauges the Poincar\'e group in gravity. Indeed, the torsion
constraint allows one to relate the local translation parameter
$\epsilon^a$ to the infinitesimal change of coordinates parameter
$\xi^\m$. Indeed, the infinitesimal diffeomorphism $x^\m\rightarrow
x^\m+\xi^\m$ acts as the Lie derivative $$\d_\xi ={\cal L}_\xi
\equiv i_\xi d+d i_\xi\,,$$ where the inner product $i$ is defined
by
$$i_\xi\equiv \xi^\mu\frac{\partial}{\partial (dx^\mu)}\,,$$
where the derivative is understood to act from the left. Any
coordinate transformation of the frame field can be written as \bea
\d_\xi e^a=i_\xi(de^a)+d(i_\xi e^a)=i_\xi
T^a+\underbrace{\epsilon^{a}{}_b
e^b+D^L\epsilon^a}_{=\d_{\epsilon}\,e^a}\,,\nonumber\eea where the
Poincar\'e gauge parameter is given by $\epsilon=i_\xi \omega$.
Therefore, when $T^a$ vanishes any coordinate transformation of the
frame field can be interpreted as a local Poincar\'e
transformation of the frame field, and reciprocally. 

To summarize, the Einstein-Cartan formulation of gravity is indeed
a fibre bundle construction where the Poincar\'e algebra
$\miso(D-1,1)$ is the fiber, $\o$ the connection and $R$ the
curvature, but, unlike for Yang-Mills theories, the equations of
motion imposes some constraints on the curvature ($T^a=0$), and
some fields are auxiliary ($\omega^{ab}$). A fully covariant
formulation is achieved in the $AdS$ case with the aid of
compensator formalism as explained in Section \ref{grav}.


\chapter{Details of the Procedure of Factoring out the Ideal ${\cal
I}[V]$}\label{App:VAB}


In this Appendix we present some of the details of the procedure of
factoring out the ideal ${\cal I}[V]$ defined in \eq{idealV} from
the enveloping algebra ${\cal U}$ of $\mso(D+1;\Comp)$ defined in
\eq{calU}.

In general, suppose that ${\cal I}[V]=V\star {\cal U}$ for an
$\mso(D+1;\Comp)$ irreducible element
$V=\l^{A_1,\dots,A_n}V_{A_1,\dots,A_n}$, where
$\l^{A_1,\dots,A_n}\in \Comp$ and $V_{A_1,\dots,A_n}$ is a (Young
projected) monomial built from $M_{AB}$ and $\eta_{AB}$, thus
obeying
\bea
[M_{BC},V_{A_1,\dots,A_n}]_\star&=&2i\eta_{A_1[C}V_{B],A_2,\dots,A_n}+\cdots+
2i\eta_{A_n[C|}V_{A_1,\dots,A_{n-1},|B]}\ .\eea
It follows that if $X\in {\cal U}$, then $X\star V=V\star X'$ for
some $X'\in {\cal U}$, so that the space ${\cal I}[V]$ is a both a
left and right ideal in ${\cal U}$, and it is equivalent to the
right ideal generated by right-multiplication by $V$, \emph{i.e.}
${\cal I}[V]=V\star {\cal U}={\cal U}\star V$. Thus, one may use the
notation
\bea X\star V\star X'&\simeq& 0\qquad\mbox{for all $X,X'\in{\cal
U}$}\ .\eea
The above considerations can be extended straightforwardly to the
case that $V$ is reducible.

Turning to the specific case of $V$ given by \eq{VAB} and
\eq{VABCD}, let us we show their equivalence to \eq{papa}. First, by
decomposing $V_{AB}\simeq 0$ and $V_{ABCD}\simeq 0$ under
$\mso(D;\Comp)$ one immediately arrives at \eq{v00}-\eq{v0abc}.
Formally, the $\mso(D+1;\Comp)$-irreducibility of $V_{AB}\simeq 0$
implies that $V_{0'a}\simeq0$ and $V_{ab}\simeq 0$ follow from
$V_{0'0'}\simeq 0$. Similarly, the constraint $V_{abcd}\simeq 0$
follows from $V_{0'abc}\simeq0$. Explicitly, from the algebra and
$V_{0'0'}\simeq 0$, and using the fact that $\mu^2$ defined by
\eq{mu2} is a commuting element, it follows that
\bea P^a\star M_{ab}\simeq M_{ba}\star P^a\simeq i(\e_0+1)P_b\
.\label{pamab2}\eea
The constraint \eq{v0a} then follows immediately. Alternatively, one
may compute $V_{0'a}=-\ft{i}4 [P_a,P^b\star P_b]_\star\simeq-\ft{i}4
[P_a,\mu^2]_\star=0$. Next, the algebra, eq. \eq{pamab2} and
$P^a\star P_a\simeq \mu^2$ imply
\bea M_{(a}{}^c\star M_{b)c}&\simeq & -P_{(a}\star
P_{b)}+\mu^2\eta_{ab}\ ,\label{macmbc}\eea
that is, $V_{ab}\simeq 0$. Similarly, the algebra implies that
$[P_{[a},V_{0'bcd]}]_\star$ is proportional to $V_{abcd}$, so that
$V_{0'abc}\simeq 0$, \emph{i.e.} $P_{[a}\star P_b\star P_{c]}\simeq
0$, implies $V_{abcd}\simeq 0$, \emph{i.e.} $M_{[ab}\star
M_{cd}]\simeq 0$. Finally, the value of $\mu^2$ is fixed from
\eq{detmu2}.

Next, let us use the contraction rules \eq{papa}, \eq{pamab} and
\eq{macmcb} to derive the lemma \eq{lemmaApp}, that is, compute the
coefficient
\bea \k_{n}\equiv \k_{n,0;1}\ .\eea
To do so, we demand that the right-hand side of \eq{lemma} is
traceless. To this end, we first expand
$$\eta^{bc}P_{(b}\star P_c\star P_{a_1}\cdots\star P_{a_{n-2})}
\ \simeq\ \e_0P_{(a_1}\cdots\star P_{a_{n-2})}$$$$ +{2\over
n(n-1)}\sum_{1\leq i<j\leq n} P_{a_1}\cdots\star P_{a_{i-1}}\star
[P_b,P_{a_{i}}\star\cdots \star P_{a_{j-2}}]_\star\star P^b\star
P_{a_{j-1}}\star \cdots \star P_{a_{n-2})}\ ,$$
where we have used \eq{papa}, and proceed by calculating the terms
in the sum up trace parts, which only affect the higher traces in
\eq{lemma}. Thus, for $i=1$ and $j=n$, and using also \eq{pamab}, we
find that
\bea &&[P_b,P_{a_{1}}\star\cdots \star P_{a_{n-2}}]_\star\star
P^b\nn\\[5pt]&=&iM_{ba_1}\star P_{a_2}\star\cdots \star P_{a_{n-2}}\star P^b
+P_{a_1}\star (iM_{ba_2})\star \cdots \star P_{a_{n-2}}\star
P^b+\cdots\nn\\[5pt] &\simeq&(\e_0+1) P_{a_1}\star\cdots \star
P_{a_{n-2}}+\eta_{ba_2}P_{a_1}\star P_{a_3}\star \cdots\star
P_{a_{n-2}}\star P^b\nn\\[5pt] &&+P_{a_2}\star (\eta_{ba_3}P_{a_2})\star
P_{a_4}\star \cdots\star P_{a_{n-2}}\star P^b+P_{a_2}\star
P_{a_3}\star (\eta_{ba_4}P_{a_3})\star P_{a_5}\star
\cdots\star P_{a_{n-2}}\star P^b+\cdots\qquad\qquad\nn\\[5pt]&&+
(\e_0+1) P_{a_1}\star\cdots \star P_{a_{n-2}}+P_{a_2}\star
(\eta_{ba_3}P_{a_2})\star P_{a_4}\star
\cdots\star P_{a_{n-2}}\star P^b+\cdots\nn\\[5pt]&&+\cdots\nn\\[5pt]&&+{\cal O}(\eta)\nn\\[5pt]
&=&\left((n-2)(\e_0+1)+\ft12(n-2)(n-3)\right)P_{a_1}\star\cdots
\star P_{a_{n-2}}+{\cal O}(\eta)\ ,\label{intermediate1}\eea
where the symmetrization on $a_1\cdots a_{n-2}$ has been suppressed.
The contributions from the terms with $j=n$ and $i=1+k$ for
$k=0,\dots,n-2$ are obtained by letting $n\rightarrow n-k$ in
\eq{intermediate1}, and their sum is given by
\bea \sum_{k=0}^{n-2} k(\e_0+1+\ft12(k-1))P_{(a_1}\star\cdots \star
P_{a_{n-2})}&=& \ft16(n-1)(n-2)(n+3\e_0) P_{(a_1}\star\cdots \star
P_{a_{n-2})}\label{intermediate3}\eea
plus trace parts. The contributions from the the terms with $j=n-k$
for $k=0,\dots,n-2$ are obtained by letting $n\rightarrow n-k$ in
\eq{intermediate3}, and their sum is given by $P_{(a_1}\star\cdots
\star P_{a_{n-2})}$ times the numerical factor
\bea \ft16 \sum_{k=0}^{n-2}(n-1-k)(n-2-k)(n-k+3\e_0) &=& \ft1{24}
n(n-1)(n-2)(n+4\e_0+1)\ .\eea
Hence,
\bea \eta^{bc}P_{(b}\star P_c\star P_{a_1}\cdots\star
P_{a_{n-2})}&\simeq&
\left(\e_0+\ft1{12}(n-2)(n+4\e_0+1)\right)P_{(a_1}\star\cdots\star
P_{a_{n-2})}+ {\cal O}(\eta)\nn\\[5pt]&=& \ft1{12}(n+1)(n+4\e_0-2)P_{(a_1}\star\cdots\star
P_{a_{n-2})}\ .\eea
On the other hand, tracing the second term on the right hand side of
\eq{lemma}, we find $\k_n$ times
\bea \eta^{bc}\eta_{(bc}P_{a_1}\star\cdots\star P_{a_{n-2})}&=&
{2(2(n-2)+2\e_0+3)\over n(n-1)}P_{(a_1}\star \cdots\star
P_{a_{n-2})}+ {\cal O}(\eta)\ .\eea
The tracelessness of the right hand side of \eq{lemma} thus requires
\bea \ft1{12}(n+1)(n+4\e_0-2)+{2(2n+2\e_0-1)\over n(n-1)}\k_n&=&0\
,\eea
which is equivalent to \eq{kn}.

Finally, let us show \eq{TPa} in the case of $s=0$, where it reads
\bea \Ac_{P_a} T_{b(n)}&=& \{P_a,T_{b(n)}\}_\star\ =\ 2T_{ab(n)}+2
l^{(0)}_n\eta_{a\{b_1}T_{b(n-1)\}}\ ,\label{TPa0}\eea
where, in the second term, the symmetric and traceless
projection
\bea \eta_{a\{b_1}T_{b(n-1)\}}&\equiv &
\eta_{a(b_1}T_{b(n-1))}+\alpha_n\eta_{(b_1 b_2}T_{b(n-2))a}\ ,\qquad
\alpha_n\ =\ \alpha_{n,0}\ =\ -{n-1\over 2(n+\e_0-\ft12)}\ ,
\label{symmtraceless}\eea
and the coefficient
\bea \lambda^{(0)}_n&=&{n(n+2\e_0-1)(n+1)\over 8(n+\e_0+\ft12)}\
.\label{ln0}\eea
Let us first compute this coefficient by imposing the trace
conditions on \eq{TPa0} using the contraction rules \eq{papa},
\eq{pamab} and \eq{macmcb}. Thus, we contract \eq{TPa0} by
$\eta^{ab_n}$, which yields
\bea \{P^a, T_{ab(n-1)}\}_\star&=& \l_n^{(0)}
\eta^{ac}\eta_{a\{b_1}T_{b(n-2)c\}}\ .\eea
On the right hand side, we use \eq{symmtraceless}, and calculate
\bea \eta^{ac}\eta_{a\{b_1}T_{b(n-2)c\}}&=&
{(n+2\e_0+2)(n+\e_0-\ft12)-n+1\over n(n+\e_0-\ft12)}
T_{b(n-1)}\nn\\[5pt]&=&{(n+\e_0+\ft12)(n+2\e_0)\over n(n+\e_0-\ft12)}
T_{b(n-1)}\ ,\eea
so that
\bea \{P^a, T_{ab(n-1)}\}_\star&=&
\l_n^{(0)}{(n+\e_0+\ft12)(n+2\e_0)\over n(n+\e_0-\ft12)} T_{b(n-1)}\
.\label{intermediate2}\eea
On the left hand side, we first use $P^a\star
T_{ab(n-1)}=T_{ab(n-1)}\star P^a$, which can be shown using the
anti-automorphism $\tau$ in \eq{tau}. To calculate $P^a\star
T_{ab(n-1)}$ we use the lemma \eq{lemma} and the ideal contraction
rules \eq{papa} and \eq{pamab}:
\bea P^a\star T_{ab(n-1)}&=& {1\over n}P^a\star\sum_{i=1}^n
P_{(b_{1}}\star\cdots
P_{b_{i-1}|}\star P_a\star P_{|b_{i}}\star\cdots\star P_{b_{n-1})}\nn\\[5pt]&&+
{2\k_n\over n}P_{(b_1}\star\cdots \star P_{b_{n-1})}+{\cal O}(\eta)\nn\\[5pt]&\simeq&
\left(\e_0+\ft12(\e_0+1)(n-1)+\ft16(n-1)(n-2)+{2\k_n\over
n}\right)P_{(b_1}\star\cdots
\star P_{b_{n-1})}+{\cal O}(\eta)\nn\\[5pt]
&=& {(n+2\e_0)(n+2\e_0-1)(n+1)\over
8(n+\e_0-\ft12)}P_{(b_1}\star\cdots \star
P_{b_{n-1})}+{\cal O}(\eta)\nn\\[5pt]
&=& {(n+2\e_0)(n+2\e_0-1)(n+1)\over 8(n+\e_0-\ft12)}T_{b(n-1)}+{\cal
O}(\eta)\ ,\eea
where we note that the trace parts (which are irrelevant for our
calculations) must cancel among themselves. Substituting back into
\eq{intermediate2}, we finally obtain \eq{ln0}.

As a check of \eq{TPa0}, one may verify the closure relation
\bea[\Ac_{P_a},\Ac_{P_b}]T_{c(n)}&=&i\widetilde{\Ad}_{M_{ab}}T_{c(n)}\
.\label{closure}\eea
Using \eq{TPa0}, the left hand side can be expanded as
\bea
&&2\widetilde{\Ad}_{P_a}T_{bc(n)}+2\l_n^{(0)}\eta_{b\{c_1|}\widetilde{\Ad}_{P_a}T_{|c(n-1)\}}-(a\leftrightarrow
b)\nn\\[5pt]
&=&4\l_{n+1}^{(0)} \eta_{a\{b}T_{c(n)\}}+4\l_{n}^{(0)}
\eta_{b\{c_1}T_{c(n-1)\}a}+4\l_n^{(0)}\l_{n-1}^{(0)}\eta_{b\{c_1|}\eta_{a|c_2}T_{|c(n-2)\}}-(a\leftrightarrow
b)\ .\label{intermediate4}\eea
The last term cancel upon the anti-symmetrization in $a$ and $b$, as
can be seen by expanding it explicitly using \eq{symmtraceless}:
\bea &&\eta_{b\{c_1|}\eta_{a|c_2}T_{|c(n-2)\}}-(a\leftrightarrow b)
\nn\\[5pt]&=&
\eta_{bc_1}\left(\eta_{ac_2}T_{c(n-2)}+\a_{n-1}\eta_{c_2c_3}T_{c(n-3)a}\right)
+\a_n\eta_{c_1c_2}\left(\eta_{a
(c_3}T_{c(n-3)b)}+\a_{n-1}\eta_{(c_3c_4}
T_{c(n-4)b)a}\right)-(a\leftrightarrow b)\nn\\[5pt]&=&\left(\a_{n-1}-
{n-2\over n-1}\a_n+{2\a_n\a_{n-1}\over
n-1}\right)\eta_{bc_1}\eta_{c_2c_3}T_{c(n-3)a}\ =\ 0\ ,\eea
where the symmetrization on $c_1,\dots,c_n$ has been suppressed.
Expanding the two first terms in \eq{intermediate4} using
\eq{symmtraceless} and \eq{ln0}, one finds
\bea &&4\l_{n+1}^{(0)} \left(\eta_{a(b}
T_{c(n))}+\a_{n+1}\eta_{(bc_1}T_{c(n-1))a}\right)+4\l_{n}^{(0)}
\left(\eta_{b(c_1}T_{c(n-1))a}+\a_n\eta_{(c_1c_2}T_{c(n-2))ba}\right)
-(a\leftrightarrow b)\nn\\[5pt]
&=&8\left({n\l_{n+1}^{(0)}\over n+1}-{2\a_{n+1}\l_{n+1}^{(0)}\over
n+1}-\l_n^{(0)}\right)\eta_{c_1[a}T_{b]c(n-1)}
\nn\\[5pt]&=&2n\eta_{c_1[a}T_{b]c(n-1)}\ ,\label{closure2}\eea
which one identifies as the right hand side of \eq{closure}.



\chapter{Quadratic and Quartic Casimir Operators}\label{App:Cas}


The quadratic and and quartic Casimir operators of $\mso(D+1;\Comp)$
are defined by
\bea C_2[\mso(D+1;\Comp)]&=& \ft12 M^{AB} \star M_{AB}\
,\label{C2}\\[5pt]
C_4[\mso(D+1;\Comp)]&=&\ft12 M_A{}^B\star M_B{}^C\star M_C{}^D\star
M_D{}^A\ .\label{C4}\eea
Acting on lowest and highest weight states $|e_0;{\bf
s}_0\rangle^\pm$, which are annihilated by $L^\mp_r$, they can be
rewritten using \eq{ll}-\eq{ml}, and the resulting values are
\bea C_2[\mso(D+1;\Comp)]|e_0;{\bf s}_0]&=&
e_0(e_0\mp(D-1))+C_2[\mso(D-1;\Comp)|{\bf s}_0]\label{C2lhws}\\[5pt]
C_4[\mso(D+1;\Comp)|e_0;{\bf s}_0]|&=&
e_0(e_0\mp(D-1))\left(e_0(e_0\mp(D-1))+\ft12
(D-1)(D-2)\right)\nn\\&&+C_4[\mso(D-1;\Comp)|{\bf
s}_0]-C_2[\mso(D-1;\Comp)|{\bf s}_0]\ ,\label{C4lhws}\eea
where the Casimir operators that are quadratic and quartic in
angular momenta are defined by
\bea C_2[\mso(D-1;\Comp)]&=&\ft12 M^{rs}M_{rs}\ ,\\[5pt]
C_4[\mso(D-1;\Comp)]&=&\ft12 M_r{}^s\star M_s{}^t\star M_t{}^u\star
M_u{}^r\ ,\eea
and given in the $\mso(D-1;\Comp)$ irrep with highest weight ${\bf
s}_0=(m_1,\dots,m_{\nu-1})$ by
\bea C_2[\mso(D-1;\Comp)|{\bf s}_0] &=&
\sum_{k=1}^{\nu-1}m_k(m_k+D_k)\ ,\label{c2d-1}\\[5pt]
C_4[\mso(D-1;\Comp)|{\bf s}_0] &=&
\sum_{k=1}^{\nu-1}m_k(m_k+D_k)\left(m_k(m_k+D_k)+\ft12(D_k)(D_k-1)+1-k\right)\
,\label{c4d-1}\eea
where $D_k=D-1-2k$. The latter two equations follow by recursive use
of \eq{C2hws} and \eq{C4hws} in the case of a highest weight state.

To evaluate the Casimir operators in $\ell$th level of the adjoint
representation, defined in \eq{Lell}, we use \eq{C2lhws} and
\eq{C2lhws} with the highest weight $(2\ell+1,2\ell+1)$, which
immediately gives \eq{c2ell} and \eq{c4ell} (with $s=2\ell+2$). The
same values follow for the massless lowest weight and highest weight
spaces $\mD^\pm(\pm(s+2\e_0);(s))$. In the twisted adjoint
representation, we first rewrite the $\pi$-twisted commutators in
terms of commutators plus terms that can be calculated directly
using the contraction rule \eq{papa}. In the case of the quadratic
Casimir we find
\bea \widetilde{\Ad}_{C_2[\mso(D+1;\Comp)]}(S)&=&
\Ad_{C_2[\mso(D;\Comp)]}(S)-\{P^a,\{P_a,S\}_\star\}_\star\nn\\[5pt]&=&
\Ad_{C_2[\mso(D;\Comp)]}(S)-2\e_0 S-2P^a\star S\star P_a\nn\\[5pt]&=&
\Ad_{C_2[\mso(D+1;\Comp)]}(S_\ell)+4P^a\star S\star P_a\ ,\eea
from which $P^a\star S\star P_a$ can be eliminated, which yields the
explicit formula
\bea
T(C_2[\mso(D+1;\Comp)])(S)&=&2\Ad_{C_2[\mso(D;\Comp)]}(S)-\Ad_{C_2[\mso(D+1;\Comp)]}(S_\ell)-4\e_0
S\ .\eea
At the $\ell$th level, defined by \eq{levelell}, the elements
$S_\ell$ carry highest weights $(s+k,s)$ and $(s+k,s+k)$ with
$s=2\ell+2$ and $k=0,1,\dots$, of (the adjoint actions) of
$\mso(D;\Comp)$ and $\mso(D+1;\Comp)$, respectively, and we find
that for all $k$
\bea
T(C_2[\mso(D+1;\Comp)])(S_\ell)&=&\left(2C_2[\mso(D;\Comp)|(s+k,s)]-C_2[\mso(D+1;\Comp)|(s+k,s+k)]-4\e_0\right)S_\ell
\nn\\[5pt]&=&\left(
2C_2[\mso(D;\Comp)|(s,s)]-C_2[\mso(D+1;\Comp)|(s,s)]-4\e_0\right)S_\ell\nn\\[5pt]&=&
C_2[\mso(D+1;\Comp)|\ell] S_\ell\ .\eea
Similarly, in the case of the quartic Casimir,
\bea
\widetilde{\Ad}_{C_4[\mso(D+1;\Comp)]}(S)&=&\Ad_{C_4[\mso(D;\Comp)]}(S)+\nn\\
&&+\ft12[M_a{}^b,[M_b{}^c,\{P_c,\{P^a,S\}_\star\}_\star]_\star]_\star
+\ft12[M_a{}^b,\{P_b,\{P^c,[M_c{}^a,S]_\star\}_\star\}_\star]_\star\nn\\&&
+\ft12\{P_a,\{P^b,[M_b{}^c,[M_c{}^a,S]_\star]_\star\}_\star\}_\star+
+\ft12\{P^a,[M_a{}^b,[M_b{}^c,\{P_c,S\}_\star]_\star]_\star\}_\star\nn\\&&
+\ft12 \{P_a,\{P^b,\{P_b,\{P^a,S\}_\star\}_\star\}_\star\}_\star +
+\ft12
\{P^a,\{P_b,\{P^b,\{P_b,S\}_\star\}_\star\}_\star\}_\star\nn\\[5pt]
&=&\Ad_{C_4[\mso(D;\Comp)]}(S)+C_+(S)+C_-(S)\nn\\[5pt]
&=& \Ad_{C_4[\mso(D+1;\Comp)]}(S)+2C_-(S)\ ,\eea
where $C_{+}(S)$ and $C_-(S)$ are the terms with an even and odd
number of translation generators standing to the right of $S$,
respectively. Eliminating $C_-(S)$ leads to
\bea
\widetilde{\Ad}_{C_4[\mso(D+1;\Comp)]}(S)&=&2\Ad_{C_4[\mso(D;\Comp)]}(S)-\Ad_{C_4[\mso(D+1;\Comp)]}(S)+2C_+(S)\
.\eea
The quantity $C_+(S)$ can be calculated using of \eq{pamab},
\eq{macmbc}, and $M_{ab}\star S\star
M^{ab}=-\Ad_{C_2[\mso(D;\Comp)]}(S)+\{M^{ab}\star M_{ab},S\}_\star$,
and one finds
\bea C_+(S)&=&\Ad_{C_2[\mso(D;\Comp)]}(S)-2\e_0(2\e_0^2-\e_0+1)S\
.\eea
Thus, using the above assignments of highest weights for $S_\ell$,
we find that
\bea
\widetilde{\Ad}_{C_4[\mso(D+1;\Comp)]}(S_\ell)&=&\left(C_4[\mso(D;\Comp)|(s+k,s)]-
C_4[\mso(D+1;\Comp)|(s+k,s+k)]\right)S_\ell\nn\\&&+
\left(C_2[\mso(D;\Comp)|(s+k,s)]-4\e_0(2\e_0^2-\e_0+1)\right)S_\ell\nn\\[5pt]
&=&C_4[\mso(D+1;\Comp)|\ell] S_\ell\ .\eea


\chapter{Computing $\Ac_{P_a}T_{b(n),c(m)}$ from the Mass
Formula}\label{App:T}


In this Appendix we shall derive the expression \eq{lambda} for the
coefficient $\l_k^{(s)}$ in \eq{TPa} using the Casimir relation
\eq{Casimirrelation}, or, equivalently, the linearized zero-form
constraint \eq{DPhicomponents} and the Weyl-tensor mass formula
\eq{msk}.

Let us begin with the case of $s=0$, where \eq{TPa} reduces to
\eq{TPa0}. Starting from the linearized zero-form master constraint
\eq{linzeroform}, \emph{i.e.} $\nabla\Phi_{(0)}-i
e^a\{P_a,\Phi_{(0)}\}_\star=0$, and expanding $\Phi_{(0)}$ using
\eq{Phis}, we obtain the component form \eq{DPhicomponents} of the
constraint for $s=0$, that is
\bea \nabla_b \Phi_{a(n)}-2n
\eta_{b\{a_1}\Phi_{a(n-1)\}}+{2\l_n^{(0)}\over n+1}\Phi_{ba(n)}&=&0\
,\label{DPhicomponents0}\eea
where we recall the definition \eq{symmtraceless} of the symmetric
and traceless projection $\eta_{b\{a_1}\Phi_{a(n-1)}$. The symmetric
traceless part of \eq{DPhicomponents0} immediately gives
\eq{Phiauxiliary} for $s=0$, while the trace part of
\eq{DPhicomponents0} can be used to derive the masses \eq{msk} for
$s=0$. to this end, one first contracts \eq{DPhicomponents0} with
$\nabla^b$, which yields
\bea
\nabla^2\Phi_{a(n)}-2n\eta^{bc}\left(\eta_{b(a_1|}\nabla_c\Phi_{|a(n-1))}+\a_n
\eta_{(a_1a_2|}\nabla_c\Phi_{|a(n-2))b}\right)+{2\l^{(0)}_{n+1}\over
n+1}\nabla^b\Phi_{ba(n)}&=&0\ .\eea
One then substitutes $\nabla_c\Phi_{a(n-1)}$ and
$\nabla_c\Phi_{a(n1)}$ using \eq{Phiauxiliary}, and takes the
symmetric and traceless projection in $a(n)$, which leads to
\bea
\nabla^2\Phi_{a(n)}+4\l_n^{(0)}\Phi_{a(n)}+4\l^{(0)}_{n+1}\eta^{bc}
\left(\eta_{c(b}\Phi_{a(n))}+\a_{n+1}\eta_{(ba_1}\Phi_{a(n-1))c}\right)&=&0\
.\eea
Performing the traces, one ends up with the following expression for
the mass:
\bea m^2_{0,n}&=&-4\l_n^{(0)}-4\l_{n+1}^{(0)}{1\over
n+1}\left(n+2\e_0+3-{n\over n+\e_0+\ft12}\right)\nn\\[5pt]&=&
-4l_n^{(0)}-4l_{n+1}^{(0)}{(n+2\e_0+1)(n+\e_0+\ft32)\over
(n+1)(n+\e_0+\ft12)}\ .\label{masses0}\eea
Inserting \eq{ln0}, this expressions can be simplified, with the
result
\bea m^2_{0,n}&=&-(n^2+(2\e_0+1)n+4\e_0)\ ,\eea
in agreement with \eq{msk} for $s=0$.

Turning to the case of general $s$, we use \eq{TPa} and the
expansion \eq{Phis} in the linearized zero-form master constraint
\eq{linzeroform}:
\bea 0&=&\nabla_c\Phi-i\{P_c,\Phi\}_\star\nn\\[5pt]
&=& \sum_{s,n}{i^n\over n!} T^{a(s+n),b(s)} \nabla_c
\Phi_{a(s+n),b(s)}-\sum_{s,n}{i^{n+1}\over n!}
2\D_{n+s,s}T_{c}{}^{a(s+n),b(s)} \Phi_{a(s+n),b(s)}\nn\\[5pt]&&-
\sum_{s,n}{i^{n+1}\over n!} 2\l_n^{(s)} \eta_{c\langle
a}T_{a(s+n-1),b(s)\rangle} \Phi^{a(s+n),b(s)}\ .\eea
To read off the corresponding component equations, we rewrite the
middle term using
\bea T_{c}{}^{a(s+n),b(s)} \Phi_{a(s+n),b(s)}&=&T^{a(s+n+1),b(s)}
\eta_{c(a}\Phi_{a(s+n)),b(s)}\ =\ T^{a(s+n+1),b(s)} \eta_{c\{
a}\Phi_{a(s+n),b(s)\}}\ ,\eea
and the last term using
\bea &&\eta_{c\{ a}T_{a(s+n-1),b(s)\}} \Phi^{a(s+n),b(s)}\ =\
\left(\eta_{c( a}T_{a(s+n-1)),b(s)} +\mbox{($\eta_{aa}$ and
$\eta_{ab}$ traces)}\right)\Phi^{a(s+n),b(s)}\nn\\[5pt]&=& \eta_{c(
a}T_{a(s+n-1)),b(s)}\Phi^{a(s+n),b(s)}\ =\
T_{a(s+n-1),b(s)}\Phi_{c}{}^{\{ a(s+n-1),b(s)\} }\ ,\eea
after which we arrive at \eq{DPhicomponents}. Contracting by
$\nabla^c$ and we get
\bea \nabla^2\Phi_{a(s+n),b(s)}&=&
2n\D_{s+n-1,s}\eta^{cd}\nabla_d\eta_{c\{ a}\Phi_{a(s+n-1),b(s)\}}-
{2\l_{n+1}^{(s)}\over n+1} \eta^{cd}\nabla_d\Phi_{c\{
a(s+n),b(s)\}}\ ,\label{intermediatemasses}\eea
where we recall that $\eta_{c\{ a}\Phi_{a(s+n-1),b(s)\}}$ is given
by \eq{hooked}. The next step is to use \eq{DPhicomponents} to
substitute the gradients on the right hand side. In the first term
we obtain
\bea &&-4\D_{s+n-1,s}\l_{n}^{(s)}\eta^{cd} \left(\eta_{c a}\Phi_{d
\langle a(s+n-1),b(s)\rangle} +\a_{s+n,s}\eta_{a(2)} \Phi_{d\langle
a(s+n-2)c,b(s)\rangle}\right.\nn\\[5pt]&&\left.+\beta_{s+n,s}\eta_{a(2)}\Phi_{d\langle
a(n+s-2)b,cb(s-1)\rangle}+\c_{s+n,s}\eta_{ab}\Phi_{d\langle
a(s+n-1),cb(s-1)\rangle}\right)\nn\\[5pt]&=&
-4\D_{s+n-1,s}\l_{n}^{(s)}\Phi_{a\langle a(s+n-1),b(s)\rangle}\nn\\[5pt]&=&
-4\D_{s+n-1,s}\l_{n}^{(s)}\widetilde \D_{s+n,s}\Phi_{a(s+n),b(s)}\
,\eea
where the intermediate $\langle\cdots\rangle$ Young projections
imposed \emph{prior} to the final symmetrization on $a$ and $b$
indices, and the coefficient $\widetilde \D_{s+n,s}$ is defined by
\bea \Phi_{a\langle a(s+n-1),b(s)\rangle}&=& \widetilde
\D_{s+n,s}\Phi_{a(s+n),b(s)}\ .\eea
To compute this coefficient, we expand the left hand side using the
definition of the Young projector:
\bea \Phi_{a\langle a(s+n-1),b(s)\rangle}&=& {(s+n-1)! s!\over
(s+n)\cdots (n+1)(n-1)\cdots 1\times s!} \sum_{k=0}^s
(-1)^k{s\choose k}\Phi_{ a(s+n-k)b(k),a(k)b(s-k)}\ .\eea
In the $k$th term, we cycle the $b$-indices back to their original
position, using $\Phi_{(a(s-k)b(k),a_1)a(k-1)b(s-k)}=0$, which
yields
\bea \Phi_{a(s-k)b(k),a(k)b(s-k)}&=&-{k\over
s+n-k+1}\Phi_{a(s-k+1)b(k-1),a(k-1)b(s-k+1)}\ =\ {(-1)^k\over
{s+n\choose k}}\Phi_{a(s),b(s)}\ .\eea
Thus,
\bea \widetilde \D_{s+n,s}&=& {n\over s+n}\sum_{k=0}^s {{s\choose
k}\over {s+n\choose k}}\ =\ {n\over s+n}{1\over
(s+1)_{n}}\sum_{k=0}^s(s+1-k)_{n}\ =\ {n\over s+n}{s+n+1\over n+1}\
.\eea
Using also \eq{Delta}, the first term on the right-hand side of
\eq{intermediatemasses} is found to be
\bea -4\D_{s+n-1,s}\l_{n}^{(s)}\widetilde \D_{s+n,s}&=&
-4\l_{n}^{(s)}\ .\eea
Turning to the second term, substitution of the gradient yields
\bea &&- 4\l_{n+1}^{(s)} \D_{s+n,s}\eta^{cd}\eta_{c\{ d}\Phi_{\{
a(s+n),b(s)\}\}}\nn\\[5pt]&=&
- 4\l_{n+1}^{(s)}\D_{s+n,s}{1\over
n+s+1}\left(s+n+2\e_0+3+2\a_{s+n+1,s}-{2\beta_{s+n+1,s}\over
s+n}+\c_{s+n+1,s}\right)\Phi_{a(s+n),b(s)}\nn\\[5pt]&=&-4\l_{n+1}^{(s)}\D_{s+n,s}
{(n+s+2\e_0)(n+s+\e_0+\ft32)(n+2s+2\e_0+1)\over (n+s+1)
(n+2s+2\e_0)(n+s+\e_0+\ft12)}\Phi_{a(s+n),b(s)}\ .\eea
Thus, upon adding the two contributions, we find
\bea m^2_{s,n}&=&-4\l_{n}^{(s)}-4\l_{n+1}^{(s)}\D_{s+n,s}
{(n+s+2\e_0)(n+s+\e_0+\ft32)(n+2s+2\e_0+1)\over (n+s+1)
(n+2s+2\e_0)(n+s+\e_0+\ft12)}\ ,\eea
where $m^2_{s,n}=-4\e_0-2s-(n+2s+2\e_0+1)n$, which serves as a
recursion relation for determining $\l_{n}^{(s)}$, given the initial
datum $\l_0^{(s)}=0$, with solution given by \eq{lambda}.


\chapter{Two-Component Spinor and Curvature Conventions}\label{App:F}


We use conventions in which the generators of the various real forms
of $SO(5;\Comp)$ obey
\be [M_{AB},M_{CD}]\ =\ 4i\y_{[C|[B}M_{A]|D]}\ ,\qquad
(M_{AB})^\dagger\ =\ \sigma(M_{AB})\ ,\label{sogena}\ee
with $\eta_{AB}=(\eta_{ab};-\lambda^2)$, where $\eta_{ab}$ specifies
the signature of the tangent space. Under $M_{AB}\rightarrow
(M_{ab},P_a)$, the commutation relations decompose into
\be [M_{ab},M_{cd}]_\star\ =\ 4i\y_{[c|[b}M_{a]|d]}\ ,\qquad
[M_{ab},P_c]_\star\ =\ 2i\y_{c[b}P_{a]}\ ,\qquad [P_a,P_b]_\star\ =\
i\lambda^2 M_{ab}\ .\label{sogenb}\ee
The corresponding oscillator realization is taken to be \eq{mab2},
which we repeat here for convenience
  \be
  M_{ab}\ =\ -\frac18 \left[~ (\s_{ab})^{\a\b}y_\a y_\b+
  (\sb_{ab})^{\ad\bd}\tilde y_{\ad}\yb_{\bd}~\right]\
,\qquad P_{a}\ =\
  \frac{\l}4 (\s_a)^{\a\bd}y_\a \yb_{\bd}\ .\label{mab}
  \ee
Here, the van der Waerden symbols obey
  \bea
   (\s^{a})_{\a}{}^{\ad}(\sb^{b})_{\ad}{}^{\b}&=&
\y^{ab}\d_{\a}^{\b}\
  +\ (\s^{ab})_{\a}{}^{\b} \ ,\qquad
  (\sb^{a})_{\ad}{}^{\a}(\s^{b})_{\a}{}^{\bd}\ =\
\y^{ab}\d^{\bd}_{\ad}\
  +\ (\sb^{ab})_{\ad}{}^{\bd} \ ,\label{so4a}\w2
  \ft12 \e_{abcd}(\s^{cd})_{\a\b}&=& \left\{\ba{ll}
  (\s_{ab})_{\a\b}\ ,&\mbox{$(4,0)$ and $(2,2)$ signature\ ,}\\[5pt]i(\s_{ab})_{\a\b}
  \ ,&\mbox{$(3,1)$ signature\ ,}\ea\right.\label{so4b}\eea
and reality conditions
\bea ((\s^a)_{\a\bd})^\dagger\ =\ \left\{
\begin{array}{ll}
-(\sb^a)^{\bd\a} \ = \ -(\s^a)^{\a\bd}&\mbox{for $SU(2)$\ ,}
\\[5pt]
(\sb^a)_{\ad\b} \ = \ (\s^a)_{\b\ad} &\mbox{for
$SL(2,\Comp)$\ ,} \\[5pt]
(\sb^a)_{\bd\a} \ = \ (\s^a)_{\a\bd} &\mbox{for $Sp(2)$\ ,}
\end{array}
\right.\eea
and
\bea ((\s^{ab})_{\a\b})^\dagger\ =\ \left\{
\begin{array}{ll}
(\s^{ab})^{\a\b} &\mbox{for $SU(2)$\ ,} \\[5pt]
(\sb^{ab})_{\ad\bd} &\mbox{for $SL(2,\Comp)$\ ,} \\[5pt]
(\s^{ab})_{\a\b} &\mbox{for $Sp(2)$\ .} \end{array} \right.\eea
The reality conditions on spinor oscillators are given in
(\ref{su2}), (\ref{sl2}) and (\ref{sp2}). Spinor indices are raised
and lowered according to the following conventions,
$A^\a=\epsilon^{\a\b}A_\b$, $A_\a=A^\b\epsilon_{\b\a}$, where
\be \e^{\a\b}\e_{\c\d} \ = \ 2 \d^{\a\b}_{\c\d} \ , \qquad
\e^{\a\b}\e_{\a\c} \ = \ \d^\b_\c \ ,\ee
and
\be (\e_{\a\b})^\dagger \ = \ \left\{ \begin{array}{ll}
\e^{\a\b}& \mbox{for $SU(2)$\ ,} \\[5pt]
\e_{\ad\bd}& \mbox{for $SL(2,\Comp)$\ ,} \\[5pt]
\e_{\a\b} & \mbox{for $Sp(2)$\ .} \end{array} \right. \ee
One may use the following convenient representations:
\bea SU(2)&: & \qquad \s^a=(i,\s^i) \ ,\qquad \sb^a=(-i,\s^i) \
,\qquad
\e=i\s^2 \ ; \\[5pt]
  SL(2,\Comp)&: & \qquad \s^a=(-i\s^2,-i\s^i\s^2) \
,\qquad
\sb^a=(-i\s^2,i\s^2\s^i) \ , \qquad\e=i\s^2 \ ; \\[5pt]
Sp(2)& : & \qquad \s^a=(1,\ts^i) \ , \qquad\sb^a=(-1,\ts^i) \
,\qquad \e=i\s^2 \ ,\eea
where in the last case $\ts^i = (\s^1,i\s^2,\s^3)$. The real form of
the $\mso(5;\Comp)$-valued connection $\O$ can be expressed, using
(\ref{em}) and (\ref{mab}), as
  \be
   \O\ =\ \frac1{4i}
  dx^\mu\left[\omega_\mu^{\a\b}~y_\a y_\b
  +\overline{\omega}_\mu{}^{\dot\a\dot\b}~{\bar
y}_{\dot\a}{\bar y}_{\dot\b}
  + 2 e_\mu^{\a\dot\b}~y_\a {\bar y}_{\dot\b}\right]\
  ,\label{Omega}
  \ee
where
  \be
  \o^{\a\b}\ =\ -\ft14(\s_{ab})^{\a\b}~\o^{ab}\ ,
  \quad
  \ob^{\dot\a\dot\b}\ =\
-\ft14({\sb}_{ab})^{\dot\a\dot\b}~\o^{ab}\ ,
  \quad
  e^{\a\dot\a}\ =\ \ft{\lambda}2(\s_{a})^{\a
\dot\a}~e^{a}\ .
  \label{convert}
  \ee
Likewise, for the curvature ${\cal R}=d\O+\O\star \O$ one finds
  \bea
  {\cal R}_{\a\b}&=& d\o_{\a\b}
+\o_{\a\c}\wedge\o_{\b}{}^{\c}+
  e_{\a\dd}\wedge e_{\b}{}^{\dd}\ ,
  \label{rab}\w2
  \overline{\cal R}_{\dot\a\dot\b}&=&
d\overline{\o}_{\ad\bd}
  +\overline{\o}_{\ad\cd}\wedge\overline{\o}_{\bd}{}^{\cd}
  +e_{\d\ad}\wedge e^{\d}{}_{\bd}\ ,
  \label{radbd}\w2
  {\cal R}_{\a\dot\b}&=& de_{\a\bd}+ \o_{\a\c}\wedge
  e^{\c}{}_{\bd}+\overline{\o}_{\bd\dd}\wedge
e_{\a}{}^{\dd}\
  ,\label{rabd}
  \eea
and
  \be
  {\cal R}^{ab}\ =\ d\o^{ab}+\o^a{}_c\wedge\o^{cb}
+\lambda^2
  e^a\wedge e^b\ ,\qquad {\cal R}^a\ =\ d
e^a+\o^a{}_b\wedge e^b\ .
  \label{curvcomp} \ee
Setting ${\cal R}=0$ gives the Riemann tensor
  \be
  R_{\m\n,\r\s}\ = \
  -\lambda^2 \left( g_{\mu\rho} g_{\nu\sigma}-
   g_{\nu\rho} g_{\mu\sigma} \right)\ ,\label{ads}\ee
corresponding to the maximally symmetric vacuum solution of gravity
with action
\bea S&=&\frac{1}{16\pi G_{4}}\int d^{4}x\sqrt{-g}(R-2\Lambda)\
,\qquad \L\ =\ -3\l^2\ .\label{pure}\eea


\chapter{Further Notation Used for the Solutions}\label{AppDef}


The gauge function $L(x;y,\yb)$ defined in \eq{wL} can be written as
\bea L &=& {2h\over 1+h} \exp (-iy a \yb)\ , \label{L}\eea
where
\bea a_{\a\ad}&=&{\lambda x_{\a\ad}\over 1+h}\ ,\qquad
x_{\a\ad}=(\s^a)_{\a\ad} x_a\ ,\label{aaad}\w2
x^2 &=&  \eta_{ab} x^a x^b\ ,\qquad  h =  \sqrt{1-\lambda^2x^2}\
.\label{aad} \eea
Useful relations that follow from these definitions are
\be a^2={1-h\over 1+h}\ ,\qquad h={1-a^2\over 1+a^2}\ .\ee
The Maurer-Cartan form based on $L$ defined in \eq{wL} yields the
the vierbein and Lorentz connection
\be e_{(0)}{}^{\a\ad}\ =\ -{\l(\s^a)^{\a\ad}dx_a\over h^2}\ ,\qquad
\o_{(0)}{}^{\a\b}\ =\ - {\l^2(\s^{ab})^{\a\b} dx_a x_b\over h^2}\
,\label{adseo}\ee
with Riemann tensor given by
\be
 R_{(0)\m\n,\r\s}\ = \
 -\lambda^2 \left( g_{(0)\mu\rho} g_{(0)\nu\sigma}-
  g_{(0)\nu\rho} g_{(0)\mu\sigma} \right)\ .
\ee
In the case of type 3 solutions, a useful definition is
\bea b_{\a\b}&=&2\l_{(\a}\m_{\b)}\ ,\qquad \l^\a \mu_\a\ =\ \ft{i}2\
. \label{bab}\eea
It obeys the relation $(b^2)_\a{}^\b\ =\ -\ft 14 \d_\a^\b$ and it
defines an almost complex structure via the relations (see, for
example, \cite{Pope:1982ad})
\be b_{\a\b}=\frac{1}{8} (\sigma^{ab})_{\a\b}\,J_{ab}\ ,\qquad
J_{ab}=(\sigma_{ab})^{\a\b}\,b_{\a\b}\ ,\qquad J_a{}^c J_c{}^b=
-\delta_a^b\ . \label{jj}\ee
Similarly, using the definition
\be {\tilde b}_{\a\b}= a^{-2} (a{\bar b} \bar a)_{\a\b}\
,\label{btilde}\ee
we have the relations
\be {\tilde b}_{\a\b}=\frac{1}{8} (\sigma^{ab})_{\a\b}\,{\tilde
J}_{ab}\ ,\qquad {\tilde J}_{ab}=(\sigma_{ab})^{\a\b}\, {\tilde
b}_{\a\b}\ ,\qquad {\tilde J}_a{}^c {\tilde J}_c{}^b= -\delta_a^b\
.\label{jtilde} \ee
Finally, we have the following definition for spinors used in
describing a Type 3 solution:
\be U_{\ad}={x^a\over \sqrt{x^2}}
\left({\bar\sigma}_a\lambda\right)_{\ad}\ , \qquad V_{\ad}=
{x^a\over \sqrt{x^2}} \left({\bar\sigma}_a \mu\right)_{\ad}\ .
\label{UV}\ee


\chapter{Weyl-ordered Projectors}\label{AppProj}


Weyl-ordered projectors $P(y,\bar y)$ can be constructed by
recombining $(y,\bar y)$ into a pair of Heisenberg oscillators
$(a_i,b^j)$ ($i,j=1,2)$ obeying
\bea [a_i,b^j]_\star&=& \d_i^j\ .\eea
For example, one can take
\bea a_1&=&u\ =\ \l^\a y_\a\ ,\qquad b^1\ =\ v\ =\ \mu^\a y_\a\
,\\[5pt]
a_2&=&\bar u\ =\ \bar\l^{\ad}\bar y_{\ad}\ ,\qquad b^2\ =\ \bar v\
=\ \bar\mu^{\ad}\bar y_{\ad}\ ,\eea
where the constant spinors are normalized as
\bea \l^\a\mu_\a&=&\ft{i}2\ ,\qquad \bar\l^{\ad}\bar\mu_{\ad}\ =\
\ft{i}2\ .\eea
The projectors, obeying the appropriate reality conditions, take the
form
\bea P&=&\sum_{n_1,n_2\in~\integ+\ft12}\th_{n_1,n_2} P_{n_1,n_2}\
,\qquad \bar P\ =\ \sum_{n_1,n_2\in~\integ+\ft12}\bar\th_{n_1,n_2}
P_{n_1,n_2}\ ,\label{generalP}\eea
where $\theta_{n_1,n_2}\in\{0,1\}$ and
$\bar\th_{n_1,n_2}\in\{0,1\}$, with
\bea \mbox{$(3,1)$ signature}&:& \th_{n_1,n_2}\ =\
\bar\th_{n_1,n_2}\ ,\\[5pt]
\mbox{$(4,0)$ and ($2,2)$ signatures}&:& \th_{n_1,n_2}\
,\bar\th_{n_1,n_2}\quad\mbox{independent}\ ,\eea
and
\bea P_{n_1,n_2}&=& 4(-1)^{n_1+n_2-\ft{\e_1+\e_2}2}e^{-2\sum_i \e_i
w_i}L_{n_1-\ft{\e_1}2}(4\e_1 w_1)L_{n_2-\ft{\e_2}2}(4\e_2 w_2)\ ,\\[5pt]
w_i&=&b^i a_i\ =\ b^i\star a_i+\ft12\ =\ a_i\star b^i-\ft12\qquad
\mbox{(no sum)}\ ,\eea
with $\e_i=n_i/|n_i|$ and $L_n(x)={1\over n!}e^x{d^n\over
dx^n}(e^{-x}x^n)$ are the Laguerre polynomials. The projector
property of $P$ and $\bar P$ follows from
\bea P_{m_1,m_2}\star P_{n_1,n_2}&=& \d_{m_1 n_1}\d_{m_2
n_2}P_{n_1,n_2}\
,\label{projprop}\\[5pt]
(w_i-n_i)\star P_{n_1,n_2}&=& 0\ ,\\[5pt] \t(P_{n_1,n_2})&=& P_{-n_1,-n_2}\ .\label{tauP}\eea
Here, $w_i-\ft12$ is the Weyl-ordered form of the number operator,
and
\bea 2(-1)^{n_i-\ft{\e_i}2} e^{-2\e_i w_i}L_{n_i-\ft{\e_i}2}(4\e_i
w_i)&=&\mx{\{}{ll}{|n_i\rangle\langle n_i|&\mbox{for $n_i>0$}\\[5pt]
(-1)^{-n_i-\ft12}|n_i\rangle\langle n_i|& \mbox{for $n_i<0$}}{.}\eea
where
$|n_i\rangle=\ft{(b^i)^{n_i-\ft12}}{\sqrt{(n_i-\ft12)!}}|0\rangle$
with $n_i>0$ belongs to the standard Fock space, built by acting
with $b^i$ on the ground state $|0\rangle$ obeying $a_i|0\rangle=0$,
while
$|n_i\rangle=\ft{(a_i)^{-n_i-\ft12}}{\sqrt{(-n_i-\ft12)!}}|\tilde
0\rangle$ for $n_i<0$ are anti-Fock space states, built by acting
with $a_i$ on the anti-ground state $|\tilde 0\rangle=0$ obeying
$b^i|\tilde 0\rangle=0$. Equation \eq{projprop} holds formally,
since the inner product between a Fock space state and an anti-Fock
space state vanishes. However, the corresponding Weyl-ordered
projectors have divergent $\star$-products, as can be seen from the
lemma
\bea e^{suv}\star e^{tuv}&=& {1\over
1+\frac{st}4}\exp\left({\frac{s+t}{1+\frac{st}4}\,uv}\right)\ .\eea
Thus, lacking, at present, a suitable regularization scheme that
does not violate associativity and other basic properties of the
$\star$-product algebra, we shall restrict our attention to
projectors that are constructed in either the Fock space or the
anti-Fock space, \emph{i.e.}
\bea \th_{n_1,n_2}&=&1 \quad \mbox{only if $(n_1,n_2)\in Q$}\
,\label{constrth}\eea
where $Q$ is anyone of the four quadrants in the $(n_1,n_2)$ plane.
From \eq{tauP}, it follows that these projectors are not invariant
under the $\tau$ map, and therefore the master fields Type 2 and
Type 3 solutions will be those of the non-minimal model, where the
$\tau$ conditions are relaxed to $\pi\bar\pi$ conditions, which are
certainly satisfied.

We also note that in order to solve the higher-spin equations it is
essential that
\bea [P,\bar P]_\star&=&\sum_{n_1,n_2}
(\th_{n_1,n_2}\bar\th_{n_1,n_2}-\bar\th_{n_1,n_2}\th_{n_1,n_2})P_{n_1,n_2}\
=\ 0\ ,\eea
which holds for independent $\th_{n_1,n_2}$ and
$\bar{\th}_{n_1,n_2}$ parameters (in the Euclidean and Kleinian
signatures). Moreover, one can work with a reduced set of
oscillators, say $a_1=u$ and $b^1=v$, by taking
$\th_{n_1,n_2}=\th(\pm n_2)\th_{n_1}$, where $\th(x)=0$ for $x<0$
and $\th(x)=1$ for $x>0$, and summing over all values of $n_2$ using
\bea \sum_{k=0}^\infty t^k L_k(x)&=&(1-t)^{-1}\exp(-xt(1-t)^{-1})\
.\label{generating}\eea
Setting $n=n_1$ and $\e=\e_1$, this leads to
\bea P&=& \sum_{n\in ~\integ+\ft12}\th_n P_n\ ,\qquad \bar P\ =\
 \sum_{n\in ~\integ+\ft12}\bar\th_n P_n\\[5pt]
P_n&=& 2(-1)^{n-\ft{\e}2} e^{-2\e uv} L_n(4\e uv)\
,\label{reduced}\eea
with suitable reality conditions on the $\th_n$ parameters. Finally,
setting $\th_n=\th(\pm n)$, and using \eq{generating} once more, one
finds that setting all $\th$-parameters equal to $1$ gives $P=1$.


\chapter{Calculation of $V=L^{-1}\star P\star L$}\label{App2}


In this Appendix we compute $V=L^{-1}\star P\star L$ where $L$ is
the gauge function given in \eq{L} and $P$ is a projector of the
form given in \eq{generalP}. Let us begin by considering the case of
$P=P_{\ft12}=2e^{-2uv}$, \emph{i.e.}
\bea V &=&{8h^2\over(1+h)^2} e^{iy a \bar y}\star e^{yby}\star
e^{-iy a\bar  y}\ ,\eea
where $y a\bar y= y^\a a_{\a}{}^{\ad}\bar y_{\ad}$ and $yby=y^\a
b_\a{}^\b y_\b$, with $a_{\a\ad}$ and $b_{\a\b}$ given by \eq{aaad}
and \eq{bab}. The first $\star$-product can be performed treating
the integration variables $(\xi_\a,\eta_\a)$ and
$(\bar\xi_{\ad},\bar\eta_{\ad})$ as separate real variables. Using
the formulae (B.1) provided in \cite{Sezgin:2005pv}, we find
\bea V &=&{8h^2\over(1+h)^2} e^{i ya\bar y+ (y-\bar y a)b(y+a\bar
y)}\star e^{-i y a\bar y}\ .\eea
The remaining $\star$-product leads to the Gaussian integral
\bea V&=&{8h^2\over(1+h)^2} \int {d^4\xi d^4\eta\over (2\pi)^4}
e^{\ft12 \Xi^I M_I{}^J \Xi_J+\Xi^I N_I +(y-\yb a)b(y+a\yb)}\
,\label{b4} \eea
where $\Xi^I=(\x^\a,\bar \x^{\ad};\eta^{\a},\bar\eta^{\ad})$ and
$\Xi_I=(\x_\a,\bar
\x_{\ad};\eta_{\a},\bar\eta_{\ad})=\Xi^J\Omega_{JI}$, with
block-diagonal symplectic metric $\Omega=\e\oplus\bar
\e\oplus\e\oplus\bar \e$, and
\bea M&=&\left[\ba{cc}A&-i\\i&B\ea\right]\ ,\\[5pt] A&=&
\left[\ba{cc}2b&ia+2ba\\ia-2ba&-2\bar aba\ea \right]
\ ,\qquad B\ =\ \left[\ba{cc}0&-ia\\-i\bar a&0\ea\right]\ ,\\[5pt] N&=&
\left[\ba{l} i(1-2ib)a\bar y +2by\\-2\bar aba\bar y +i\bar
a(1+2ib)y\\-ia\bar y\\-i\bar ay\ea \right]\ .\eea
The Gaussian integration gives
\bea V &=&{8h^2\over(1+h)^2\sqrt{\det M}}~e^{\ft12 N^I(M^{-1})_I{}^J
N_J  + (y-\bar y a)b(y+a\bar y)}\ .\eea
From $\det M= \det (1+AB)$, and noting that the matrices defined as
\bea C&\equiv & {BA-a^2\over 2i}\ =\ \left[\ba{cc}a^2b&a^2ba\\-\bar
ab&-\bar aba\ea\right]\ ,\qquad \tilde C\ \equiv \ {AB-a^2\over 2i}\
=\ \left[\ba{cc}-a^2b&-ba\\\bar aba^2&\bar aba\ea\right]\ ,\eea
are nilpotent, i.e. $C^2 = \tilde C^2\ =\ 0$, one finds
\bea\det M&=& (1-a^2)^4\ ,\eea
and, using $1-a^2=2h/(1+h)$, the pre-factor in $V$ is thus given by
\bea {8h^2\over(1+h)^2\sqrt{\det M}}&=&2\ .\eea
Next, using geometric series expansions, one finds
\bea M^{-1}&=& {i\over (1-a^2)}\left[\ba{cc}i(1-a^2)B+2B\tilde C&
-(1-a^2)-2iC\\ 1-a^2+2i\tilde C&i(1-a^2)A+2AC\ea\right]\ ,\eea
and
\bea \ft12 N^I(M^{-1})_I{}^J N_J&=& {4a^2 yby +2(1+4a^2-a^4) yba\yb
-(3-a^2)(1+a^2) \bar y \bar aba\bar y\over (1-a^2)^2}\ .\eea
Adding the classical term in the exponent in \eq{b4} yields the
final result
\bea V &=& 2 \exp \left(-{ [2\yb \bar a -(1+a^2)y]\,b\,[2a\yb
+(1+a^2)y]\over (1-a^2)^2}\right)\ .\label{Vrho0}\eea
The projector property $V\star V=V$ follows manifestly from
\bea V&=& 2\exp (-2\tilde u\tilde v)\ ,\qquad [\tilde u,\tilde
v]_\star\ =\ 1\ ,\eea
where
\bea \tilde u&=&\l^\a\eta_\a\ ,\qquad \tilde v\ =\ \mu^\a\eta_\a\
,\eea
with
\bea \eta_\a&=& {[(1+a^2)y+2a\yb]_\a\over 1-a^2}\ ,\qquad
[\eta_\a,\eta_\b]_\star\ =\ 2i\e_{\a\b}\ .\eea
Thus, the net effect of rotating the projector $P_{\ft12}(u,v)$
given in \eq{reduced} is to replace the oscillators $u$ and $v$ by
their rotated dittos $\tilde u$ and $\tilde v$. We claim, without
proof, that this generalizes to any $n$, \emph{viz.}
\bea L^{-1}\star P_n(u,v)\star L&=& P_n(\tilde u,\tilde v)\ .\eea
Similarly, for $P_n(\bar u,\bar v)$ we have
\bea L^{-1}\star P_n(\bar u,\bar v)\star L&=& P_n(\tilde{\bar
u},\tilde {\bar v})\ ,\eea
where
\bea \tilde{ \bar u}&=&\bar\l^{\ad}\bar \eta_{\ad}\ ,\qquad \tilde{
\bar v}\ =\ \bar\mu^{\ad}\bar\eta_{\ad}\ ,\eea
with
\bea \bar\eta_{\ad}&=& {[(1+a^2)\yb+2\bar a y]_{\ad}\over 1-a^2}\
,\qquad [\bar\eta_{\ad},\bar\eta_{\bd}]_\star\ =\ 2i\e_{\ad\bd}\
.\eea
Finally, using $[\eta_{\a},\bar\eta_{\ad}]_\star=0$, we deduce that
\bea V&=&L^{-1}\star P\star L\ =\ \sum_{n_1,n_2}\th_{n_1,n_2}
P_{n_1,n_2}(\tilde u,\tilde v;\tilde{\bar u},\tilde{\bar v})\
.\label{generalV}\eea
%


\chapter{Traces and Projectors in Oscillator
Algebras}\label{App:G}


In this appendix we collect some basic properties of the
representation theory of a single oscillator of importance to the
doublet-oscillator realization of $\mso(5;\Comp)$ in Section
\ref{supersing}.

\begin{center}{\it Trace and Supertrace}\end{center}

We start from the complexified Heisenberg algebra
\bea u\star v-v\star u&=&1\ ,\label{Heisenberg}\eea
generates an associative $\star$-product algebra of Weyl-ordered
functions $f(u,v)$ with product
\bea f\star g&=& \int_{\Comp\times \Comp}~{d\xi d\bar\xi d\eta
d\bar\eta\over \pi^2} ~e^{2i(\bar\xi\eta+\bar\eta\xi)}
f(u+\xi,v+\bar\xi)g(u+i\eta,v-i\bar\eta)\ ,\eea
where $d\xi d\bar \xi=2d({\rm Re}\xi)d({\rm Im}\xi)$. The algebra
admits two inequivalent hermitian conjugation rules,
\bea u^\dagger&=&v\ ,\qquad v^\dagger\ = u\ ,\label{dag}\\[5pt] u^\ddag&=& -v\ ,\qquad
v^\ddag\ =\ -u\ .\label{ddag}\eea
It also admits two inequivalent traces, namely, the cyclic trace
\bea \mathrm{Tr}_+(f)&=& \int_{\Comp} {du d\bar u\over 2\pi}
f(u,\bar u)\ ,\label{trplus}\eea
obeying
\bea \Tr_+(f\star g)&=& \Tr_+(fg)\ =\ \Tr_+(g\star f)\
,\label{cyclic}\eea
up to boundary terms, and the graded-cyclic trace
\bea \mathrm{Tr}_-(f)&=&{f(0,0)\over 2}\ ,\eea
obeying
\bea \mathrm{Tr}_-(f\star g)\ =\
(-1)^{\e(f)\e(g)}\mathrm{Tr}_-(g\star f)\ , \label{grcicl}\eea
for functions $f$ and $g$ with definite parity,
$f(-u,-v)=(-1)^{\e(f)}f(u,v)$ \emph{idem} $g$. Note that the
functions $f$ and $g$ in the argument of the traces are always
supposed to be Weyl-ordered.

Let us show that
\bea \Tr_\pm(f)&=&\Tr_\mp((-1)^{N}_\star\star f)\ ,\qquad N\ =\
v\star u\ ,\label{thm}\eea
where we use the notation
\bea x^A_\star&=&\exp_\star (A\ln x)\ ,\qquad \exp_\star A\ =\
\sum_{n=0}^\infty {A^{\star n}\over n!}\ ,\qquad A^{\star n}\ =\
\underbrace{A\star\cdots \star A}_{\mbox{$n$ times}}\ .\eea
To this end, we begin by deriving the lemma
\bea \exp_\star (\a w)&=& {\exp({2w\tanh\ft\a2})\over \cosh{\a\over
2}}\ ,\qquad w\ =\ N+\ft12\ =\ uv\ ,\label{lemma}\eea
with $\a\in\Comp\backslash\{\pm i\pi,\pm 3i\pi,\dots\}$. This
follows by acting on both sides of \eq{lemma} with
$\partial/\partial\a$ and using
\bea w\star f(w)&=&\left(w-{1\over 4} {\partial\over \partial
w}-{1\over 4} w{\partial^2\over \partial w^2}\right)f(w)\
.\label{wstarf}\eea
Thus, setting $\exp_\star \a w =r(\a)\exp(s(\a) w)$, one finds
$r'=-rs/4$ and $s'=1-s^2/4$, subject to $r(0)=1$ and $s(0)=0$, with
the solution $r^{-1}=\cosh(\a/2)$ and $s=2\tanh(\a/2)$. To proceed
with the proof of \eq{lemma}, we need to examine the nature of
$\exp_\star (i(\pi+\e)N)$ more carefully in the singular limit
$\eta=-\sin(\e/2)\rightarrow 0$. Here,
\bea \exp_\star(i(\pi+\e)N)\sim -i {\exp{2iuv\over\eta}\over \eta}\
,\eea
so that, using \eq{cyclic},
\bea \lim_{\e\rightarrow 0}\Tr_+(\exp_\star(i(\pi+\e)N)\star f)\ =\
-i\lim_{\eta\rightarrow 0}\int_{\Comp}{dud\bar u\over 2\pi}
{\exp{2iu\bar u\over\eta}\over \eta} f(u,\bar u)\ =\ {f(0,0)\over
2}\ .\eea
Moreover, from $\exp_\star (\a N)\star \exp_\star(\b
N)=\exp_\star((\a+\b)N)$, it follows, using the regularized
Weyl-ordered form, that
\bea \lim_{\e\rightarrow 0}
\left[\exp_\star(i(\pi+\e)N)\right]^{\star 2} &=&\lim_{\e\rightarrow
0} \exp_\star(2i(\pi+\e)N)\\[5pt]& =&\lim_{\e\rightarrow 0}{\exp(2\tanh
(i(\pi+\e))N -i(\pi+\e))\over \cosh i(\pi+\e)}\ =\ 1\ ,\eea
in agreement with $(-1)_\star^{N}\star
(-1)_\star^N=(-1)_\star^{2N}=1$.

\begin{center}{\it Inner Products and Projectors}\end{center}

The two inequivalent traces $\Tr_\pm$ are related to two
inequivalent inner products ${}_\pm\langle\Psi|\Psi'\rangle \equiv
I_\pm(\ket{\Psi},\ket{\Psi'})$ on the Fock space
\bea {\cal F}&=&\span{\Comp}\left\{ \ket{n}\ =\ {v^n\over
\sqrt{n!}}~\ket{0}\right\}_{n=0}^\infty\ ,\qquad u\ket{0}\ =\ 0\
,\eea
defined by
\bea I_\pm (\m \ket{m},\n \ket{n})&=& \bar\mu \nu (\pm 1)^m\d_{mn}\
,\qquad \mu,\nu\in \Comp\ .\eea
The relation is
\bea I_\pm (\ket{m},\ket{n})&=& \Tr_\pm(P_{n,m})\
,\label{relation}\eea
where
\bea P_{n,m}&=& {1\over \sqrt{m!n!}}~v^n\star P_{0,0}\star u^m\
,\qquad P_{0,0}\ =\ 2e^{-2w}\ ,\label{Pmn}\eea
can be identified as
\bea P_{n,m}&=&\ket{n}{}_+\bra{m}\ =\ \ket{n}{}\bra{m}\
,\label{Pmn2}\eea
where we use the convention that $\bra{\Psi}={}_+\bra{\Psi}$. To
show \eq{relation}, we use the cyclicity properties of $\Tr_\pm$ and
\bea u\star P_{0,0}&=&P_{0,0}\star v\ =\ 0\ ,\label{P00}\eea
to derive $\Tr_\pm(P_{n,m})=(\pm 1)^n\d_{mn}\Tr_\pm(P_{0,0})$ where
$\Tr_\pm(P_{0,0})=1$, which yields the desired result. As a
byproduct of \eq{P00} it follows that
\bea P_{m,n}\star P_{p,q}&=&\d_{np}P_{m,q}\ .\eea
The relation \eq{thm}, between the traces, implies
\bea {}_\pm\langle\Psi|\Psi'\rangle &=&{}_\mp\langle\Psi|
(-1)_\star^N\vert \Psi'\rangle\ ,\eea
inducing the hermitian conjugation rules \eq{dag} and \eq{ddag}, as
follows
\bea I_\pm(f
\ket{\Psi},\ket{\Psi'})&=&I_\pm(\ket{\Psi},g\ket{\Psi'})\ ,\qquad g\
=\ \left\{\ba{ll}
f^\dagger&\mbox{for $I_+$}\\[5pt] f^\ddag&\mbox{for $I_-$}\ea\right.\eea
The operator $(-1)_\star^N$ has a finite normal-ordered
representation,
\bea (-1)_\star^N&=& :e^{-2N}:\ ,\eea
as can be seen by computing the projectors on even and odd states,
\bea P_{(\pm)}&=& \sum_{n=0}^\infty \ft12(1\pm(-1)^n) \vert
n\rangle\langle n\vert\ =\ \ft12:(e^N\pm e^{-N})\vert
0\rangle\langle 0\vert:\ ,\eea
adding them, \emph{viz.} $P_{(+)}+P_{(-)}=:e^{N}\vert
0\rangle\langle 0\vert:=1$, to obtain $\vert 0\rangle\langle
0\vert=:e^{-N}:$, from which it follows that
\bea P_{(\pm)}&=&\ft12 (1\pm:e^{-2N}:)\ .\eea
Finally, as an explicit check of $\Tr_\pm(P_{n,n})=(\pm 1)^n$ one
can compute explicitly
\bea P_{n,n}&=&\ket{n}\bra{n}\ =\ 2(-1)^n e^{-2w}L_n(4w)\ ,\eea
either by direct evaluation of the $\star$-products in \eq{Pmn}, or,
by starting from \eq{Pmn2} and using
\bea :e^{a u+ b v}:&=& e^{au}_\star\star e^{bv}_\star\ =\ e_\star^{a
u+ bv+\ft12 ab}\ =\ e^{au+bv+\ft 12 ab}\ ,\eea
and $\ket{0}\bra{0}=:\exp(-uv):$, together with Fourier
transformation techniques. This enables one to calculate
\bea P_{n,n}&=&\ket{n}\bra{n}\ =\ {1\over n!} :v^n e^{-vu} u^n:\nn\\[5pt]
&=&\int {dkd\bar k\over 2\pi}:e^{-i(\bar k u+k v)-\bar k
k}:L_n(\bar k k)\nn\\[5pt]
&=&\sum_{p=0}^n{n\choose n-p} {1\over
p!}(\partial_u\partial_v)^p\int {dkd\bar k\over 2\pi}e^{-i(\bar k
u+k v)-\ft12 \bar k k}\nn\\[5pt]&=&2\sum_{p=0}^n{n\choose n-p} (-2)^p e^{-2w}L_p(2w)\nn\\[5pt]&=&
2(-1)^n e^{-2w} L_n(4w)\ ,\eea
where
\bea L_n(x)&=&{1\over n!}e^{x}{d^n\over dx^n} (e^{-x} x^n)\ =\
\sum_{p=0}^n{n\choose n-p} {(-1)^p\over p!} x^p\ ,\eea
and the last step in the calculation follows from the identities
\bea \sum_{n=0}^\infty (-z)^n L_n(2x)&=& {e^{2xz\over 1+z}\over
1+z}\ =\ \sum_{n=0}^\infty\sum_{p=0}^n {n\choose n-p}(-2)^pL_p(x)
z^n\ .\eea

\begin{center}{\it Anti-Fock Space and Discrete Maps}\end{center}

Next, we introduce the anti-Fock space
\bea {\cal F}^- &=& \left\{\ket{n}^-\ =\
\frac{u^n}{\sqrt{n!}}~\ket{0}^-\right\}_{n=0}^\infty \ , \qquad
v\ket{0}^-\ =\ 0 \ ,\eea
and define its two inequivalent duals ${\cal F}^{\star -}_\pm$ by
\bea {}_\pm^-\bra{m}n\rangle^-&=& (\mp 1)^m\d_{mn}\ .\eea
One can identify ${}_\pm^-\bra{m}n\rangle^-=\Tr_\pm(P^-_{n,m})$ with
\bea P^-_{n,m}&=& \ket{n}^-{}^-\bra{m}\ =\ {1\over
\sqrt{n!m!}}~u^n\star P^-_{0,0}\star v^m\ ,\qquad P^-_{0,0}\ =\
2e^{2w}\ ,\eea
where we note that $P^-_{m,n}\star
P^-_{p,q}=(-1)^n\delta_{np}P^-_{m,q}$ and 
\bea P^-_{n,n}&=&2 e^{2w}L_n(-4w)\ .\eea
We also set ${\cal F}^+={\cal F}$ and $P^+_{n,m}=P_{n,m}$.

The $\star$-product algebra admits the automorphism $\pi$ and
anti-automorphism $\tau$ given by
\bea \pi(f(u,v))&=& f(i v,iu)\ ,\qquad \tau(f(u,v))\ =\ f(i u,i v)\
.\eea
These exchange the $P^\pm$ projectors as follows
\bea\pi(P^\pm_{n,m})&=&i^{m+n}P^\mp_{n,m}\ ,\qquad
\tau(P^\pm_{n,m})\ =\ i^{m+n}P^\mp_{m,n}\ .\label{piPmn}\eea
Their self-compositions are given by
\bea \pi\circ \pi&=&\tau\circ \tau\ =\ \Ad_{(-1)_\star^{N'}}\
,\qquad N'\ =\ \sum_{n=0}^\infty\left((w-\ft12)\star
P_{n,n}+(w+\ft12)\star P^-_{n,n}\right)\ ,\label{squares}\eea
where $\Ad_X(Y)=X\star Y\star X^{-1}$, while their mutual
composition defines the reflector
\bea R&=&\pi\circ \t\ ,\qquad R(f(u,v)) =\ f(-v,-u)\ ,\qquad R\circ
R\ =\ \mathrm{Id}\ .\eea
We note that $(w\mp 1/2)\ket{n}^\pm=\pm n \ket{n}^\pm$ and
${}^\pm\bra{n}(w\mp 1/2)=\pm n {}^\pm\bra{n}$, while $w\ket{n}^\pm=
\pm(n+\ft12)\ket{n}^\pm$ and ${}^\pm\bra{n}w=
\pm(n+\ft12){}^\pm\bra{n}$. The above maps extend naturally to
\bea \pi&:&{\cal F}^\pm\oplus {\cal F}^{\star \pm}\rightarrow {\cal
F}^\mp\oplus{\cal F}^{\star\mp}\ ,\\[5pt] \tau&:&{\cal F}^{\pm}\rightarrow
{\cal F}^{\star\mp}\ ,\\[5pt] R&:&{\cal F}^\pm\rightarrow{\cal F}^{\star\pm}\ ,\eea
by defining
\bea \pi(\ket{0}^\pm)&=& \ket{0}^\mp\ ,\qquad
\pi({}^\pm\bra{0})\ =\ {}^\mp\bra{0}\ ,\\[5pt] \tau(\ket{0}^\pm)&=&
{}^\mp\bra{0}\ ,\qquad \tau({}^\pm\bra{0})\ =\ \ket{0}^\pm\ , \eea
such that \eq{squares} hold in the generalized sense that
$\Ad_X(\ket{\Psi})=X\ket{\Psi}$ and
$\Ad_X(\bra{\Psi})=\bra{\Psi}X^{-1}$. It follows that
\bea R(\ket{0}^\pm)&=&{}^\pm\bra{0}\ ,\qquad R({}^\pm\bra{0})\ =\
\ket{0}^\pm\ .\eea
We see that the reflector acts in the real basis
$\{\ket{n}^\pm,{}^\pm\bra{n}\}$ as the hermitian conjugation
\eq{ddag}. This is to be compared with the case of
doublet-oscillators where the reflector is modified by the
requirement thst it should preserve $SU(2)$ quantum numbers (spatial
spins) while flipping $U(1)$ quantum numbers (the energy).

\begin{center}{\it Projectors and $w$-Invariants}\end{center}

The projectors $P^\pm_{n,n}$ are special cases of the more general
functions $M_\k=M_\k(w)$ obeying the $w$-invariance
condition 
\bea (w-\kappa)\star M_\kappa&=&0\ \label{Mkappa}\eea
for $\kappa\in \Comp$. These functions can be written as
\bea M_\kappa&=& {\cal N}_\kappa\oint_{C} {d\alpha\over 2\pi i}
g^{(\kappa)}(\alpha)\ ,\eea
where ${\cal N}_\kappa\in\Comp$;
\bea g^{(\kappa)}(\alpha)&=&e_\star^{\alpha(w-\kappa)}\ =\
(1+\ft{s}2)^{\ft12-\kappa}(1-\ft{s}2)^{\ft12+\kappa}e^{s w}\ ,\qquad
s\ =\ 2\tanh\ft{\alpha} 2\ ;\eea
and $C$ a contour encircling $i\pi$ clockwise, so that its image
$\Gamma=s(C)$ encircles $[-2,2]$ counterclockwise. Taking $C$ to be
a small circle, $\Gamma$ becomes a large contour. Enlarging $C$ to
the ''box'' $-C=\{i\e+x:-L\leq x\leq L\}\cup \{L+ix:\e\leq x\leq
2\pi-\e\}\cup\{i(2\pi-\e)-x:-L\leq x\leq
L\}\cup\{-L+i(2\pi-x):\e\leq x\leq 2\pi-\e\}$, $\Gamma$ becomes a
''dogbone'' encircling $[-2,2]$. The normal-ordered form of the
group elements reads
\bea g^{(\kappa)}(\alpha)&=& (1+\lambda)^{\ft12-\kappa}:e^{\lambda
w}:\ ,\qquad \lambda\ =\ {s\over 1-\ft s2}\ =\ e^\a-1\ .\eea
The composition rule
\bea e^{sw}\star e^{s' w}&=& {1\over 1+\ft{ss'}4}e^{s'' w}\ ,\qquad
s''\ =\ {s+s'\over 1+\ft {ss'}4}\ ,\\[5pt] :e^{\l w}:\star :e^{\l'
w}&=& :e^{\l''w}:\ ,\qquad \l''\ =\ \l+\l'+\l\l'\ ,\eea
implies
\bea g^{(\kappa)}(\a)\star g^{(\kappa)}(\a')&=&
g^{(\kappa)}(\a+\a')\ =\ g^{(\kappa)}(\a(s''))\ =\
g^{(\kappa)}(\a(\l''))\ .\eea
Changing variables, one finds
\bea M_\kappa&=&{\cal N}_\kappa\oint_\Gamma {ds\over 2\pi i
(1-\ft{s^2}4)} g^{(\kappa)}(\alpha(s))\ =\ {\cal
N}_\kappa\oint_{\Gamma'}{d\lambda\over 2\pi i(1+\lambda)}
g^{(\kappa)}(\a(\l))\ .\eea
Using the Weyl-ordered form and \eq{wstarf}, one can then easily
verify \eq{Mkappa}. There exist other choices of contour $C$ leading
to $w$-invariants, such as $C=i[-\pi,\pi]$ and $C=\Real$ leading to
integrals over $U(1)$ and $GL(1;\Real)$, respectively, but we shall
not consider them further here.

For $\kappa=\pm(n+1/2)$, $\Gamma$ collapses to a circle around the
pole at $s=\mp 2$, so that
\bea M_{\pm(n+\ft12)}&=& 2(\mp 1)^n e^{\mp 2w}L_n(\pm 4w)\ =\
P^\pm_{n,n}\quad\mbox{for ${\cal N}_{n+\ft12}=1$, ${\cal
N}_{-n-\ft12}=(-1)^{n+1}$}\ .\eea
Using the change of variables found in \cite{Sagnotti:2005ns}, and
making use of analytical continuation and deformations of contours,
one can calculate
\bea (M_{\pm(n+\ft12)})^{\star 2}&\!\!=\!\!& {\cal
N}_{\pm(n+\ft12)}\oint_{\mp 2}{ds\over 2\pi i (1-\ft{s^2}4)}
M_{\pm(n+\ft12)}\!\!\ =\!\!\ \pm {\cal
N}_{\pm(n+\ft12)}M_{\pm(n+\ft12)}\!\!\ =\!\!\ (-1)^n
M_{\pm(n+\ft12)}\ ,\qquad\qquad\eea
in agreement with $(P^\pm_{n,n})^{\star 2}=(-1)^n P^\pm_{n,n}$. For
$\kappa\notin(\integ+1/2)$, the branch cut prevents $\Gamma$ from
collapsing, so that deforming back the contour it now encircles both
$s=-2$ and $s=2$, with the result that
\bea M_\kappa^{\star 2}&=& {\cal N}_\kappa \oint_{\Gamma}{ds\over
2\pi i (1-\ft{s^2}4)} M_{\kappa}\ =\ 0\ .\eea
Interestingly, the case of $\kappa=0$ is closely related to the
dressing function of the 5D higher-spin gauge theory based on spinor
oscillators, introduced in \cite{5d} and later analyzed in more
detail in \cite{Sagnotti:2005ns}. The above analysis suggests that a
suitable regularization of the dressing function has a well-defined
\emph{vanishing} self-composition, which would greatly simplify the
perturbative weak-field expansion.

\begin{center}{\it Fermionic Oscillators}\end{center}

It is also interesting to look along similar lines at the
complexified Clifford algebra
\bea \{\c,\d\}_\star \ \equiv \ \c\star\d+\d\star\c \ = \ 1 \ ,
\label{anticomm}\eea
with Weyl-ordering defined by $\c\star\d=\c\d+1/2$ where $\
\c\d=[\c,\d]_\star/2=-\d\c$. The corresponding Fock space consists
of a vacuum state $\ket{0}$ obeying $\c\ket{0}=0$ and an excited
state $\ket{1}=\d\ket{0}$. The algebra \eq{anticomm} admits two
inequivalent hermitian conjugation rules,
\bea \c^\dagger&=&\d\ ,\qquad \d^\dagger\ = \ \c\ ,\\[5pt] \c^\ddag&=& -\d\ ,\qquad
\d^\ddag\ =\ -\c\ .\label{ddag}\eea
corresponding to the inner products
\bea I_\pm(f\star
\ket{\Psi},\ket{\Psi'})&=&I_\pm(\ket{\Psi},g\star\ket{\Psi'})\
,\qquad g\ =\ \left\{\ba{ll}
f^\dagger&\mbox{for $I_+$}\\[5pt] f^\ddag&\mbox{for $I_-$}\ea\right.\ ,\eea
related by
\bea {}_\pm\langle\Psi|\Psi'\rangle &=&{}_\mp\langle\Psi|\star
(-1)_\star^F\star \Psi'\rangle\ ,\qquad F\ =\ \d\star\c\ .\eea
We note that ${}_\pm\langle 1|1\rangle=\pm {}_{\pm}\langle
0|0\rangle$, and we choose ${}_{\pm}\langle 0|0\rangle=1$. The
description in terms of states and inner products can be replaced by
the dual $\star$-algebra picture that makes use of the projectors
\bea P_0 &=&\ket{0}\bra{0} \ = \ 1-\d\star\c\ =\ \ft12
(1-2\d\c)\ ,\\[5pt] \qquad P_1&=&\ket{1}\bra{1} \ = \
\d\star\c \ =\ \ft12(1+2\d\c)\ ,\eea
and the trace operations
\bea \Tr_+(f) & = & 2\,f(0,0) \ ,\label{ftrace1}\\[5pt]
\Tr_-(f) & = & -\int d\c d\bar{\c}\,f(\c,\bar{\c}) \
,\label{ftrace2}\eea
where $f=f(\c,\d)$ is Weyl-ordered and we use the Berezin
integration rule $\int d\c d\bar{\c}~\bar\c \c=1$. Interestingly
enough, comparing to bosons, the definitions of the traces $\Tr_+$
and $\Tr_-$ are interchanged. For example, we have
$\Tr_+(P_0)=\Tr_+(P_1)=1$ while $\Tr_-(P_0)=-\Tr_-(P_1)=1$. Finally,
one can show that
\bea \Tr_\pm((-1)_\star^F\star f) \ =\ \Tr_\mp(f) \
.\label{ftraces}\eea
To this end, we first simplify
\bea \int d\c d\bar{\c}\,f\star g \ = \ \int d\c d\bar{\c}\,fg \ ,
\label{addexp}\eea
using the fact that the terms in $f\star g$ involving at least one
contraction do not survive the integration, as can be seen directly
by expanding $f=f_0+f_1\c+f_2\bar{\c}+f_3\bar{\c}\c$, \emph{idem}
$g$. Then, from $F=P_1$ it follows that
\bea (-1)_\star^F\ =\ \exp_\star(i\pi F)\ = \ 1+\sum_{n=1}^\infty
\frac{(i\pi)^n}{n!}P_1 \ = \ 1+(e^{i\pi}-1)P_1 \ = \ 1-2P_1 \ =\
-2\bar{\c}\c\ ,\eea
so that, using \eq{addexp},
\bea \Tr_-((-1)_\star^F\star f) \ = \ -\int d\c
d\bar{\c}\,(-2\bar{\c}\c)f(\c,\bar{\c}) \ =\ 2f(0,0) \ =\ \Tr_+(f)\
\ ,\eea
where integration is performed using $\bar{\c}\c=\d(\bar \c)\d(\c)$.
The converse, \emph{i.e.} $\Tr_+((-1)_\star^F\star f) =\Tr_-(f)$,
then follows by letting $f\rightarrow (-1)_\star^F\star f$ and using
$(-1)_\star^F\star(-1)_\star^F=(-2\bar{\c}\c)\star(-2\bar{\c}\c)=1$.


\end{appendix}


\end{document}